\documentclass[conference,onecolumn]{sty/IEEEtran}
\IEEEoverridecommandlockouts
\usepackage[utf8]{inputenc}
\usepackage{stackengine}
\usepackage{amssymb}
\usepackage{amsmath}
\usepackage{mathrsfs}
\usepackage{ulem}
\usepackage{epsf,epsfig}
\usepackage{cite}
\usepackage{color,xcolor}
\usepackage{dsfont}
\usepackage{bbm}
\usepackage{amsthm}
\usepackage{physics}
\usepackage{tcolorbox}
\usepackage{todonotes}
\usepackage{wrapfig}
\usepackage{url}
\usepackage{graphicx}
\usepackage{eurosym}
\usepackage{enumitem}
\usepackage{nopageno}
\usepackage{fancyhdr}
\usepackage{mathtools}
\usepackage{stmaryrd}

\def\naturals{\mathbb{N}}
\def\reals{\mathbb{R}}
\def\complex{\mathbb{C}}

\def\fieldq{\mathcal{F}_{q}}
\def\Expectation{\mathbb{E}}

\newcommand{\msout}[1]{\text{\sout{\ensuremath{#1}}}}

\definecolor{darkgreen}{rgb}{0,0.5,0}
\definecolor{darkmagenta}{rgb}{.5,0,.5}
\definecolor{darkblue}{rgb}{0,0,0.5}

\def\CalA{\mathcal{A}}
\def\CalB{\mathcal{B}}
\def\CalC{\mathcal{C}}
\def\CalD{\mathcal{D}}
\def\CalE{\mathcal{E}}
\def\CalF{\mathcal{F}}

\def\CalH{\mathcal{H}}
\def\CalI{\mathcal{I}}
\def\CalJ{\mathcal{J}}
\def\CalK{\mathcal{K}}
\def\CalL{\mathcal{L}}

\def\CalN{\mathcal{N}}

\def\CalP{\mathcal{P}}
\def\CalQ{\mathcal{Q}}

\def\CalS{\mathcal{S}}
\def\CalT{\mathcal{T}}
\def\CalU{\mathcal{U}}
\def\CalV{\mathcal{V}}
\def\CalW{\mathcal{W}}
\def\CalX{\mathcal{X}}
\def\CalY{\mathcal{Y}}

\def\ulineA{\underline{A}}
\def\ulineB{\underline{B}}
\def\ulineC{\underline{C}}

\def\ulineE{\underline{E}}

\def\ulineK{\underline{K}}

\def\ulineM{\underline{M}}

\def\ulineR{\underline{R}}

\def\ulineW{\underline{W}}

\def\ulineb{\underline{b}}

\def\ulineg{\underline{g}}

\def\ulinek{\underline{k}}

\def\ulinem{\underline{m}}

\def\ulinew{\underline{w}}

\def\uliney{\underline{y}}

\def\hatz{\hat{z}}

\def\CalA{\mathcal{A}}
\def\CalB{\mathcal{B}}
\def\CalC{\mathcal{C}}
\def\CalD{\mathcal{D}}
\def\CalE{\mathcal{E}}
\def\CalF{\mathcal{F}}

\def\CalH{\mathcal{H}}
\def\CalI{\mathcal{I}}
\def\CalJ{\mathcal{J}}
\def\CalK{\mathcal{K}}
\def\CalL{\mathcal{L}}

\def\CalN{\mathcal{N}}

\def\CalP{\mathcal{P}}
\def\CalQ{\mathcal{Q}}

\def\CalS{\mathcal{S}}
\def\CalT{\mathcal{T}}
\def\CalU{\mathcal{U}}
\def\CalV{\mathcal{V}}
\def\CalW{\mathcal{W}}
\def\CalX{\mathcal{X}}
\def\CalY{\mathcal{Y}}

\def\hatY{\hat{Y}}

\def\ulineCalE{\underline{\mathcal{E}}}

\def\ulineCalI{\underline{\mathcal{I}}}

\def\ulineCalW{\underline{\mathcal{W}}}
\def\ulineCalX{\underline{\mathcal{X}}}

\def\parsec{\par\noindent}
\def\med{\medskip\parsec}

\def\define{\mathrel{\ensurestackMath{\stackon[1pt]{=}{\scriptstyle\Delta}}}}

\def\boldCalQ{\boldsymbol{\mathcal{Q}}}

\def\boldCalW{\boldsymbol{\mathcal{W}}}

\def\boldCalY{\boldsymbol{\mathcal{Y}}}

\def\boldA{\boldsymbol{A}}
\def\boldB{\boldsymbol{B}}

\def\boldE{\boldsymbol{E}}

\def\boldR{\boldsymbol{R}}

\def\boldT{\boldsymbol{T}}

\def\boldW{\boldsymbol{W}}

\def\boldi{\boldsymbol{i}}

\def\boldp{\boldsymbol{p}}

\def\boldw{\boldsymbol{w}}

\def\boldy{\boldsymbol{y}}

\def\boldR{\boldsymbol{R}}

\def\boldy{\boldsymbol{y}}

\def\boldA{\boldsymbol{A}}
\def\boldB{\boldsymbol{B}}

\def\boldW{\boldsymbol{W}}
\def\boldw{\boldsymbol{w}}

\def\boldlambda{\boldsymbol{\lambda}}

\def\boldrho{\boldsymbol{\rho}}

\def\olinex{\overline{x}}

\def\ScrA{\mathscr{A}}

\def\ScrE{\mathscr{E}}

\def\ScrO{\mathscr{O}}

\def\ScrT{\mathscr{T}}

\def\sfa{\mathsf{a}}
\def\sfb{\mathsf{b}}
\def\sfc{\mathsf{c}}

\def\sfm{\mathsf{m}}

\def\sfs{\mathsf{s}}

\def\sfA{\mathsf{A}}
\def\sfB{\mathsf{B}}
\def\sfC{\mathsf{C}}

\def\sfL{\mathsf{L}}

\def\sfV{\mathsf{V}}

\def\tildeb{\tilde{b}}

\def\dulinek{\doubleunderline{k}}
\def\dulinel{\doubleunderline{l}}
\def\dulinem{\doubleunderline{m}}

\def\dulines{\doubleunderline{s}}

\newtheorem{remark}{Remark}
\newtheorem{theorem}{Theorem}

\newtheorem{definition}{Definition}
\newtheorem{lemma}{Lemma}
\newtheorem{example}{Example}
\newtheorem{proposition}{Proposition}
\newtheorem{fact}{Fact}
\newtheorem{note}{{Note}}

\def\span{\mbox{span}}
\def\supp{\mbox{supp}}

\newcommand{\colBlue}[1]{\textcolor{blue}{#1}}
\newcommand{\colRed}[1]{\textcolor{red}{#1}}

\newcommand{\colGreen}[1]{\textcolor{darkgreen}{#1}}

\newtheorem{notation}{Notation}

\usepackage{cite}
\usepackage{amsmath,amssymb,amsfonts}
\usepackage{algorithmic}
\usepackage{graphicx}
\usepackage{textcomp}
\usepackage{xcolor}
\def\BibTeX{{\rm B\kern-.05em{\sc i\kern-.025em b}\kern-.08em
    T\kern-.1667em\lower.7ex\hbox{E}\kern-.125emX}}

    \def\doubleunderline#1{\underline{\underline{#1}}}

\def\dulinel{\doubleunderline{l}}
\def\dulines{\doubleunderline{s}}

\def\olinescre{\overline{\mathscr{E}}}

\def\Bin{B}

\def\brho{\boldsymbol{\rho}}
\def\boldrho{\boldsymbol{\rho}}

\def\brho{\boldsymbol{\rho}}
\def\boldrho{\boldsymbol{\rho}}
\def\boldtheta{\boldsymbol{\theta}}
\def\boldsigma{\boldsymbol{\sigma}}

\newcommand{\mt}{\sfm}

\newcommand{\ulinesfc}{\underline{\sfc}}
\newcommand{\ulinesfC}{\underline{\sfC}}

\newcommand{\tinyoplus}{{\scriptscriptstyle{\oplus}}}
\newcommand{\boplus}{b^{\tinyoplus}}
\newcommand{\tildeboplus}{\tildeb^{\tinyoplus}}

\begin{document}
\setlength{\abovedisplayskip}{4pt plus 2pt minus 1pt}
\setlength{\belowdisplayskip}{4pt plus 2pt minus 1pt}
\setlength{\abovedisplayshortskip}{2pt plus 1pt minus 1pt}
\setlength{\belowdisplayshortskip}{3pt plus 1pt minus 1pt}
\setlength{\jot}{2pt}
\emergencystretch=1.5em

\title{Efficient One-Shot Simulation of Network Instruments and Multi-party Purity Distillation
}
\author{\IEEEauthorblockN{Igor BERNARD}
\and
\IEEEauthorblockN{Arun PADAKANDLA}
}

\maketitle

\begin{abstract}
We undertake an information theoretic study of instrument compression problems (ICPs) in {multiple network} configurations, wherein the goal is to design and analyze protocols that simulate specific quantum classical (QC) correlation amidst distributed parties - some holding components of the quantum state and the rest holding classical registers - with minimum communication. We present a \textit{unified framework} of protocols based on likelihood instruments whose analysis yields new inner bounds in the one-shot and asymptotic regimes for the different network configurations we study. Our framework comprises of a common template for designing the instrument simulation protocol and analyzing its information theoretic performance. We also address a related problem of multi-party purity distillation wherein three distributed parties have to distil pure states efficiently via local operations and minimal communication. In regards to the tools, in addition to one-shot purity concentration protocol of \cite{202203arXiv_ChaPadSen} and one-shot quantum simultaneous decoder of Sen \cite{202102SAD_Sen}, we leverage Sen's fully smooth multi-partite covering \cite{202504arXiv_Sen} for all of our covering bounds.
\end{abstract}

\section{Introduction}
\label{Sec:Introduction}

The outcome of a measurement being inherently random, a question of {fundamental} interest \cite{197109IJP_Gro,200805PRL_BusHayHor,201210JPA_WilHayBusHsi} is to quantify the amount of \textit{information} it contains about the {(}quantum{)} state being measured. Despite its simplicity, a satisfying mathematical formulation of this question remained elusive since the early work of \cite{197109IJP_Gro}. Building on \cite{200005PhyA_MasPop,200101PhyA_WinMas}, Winter noted\cite{200401CommMathPhy_Win} that the effect of a measurement is to induce a specific \textit{quantum-classical} (QC) \textit{correlation} between the post-measurement state and the (classical) register holding the outcome (Fig.~\ref{Fig:SnglEncSnglDecMsmtSimulation}). What is the {minimum communication} (resources) necessary to \textit{simulate} the \underline{same} QC correlation if the state and the register were \underline{distributed}? Modeling this communication problem in an information-theoretic {(info-theoretic)} framework, Winter formulated this question into what is referred to as the \textit{{m}easurement {c}ompression {p}roblem} (MCP). Designing an alternate \textit{{m}easurement {s}imulation {p}rotocol} (MSP) that simulates the same QC correlation with minimal communication, Winter \cite{200401CommMathPhy_Win} quantified the information content in a measurement's outcome.

\begin{figure}
\vspace{-0.1in}
\includegraphics[width=7.2in]{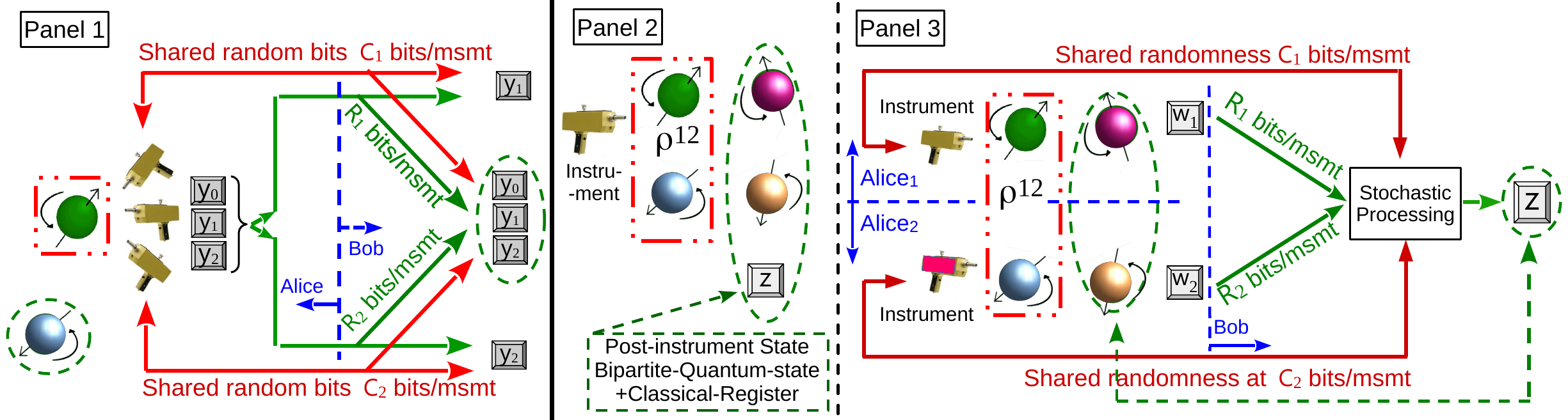}
\vspace{-0.25in}
\caption{Throughout, the (i) (quantum) state enclosed in the box bordered by \colRed{$-\cdot\cdot-\cdot\cdot$} depicts state prior to instrument action, and (ii) \{state+classical register\} enclosed within green \colGreen{$--$} ellipse and/or circle depicts joint quantum+classical state post instrument action. {\underline{\textbf{Multiple-Description ICP} (Panel 1)}: Consider a high resolution instrument with outcome $x \in \CalX$ having three coordinate outcomes, i.e., $x=(y_{0},y_{1},y_{2}) \in \CalX = \CalY_{0}\times \CalY_{1}\times \CalY_{2}$. Associate the individual co-ordinate outcomes $y_{1} \in \CalY_{1}$ and $y_{2} \in \CalY_{2}$ with two lower resolution instruments. Alice holds the state and is connected to Bob via two links, each providing shared randomness and bit communication. The goal is to quantify the minimum rates of these links so that (i) individually they can simulate the action of respective co-ordinate instruments and (ii) together they can simulate the high resolution instrument's action. } \underline{\textbf{Distributed ICP} (Panels 2,3)}: {Panel 2} depicts the action of a specified instrument on a bipartite state. Panel 3 depicts the distributed ICP scenario wherein the two components of the bipartite state are distributed amidst separated parties Alice$_{1}$ and Alice$_{2}$. Alice$_{i}$ can transfer $R_{i}$ bits to Bob with whom she shares $C_{i}$ bits of common randomness. The goal is to identify the set of of all $(R_{1},R_{2},C_{1},C_{2})$ quadruples for which the same post-instrument {state+classical register} can be simulated distributedly.}
\vspace{-0.25in}
\label{Fig:DistISAndMultDescIS}
\end{figure}

{Winter's elegant formulation and solution yielded an answer that is both intuitive and mathematically rigorous. More importantly, his pursuit of MCP led to the study and design of a minimum communication protocol to simulate specific QC correlations - a tool proven to be of much utility in tackling several fundamental problems in quantum information theory (QIT).} Indeed, designing communication efficient protocols that simulate specific QC correlation forms a central primitive in solving problems such as quantum-quantum rate distortion\cite{201301TIT_DatHsiWil}, reverse Shannon theorem\cite{201405TIT_BenDevHarShoWin} and remote state preparation \cite{200107PRL_BenDiVShoSmoTerWoo}. Furthermore, such protocols are key to accomplishing entanglement \cite{200501MPES_DevWin} and purity distillation (ED and PD) \cite{200506PRA_Dev,200707PRA_KroDev,202206ISIT_AtiPra,202306ISIT_ChaNemBus,202309ALL_ChaJaiSen} - important tasks at the intersection of QIT, thermodynamics and resource theory. These connections and the broader applicability of tools developed in solving MCP have motivated extensive MCP studies \cite{201210JPA_WilHayBusHsi,201412TIT_BerRenWil,201909TIT_AnsJaiWar,202202TIT_AtiHeiPra,202206ISIT_ChaPadSen,202107ISIT_AtiPra,201907ISIT_HeiAtiPra,202407ISIT_PadWar} in diverse network configurations that continue till date \cite{2026MMTIT_Pad,202504TIT_AtiPra}. While initial works \cite{200401CommMathPhy_Win,201210JPA_WilHayBusHsi,201412TIT_BerRenWil} considered a \textit{single} component quantum state yielding \textit{one} classical outcome in the asymptotic regime, MCP studies involving quantum side information (QSI) \cite{201210JPA_WilHayBusHsi,201909TIT_AnsJaiWar}, a bipartite state with distributed components \cite{202202TIT_AtiHeiPra,201907ISIT_HeiAtiPra,202504TIT_AtiPra,202206ISIT_AtiPra,2026MMTIT_Pad} and/or multiple outcomes (Fig.~\ref{Fig:DistISAndMultDescIS}) \cite{202206ISIT_ChaPadSen} have either exposed challenges \cite{201909TIT_AnsJaiWar} when tackled in the more general one-shot regime spurring the development of refined analytical tools \cite{201001TIT_HarJaiMcaRad,201709PRL_AnsDevVamJan,201909TIT_AnsJaiWar,201709QIP_Wil,202306ISIT_CheGao}, or motivated newer strategies \cite{202504TIT_AtiPra} in the asymptotic regime. {The centrality of MCP studies in QIT motivate our pursuit.}

In addition to tackling MCPs, the above discussions indicate that design and analysis of protocols that simulate target QC correlations with minimal communication has greater utility. Indeed, while solving an MCP entails the design of a protocol that only preserves QC correlation between the inaccessible reference component and classical outcome while ignoring (tracing out) the accessible post-measurement component, tasks such as ED, PD require protocols that must simulate the \underline{entire} QC correlation including \underline{both} quantum components. Taking this cue, we formulate all our problems in terms of \textit{instruments} (Defn.~\ref{Defn:QuantumInstrument}) and address \textit{instrument compression problems} (ICPs) wherein the goal is to simulate the action of instruments with minimal communication. {Instruments being the most general measurements, our formulations subsume corresponding MCP formulations (Rem.~\ref{Rem:ICPsRecoverMCPs}) and our findings apply to the latter.} {Rest of our discussion is therefore in terms of instruments and ICPs.}

In this work, we undertake a comprehensive Shannon theoretic study of ICP in \textit{multiple network} configurations in both one-shot and asymptotic regimes. We note that any such study of an ICP involves characterizing lower and upper bounds, i.e., outer and inner bounds, on the minimum communication resources necessary to establish a specific QC correlation. Our focus throughout will be on characterizing inner bounds. This involves designing an \textit{instrument simulation protocol} (ISP) and characterizing its {info-theoretic} performance using an appropriate set of analytical tools. Our goal in this work is to (i) identify a \textit{unified framework} of analytical tools and techniques that can (ii) lead us to prove achievability of new inner bounds to the ICP \textit{both} in the \textit{one-shot} \cite{201909TIT_AnsJaiWar,202407ISIT_PadWar} and \textit{asymptotic} \cite{200401CommMathPhy_Win,201210JPA_WilHayBusHsi,201412TIT_BerRenWil,202202TIT_AtiHeiPra,202504TIT_AtiPra} regimes for \textit{different network configurations}. Towards achieving this goal in this second part manuscript, we build on our first part \cite{2026MMTIT_Pad} findings wherein we proposed a new framework of techniques based on \textit{likelihood POVMs} (positive operator valued measurements). {The proposed framework was proven \cite{2026MMTIT_Pad} to be optimal for all scenarios involving a single encoder and a single decoder with and without QSI, henceforth referred to as \textit{point-to-point} (PTP) scenarios, in both one-shot and asymptotic regimes. Going further, this unified likelihood POVM based framework yielded {new strictly larger one-shot inner bound} \cite[Thm.~3]{2026MMTIT_Pad} for the network MCP involving distributed entangled components, also referred to herein as the distributed MCP.} Having thus proved its utility in \cite{2026MMTIT_Pad}, our first {contribution} here is to enhance this framework in the following {three} aspects: (i) design and analyze a single high resolution likelihood instrument that can efficiently compress outcomes of multiple \textit{compatible} instruments (Sec.~\ref{Sec:MultiLinkIS}), (ii) incorporating algebraic structure into instruments to design \textit{structured likelihood instruments} (Sec.~\ref{Sec:CosetDIS}), {and} (iii) {adopt} quantum simultaneous decoding in the one-shot regime (Sec.~\ref{Sec:QSIDistributedIS}). This enhancement fuels our second finding which concerns characterizing inner bounds for network ICPs. We characterize \textit{new inner bounds} to (a) the multiple-description ICP (Thm.~\ref{Thm:OneShotMLIS}., Rem.~\ref{Rem:MLComparisonPriorOneShot}), (b) multi-party purity distillation (Thm.~\ref{Thm:PurityAchievability}, Rem.~\ref{Rmk:AsymptoticPurityKappa}) and (c) distributed ICP with QSI (Thm.~\ref{Thm:DistInstSimltnOneShotQSI}, Rem.~\ref{Rem:QSIComparison}). Our proposed framework is also proven to achieve the asymptotic inner bound using structured codes for the distributed MCP in Sec.~\ref{Sec:DISCosetCodes}. {In conjunction with \cite{2026MMTIT_Pad}, this establishes that the unified framework of ISPs based on  likelihood POVMs achieve the current known largest inner bounds for all network ICPs in both the one-shot and asymptotic regimes.} We now elaborate on our specific findings for the different network ICPs.


\med\textbf{Multiple-Description (MD) ICP}: A broad question of interest is to quantify the \textit{incremental information} contained in higher resolution instruments. As a simple example, consider a high resolution POVM $\lambda_{\CalX}$ with outcome set $\CalX$. Consider a lower resolution POVM $\theta_{\CalY_{1}}$ where each outcome $y_{1} \in \CalY_{1}$ is associated with a subset $A_{y_{1}} \subseteq \CalX$ such that the collection $A_{y_{1}} : y_{1} \in \CalY_{1}$ of subsets forms a \textit{partition} of $\CalX$. How much \textit{incremental} information does outcome of $\lambda_{\CalX}$ provide, beyond and above that available from the outcome of $\theta_{\CalY_{1}}$? One can go a step further to consider two lower resolution POVMs $\theta_{\CalY_{1}},\theta_{\CalY_{2}}$ and enquire the amount of \textit{incremental} information contained in the outcome of $\lambda_{\CalX}$, beyond and above that available in (i) only $\theta_{\CalY_{1}}$'s outcome, (ii) only $\theta_{\CalY_{2}}$'s outcome, and (iii) both of $\theta_{\CalY_{1}}$, $\theta_{\CalY_{2}}$'s outcomes. Note that unless one of $\theta_{\CalY_{1}}$, $\theta_{\CalY_{2}}$ is of a lower resolution than the other, (iii) cannot be answered by solving the earlier simpler problem involving just one lower resolution POVM. From our modeling it will be evident that the question we address can also be viewed alternatively as: How does one optimally split the information needed to simulate a high resolution POVM $\lambda_{\CalX}$ into three parts, so that two of the parts can individually simulate the action of two lower resolution POVMs $\theta_{\CalY_{1}},\theta_{\CalY_{2}}$, while all three together can simulate $\lambda_{\CalX}$?

In Sec.~\ref{Sec:MultiLinkIS}, we formulate and address a \textit{multiple-description (MD) ICP} to answer the above question both in the one-shot and asymptotic regimes. See Fig.~\ref{Fig:DistISAndMultDescIS}. Our formulation is inspired by the well-studied classical problem of \textit{multiple-description coding} (MDC) \cite[Ch.~13]{BkNITElGamalKim_2011}, \cite{198707TIT_ZhaBer,198007BSTJ_Wit,198112BSTJ_WitWyn,198211TIT_ElGCov}, wherein the goal is to encode a source in multi-resolution fashion so that certain segments of the encoded bits can yield descriptions of the source to specified distortion levels. We design and analyze a new two-layer ISP based on likelihood instruments to simulate the desired QC correlations in a multi-resolution fashion. A recurring challenge (Rem.~\ref{Rem:MDICPProxyStates}, \ref{Rem:OneShotDIS-COSTrick}) encountered in analyzing likelihood instrument based ISPs throughout is the presence of inverted random operators for which we cannot prove concentration. Our broad approach is to neutralize these inverted random operators by operating the ISP on \textit{proxy states}, a technique devised in \cite{2026MMTIT_Pad}. Here, our formulation with a third highest resolution instrument adds new complexity to this challenge forcing us to consider two sets of \textit{proxy states}. See Rem.~\ref{Rem:MDICPSecondPrxyStW0layer}. This is also linked to the challenge that the smoothed multi-party covering \cite{202504arXiv_Sen} does not readily handle a non-product distribution of the multi-party code (Rem.~\ref{Rem:MDICP:ASecondSplitInCov}), requiring us to do a two step covering. Identifying appropriate tools, we analyze our proposed ISP to characterize a new inner bound (Thm.~\ref{Thm:OneShotMLIS}) in the one-shot regime. For the asymptotic regime, we go an extra step. Zhang and Berger \cite{198707TIT_ZhaBer} have proven that a \textit{three-layer} code with a common base layer can yield strictly larger inner bounds for the classical MD problem. Adopting this, we design and analyze a three-layer likelihood instrument based ISP to derive an inner bound (Thm.~\ref{Thm:AsymptoticZBMLIS}) that subsumes the natural generalization of our one-shot inner bound, and one that can be strictly larger (Rem.~\ref{Rmk:ZBBaseLayerGain}). We have not studied a three-layer ISP in the one-shot regime in the interest of brevity and a simplified exposition. Indeed, our technique of employing two sets of proxy states can be extended with a third set to adopt Zhang-Berger's coding \cite{198707TIT_ZhaBer} in one-shot.

Our findings go well beyond \cite{202206ISIT_ChaPadSen} which is, to our best knowledge, the only prior work on multiple-description ICP. \cite{202206ISIT_ChaPadSen} studies simulation wrt only 2 instruments, whereas we demand simulation wrt to a third highest resolution instrument, lending our formulation more generality with additional complexity. In fact, it is this third instrument that forces the use of two proxy states and bumps up the sum rate (Rem.~\ref{Rem:MLICPBumpInSumRate}). Due to above mentioned challenges, the ISP in \cite{202206ISIT_ChaPadSen} is inapplicable to the problem studied herein. But for the difference in the one-shot information quantities, our one-shot inner bound subsumes that proven achievable in \cite{202206ISIT_ChaPadSen}. See Rem.~\ref{Rem:MLComparisonPriorOneShot}. Owing to our enhanced three-layer ISP, our asymptotic-regime inner bound (Thm.~\ref{Thm:AsymptoticZBMLIS}) is strictly larger than the natural generalization of \cite{202206ISIT_ChaPadSen}'s one-shot inner bound, thereby yielding new asymptotic regime inner bounds.

\begin{figure}
\vspace{-0.1in}
\includegraphics[width=7.2in]{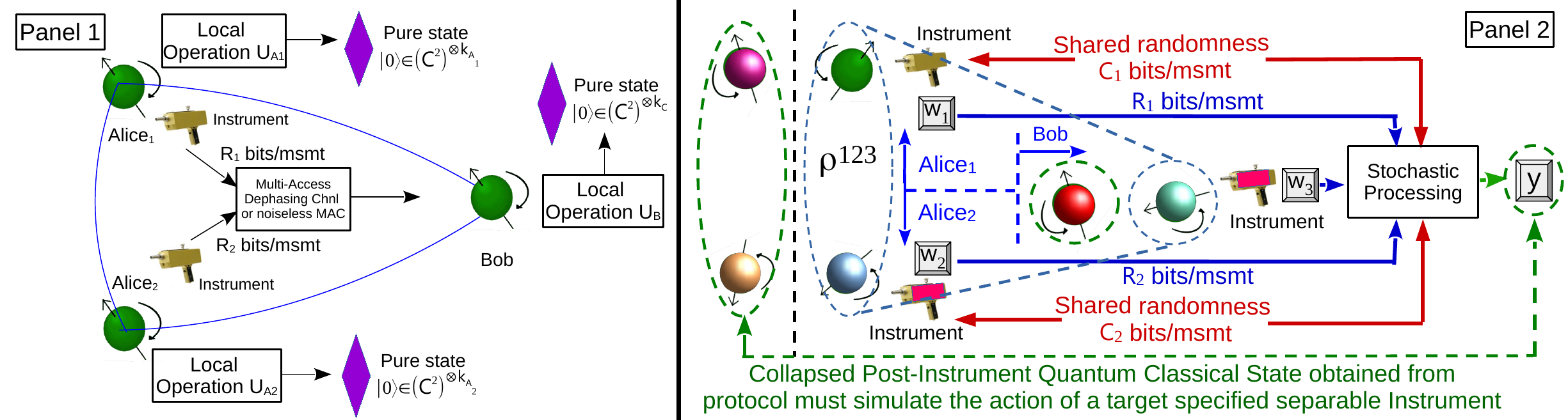}
\vspace{-0.25in}
\caption{{\textbf{\underline{Multi-party Purity Distillation}} (Panel 1): Distributed parties Alice$_{1}$, Alice$_{2}$ have access to their respective components of a tripartite entangled state and can communicate to the third party Bob, holding the third component, on a rate limited dephasing MAC. The goal is to design a protocol where each party can perform local unitaries on their respective components of the mixed state and utilize the dephasing MAC to distil local purity. Here, the violet diamond depicts pure states. \underline{\textbf{Distributed ICP with QSI} (Panel 2)}: Distributed parties Alice$_{1}$, Alice$_{2}$ and Bob have to distributively simulate the action of a separable instrument acting only on the components held by the former two parties. The goal is to quantify the minimum communication resources needed.}}
\vspace{-0.25in}
\label{Fig:QSIAndPD}
\end{figure}

\med\textbf{Multi-Party Purity Distillation}: Local pure states form a ubiquitous resource in quantum information processing~\cite{200306PRA_HorHorOpp}. In contrast to a popular belief that pure states are freely `available', thermodynamic considerations reveal \cite{196107IBM_Lan,200303PRL_HorHorHorHorOppSenSen} that work has to be done to distil pure states. Multiple distributed parties having access to components of a larger entangled state needing to distil local pure states is a naturally occurring scenario. Motivated by this, we study a multi-party PD \cite{202206ISIT_AtiPra} scenario (Fig.~\ref{Fig:QSIAndPD}) involving {three} distributed parties - Alice$_{1}$, Alice$_{2}$ and Bob - with each having access to one component of a tripartite state, and capable of performing local operations. Moreover, Alice$_{1}$ and Alice$_{2}$ can communicate to Bob via a limited rate dephasing multiple-access channel (MAC). What is the total number of pure states that the three parties can distil via any protocol?

{Devetak and Krovi's study \cite{200707PRA_KroDev} of PD in a PTP scenario involving two parties motivated Atif and Pradhan \cite{202206ISIT_AtiPra} to formulate and investigate this multi-party PD question. Building from \cite{200707PRA_KroDev}, it is natural to see that Alice$_{1}$, Alice$_{2}$ have to perform local measurements and communicate their outcomes to Bob to enable him efficiently distil his share of purity. The limited rate dephasing MAC forces them to compress their measurement outcomes and optimize their communication. One is naturally led to the problem of \textit{distributed ICP}, a scenario of much interest in this work and one that we briefly discuss now.}

Consider three distributed parties with Alice$_{1}$, Alice$_{2}$ having access to one component each of a possibly entangled bipartite state $\rho^{A_{1}A_{2}}$ and Bob holding a classical register. See Panel 2 in Fig.~\ref{Fig:DistISAndMultDescIS}. Given an instrument acting on the bipartite state that can be simulated distributedly, formalized as a \textit{separable instrument} in Defn.~\ref{Defn:SeparableInstrument}, the distributed ICP seeks to quantify the minimum communication resources necessary for the three parties to simulate the QC correlation induced by the given instrument with Bob holding the the classical outcome and Alice$_{1}$, Alice$_{2}$ holding the respective quantum components. As we elaborate in the sequel, the distributed ICP is a considerably richer problem than PTP scenarios, i.e. those involving a single encoder-decoder pair. The distributed ICP was first formulated \cite{201907ISIT_HeiAtiPra} and tackled \cite{202202TIT_AtiHeiPra} in the context of POVMs through a series of works \cite{202107ISIT_AtiPra,202504TIT_AtiPra}, wherein asymptotic regime inner bounds were characterized. Leveraging their inner bounds \cite{202202TIT_AtiHeiPra,202504TIT_AtiPra} for the distributed MCP, Atif and Pradhan \cite{202206ISIT_AtiPra} have characterized the current known largest inner bound for the multi-party PD in the asymptotic regime. However, the distributed nature of the components, quantum non-commutativity and entanglement impose a specific challenge that forces Pradhan et.~al.~\cite{201907ISIT_HeiAtiPra,202202TIT_AtiHeiPra,202504TIT_AtiPra} to restrict attention to time sharing based distributed MSPs. Since these protocols are at the heart of their multi-party PD protocols, Atif and Pradhan's \cite{202206ISIT_AtiPra} study and inner bounds for the multi-party PD are also restricted to the asymptotic regime. The recent work of Chakraborty et.~al.~\cite{202306ISIT_ChaNemBus} for one-shot PD is restricted to a PTP scenario since they cannot handle a separable measurement/instrument on distributed entangled components.

The one-shot multi-party PD thus being open, we first address the distributed ICP in Sec.~\ref{Sec:DistInstrSimOneShot}, wherein we develop a likelihood POVM based one-shot protocol. Building on \cite{2026MMTIT_Pad}, we enhance our compatibility operator sliding (COS) to characterize an one-shot inner bound (Thm.~\ref{Thm:DistInstSimltnOneShot}), analogous to that in \cite{2026MMTIT_Pad}, which naturally enlarges to the asymptotic inner bound characterized in \cite{202202TIT_AtiHeiPra} for the distributed MCP. Leveraging this protocol and the one-shot purity concentration protocol of Chakraborty et.~al.~\cite{202306ISIT_ChaNemBus}, we devise a new one-shot multi-party PD protocol in Sec.~\ref{Sec:PurityDistillation}. Analyzing its performance, we characterize a new one-shot inner bound for the multi-party PD in Thm.~\ref{Thm:PurityAchievability}. The natural extension of this inner bound to the asymptotic regime yields an inner bound that can be strictly larger than that characterized in \cite{202206ISIT_AtiPra} - the only known prior work on multi-party PD. See Rem.~\ref{Rmk:AsymptoticPurityKappa}.

\med\textbf{Structured Instruments for Distributed ICP}: In contrast to point-to-point (PTP) ICPs \cite{200401CommMathPhy_Win,201210JPA_WilHayBusHsi,201909TIT_AnsJaiWar} with or without QSI, distributed ICP is a much richer problem since it involves distributed simultaneous compression of two correlated outcomes. Indeed, firstly while the PTP ICPs admit characterization of {\textit{matching} inner and outer bounds}, the distributed ICP remains open, requiring newer ideas. Secondly, while ISPs based on conventional \textit{unstructured independent identically distributed} (IID) codes are optimal for PTP ICPs, strictly larger inner bounds can be achieved \cite{202504TIT_AtiPra} via protocols based on \textit{jointly designed} codes possessing \textit{algebraic closure} properties, henceforth referred to as structured codes. Having focussed on unstructured code based ISPs in Sec.~\ref{Sec:DistInstrSimOneShot}, we incorporate algebraic structure to design \textit{coset code} based ISPs in Sec.~\ref{Sec:DISCosetCodes}, analyze the same and prove achievability of the current known best inner bound \cite{202504TIT_AtiPra} achievable via structured measurements. Here, we note that analysis of coset code based ISPs are fundamentally different from those based on conventional IID codes. Indeed, the codewords of the former are uniformly distributed and only pairwise independent (Rem.~\ref{Rem:DISCosetCdsMeasureTransformTerm}) precluding the use of concentration bounds such as the operator Chernoff bound when analyzing the corresponding instruments. Despite these new elements that complicate analysis, our design of the ISP, the corresponding instruments, analysis steps follow the same template as in other proofs highlighting the unified structure of the likelihood instrument based framework. Our primary motivation in designing and analyzing structured likelihood instruments for distributed ICP despite not proving new inner bounds is to provide a first step towards a broader goal of designing and analyzing one-shot ISPs for distributed ICP based on structured likelihood instruments. We elaborate on this in Rem.~\ref{Rem:DISCCBroaderGoalOfOneShot}.

\textbf{Distributed ICP with QSI}: With an intent to comprehensively address all challenges encountered in network ICPs, we address the distributed ICP with QSI depicted in Fig.~\ref{Fig:QSIAndPD}, wherein decoder Bob has access to QSI. Beyond distributed ICP, the presence of QSI enables the encoders Alice$_{1}$ and Alice$_{2}$ to reduce their communication rates via further enhanced binning (see~\cite{200310PRA_DevWin,201203TIT_RenRen} for classical data compression with QSI at the decoder). While conceptually this is analogous to PTP ICP with QSI, the presence of two encoders accessing distributed components implies the the decoder Bob has to perform simultaneous decoding of the encoder's classical outcomes via a POVM applied on his QSI component. As evidenced through a series \cite{201206TIT_FawHaySavSenWil,202103SAD_Sen,202102SAD_Sen} of works, starting from Winter's initial recognition \cite{200107TIT_Win}, it has been observed that design and analysis of a simultaneous quantum decoder is challenging. In a significant recent finding, Sen has developed techniques of tilting smoothing and augmentation to design and analyze a one-shot quantum simultaneous decoder~\cite{202103SAD_Sen,202102SAD_Sen}. In Sec.~\ref{Sec:QSIDistributedIS}, we demonstrate that our likelihood POVM based protocol for distributed ICP is versatile enough to seamlessly incorporate Sen's simultaneous quantum decoder and yield a new one-shot inner bound for the distributed ICP with QSI which naturally generalizes to the current known best inner bound in the asymptotic regime. In the interest of brevity, we refer to \cite{} for the proofs.

A brief summary mentioning the common threads, challenges and tools we employ is in order. All our ISPs have a common likelihood structure \cite{2026MMTIT_Pad} whose roots can be traced back to \cite[Sec.~V-D]{201311TIT_Cuf}. Likelihood encoders involve (Rems.~\ref{Rem:MDICPProxyStates} , \ref{Rem:DISInverseOfRandomVariables}) random inverse operators. Lack of concentration bounds for the latter leads us to adopt a technique based on proxy states (Rems.~\ref{Rem:DISOneShotProofStepsAndProxyStates} and~\ref{Rem:MDICPNewApproachPurificationBounding}) wherein we analyze the ISP on a proxy state instead of the original state and eventually prove the two states are close. The different network configurations impose specific challenges to this common sequence of proof steps. While we need multiple proxy states (Rems.~\ref{Rem:MDICPSecondPrxyStW0layer}, \ref{Rem:MDICP:ASecondSplitInCov}) for multiple description ICP, entanglement in distributed ICP pushes us to carefully design compatible operators (Rem.~\ref{Rem:OneShotDIS-COSTrick}) that can be commuted. In addition to new inner bounds for multiple description ICP (Thms.~\ref{Thm:OneShotMLIS}  and \ref{Thm:AsymptoticZBMLIS}) , multi-party PD (Thms.~\ref{Thm:PurityAchievability} and \ref{Thm:AsymptoticPurityAchievability}) and distributed ICP with QSI (Thm.~\ref{Thm:DistInstSimltnOneShotQSI}), the techniques we have developed to analyze these different configurations using a unified structure of the likelihood encoder maybe regarded as one of our (technical) findings. In regards to tools, Sen's multi-party fully smooth covering~\cite{202504arXiv_Sen} is used through all our proofs. In addition, the purity concentration protocol of \cite{202306ISIT_ChaNemBus} and Sen's simultaneous quantum decoder~\cite{202103SAD_Sen,202102SAD_Sen} are employed for multi-party PD and distributed ICP with QSI respectively. Lastly, in regards to our exposition, we have provided detailed proofs for all our theorems except Thm.~\ref{Thm:AsymptoticZBMLIS} and Thm.~\ref{Thm:DistInstSimltnOneShotQSI}, for which we have provided substantial outlines. More steps in the proof of the latter two theorems can be found in \cite{}. 

\begin{table*}
\centering
\small
\setlength{\tabcolsep}{3pt}
\begin{tabular}{|c|l|c|l|c|l|}
\hline
Abbreviation & Expansion & Abbreviation & Expansion & Abbreviation & Expansion \\
\hline

POVM & Positive Operator Valued Measurement
& PMF & Probability Mass Function
& PTP & Point-To-Point
\\\hline

QIT & Quantum Information Theory
& QSI & Quantum Side Information
& TP & Trace-Preserving
\\\hline

MSP & Measurement Simulation Protocol
& q-system & Quantum System
& QC & Quantum Classical
\\\hline

ICP & Instrument Compression Problem
& PSD & Positive Semi-Definite
& ONB & Ortho-Normal Basis
\\\hline

MCP & Measurement Compression Problem
& TNI & Trace Non-Increasing
& CP & Completely Positive
\\\hline

ISP & Instrument Simulation Protocol
& Info-theoretic & Information Theoretic
& PD & Purity Distillation
\\\hline

\end{tabular}
\vspace{0.1cm}
\caption{Abbreviations used in this paper.}
\label{Table:Abbreviations}
\vspace{-0.9cm}
\end{table*}

\section{Preliminaries and Point-to-Point Instrument Compression Problem}
\label{Sec:ProblemStatements}

After specifying notation in Sec.~\ref{Sec:PrelimsProblemStatement} that is used throughout, we describe formulation of the point-to-point ICP in Sec.~\ref{SubSec:PTP-ICProblemStatement}.

\subsection{Notation, One-shot Information quantities and their properties}
\label{Sec:PrelimsProblemStatement}
We supplement standard notation \cite{BkWilde_2017} adopted in quantum information theory with the following. For a positive integer $n$, we let $[n] \define \{1,\cdots,n\}, [\overline{n}]\define\{0\}\cup [n]$. When $i \in \{1,2\}$, we let $\msout{i}$ to denote the complement index, i.e., $\{i,\msout{i}\} = \{1,2 \}$. All Hilbert (state) spaces are assumed to be finite dimensional. $\mathcal{L}(\mathcal{H})$, $\mathcal{P}(\mathcal{H})$, $\mathcal{D}(\mathcal{H})$ denote the collection of linear, positive semi-definite (PSD) and density operators acting on Hilbert space $\mathcal{H}$ respectively. $\alpha^{*} \in \complex$ and $T^{\dagger} \in \CalL(\CalH)$ denote complex conjugate of $\alpha\in\complex$ and adjoint of $T \in \CalL(\CalH)$ respectively. For any Hilbert space $\CalH$ of dimension $d \in \naturals$, we let $\{\ket{\mathsf{b}_{1}},\cdots,\ket{\mathsf{b}_{d}}\} \subseteq \CalH$ denote an arbitrary but \underline{fixed} orthonormal basis (ONB) and define conjugation maps wrt this fixed ONB. Specifically, for any $\ket{x} =\sum_{i=1}^{d}\alpha_{i}\ket{\mathsf{b}_{i}} \in \CalH$, we let $\ket{\olinex}\define \sum_{i=1}^{d}\alpha_{i}^{*}\ket{\mathsf{b}_{i}}\in \CalH$ denote its conjugation. For $A \in \CalL(\CalH)$, we let $A^{*}$ denote the unique operator satisfying $\mel{\sfb_{r}}{A^{*}}{\sfb_{s}} = \left(\!\mel{\sfb_{r}}{A}{\sfb_{s}}\!\right)^{*}$ for all $r,s \in [d]$. $\tau\in\CalP(\CalH)$ is subnormalized if $\Tr\tau\le1$.

$\CalH_{A}$ and $H_{R}$ will denote state spaces of the accessible and inaccessible (or reference) components of our quantum systems respectively. For state $\rho \in \CalD(\CalH_{A})$, $\ket{\phi_{\rho}} \in \CalH_{R} \otimes \CalH_{A}$ denotes a purification of $\rho$ and $\ket{\varphi_{\rho}} = \sum_{x}\ket{\overline{\eta}_{x}}\otimes \sqrt{\rho}\ket{\eta_{x}} \in \CalH_{R} \otimes \CalH_{A}$, where $\{\ket{\eta_{x}} : x \in \CalX\} \subseteq \CalH_{A}$ is any orthonormal basis, denotes the \textit{canonical purification} of $\rho \in \CalD(\CalH_{A})$. In the latter case, $\CalH_{R} = \CalH_{A}$ are isomorphic. We let $\phi_{\rho}\define \ketbra{\phi_{\rho}}$ and $\varphi_{\rho}\define \ketbra{\varphi_{\rho}}$. We extend definition of canonical purification to the space $\CalP(\CalH)$ of PSD operators. With this, we note $\Tr\varphi_\tau=\|\ket{\varphi_\tau}\|_2^2=\Tr\tau$ for any $\tau \in \CalP(\CalH)$ and $\Tr(\varphi_\rho)=1$ if $\rho \in \CalD(\CalH)$. $\norm{A}_{1}\define \tr(\sqrt{A^{\dagger}A})$ denotes trace norm. $D(\rho||\sigma)\define \tr\left\{\rho\log \rho - \rho\log \sigma \right\}$ denote the relative entropy between $\rho,\sigma \in \CalD(\CalH)$.

All sets are assumed to be of finite cardinality. $\CalD(\CalX)$ denote the set of all probability mass functions (PMFs) on $\CalX$ and $\complex^{|\CalX|}$ denote the state space of the classical register taking values in $\CalX$. For a stochastic matrix $p_{Y|W}:\CalW \rightarrow \CalY$, we let $\olinescre^{{Y|W}}_{p}(\cdot) : \CalL(\complex^{|\CalW|}) \rightarrow \CalL(\complex^{|\CalW|} \otimes \complex^{|\CalY|} )$ and $\ScrE^{{Y|W}}_{p}(\cdot): \CalL(\complex^{|\CalW|}) \rightarrow \CalL(\complex^{|\CalY|})$ denote the CPTP maps
\begin{eqnarray}
\label{Eqn:ClassicalStochasticMaps}
\olinescre^{{Y|W}}_{p}(a) \define\!\!\!\!\!\!\!\! \sum_{(w,y) \in \mathcal{W}\times \mathcal{Y}}\!\!\!\!p_{Y|W}(y|w)\ket{w}\!\mel{w}{a}{w}\!\bra{w}\otimes \ketbra{y}\mbox{ and }\ScrE^{{Y|W}}_{p}(a)\define\!\!\!\! \sum_{(w,y) \in \mathcal{W}\times \mathcal{Y}}\!\!\!\!p_{Y|W}(y|w)\mel{w}{a}{w}\ketbra{y}.
\end{eqnarray}

\med\textit{\underline{Underline} Notation}: We let an \underline{underline} denote an appropriate aggregation of related objects. For example, with (1) sets $\CalX_{1},\CalX_{2}$, we let $\ulineCalX \define \CalX_{1}\times \CalX_{2}$, (2) Hilbert spaces, $\CalH_{A_1},\CalH_{A_2}$, we let $\CalH_{\ulineA}\define\CalH_{A_1}\!\otimes\CalH_{A_2}$, (3) indices $k_{i} \in [K_{i}]:i \in [2]$, we let $\ulinek\define(k_1,k_2)$, (4) stochastic matrix, we let $p_{Y|\ulineW}\define p_{Y|W_1W_2}$. The appropriate aggregation will be clear from context.

\med\textbf{Bold} \textit{for $n-$letter} objects: In addition to Sec.~\ref{Sec:DISCosetCodes} which deals exclusively in the asymptotic regime, we characterize inner bounds in the asymptotic regime as corollaries of our one-shot inner bounds. To denote $n-$letter objects, we employ a \textbf{bold} notation. For state space $\CalH_{A}$, set $\CalY$, measurement $\lambda$ or $\lambda_{\CalY}$ PMF $p_{W} \in \CalD(\CalW)$ and $\rho \in \CalD(\CalH_{A})$, we let $\CalH_{\boldA} \define \CalH_{A}^{\otimes n}$, $\boldCalY \define \CalY^{n}$, $\boldlambda \define\boldlambda_{\boldCalY} \define \{ \boldlambda_{\boldy} \define \lambda_{y_{1}}\otimes \cdots \otimes \lambda_{y^{n}}: \boldy \define (y_{1},\cdots,y_{n}) \in \boldCalY \define \CalY^{n}\}$, $\boldp_{W}=p_{W}^{n}=\prod_{t=1}^{n}p_{W}$ and $\brho\define \rho^{\otimes n}$ respectively.

For a state $\sigma^{AB} \in \CalD(\CalH_{A} \otimes \CalH_{B})$, we let $I(A:B)_{\sigma} \define H(A)_\rho+H(B)_\rho-H(AB)_\rho
=D(\rho^{AB}\Vert\rho^A\otimes\rho^B)$, where $\sigma^{A} = \tr_{B}(\sigma^{AB}), \sigma^{B} = \tr_{A}(\sigma^{AB})$ and $H(A)_\rho\define-\Tr[\rho^A\log\rho^A]$ denotes the Von Neumann entropy. All logarithms are wrt base $2$. Our one-shot performance limits are characterized via broadly three families of information measures. Our primary covering tools are the fully smooth multi-partite convex-split lemmas of Sen~\cite{202504arXiv_Sen}. For purity concentration, we also use the smoothed support max-entropy of~\cite{202208arXiv_ChaNemBus}, defined next. For $\rho^A\in\CalD(\CalH_A)$ and $\varepsilon\in[0,1]$, we let
\begin{equation}
\label{Eqn:TildeHmaxCondY}
\widetilde H_{\max}^{\varepsilon}(A)_{\rho}
\define
\min_{\Pi_A^2=\Pi_A=\Pi_A^\dagger,\ \Pi_A\neq0:\,\Tr[\Pi_A\rho^A]\ge1-\varepsilon}
\log\Tr[\Pi_A]~\mbox{ and }~\widetilde H_{\max}^{\varepsilon}(A|Y)_\rho
\define
\min_{\mathcal G\subseteq\mathcal Y:\,p_Y(\mathcal G)>1-\sqrt{\varepsilon}}
\max_{y\in\mathcal G}\widetilde H_{\max}^{\varepsilon}(A)_{\rho_y},
\end{equation}
where the latter quantity is defined for a cq-state
$\rho^{YA}=\sum_{y\in\CalY}p_Y(y)\ketbra{y}_Y\otimes\rho_y^A$ and $\varepsilon\in(0,1]$. Note that $\widetilde H_{\max}^{\varepsilon}(A)_{\rho}$ is the logarithm of the
smallest dimension of a subspace carrying probability at least $1-\varepsilon$ under
$\rho^A$. We will also have opportunity to leverage sandwiched R\'enyi divergences \cite{201312JMP_MulDupSzefehTom,201407CMP_WilWinYan} of order $2$, defined next. For $\alpha,\beta \in \CalP(\CalH)$ with $\alpha$ subnormalized, we let
\begin{eqnarray}
D_2(\alpha\Vert\beta)\define
\begin{cases}
2\log\|\beta^{-1/4}\alpha\beta^{-1/4}\|_2,
& \text{if }\supp(\alpha)\subseteq\supp(\beta),\\
+\infty, & \text{otherwise},
\end{cases}\mbox{ and }D_2^\varepsilon(\alpha\Vert\beta)
\define
\min_{\alpha'\ge0:\,\|\alpha'-\alpha\|_1\le\varepsilon\Tr\alpha}
D_2(\alpha'\Vert\beta)\mbox{ for }\varepsilon\in(0,1),
\nonumber
\end{eqnarray}
where the minimization is over subnormalized operators $\alpha'$. Next, our packing bounds are characterized via hypothesis-testing divergence \cite{201205PRL_WanRen}. For state $\rho \in \CalD(\CalH)$, PSD $\sigma \in \CalP(\CalH)$ and $\varepsilon\in(0,1)$, the hypothesis-testing divergence  $D_H^\varepsilon(\rho\Vert\sigma)$ is defined as
\begin{eqnarray}
\label{Eqn:NotationHypoTestDiv}
D_H^\varepsilon(\rho\Vert\sigma)
\define-\log\inf_{0\le Q\le I:\,\Tr[Q\rho]\ge1-\varepsilon}\Tr[Q\sigma].\mbox{ Finally, collating the associated one-shot mutual information quantities, we let}
\nonumber\\
I_2^\varepsilon(A:B)_\rho
\define D_2^\varepsilon(\rho^{AB}\Vert\rho^A\otimes\rho^B),
~~~
I_H^\varepsilon(A:B)_\rho
\define D_H^\varepsilon(\rho^{AB}\Vert\rho^A\otimes\rho^B) \mbox{ for any bipartite state }\rho^{AB} \in \CalD(\CalH_{A}\otimes \CalH_{B}).
\nonumber
\end{eqnarray}
\begin{lemma}
\label{Lem:One-ShotInfoQuantConvergence}
Let
$(\varepsilon_n)_{n\ge1}$ satisfy $\varepsilon_n\downarrow0$ and
$\frac1n\log\frac1{\varepsilon_n}\to0$. For $\rho,\sigma \in \CalD(\CalH)$ with $\supp(\rho)\subseteq\supp(\sigma)$ and $\theta_{AB} \in \CalD(\CalH_{A}\otimes \CalH_{B})$
\[
\lim_{n\to\infty}\frac1n
D_2^{\varepsilon_n}(\boldrho\Vert\boldsigma)
=
\lim_{n\to\infty}\frac1n
D_H^{\varepsilon_n}(\boldrho\Vert\boldsigma)
=
D(\rho\Vert\sigma), \qquad \lim_{n\to\infty}\frac1n I_2^{\varepsilon_n}(A^n:B^n)_{\boldtheta}
=
\lim_{n\to\infty}\frac1n I_H^{\varepsilon_n}(A^n:B^n)_{\boldtheta}
=
I(A:B)_\rho.
\]
\end{lemma}
\noindent See Lemma \ref{Lem:UpperAEPConditionalTildeHmax} for convergence of the conditional quantity in~\eqref{Eqn:TildeHmaxCondY}. Convergence proofs of the rest is standard; see \cite{2016BOOK_Tom,201205PRL_WanRen}.

\subsection{Formulation and Statement of Point-to-point Instrument Simulation Problem}
\label{SubSec:PTP-ICProblemStatement}

For pedagogical reasons, we now discuss formulation of a point-to-point (PTP) MCP - the simplest ICP involving measurement action on a single component quantum state yielding one outcome. Following this, we recall the notion of instruments, provide a formal statement of PTP ICP in terms of instruments and state the solution, i.e. characterize the minimal necessary communication resources. This assists our reader reason our problem statements of network ICPs in subsequent sections.

\begin{figure}
    \vspace{-0.25in}
    \includegraphics[width=6.85in]{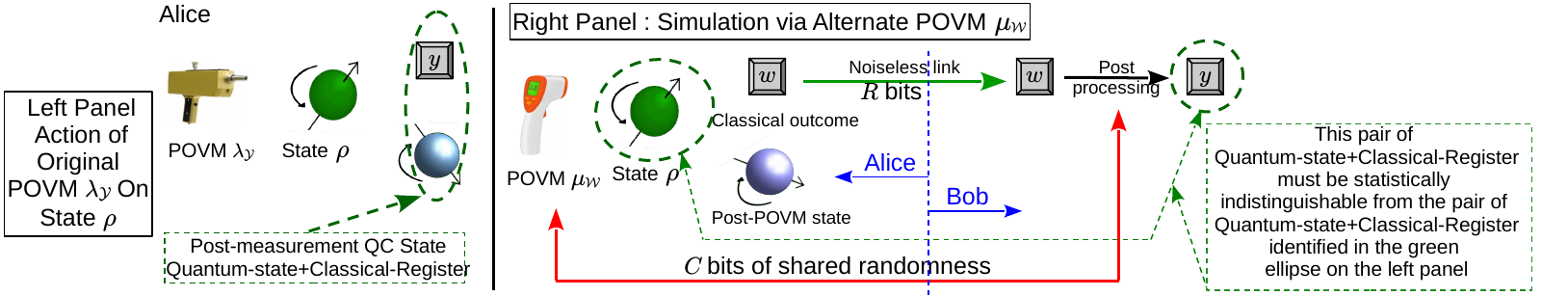}
    \vspace{-0.15in}
    \caption{When Alice performs a measurement on a quantum state, a specific quantum correlation (QC) correlation is induced between the post-measurement state and her classical register that holds the outcome. See green ellipse in the left panel. What is the minimum number $R$ of bits that Alice must communicate to a distributed party Bob with whom she shares $C$ bits of common random bits to induce the same correlation between his register and the post-measurement state she holds? Broadly stated, the MCP investigates what resources, in terms of rates of the noiseless link ($R$) and common random bits ($C$), are needed to simulate this same correlation distributedly. See Sec.~\ref{SubSec:PTP-ICProblemStatement} for a description and a precise formulation.}
    \label{Fig:SnglEncSnglDecMsmtSimulation}
    \vspace{-0.2in}
\end{figure}

When Alice performs a measurement $\lambda_{\CalY}$ on state $\rho \in \CalD(\CalH_{A})$, a specific QC correlation is induced between the post-measurement state and her classical register that holds the outcome (Fig.~\ref{Fig:SnglEncSnglDecMsmtSimulation}). What is the minimum number $R$ of bits that Alice must communicate to Bob to induce the same QC correlation between the the post-measurement state she possesses and the contents of his classical register? In answering this, Winter \cite{200401CommMathPhy_Win} made the crucial observation that some of the randomness necessary to induce the necessary QC correlation can be sourced from an external source unrelated to both the q-system and the measurement process, so long as it \textit{shared} between the two parties. This led him to design a \textit{measurement simulation protocol} (MSP) employing an alternate $n-$letter measurement, which when executed on the tensor state $\boldrho\define \rho^{\otimes n}$ results in an identical post-measurement QC state as the original measurement $\boldlambda_{\boldCalY} \define \lambda_{\CalY}^{\otimes n} $, but with the desirable property that the classical outcome of the protocol requires Alice to communicate fewer bits. By communicating these fewer bits to Bob, the simulation protocol can potentially establish the same QC correlation we sought. We are thus led to the following.

Alice holding $n$ copies of the state, i.e $\boldrho \in \CalD(\CalH_{\boldA})$, and an agnostic distributed observer Bob holding the classical register are both provided $C$ bits of (external) shared random bits per measurement, which are statistically independent of the state $\boldrho$. These can be utilized by Alice to potentially choose a measurement at random from a bank of designed measurements. Alice can communicate $R$ bits/msmt to Bob. Bob can then utilize the $nC$ bits of shared randomness to post process the information he receives in the $nR$ bits, to populate his classical register. We require that the QC outcome of this protocol - the post-measurement state at Alice and Bob's classical register contents - to be statistically indistinguishable from that of the original measurement $\boldlambda_{\boldCalY}$. Since the $C$ shared random bits/msmt are statistically independent of the state $\rho$, Winter identified this as \textit{extrinsic randomness} and the $R$ bits/msmt communicated by Alice as the \textit{intrinsic information} of the state $\rho$ in the outcome of measurement $\lambda_{\CalY}$. The minimum value of $R$ for which, there exists a $C$ and a corresponding measurement simulation protocol that can accomplish this, is therefore the information content in the outcome of the measurement $\lambda_{\CalY}$ on $\rho$. Going further, the MCP seeks to characterize all such rate pairs $(R,C)$ for which such a simulation protocol exists, thereby quantifying the entire trade-off between intrinsic information and extrinsic randomness in the outcome of measurement $\lambda_{\mathcal{Y}}$ on state $\rho$.

Two remarks are in order. Firstly, in his formulation Winter \cite{200401CommMathPhy_Win} modeled the measurement process in the conventional {info-theoretic} setting of asymptotically large number of identical measurements over identical copies of the q-system and characterized his rates in the asymptotic regime. Secondly, since his \cite{200401CommMathPhy_Win} goal was to only quantify the information content in the measurement's outcome he restricted attention to modeling the action of POVMs and focussed only on the QC correlation induced between the inaccessible reference and the outcome by discarding (tracing out) the accessible post-measurement state.

Through most of our work, except for findings presented in Sec.~\ref{Sec:DISCosetCodes}, we focus on both one-shot and asymptotic regimes. Secondly, in addition to characterizing rates of communication and common randomness, our study of PD will require us to simulate specific QC correlation induced by arbitrary measurements, not necessarily POVMs, between the post-measurement state and the classical outcome. In other words, this precludes us from ignoring the effect of the measurement on the accessible component and are thus prevented from tracing out the latter in the analysis. Our study takes this opportunity to consider \textit{instruments} which, as defined below, are most general measurements yielding both classical and quantum outcomes.

\begin{definition}
\label{Defn:QuantumInstrument}
A quantum instrument with input system $A$, output system $B$, and finite outcome alphabet $\CalY$ is a collection
$\CalJ=\{\CalJ_y:\CalL(\CalH_A)\to\CalL(\CalH_B)\}_{y\in\CalY}$ of completely positive trace-nonincreasing (CP TNI) maps such that $\sum_{y\in\CalY}\CalJ_y$ is trace preserving. We identify the instrument with the CPTP map
$\CalJ(\rho)\define\sum_{y\in\CalY}\CalJ_y(\rho)\otimes\ketbra{y}_Y$.
\end{definition}

Clearly, the class of instruments subsume both POVMs and measurements. From Defn.~\ref{Defn:QuantumInstrument}, we also note that the above theme motivating formulation of MCP can be naturally extended to formulating an ICP. Indeed, an instrument too induces a specific QC correlation between the post-measurement q-system and the classical register, albeit the q-system potentially lives in a different state space. Nevertheless, one may design an analogous instrument simulation protocol (ISP) comprising of a \textit{bank of encoder instruments} and a corresponding \textit{bank of post-processing decoder maps} to efficient utilize the communication resources - external shared common random bits and noiseless bit pipe from Alice to Bob. We are led to the following.

\begin{definition}[Point-to-point one-shot instrument simulation protocol]
\label{Defn:OneShotPTPIS}
A $(K,M)$ point-to-point one-shot instrument simulation protocol for $\CalJ$ on $\rho$ consists of:
\begin{itemize}
\item a bank of encoding $K$ instruments
\[
\CalE_k:\CalL(\CalH_A)\longrightarrow \CalL(\CalH_B\otimes\mathbb{C}^{M}),
\qquad k\in[K],
\]
each indexed by the common randomness index $k$
\item $K$ decoding CPTP maps, one for each common-randomness index,
\[
\CalD_k:\CalL(\mathbb{C}^{M})\longrightarrow \CalL(\mathbb{C}^{\CalY}),
\qquad k\in[K].
\]
\end{itemize}
\end{definition}
\begin{definition}[Point-to-point instrument simulation criterion]
\label{Defn:PTPISCriterion}
For $\eta>0$, the action of $\CalJ$ on $\rho$ can be $\eta$-simulated with communication cost $R$ and common-randomness cost $C$ if there exists a $(K,M)$ point-to-point one-shot instrument simulation protocol satisfying $\log M\leq R$, $\log K\leq C$, and
\begin{equation*}
\left\|
\frac{1}{K}\sum_{k=1}^{K}
\Bigl(\mathrm{id}_{BR}\otimes\CalD_k\Bigr)\circ
\Bigl(\mathrm{id}_R\otimes\CalE_k\Bigr)(\varphi_\rho)
-
(\mathrm{id}_R\otimes \CalJ)(\varphi_\rho)
\right\|_1
\leq \eta.
\end{equation*}
The action of $\CalJ$ on $\rho$ is asymptotically simulable with resource costs $(R,C)$ if, for every $\eta>0$, there exists $N_\eta$ such that for all $n\geq N_\eta$, the action of $\CalJ^{\otimes n}$ on $\rho^{\otimes n}$ can be $\eta$-simulated with communication cost $nR$ and common-randomness cost $nC$.
\end{definition}

Winter \cite{200401CommMathPhy_Win} characterized the entire trade-off of $(R,C)$ pairs that permits asymptotic simulability for the particular case of POVMs under the requirement that both Alice and Bob has to possess the classical outcome. The point-to-point ICP when restricted to POVMs was first studied by Anshu, Jain and Warsi in \cite{201909TIT_AnsJaiWar} who developed a one-shot measurement-compression protocol with quantum side information based on convex splitting and position-based decoding, and characterized its communication and shared-randomness costs in terms of smooth max- and hypothesis-testing relative entropies. Among the works that have followed \cite{201909TIT_AnsJaiWar}, we note the recent work of Cheng and Gao \cite[Sec.~5.4]{202305arXiv_CheGao} that derives error exponents for the point-to-point ICP, once again when restricted to POVMs. For the asymptotic setting, Wilde et al.~\cite[Thms.~8 and~9]{201210JPA_WilHayBusHsi} discuss instrument compression and the communication--randomness tradeoff for non-feedback measurement compression.\footnote{The control state in~\cite[Thm.~9, Eq.~(49)]{201210JPA_WilHayBusHsi} is written after tracing out the accessible quantum output. Definition~\ref{Defn:PTPISCriterion} here explicitly requires preservation of that output, denoted by $B$, which is therefore retained in the information quantity in our bound below.} \cite{202206ISIT_ChaPadSen} studies instrument simulation in the one-shot setting and derives an achievable communication--randomness tradeoff, including the point-to-point case. Achievability of the following inner bound can be obtained as a particular case of the proofs provided for either Thm.~\ref{Thm:OneShotMLIS} or Thm.~\ref{Thm:DistInstSimltnOneShot}.

\begin{theorem}[Point-to-point instrument simulation]
\label{Thm:PTPInstrumentSimulation}
Let $\CalJ=\{\CalJ_y:y\in\CalY\}$ be an instrument and let $\rho\in\CalD(\CalH_A)$. Suppose there exist a finite alphabet $\CalW$, an instrument $\CalI=\{\CalI_w:w\in\CalW\}$, and a stochastic matrix $p_{Y|W}$ such that
\[
\CalJ_y=\sum_{w\in\CalW}p_{Y|W}(y|w)\CalI_w,\qquad y\in\CalY.
\]
Let $\sigma^{RBWY}\define
\bigl(\mathrm{id}_{RB}\otimes\olinescre^{Y|W}_p\bigr)
\circ (\mathrm{id}_R\otimes\CalI)(\varphi_\rho).$
Then the action of $\CalJ$ on $\rho$ is asymptotically simulable with communication rate $R$ and common-randomness rate $C$ whenever
\begin{equation*}
R>I(W:R)_\sigma,\qquad R+C>I(W:YBR)_\sigma.
\end{equation*}
\end{theorem}

\begin{remark}
\label{Rem:ICPsRecoverMCPs}
One can recover the original measurement compression rates as a special case of our setup.
Indeed, the standard setting corresponds to the situation where we do not require preserving the
post-measurement state on Alice's system. Concretely, given a POVM
$\{\Lambda_y\}_{y \in \mathcal{Y}}$, it is enough to simulate the instrument
$\mathcal{I}:\mathcal{L}(\mathcal{H}_A)\to \mathcal{L}(\mathbb{C}^{|\mathcal{Y}|})$ defined by
\[
\mathcal{I}(\rho)
= \operatorname{Tr}_A\left[\sum_{y\in\mathcal{Y}}
M_y\,\rho\,M_y^\dagger \otimes |y\rangle\langle y|\right],
\]
where $\{M_y\}_{y\in\mathcal{Y}}$ are Kraus operators acting on $\mathcal{H}_A$, i.e.,
$M_y:\mathcal{H}_A\to \mathcal{H}_A$, and satisfy $M_y^\dagger M_y = \Lambda_y$ for all
$y\in\mathcal{Y}$. More generally, the same reduction applies to all the instrument-simulation
problems studied in this paper: the corresponding measurement-simulation setting is recovered by
discarding the post-measurement quantum output systems and retaining only the classical outcomes.
This form of simulation is sufficient for some applications; however, in other tasks such as purity
distillation, it is crucial to preserve the post-measurement state as well.
\end{remark}

\section{Multiple-Description Instrument Simulation}
\label{Sec:MultiLinkIS}
Alluding back to the discussion in Sec.~\ref{Sec:Introduction}, we begin by formalizing the notions of higher resolution (fine-grained) and lower resolution (coarse-grained).

\begin{definition}
\label{Defn:FineAndCoarseGranings}
Let
$\CalJ^{\CalX}
=
\{\CalJ_x:\CalL(\CalH_A)\to\CalL(\CalH_B):x\in\CalX\}$ and
$\CalK^{\CalY}
=
\{\CalK_y:\CalL(\CalH_A)\to\CalL(\CalH_B):y\in\CalY\}$ be two instruments operating with the same input, output quantum spaces. We say that $\CalK^{\CalY}$ is a \emph{coarse-graining} of
$\CalJ^{\CalX}$ if there exists a deterministic map
$f:\CalX\to\CalY$ such that
\begin{equation}
\label{Eqn:InstrumentCoarseGraining}
\CalK_y
=
\sum_{x:\,f(x)=y}\CalJ_x,
\qquad y\in\CalY.
\end{equation}
In this case, $\CalJ^{\CalX}$ is said to be a \emph{fine-graining} of
$\CalK^{\CalY}$. Equivalently, the fibers
$\{f^{-1}(y):y\in\CalY\}$ partition the outcome alphabet $\CalX$, and the lower-resolution outcome $y$ identifies
only the cell of this partition containing the higher-resolution outcome $x$.
\end{definition}
We now describe the scenario of interest in our multiple-description ICP. As depicted in Panel 3 of Fig.~\ref{Fig:DistISAndMultDescIS}, consider a single component quantum state $\rho \in \CalD(\CalH_{A})$ accessible to Alice. Alice is connected to a distributed party Bob who holds classical registers via two links - Link$_{1}$ and Link$_{2}$ - of communication resources. Link$_{i}$ enables (i) Alice to transfer $R_{i}$ bits to Bob, and (ii) both parties to share $C_{i}$ bits of common randomness, both metered per instrument action. Consider two coarse-grained instruments $\CalJ^{Y_{1}},\CalJ^{Y_{2}}$ of $\CalJ^{\CalX}$. The multiple-description ICP comes with three requirements. Firstly, with Link$_{1}$ alone being active, Alice and Bob must be able to simulate the action of instrument $\CalJ^{\CalY_{1}}$. Secondly, with Link$_{2}$ alone being active, Alice and Bob must be able to simulate the action of instrument $\CalJ^{\CalY_{1}}$. Using both Link$_{1}$ and Link$_{2}$, Alice and Bob need to simulate the fine grained instrument $\CalJ^{\CalX}$.

Before we address the problem, our first step is to prove that it suffices to consider a simpler form for the three instruments. Essentially, the following lemma proves that we could have equivalently stated our problem in terms of a single fine grained instrument $\CalJ^{\CalY_{0}\times \CalY_{1} \times \CalY_{2}}$ having an outcome alphabet $\CalY_{0}\times \CalY_{1} \times \CalY_{2}$ that is a Cartesian product of three alphabets $\CalY_{i} : i \in \{0,1,2\}$ and demanding that (i) each Link$_{i}$ individually enable the parties simulate the co-ordinate outcome in $\CalY_{i}$ for $i=1,2$, and (ii) both Link$_{1}$ and Link$_{2}$ enables Alice and Bob simulate all three coordinates.

\begin{lemma}
\label{Lem:ThreeRegisterRepresentation}
Let $\CalJ^{\CalX}
=
\{\CalJ_x:x\in\CalX\}$ be an instrument, and for $i\in[2]$ let $\CalJ^{(i)}
=
\{\CalJ^{(i)}_{y_i}:y_i\in\CalY_i\}$ be a coarse-graining of $\CalJ^{\CalX}$ induced by a map
$f_i:\CalX\to\CalY_i$, i.e., $\CalJ^{(i)}_{y_i}
=
\sum_{x:\,f_i(x)=y_i}\CalJ_x, y_i\in\CalY_i, i\in[2].$ Then there exist a finite alphabet $\CalY_0$ and an instrument $\widehat{\CalJ}
=
\{
\widehat{\CalJ}_{y_0,y_1,y_2}:
(y_0,y_1,y_2)
\in
\CalY_0\times\CalY_1\times\CalY_2
\}$ which is equivalent to $\CalJ^{\CalX}$ up to a relabeling of its
classical outcome, and such that
\begin{equation}
\label{Eqn:ThreeRegisterMarginals}
\CalJ^{(1)}_{y_1}
=
\sum_{y_0,y_2}
\widehat{\CalJ}_{y_0,y_1,y_2},
\qquad
\CalJ^{(2)}_{y_2}
=
\sum_{y_0,y_1}
\widehat{\CalJ}_{y_0,y_1,y_2}.
\end{equation}
Conversely, any instrument
$\widehat{\CalJ}=\{\widehat{\CalJ}_{y_0,y_1,y_2}\}$
with classical outcome $(Y_0,Y_1,Y_2)$ defines a fine-grained
instrument whose $Y_1$- and $Y_2$-marginals in
\eqref{Eqn:ThreeRegisterMarginals} are two of its coarse-grainings.
Hence, specifying a fine-grained instrument together with two of its
coarse-grainings is equivalent, up to a relabeling of the fine outcome,
to specifying a single instrument with three classical outcome registers
$(Y_0,Y_1,Y_2)$.
\end{lemma}

\begin{proof}
For $(y_1,y_2)\in\CalY_1\times\CalY_2$, let
$\CalF_{y_1,y_2}\define\{x\in\CalX:f_1(x)=y_1,\ f_2(x)=y_2\}$ and
$L\define\max_{y_1,y_2}|\CalF_{y_1,y_2}|$. Set $\CalY_0=[L]$ and, for each nonempty
$\CalF_{y_1,y_2}$, choose an injection
$\ell_{y_1,y_2}:\CalF_{y_1,y_2}\to\CalY_0$. Then $\iota(x)\define
\bigl(\ell_{f_1(x),f_2(x)}(x),f_1(x),f_2(x)\bigr)$ defines an injection
$\iota:\CalX\to\CalY_0\times\CalY_1\times\CalY_2$. Define
\begin{equation*}
\widehat{\CalJ}_{y_0,y_1,y_2}
\define
\begin{cases}
\CalJ_x, & \text{if }(y_0,y_1,y_2)=\iota(x)\text{ for some }x\in\CalX,\\
0, & \text{otherwise}.
\end{cases}
\end{equation*}
Injectivity of $\iota$ gives
$\sum_{y_0,y_1,y_2}\widehat{\CalJ}_{y_0,y_1,y_2}
=\sum_{x\in\CalX}\CalJ_x$, furthermore $\sum_{y_0,y_2}\widehat{\CalJ}_{y_0,y_1,y_2}
=\sum_{x:f_1(x)=y_1}\CalJ_x
=\CalJ^{(1)}_{y_1}$ and same identity hold for $\CalJ^{(2)}_{y_2}$ so $\widehat{\CalJ}$ is an instrument whose $Y_1$- and $Y_2$-marginals are the prescribed coarse-grainings. Conversely, for any
instrument $\widehat{\CalJ}=\{\widehat{\CalJ}_{y_0,y_1,y_2}\}$, the coordinate
projections $(y_0,y_1,y_2)\mapsto y_i$, $i\in[2]$, are deterministic maps, so
the marginals in \eqref{Eqn:ThreeRegisterMarginals} are coarse-grainings
according to \eqref{Eqn:InstrumentCoarseGraining}.
\end{proof}
\subsection{Problem Statement}
\label{SubSec:MultiLinkISProblemStatement}

We begin with some remarks on notation that we will utilize throughout this section.

\begin{notation}
\label{Rem:MLNotation}
Throughout this section, we use underlines to aggregate indices and random variables. A single \underline{underline} aggregates the \textit{two} link-specific components, while a double \doubleunderline{underline} aggregates \textit{all three}
components (including the common component indexed by $0$). For instance, we write $\underline{W}\define (W_1,W_2)$ and $\underline{k}\define(k_1,k_2)$, whereas
$\doubleunderline{m}\define(m_0,m_1,m_2)$ and $\doubleunderline{W}\define(W_0,W_1,W_2)$. In the following, when $i \in \{1,2\}$, we let $\msout{i}$ to denote the complement index, i.e., $\{i,\msout{i}\} = \{1,2 \}$. We note here that we will also be operating with indices in \{0,1,2\}. We employ the $\msout{i}$ notation only when there is no ambiguity.
\end{notation}

We refer the reader to Panel 1 in Fig.~\ref{Fig:DistISAndMultDescIS} wherein the scenario of Multiple-Description ICP is depicted. As justified in Lemma \ref{Lem:ThreeRegisterRepresentation}, we can restrict attention to an instrument whose classical outcome takes values in a Cartesian product of three sets. Fix finite alphabets $\CalY_0,\CalY_1,\CalY_2$ and consider an instrument with three classical outputs
\begin{equation*}
\CalJ:\ \CalL(\CalH_A)\longrightarrow
\CalL\!\bigl(\CalH_B\otimes \mathbb{C}^{\CalY_0}\otimes \mathbb{C}^{\CalY_1}\otimes \mathbb{C}^{\CalY_2}\bigr),
\qquad
\CalJ(\rho)
=\!\!\!\!\!\!\!\!\!
\sum_{(y_0,y_1,y_2)\in\CalY_0\times\CalY_1\times\CalY_2}\!\!\!\!\!\!\!\!\!\!\!\!
\CalJ_{y_0,y_1,y_2}(\rho)\otimes \ketbra{y_0}_{Y_0}\otimes \ketbra{y_1}_{Y_1}\otimes \ketbra{y_2}_{Y_2},
\end{equation*}
where $\{\CalJ_{y_0,y_1,y_2}\}_{y_0,y_1,y_2}$ are completely positive maps such that
$\sum_{y_0,y_1,y_2}\CalJ_{y_0,y_1,y_2}$ is CPTP. Let $\rho\in\CalD(\CalH_A)$ and define the ideal post-instrument states
\begin{equation}
\label{Eqn:IdealPostInstrumentMultiLink}
\CalT_{012}\ \define\ (\mathrm{id}_R\otimes \CalJ)(\varphi_\rho),
\qquad
\CalT_1\ \define\ \Tr_{Y_0Y_2}[\CalT_{012}],
\qquad
\CalT_2\ \define\ \Tr_{Y_0Y_1}[\CalT_{012}].
\end{equation}
We are now set to define a multiple-description ISP.

\begin{definition}[One-shot multiple-description ISP]
\label{Defn:OneShotMultiLinkIS}
A $(K_1,K_2,M_1,M_2)$ one-shot multiple-description ISP for $\CalJ$ on $\rho$ consists of
\begin{itemize}
\item A single encoding instrument: For $i \in \{1,2\}$ and each $(k_1,k_2)\in[K_1]\times[K_2]$ let,
\begin{equation*}
\CalE_{k_1,k_2}: \CalL(\CalH_A)\longrightarrow
\CalL\!\bigl(\CalH_B\otimes \mathbb{C}^{M_1}\otimes \mathbb{C}^{M_2}\bigr)\mbox{ and define }\CalE_{k_i}^{(i)}: \CalL(\CalH_A)\longrightarrow \CalL\!\bigl(\CalH_B\otimes \mathbb{C}^{M_i}\bigr)\mbox{ as }\CalE_{k_i}^{(i)} \define \!\!\!\!\sum_{k_\msout{i}\in[K_\msout{i}]}\!\!\!\!\frac{\Tr_{M_\msout{i}}\!\circ \CalE_{k_1,k_2}}{K_\msout{i}}.
\end{equation*}
\item Single-link decoding CPTP maps at Bob: For each $i\in[2]$ and each $k_i\in[K_i]$, let
\begin{equation*}
\CalD_{k_i}^{(i)}:\ \CalL(\mathbb{C}^{M_i})\longrightarrow \CalL(\mathbb{C}^{\CalY_i}).
\end{equation*}
And a joint decoding CPTP map : For each $(k_1,k_2)$, let
\begin{equation*}
\CalD_{k_1,k_2}^{(12)}:\ \CalL(\mathbb{C}^{M_1}\otimes \mathbb{C}^{M_2})\longrightarrow
\CalL(\mathbb{C}^{\CalY_0}\otimes \mathbb{C}^{\CalY_1}\otimes \mathbb{C}^{\CalY_2}).
\end{equation*}
\item Finally, define the protocol induced states : For $i \in [2]$, let
\begin{equation}
\label{Eqn:InducedStatesMultiLink}
\widetilde \CalT_i
\define
\frac{1}{K_i}\sum_{k_i\in[K_i]}
\Bigl(\mathrm{id}_{BR}\otimes \CalD_{k_i}^{(i)}\Bigr)\circ
\Bigl(\mathrm{id}_R\otimes \CalE_{k_i}^{(i)}\Bigr)(\varphi_\rho),
~~
\widetilde \CalT_{012}
\define
\frac{1}{K_1K_2}\sum_{k_1\in[K_1]}\sum_{k_2\in[K_2]}
\Bigl(\mathrm{id}_{BR}\otimes \CalD_{k_1,k_2}^{(12)}\Bigr)\circ
\Bigl(\mathrm{id}_R\otimes \CalE_{k_1,k_2}\Bigr)(\varphi_\rho).
\end{equation}
\end{itemize}
\end{definition}

\begin{definition}[Simulation criterion]
\label{Defn:MultiLinkSimulationCriterion}
For $\eta>0$, we say that $\CalJ$ can be $\eta$-simulated on $\rho$ in the multiple-description sense with communication rates $(R_1,R_2)$ and common-randomness rates $(C_1,C_2)$ if there exists a $(K_1,K_2,M_1,M_2)$ one-shot multiple-description ISP with
\[
\log K_i\le C_i,~~ \log M_i\le R_i,~~ i\in[2],
\qquad
\text{and satisfying}
\qquad
\|\widetilde \CalT_i-\CalT_i\|_1\le \eta,~~
\|\widetilde \CalT_{012}-\CalT_{012}\|_1\le \eta.
\]
where $(\CalT_{012},\CalT_1,\CalT_2)$ are defined in \eqref{Eqn:IdealPostInstrumentMultiLink}. We say that the action of $\CalJ$ on $\rho$ is asymptotically simulable in the multiple-description sense with communication rates $(R_1,R_2)$ and common-randomness rates $(C_1,C_2)$ if, for every $\eta>0$, there exists $N_\eta\in\mathbb N$ such that for all $n\ge N_\eta$, the action of $\CalJ^{\otimes n}$ on $\rho^{\otimes n}$ can be $\eta$-simulated in the multiple-description sense with communication rates $(nR_1,nR_2)$ and common-randomness rates $(nC_1,nC_2)$.
\end{definition}

Our main goal in regards to the multiple-description ICP is to characterize sufficient conditions that the pair $(R_{1},R_{2})$ of communication rates and $(C_{1},C_{2})$ of common-randomness rates must jointly satisfy to ensure $\eta-$simulability of $\CalJ$ on $\rho$ in the multiple-description sense. The following remark in conjunction with arguments in App.~\ref{AppSec:StochGrainingNotMoreGeneral} establishes the generality of our formulation.

\begin{remark}
\label{Rem:StochCompatibleIsSubsumed}
In our formulation at the beginning of this Sec.~\ref{Sec:MultiLinkIS}, we considered simulation of fine- and coarse-grained instruments (Defn.~\ref{Defn:FineAndCoarseGranings}). A reader might wonder whether allowing for randomized coarse-graining, as a stochastic post-processing of measurement outcomes~\cite{198503RMP_Hol,BkHolevo_2019} would result in a more general problem statement. In App.~\ref{AppSec:StochGrainingNotMoreGeneral}
\end{remark}

\subsection{Inner bound for Multiple-Description ICP in the One-Shot Setting}
\label{Sec:MLProtocol}

We now proceed towards characterizing our new one-shot inner bound. We begin by specifying our valid `test-channels'.

\begin{definition}[Multiple-description test channels]
\label{Defn:MultiLinkTestChannels}
Let $\ScrT_{\mathrm{ML}}(\CalJ)$ denote the collection of all quintuples
\[
(\doubleunderline{\CalW},\CalI,\allowbreak p_{Y_0|\doubleunderline{W}\,\underline{Y}},\allowbreak p_{Y_1|W_1},\allowbreak p_{Y_2|W_2}),
\qquad
\doubleunderline{\CalW}\define \CalW_0\times\CalW_1\times\CalW_2,
\qquad
\underline{Y}\define(Y_1,Y_2),
\]
such that:
\begin{itemize}
\item $\CalI:\CalL(\CalH_A)\to \CalL(\CalH_B\otimes \mathbb{C}^{\CalW_0}\otimes \mathbb{C}^{\CalW_1}\otimes \mathbb{C}^{\CalW_2})$
is an instrument specified by CP TNI maps
$\{\CalI_{\doubleunderline{w}}:\CalL(\CalH_A)\to\CalL(\CalH_B)\}_{\doubleunderline{w}\in\doubleunderline{\CalW}}$,
\item $p_{Y_1|W_1}$, $p_{Y_2|W_2}$, and $p_{Y_0|\doubleunderline{W}\,\underline{Y}}$ are stochastic matrices such that,
for every $(y_0,y_1,y_2)\in\CalY_0\times\CalY_1\times\CalY_2$,
\begin{equation*}
\CalJ_{y_0,y_1,y_2}
=
\sum_{\doubleunderline{w}=(w_0,w_1,w_2)\in\doubleunderline{\CalW}}
p_{Y_1|W_1}(y_1|w_1)\,
p_{Y_2|W_2}(y_2|w_2)\,
p_{Y_0|\doubleunderline{W}\,\underline{Y}}(y_0|\doubleunderline{w},y_1,y_2)\,
\CalI_{\doubleunderline{w}}.
\end{equation*}
\end{itemize}
For a quintuple $(\doubleunderline{\CalW},\CalI,\allowbreak p_{Y_0|\doubleunderline{W}\,\underline{Y}},\allowbreak p_{Y_1|W_1},\allowbreak p_{Y_2|W_2})\in\ScrT_{\mathrm{ML}}(\CalJ)$ define the cq-states
\begin{eqnarray}
\label{Eqn:MLStatesSigmaOmega}
&&
\omega_0 = \omega^{RB\doubleunderline{W}}
\define
(\mathrm{id}_R\otimes \CalI)(\varphi_\rho),
\qquad
\omega_1 = \omega^{RB\doubleunderline{W}Y_0Y_1Y_2}
\define
\Bigl(\mathrm{id}_{RB}\otimes \olinescre^{Y_0|\doubleunderline{W} \, \underline{Y}}_{p}\Bigr)
\bigl(\omega^{RB\doubleunderline{W}Y_1Y_2}\bigr),
\\
&&
\omega^{RB\doubleunderline{W}Y_1Y_2}
\define
\Bigl(\mathrm{id}_{RBW_0}\otimes \olinescre^{Y_1|W_1}_{p}\otimes \olinescre^{Y_2|W_2}_{p}\Bigr)
\bigl(\omega^{RB\doubleunderline{W}}\bigr),
\nonumber
\end{eqnarray}
where the maps $\olinescre_p$ are those defined in \eqref{Eqn:ClassicalStochasticMaps}; in particular, they append the corresponding $Y$ register while retaining the conditioning classical registers.
\end{definition}

\begin{definition}[One-shot Inner Bound]
\label{Defn:MLPreFMEInfoQuantities}
Fix
$(\doubleunderline{\CalW},\CalI,
p_{Y_0|\doubleunderline{W}\,\underline{Y}},
p_{Y_1|W_1},p_{Y_2|W_2})
\in\ScrT_{\mathrm{ML}}(\CalJ)$
and $0<\varepsilon<1$, and let
$\varepsilon_5\define 2\log(1/\varepsilon)$.
Associate with this test channel the set
$\widehat{\ScrA}^{\mathrm{ML}}_{\varepsilon}
(\doubleunderline{\CalW},\CalI,
p_{Y_0|\doubleunderline{W}\,\underline{Y}},
p_{Y_1|W_1},p_{Y_2|W_2})$
of all quadruples
$(R_1,R_2,C_1,C_2)\in\mathbb{R}_{\geq0}^4$
for which there exist
$\bar R_i,\bar C_i,R_{0,i},C_{0,i}\geq0$, $i\in\{1,2\}$,
such that \begin{eqnarray*}
    \begin{array}{ccccc}
    R_i=\bar R_i+R_{0,i}: i \in [2],&\bar R_1>\mathsf{a}_1^\varepsilon  ,&\bar R_1+\bar R_2>\mathsf{a}_{12}^\varepsilon   ,& R_{0,1}+R_{0,2}>\mathsf{u}^\varepsilon,&R_{0,1}+R_{0,2}+C_{0,1}+C_{0,2}>\mathsf{v}^\varepsilon\\
    C_i=\bar C_i+C_{0,i} : i \in [2],&\bar R_2>\mathsf{a}_2^\varepsilon  ,& \bar R_1+\bar C_1>\mathsf{b}_1^\varepsilon ,& \bar R_2+\bar C_2>\mathsf{b}_2^\varepsilon,& \bar R_1+\bar R_2+\bar C_1+\bar C_2>\mathsf{b}_{12}^\varepsilon
        \end{array},
\end{eqnarray*}
\begin{equation}
\label{Eqn:MLBep}
\mbox{where }~~
\begin{gathered}
\mathsf{a}_1^{\varepsilon}\define I_2^{\varepsilon}(W_1:R)_{\omega_0}+\varepsilon_5,
\qquad
\mathsf{a}_2^{\varepsilon}\define I_2^{\varepsilon}(W_2:R)_{\omega_0}+\varepsilon_5,
\qquad
\mathsf{u}^{\varepsilon}\define I_{2,\mathrm{wc}}^{\varepsilon,\varepsilon}
(W_0:R\,|\,\underline{W})_{\omega_0}+\varepsilon_5,
\\
\mathsf{b}_1^{\varepsilon}\define I_2^{\varepsilon}(W_1:\doubleunderline{Y}BR)_{\omega_1}+\varepsilon_5,
\qquad
\mathsf{b}_2^{\varepsilon}\define I_2^{\varepsilon}(W_2:\doubleunderline{Y}BR)_{\omega_1}+\varepsilon_5,
\qquad
\mathsf{v}^{\varepsilon}\define I_{2,\mathrm{wc}}^{\varepsilon,\varepsilon}
(W_0:\doubleunderline{Y}BR\,|\,\underline{W})_{\omega_1}+\varepsilon_5,
\\
\mathsf{a}_{12}^{\varepsilon}\define
D_{2}^{\varepsilon}\!\left(
\omega^{R\underline{W}}
\,\middle\|\,
\omega^{W_1}\!\otimes\!\omega^{W_2}\!\otimes\!\omega^R
\right)_{\omega_0}+\varepsilon_5,
\qquad
\mathsf{b}_{12}^{\varepsilon}\define
D_{2}^{\varepsilon}\!\left(
\omega^{RB\underline{W}\doubleunderline{Y}}
\,\middle\|\,
\omega^{W_1}\!\otimes\!\omega^{W_2}\!\otimes\!\omega^{\doubleunderline{Y}BR}
\right)_{\omega_1}+\varepsilon_5.
\end{gathered}
\end{equation}
\end{definition}
In the following theorem, we shall prove that any quadruple $(R_1,R_2,C_1,C_2) \in \widehat{\ScrA}^{\mathrm{ML}}_{\varepsilon}
(\doubleunderline{\CalW},\CalI,
p_{Y_0|\doubleunderline{W}\,\underline{Y}},
p_{Y_1|W_1},p_{Y_2|W_2})$ is achievable. To assist a reader's intuition in regards to the above characterization, we note the quantities
$\mathsf{a}_1^{\varepsilon}$,
$\mathsf{a}_2^{\varepsilon}$, and
$\mathsf{a}_{12}^{\varepsilon}$
correspond to covering of the link-specific descriptions
$W_1,W_2$, while $\mathsf{u}^{\varepsilon}$ corresponds to covering
of the central refinement $W_0$. The quantities
$\mathsf{b}_1^{\varepsilon}$,
$\mathsf{b}_2^{\varepsilon}$,
$\mathsf{b}_{12}^{\varepsilon}$, and
$\mathsf{v}^{\varepsilon}$ are the analogous quantities arising when
the full post-instrument state including both quantum and classical components is reproduced.

\begin{theorem}[One-shot achievability]
\label{Thm:OneShotMLIS}
For $0<\eta<2$, the action of $\CalJ$ on $\rho$ can be $\eta$-simulated in the multiple-description sense with communication rates $(R_1,R_2)$ and common-randomness rates $(C_1,C_2)$ if there exists a
$(\doubleunderline{\CalW},\CalI,\allowbreak p_{Y_0|\doubleunderline{W}\,\underline{Y}},\allowbreak p_{Y_1|W_1},\allowbreak p_{Y_2|W_2})\in\ScrT_{\mathrm{ML}}(\CalJ)$
such that $(R_1,R_2,C_1,C_2)\in
\widehat{\ScrA}^{\mathrm{ML}}_{\widetilde\eta}(\doubleunderline{\CalW},\CalI,\allowbreak p_{Y_0|\doubleunderline{W}\,\underline{Y}},\allowbreak p_{Y_1|W_1},\allowbreak p_{Y_2|W_2}),~
\widetilde\eta\ \define\ \frac{\eta^2}{32^2}$.

\end{theorem}

\begin{remark}
\label{Rem:MLComparisonPriorOneShot} {How does the above inner bound compare with that characterized in \cite{202206ISIT_ChaPadSen}? It is important to note that the setting of \cite{202206ISIT_ChaPadSen} includes QSI, which is not included in the present formulation, while the present model includes the additional central outcome $Y_0$. Apart from this QSI component, which can be handled via an additional binning, the asymptotic generalization of the above inner bound  in Thm.~\ref{Thm:AsymptoticMLIS} recovers the corresponding multiple-description instrument simulation inner bound in the present, more general three-outcome formulation.}
\end{remark}

\begin{proof} 
We recall the network scenario under consideration described after Defn.~\ref{Defn:FineAndCoarseGranings}. Let us begin with an outline of the proof strategy. We first introduce a temporary three-description architecture in which $W_0$ is a virtual central description, whose communication and common-randomness resources will ultimately be redistributed across the two physical links. We construct and analyze first the joint encoder corresponding to the activation pattern in which both links are available. The analysis then proceeds through a sequence of add-and-subtract decompositions and applications of the triangle inequality. By introducing suitable {proxy and intermediate states}, these decompositions isolate the covering constraints associated with the two link-specific descriptions $W_1,W_2$ and the central refinement $W_0$. Finally, after redistributing the resources associated with $W_0$, we verify that the resulting construction satisfies the single-link simulation requirements and the consistency constraints. 

Choose a generic test channel
$(\doubleunderline{\CalW},\CalI,\allowbreak p_{Y_0|\doubleunderline{W}\,\underline{Y}},\allowbreak p_{Y_1|W_1},\allowbreak p_{Y_2|W_2})\in\ScrT_{\mathrm{ML}}(\CalJ)$ and fix the same throughout this proof. Choose nonnegative rates
$\bar R_i,\bar C_i,R_{0,i},C_{0,i}$, $i\in\{1,2\}$, witnessing membership in
$\widehat{\ScrA}^{\mathrm{ML}}_{\varepsilon}$.
Set $R_0\define R_{0,1}+R_{0,2}$ and $C_0\define C_{0,1}+C_{0,2}$. For the achievability proof, we temporarily allow this additional resource pair $(K_0,M_0)$,
which will later be redistributed across the two original links.
(Throughout this subsection we keep the notation $C_0=\log K_0$ and $R_0=\log M_0$.) We begin by first specifying our encoding instrument.

\med\textbf{Design of Encoding Instrument}: The design of our encoding instrument $\CalE_{\doubleunderline{k}}^{(012)}$ will be based on three codes, one built each on $\CalW_{0},\CalW_{1}$ and $\CalW_{2}$. Towards specifying the codes and the encoding instrument, let us identify a Kraus representation of $\CalI$.

\med\noindent\textit{\underline{Kraus representation of $\CalI$}}:
Fix a Kraus representation
\begin{eqnarray}
\label{Eqn:MLQuantInstrumentI}
\CalI(\rho)
=
\Tr_E\!\left[
\sum_{\doubleunderline{w}=(w_0,w_1,w_2)\in\doubleunderline{\CalW}}
V_{\doubleunderline{w}}\, \rho\, V_{\doubleunderline{w}}^\dagger
\otimes \ketbra{w_0}_{W_0}\otimes \ketbra{w_1}_{W_1}\otimes \ketbra{w_2}_{W_2}
\right],
\nonumber
\end{eqnarray}
where the operators $V_{\doubleunderline{w}}:\CalH_A\to \CalH_B\otimes \CalH_E$ satisfy
$\sum_{\doubleunderline{w}} V_{\doubleunderline{w}}^\dagger V_{\doubleunderline{w}}=I_A$. For the input state $\rho$ and each $\doubleunderline{w}$, define
\begin{equation*}
p_{\doubleunderline{W}}(\doubleunderline{w})
\define
\Tr[\rho\,\mu_{\doubleunderline{w}}],
\qquad
p_{W_0|\underline W}(w_0|\underline w)
\define
\frac{p_{\doubleunderline W}(w_0,\underline w)}
{p_{\underline W}(\underline w)},
\qquad
\mu_{\doubleunderline{w}}\ \define\ V_{\doubleunderline{w}}^\dagger V_{\doubleunderline{w}}\ \in\ \CalP(\CalH_A),
\end{equation*}
where $p_{W_1}$, $p_{W_2}$, and $p_{\underline W}$ denote the
corresponding marginals of $p_{\doubleunderline W}$. The conditional
distribution $p_{W_0|\underline W}$ is defined by the above ratio on
the support of $p_{\underline W}$ and extended arbitrarily outside
this support, with full support on $\CalW_0$.

\medskip
\noindent\textit{\underline{Three Codes, their structure and Randomization}}: Let $K_0=2^{C_0}$ and $M_0=2^{R_0}$, and let $K_i=2^{\bar C_i}$ and $M_i=2^{\bar R_i}$ for $i\in\{1,2\}$.
Generate two independent random codebooks
\begin{equation*}
\mathcal{C}_1=\bigl\{w_1(k_1,m_1):\ k_1\in[K_1],\ m_1\in[M_1]\bigr\},
\qquad
\mathcal{C}_2=\bigl\{w_2(k_2,m_2):\ k_2\in[K_2],\ m_2\in[M_2]\bigr\},
\end{equation*}
by drawing all entries i.i.d. according to $p_{W_1}$ and $p_{W_2}$, respectively.
Conditioned on $(\mathcal{C}_1,\mathcal{C}_2)$, generate a third (conditional) codebook
\begin{equation*}
\mathcal{C}_0=\bigl\{w_0(\doubleunderline{k},\doubleunderline{m}):\ k_0\in[K_0],\ \underline{k}\in[K_1]\times[K_2],\ m_0\in[M_0],\ \underline{m}\in[M_1]\times[M_2]\bigr\},
\end{equation*}
where, conditioned on $(\mathcal{C}_1,\mathcal{C}_2)$, the entries are generated independently and, for every $(\doubleunderline{k},\doubleunderline{m})$,
\begin{eqnarray}
\label{Eqn:MDICPWoCodebookPicking}
w_0(\doubleunderline{k},\doubleunderline{m})
\sim
p_{W_0|\underline{W}}\!\Bigl(\cdot\ \big|\ w_1(k_1,m_1),w_2(k_2,m_2)\Bigr).
\end{eqnarray}
Set $\mathcal{C}\define(\mathcal{C}_0,\mathcal{C}_1,\mathcal{C}_2)$ and write $\mathbb{E}_{\mathcal{C}}[\cdot]$
for expectation over $\mathcal{C}$.

\medskip
\noindent\textit{{\underline{Encoding instrument} $\CalE_{\ulinek}^{012}$}}: For each $(\doubleunderline{k},\doubleunderline{m})$, define
\begin{equation*}
V_{\doubleunderline{k},\doubleunderline{m}}
\ \define\
V_{\,w_0(\doubleunderline{k},\doubleunderline{m}),\,w_1(k_1,m_1),\,w_2(k_2,m_2)},
\qquad
\mu_{\doubleunderline{k},\doubleunderline{m}}
\ \define\
\mu_{\,w_0(\doubleunderline{k},\doubleunderline{m}),\,w_1(k_1,m_1),\,w_2(k_2,m_2)}.
\end{equation*}
Define the (random) operators
\begin{equation}
\label{Eqn:MLDistLikelihoodPOVMdefn_012}
\begin{aligned}
S_{\doubleunderline{k}}
&\define
\sum_{m_0=1}^{M_0}\sum_{m_1=1}^{M_1}\sum_{m_2=1}^{M_2}
\frac{\sqrt{\rho}\ \mu_{\doubleunderline{k},\doubleunderline{m}}\ \sqrt{\rho}}
{M_0M_1M_2\,
p_{W_1}(w_1(k_1,m_1))\,
p_{W_2}(w_2(k_2,m_2))\,
p_{W_0|\underline{W}}\!\Bigl(w_0(\doubleunderline{k},\doubleunderline{m})\ \big|\ w_1(k_1,m_1),w_2(k_2,m_2)\Bigr)},
\\
z_{\doubleunderline{k}}
&\define
\Pi_{\supp(S_{\doubleunderline{k}})},
\qquad
\pi_{\doubleunderline{k}}\ \define\ I_A-z_{\doubleunderline{k}}.
\end{aligned}
\end{equation}
For $\doubleunderline{m}\in[M_0]\times[M_1]\times[M_2]$, define
\begin{equation}
\label{Eqn:MLThetaDefn_012}
\theta^{\doubleunderline{m}}_{\doubleunderline{k}}
\define
\frac{
V_{\doubleunderline{k},\doubleunderline{m}}\ \sqrt{\rho}\ S_{\doubleunderline{k}}^{-1/2}}
{\sqrt{M_0M_1M_2\,
p_{W_1}(w_1(k_1,m_1))\,
p_{W_2}(w_2(k_2,m_2))\,
p_{W_0|\underline{W}}\!\Bigl(w_0(\doubleunderline{k},\doubleunderline{m})\ \big|\ w_1(k_1,m_1),w_2(k_2,m_2)\Bigr)}}.
\end{equation}
Since $\pi_{\doubleunderline{k}}$ is a projector, fix an isometry $Q:\CalH_A\to \CalH_B\otimes \CalH_E$ and define the $0$-th operator
\begin{equation*}
\theta^0_{\doubleunderline{k}}\ \define\ Q\,\pi_{\doubleunderline{k}},
\qquad\text{so that}\qquad
(\theta^0_{\doubleunderline{k}})^\dagger\theta^0_{\doubleunderline{k}}=\pi_{\doubleunderline{k}},
\qquad\text{and}\qquad
(\theta^0_{\doubleunderline{k}})^\dagger\theta^0_{\doubleunderline{k}}
+
\sum_{\doubleunderline{m}}(\theta^{\doubleunderline{m}}_{\doubleunderline{k}})^\dagger\theta^{\doubleunderline{m}}_{\doubleunderline{k}}
= I_A,
\end{equation*}
hence these operators define an instrument. The $0$-th operator is kept as an internal Kraus branch, but it is not assigned an additional communication symbol. Instead, we assign it the message $\doubleunderline{m}^{\star}\define(1,1,1)$. Thus the classical message alphabet remains exactly $[M_0]\times[M_1]\times[M_2]$. We define Alice's joint encoding instrument (for shared randomness $\doubleunderline{k}$) as
\begin{equation}
\label{Eqn:MLJointEncodingInstrument012}
\CalE_{\doubleunderline{k}}^{(012)}(S)=\Tr_E\!\Bigg[\sum_{\doubleunderline{m}}\theta^{\doubleunderline{m}}_{\doubleunderline{k}}\,S\,(\theta^{\doubleunderline{m}}_{\doubleunderline{k}})^\dagger\otimes\ketbra{m_0}_{M_0}\otimes\ketbra{m_1}_{M_1}\otimes\ketbra{m_2}_{M_2}+\theta^0_{\doubleunderline{k}}\,S\,(\theta^0_{\doubleunderline{k}})^\dagger\otimes\ketbra{1}_{M_0}\otimes\ketbra{1}_{M_1}\otimes\ketbra{1}_{M_2}\Bigg].
\end{equation}
\begin{remark}
\label{rem:MDICPMeasureTransformIsInvisible}
Our definition of $S_{\dulinek}$ as characterized in \eqref{Eqn:MLDistLikelihoodPOVMdefn_012} suppresses the presence of a \textit{measure transformation} factor
\begin{eqnarray}
\label{Eqn:MLLikelihoodRatio}
\mt( \ulinew(\ulinek,\ulinem))\define
\frac{p_{\underline W}(\underline w)}{p_{W_1}(w_1)p_{W_2}(w_2)}.
\end{eqnarray}
Indeed, note that
\begin{eqnarray*}
 S_{\doubleunderline{k}}
\define
\sum_{m_0=1}^{M_0}\sum_{m_1=1}^{M_1}\sum_{m_2=1}^{M_2}
\frac{\mt( \ulinew(\ulinek,\ulinem))\sqrt{\rho}\ \mu_{\doubleunderline{k},\doubleunderline{m}}\ \sqrt{\rho}}
{M_0M_1M_2\,
p_{W_{1}W_{2}W_{0}}(w_1(k_1,m_1),w_{2}(k_{2},m_{2}),w_0(\doubleunderline{k},\doubleunderline{m}))}.
\end{eqnarray*}
This measure transformation factor highlights the mismatch between the chosen single-letter test channel and the distribution wrt which the two $W_{1}-$, $W_{2}-$codes are picked. This indeed also results in a bump of the sum rate by $I(W_{1}:W_{2})$ (see Rem.~\ref{Rem:MLICPBumpInSumRate}). As the reader will note in our proof of Thm.~\ref{Thm:DISCosetCdsInnBound}, we explicitly state and carry this measure transformation factor throughout the proof without suppressing it. See Rem.~\ref{Rem:DISCosetCdsMeasureTransformTerm}. However, in this proof we suppress the same and employ it only when needed at the end in analyzing the final term $\|\CalT_{012}^\star-\CalT_{012}\|_1$ through \eqref{Eqn:MLZetaDef} to \eqref{Eqn:MLBipCoveringRatesSecondTerm}.
\end{remark}

\medskip
\noindent\textbf{Decoding and stochastic post-processing.}
Upon receiving $m_1$ and knowing $k_1$, Bob reconstructs $\widehat w_1=w_1(k_1,m_1)$ and outputs
$\widehat y_1\sim p_{Y_1|W_1}(\cdot|\widehat w_1)$.
Upon receiving $m_2$ and knowing $k_2$, Bob reconstructs $\widehat w_2=w_2(k_2,m_2)$ and outputs
$\widehat y_2\sim p_{Y_2|W_2}(\cdot|\widehat w_2)$.
In the temporary three-description protocol, Bob reconstructs $\widehat w_0=w_0(\doubleunderline{k},\doubleunderline{m})$ and outputs
\[
\widehat y_0\sim p_{Y_0|\doubleunderline{W}\,\underline{Y}}(\cdot|\widehat w_0,\widehat w_1,\widehat w_2,\widehat y_1,\widehat y_2),
\qquad
\text{with }(\widehat y_1,\widehat y_2)\ \text{as above.}
\]
If the internal $0$-th Kraus branch occurs, the communicated message is $\doubleunderline{m}^{\star}=(1,1,1)$ and Bob applies the same decoding rule as for the ordinary message $\doubleunderline{m}^{\star}$. This branch vanishes identically on the proxy input $\varphi_{S_{\doubleunderline{k}}}$, so the proxy-state and soft-covering calculations below are unchanged.

\medskip
\noindent\textbf{Analysis}: On using the above designed encoding instrument $\CalE_{\ulinek}^{(012)}$ and the decoder CPTP map $\CalD_{\ulinek}^{(012)}$, the post-instrument state induced by the protocol is
\begin{equation*}
\widetilde \CalT_{012}
\define
\frac{1}{K_0K_1K_2}\sum_{\doubleunderline{k}}
\Bigl(\mathrm{id}_{BR}\otimes \CalD_{\doubleunderline{k}}\Bigr)\circ
\Bigl(\mathrm{id}_R\otimes \CalE_{\doubleunderline{k}}\Bigr)(\varphi_\rho), \mbox{ and }\CalT_{012}\ \define\ (\mathrm{id}_R\otimes \CalJ)(\varphi_\rho)
\end{equation*}
is the ideal (or target) post-instrument state obtained through action of the original instrument. Recalling \eqref{Eqn:IdealPostInstrumentMultiLink} and \eqref{Eqn:InducedStatesMultiLink}, our goal is to bound on the above $\mathbb{E}_{\mathcal{C}}\left[\norm{\CalT_{012}- \widetilde\CalT_{012}}_{1}\right]$, $\mathbb{E}_{\mathcal{C}}\left[\norm{\CalT_{i}- \widetilde\CalT_{i}}_{1}\right]$ for $i \in [2]$. As will be evident at the end, an upper bound on the first error will imply an upper bound on the latter two error terms. We therefore focus on $\mathbb{E}_{\mathcal{C}}\left[\norm{\CalT_{012}- \widetilde\CalT_{012}}_{1}\right]$ through much of the following analysis, save for the end. Essentially, this corresponds to both links being active and hence Bob must reconstruct the triple $(Y_0,Y_1,Y_2)$.
Throughout this analysis we work with the temporary additional resource pair $(K_0,M_0)$ introduced above, and we postpone the elimination and redistribution of $(R_0,C_0)$ across links~$1$ and~$2$ to the end. For brevity, write $\CalE_{\doubleunderline{k}}\define \CalE_{\doubleunderline{k}}^{(012)},
\CalD_{\doubleunderline{k}}\define \CalD_{\doubleunderline{k}}^{(012)},$ where $\CalE_{\doubleunderline{k}}^{(012)}$ is defined in \eqref{Eqn:MLJointEncodingInstrument012},
and $\CalD_{\doubleunderline{k}}^{(012)}$ denotes the corresponding joint decoding map that outputs
$Y_0Y_1Y_2$ by reconstructing the codewords and applying the stochastic maps
$p_{Y_0|\doubleunderline{W}\,\underline{Y}}$, $p_{Y_1|W_1}$, $p_{Y_2|W_2}$.

\begin{remark}
    \label{Rem:MDICPProxyStates}
    Referring to $\theta_{\dulinek}^{\dulinem}$ in \eqref{Eqn:MLThetaDefn_012}, we note the presence of inverted random operator $S_{\dulinek}^{-\frac{1}{2}}$. Indeed $S_{\dulinek}$ is a random operator as evident from \eqref{Eqn:MLDistLikelihoodPOVMdefn_012}. Now, from the definition of $\CalE_{\dulinek}$ it is evident that the random operator $\widetilde \CalT_{012}$ has inverted random operators $S_{\dulinek}^{-\frac{1}{2}}$. Proving concentration of inverted random variable, let alone operators, is very involved and requires strong hypothesis. The novelty of our approach is operate $\CalE_{\dulinek}$ on a \textit{\underline{proxy state}} instead of the original state $\varphi_{\rho}$. This enables us neutralize these inverted random operators. We refer to the result of operating the simulation protocol  $\CalD_{\dulinek} \cdot \CalE_{\dulinek} $ on the proxy state as \underline{intermediate states}.
\end{remark}

\medskip
\noindent\textit{\underline{Proxy States and Intermediate Terms}}: Taking a cue from \cite{2026MMTIT_Pad}, we let $S_{\dulinek}$, and specifically, its \textit{canonical} purification $\varphi_{S_{\dulinek}}$ be the \textit{proxy state} on which we apply the simulation protocol to get the \textit{intermediate state}
\begin{equation*}
\CalT_{012}^\star
\define
\frac{1}{K_0K_1K_2}\sum_{\doubleunderline{k}}
\Bigl(\mathrm{id}_{BR}\otimes \CalD_{\doubleunderline{k}}\Bigr)\circ
\Bigl(\mathrm{id}_R\otimes \CalE_{\doubleunderline{k}}\Bigr)(\varphi_{S_{\doubleunderline{k}}}).
\end{equation*}
By the triangle inequality,
\begin{equation}
\label{Eqn:MLTri012}
\bigl\|\widetilde \CalT_{012}-\CalT_{012}\bigr\|_1
\le
\bigl\|\widetilde \CalT_{012}-\CalT_{012}^\star\bigr\|_1
+
\bigl\|\CalT_{012}^\star-\CalT_{012}\bigr\|_1.
\end{equation}

\medskip
\noindent\textit{\underline{First term: bounding} $\|\widetilde \CalT_{012}-\CalT_{012}^\star\|_1$} (Communication-only Constraints):
Since $(\mathrm{id}_{BR}\otimes \CalD_{\doubleunderline{k}})\circ(\mathrm{id}_R\otimes \CalE_{\doubleunderline{k}})$ is CPTP, trace distance is contractive, we have
\begin{equation}
\label{Eqn:MLTerm1Contract012}
\bigl\|\widetilde \CalT_{012}-\CalT_{012}^\star\bigr\|_1
\le
\frac{1}{K_0K_1K_2}\sum_{\doubleunderline{k}}
\bigl\|\varphi_{\rho}-\varphi_{S_{\doubleunderline{k}}}\bigr\|_1 .
\end{equation}

\begin{remark}
Note the the RHS in \eqref{Eqn:MLTerm1Contract012} is trace distance between purifications of $S_{\dulinek}$ and $\rho$, not the states themselves. This adds complexity to the analysis. Here, our specific choice of \underline{canonical} purification aids us in the following sequence of steps.
\end{remark}
Combining this with Lemma~\ref{lem:canonical-purif-stability} gives
\begin{equation*}
\bigl\|\widetilde \CalT_{012}-\CalT_{012}^\star\bigr\|_1
\le
\frac{1}{K_0K_1K_2}\sum_{\doubleunderline{k}}
\Bigl(1+\sqrt{\Tr S_{\doubleunderline{k}}}\Bigr)
\sqrt{\bigl\|\rho-S_{\doubleunderline{k}}\bigr\|_1}.
\end{equation*}
For every fixed $\doubleunderline{k}$, the codebook construction implies
$\mathbb E_{\mathcal C}[S_{\doubleunderline{k}}]=\rho$ and hence
$\mathbb E_{\mathcal C}[\Tr S_{\doubleunderline{k}}]=1$. Therefore, by Cauchy--Schwarz and Jensen's inequality,
\begin{equation}
\label{Eqn:MLTerm1PurifFinal012}
\mathbb E_{\mathcal C}\!\left[
\bigl\|\widetilde \CalT_{012}-\CalT_{012}^\star\bigr\|_1
\right]
\le
2\left(
\mathbb E_{\mathcal C}\!\left[
\frac{1}{K_0K_1K_2}\sum_{\doubleunderline{k}}
\bigl\|\rho-S_{\doubleunderline{k}}\bigr\|_1
\right]
\right)^{\!1/2}.
\end{equation}

\begin{remark}
\label{Rem:MDICPNewApproachPurificationBounding}
In \cite{2026MMTIT_Pad}, a different sequence of steps was identified to derive a bound on $\norm{\varphi_{S_{\dulinek}} - \varphi_{\rho}}_{1}$. However, that approach gives a bound in terms of $\sqrt[4]{\norm{{S_{\dulinek}} - {\rho}}_{1}}$. Note that, our new approach encapsulated in Lemma \ref{lem:canonical-purif-stability} yields an improved upper bound in terms of $\sqrt{\norm{{S_{\dulinek}} - {\rho}}_{1}}$ - the second root of the norm in question. In the present construction, $\Tr(S_{\ulinek})$ is not necessarily equal to one because of the measure transformation factor (Rem.~\ref{rem:MDICPMeasureTransformIsInvisible}). Lemma \ref{lem:canonical-purif-stability} thus provides a direct bound on the RHS of \eqref{Eqn:MLTerm1Contract012} requiring only $\Expectation_{\CalC}\{\Tr(S_{\ulinek})\}=1$ without needing  $\Tr(S_{\ulinek})=1$.
\end{remark}

Our next task is to bound $\|\rho-S_{\doubleunderline{k}}\|_1$ on the above. We are led to the following.
\begin{remark}
\label{Rem:MDICPSecondPrxyStW0layer}
To bound $\|\rho-S_{\doubleunderline{k}}\|_1$ on the above, it is instinctive to employ Sen's multi-party smooth covering \cite{202504arXiv_Sen} (Prop.~\ref{Prop:SmoothMultipartiteSoftCovering}). However, the $W_{0},W_{1},W_{2}-$codebooks are not picked independently, as required by the hypothesis in Prop.~\ref{Prop:SmoothMultipartiteSoftCovering}. Specifically, $W_{0}-$codebook is picked conditionally. See \eqref{Eqn:MDICPWoCodebookPicking}. Prohibited from leveraging Prop.~\ref{Prop:SmoothMultipartiteSoftCovering}, we introduce a second proxy state - $S_{\ulinek}$ below - specifically to enable the $W_{0}-$codebook layer. Furthermore, we leverage Lemma~\ref{Lem:ConditionalSoftCovering} to bound the trace distance $\norm{S_{\ulinek} - S_{\dulinek}}_{1}$ between the two proxy states $S_{\ulinek},S_{\dulinek}$ below.
\end{remark}

\medskip
\noindent\textit{\underline{Bipartite and conditional soft covering}}: Define the marginal POVM elements and the corresponding two-description covering operator by
\begin{equation*}
\mu_{\underline{w}}\define\sum_{w_0\in\CalW_0}\mu_{w_0,\underline{w}},
\qquad
S_{\underline{k}}
\define
\sum_{m_1,m_2}
\frac{\sqrt{\rho}\,\mu_{\underline{w}(\underline{k},\underline{m})}\sqrt{\rho}}
{M_1M_2\,p_{W_1}(w_1(k_1,m_1))\,p_{W_2}(w_2(k_2,m_2))},
\end{equation*}
where $\underline{w}(\underline{k},\underline{m})=(w_1(k_1,m_1),w_2(k_2,m_2))$. Since $S_{\underline{k}}$ is independent of $k_0$,
\begin{equation}
\label{Eqn:MLSplitAverage}
\mathbb{E}_{\mathcal{C}}\!\left[\frac{1}{K_0K_1K_2}\sum_{\doubleunderline{k}}\|\rho-S_{\doubleunderline{k}}\|_1\right]
\le
\mathbb{E}_{\mathcal{C}}\!\left[\frac{1}{K_1K_2}\sum_{\underline{k}}\|\rho-S_{\underline{k}}\|_1\right]
+
\mathbb{E}_{\mathcal{C}}\!\left[\frac{1}{K_0K_1K_2}\sum_{\doubleunderline{k}}\|S_{\underline{k}}-S_{\doubleunderline{k}}\|_1\right].
\end{equation}
The first term is the standard bipartite soft-covering deviation for the independent codebooks $(\mathcal C_1,\mathcal C_2)$. Proposition~\ref{Prop:SmoothMultipartiteSoftCovering} gives
\begin{equation}
\label{Eqn:MLBipSoftCoveringBound_FirstTerm}
\mathbb{E}_{\mathcal{C}}\!\left[
\frac{1}{K_1K_2}\sum_{\underline{k}}\bigl\|\rho-S_{\underline{k}}\bigr\|_1
\right]
\le 16\varepsilon,
\end{equation}
provided that
\begin{equation*}
\bar R_1>I_2^{\varepsilon}(W_1:R)_{\omega_0}+\varepsilon_5,
\qquad
\bar R_2>I_2^{\varepsilon}(W_2:R)_{\omega_0}+\varepsilon_5,
\qquad
\bar R_1+\bar R_2>
D_{2}^{\varepsilon}\Bigl(\omega^{R\underline{W}}\,\Big\|\,
\omega^{W_{1}}\otimes \omega^{W_{2}}\otimes\omega^{R}\Bigr)_{\omega_0}+\varepsilon_5.
\end{equation*}
As a passing remark, we note that although the covering operators act on $A$, the information quantities are written on the canonical reference $R$; the two representations are related by transposition in the canonical basis, which preserves both the trace norm and $D_2^{\varepsilon}$. Returning back to the second term in \eqref{Eqn:MLSplitAverage}, whenever $p_{\doubleunderline W}(w_0,\underline w)>0$, define
\begin{equation*}
\rho^{w_0,\underline{w}}
\define
\frac{\sqrt{\rho}\,\mu_{w_0,\underline{w}}\sqrt{\rho}}
{p_{\doubleunderline W}(w_0,\underline w)},
\qquad
\overline{\rho}^{\underline{w}}
\define
\sum_{w_0}p_{W_0|\underline W}(w_0|\underline w)\rho^{w_0,\underline w}.
\end{equation*}
For zero-probability pairs the conditional state may be chosen arbitrarily. The identity $\frac{\sqrt{\rho}\,\mu_{w_0,\underline{w}}\sqrt{\rho}}
{p_{W_0|\underline W}(w_0|\underline w)}
=
p_{\underline W}(\underline w)\rho^{w_0,\underline w}$ 
shows directly that, after grouping the sum over $m_0$,
\begin{equation}
\label{Eqn:MLDifferenceFactorized}
S_{\underline{k}}-S_{\doubleunderline{k}}
=
\sum_{m_1,m_2}
\frac{p_{\underline W}(\underline{w}(\underline{k},\underline{m}))}
{M_1M_2\,p_{W_1}(w_1(k_1,m_1))\,p_{W_2}(w_2(k_2,m_2))}
\left(
\overline{\rho}^{\,\underline{w}(\underline{k},\underline{m})}
-
\frac1{M_0}\sum_{m_0}
\rho^{w_0(\doubleunderline{k},\doubleunderline{m}),\,\underline{w}(\underline{k},\underline{m})}
\right).
\end{equation}
Thus the refinement error is an average of ordinary soft-covering deviations for the conditional ensemble
$\{p_{W_0|\underline W}(\cdot|\underline w),\rho^{w_0,\underline w}\}$. For every fixed $(m_1,m_2)$, the pair $\underline w(\underline k,\underline m)$ is distributed according to $p_{W_1}p_{W_2}$; the likelihood factor in \eqref{Eqn:MLDifferenceFactorized} therefore converts this outer average into $p_{\underline W}$. Averaging over $(\mathcal C_1,\mathcal C_2)$ and then applying Lemma~\ref{Lem:ConditionalSoftCovering} to $\mathcal C_0$ gives
\begin{equation}
\label{Eqn:MLCondSoftCoveringBound}
\mathbb{E}_{\mathcal C}\!\left[
\frac{1}{K_0K_1K_2}\sum_{\doubleunderline{k}}
\bigl\|S_{\underline{k}}-S_{\doubleunderline{k}}\bigr\|_1
\right]
\le 6\varepsilon
\qquad\text{if}\qquad
R_0>I_{2,\mathrm{wc}}^{\varepsilon,\varepsilon}(W_0\!:\!R\,|\,\underline W)_{\omega_0}+\varepsilon_5.
\end{equation}
Combining \eqref{Eqn:MLSplitAverage}, \eqref{Eqn:MLBipSoftCoveringBound_FirstTerm}, and \eqref{Eqn:MLCondSoftCoveringBound}, and then using \eqref{Eqn:MLTerm1PurifFinal012}, yields
\begin{equation}
\label{Eqn:MLFullSApproximation}
\mathbb E_{\mathcal C}\!\left[
\frac{1}{K_0K_1K_2}\sum_{\doubleunderline{k}}\|\rho-S_{\doubleunderline{k}}\|_1
\right]\le22\varepsilon,
\qquad
\mathbb E_{\mathcal C}\!\left[\|\widetilde\CalT_{012}-\CalT_{012}^\star\|_1\right]
\le2\sqrt{22\varepsilon}.
\end{equation}

\medskip
\noindent\textit{\underline{Second term Bounding} $\|\CalT_{012}^\star-\CalT_{012}\|_1$} (Constraints on Communication+Common-randomness): The difficulty in this term is the random inverse $S_{\doubleunderline{k}}^{-1/2}$ appearing in each likelihood Kraus operator. The clever choice of the proxy state $S_{\dulinek}$ guarantees neutralization of this random inverse operator $S_{\doubleunderline{k}}^{-1/2}$. As we will see below, the canonical purification $\varphi_{S_{\doubleunderline{k}}}$ contains a factor $\sqrt{S_{\doubleunderline{k}}}$ on the accessible system, and
$S_{\doubleunderline{k}}^{-1/2}\sqrt{S_{\doubleunderline{k}}}=\Pi_{\supp(S_{\doubleunderline{k}})}$.
Since the relevant branch operators are supported on $\supp(S_{\doubleunderline{k}})$, this projector can then be removed; moreover the $0$-th branch vanishes because $\pi_{\doubleunderline{k}}\sqrt{S_{\doubleunderline{k}}}=0$. Therefore, let us define the conditional post-instrument states $\zeta^{BR}_{\doubleunderline w}$ through
\begin{equation}
\label{Eqn:MLZetaDef}
p_{\doubleunderline W}(\doubleunderline w)\zeta^{BR}_{\doubleunderline w}
\define
\Tr_E\!\left[
(\mathrm{id}_R\otimes V_{\doubleunderline w})\varphi_\rho
(\mathrm{id}_R\otimes V_{\doubleunderline w})^\dagger
\right].
\end{equation}
We now evaluate a regular likelihood branch on the proxy input. Using
the canonical-purification representation $\varphi_{S_{\doubleunderline{k}}}
=
\sum_{x,x'}
\ket{x}\!\bra{x'}_R
\otimes
\sqrt{S_{\doubleunderline{k}}}\ket{x}\!\bra{x'}
\sqrt{S_{\doubleunderline{k}}},$ together with the definition of
$\theta^{\doubleunderline{m}}_{\doubleunderline{k}}$, the factor
$S_{\doubleunderline{k}}^{-1/2}$ cancels against the
$\sqrt{S_{\doubleunderline{k}}}$ appearing in the proxy purification. Since each
$\sqrt{\rho}\,\mu_{\doubleunderline{k},\doubleunderline{m}}\sqrt{\rho}$
is supported on $\supp(S_{\doubleunderline{k}})$, we have
$V_{\doubleunderline{k},\doubleunderline{m}}\sqrt{\rho}\,
\Pi_{\supp(S_{\doubleunderline{k}})}
=
V_{\doubleunderline{k},\doubleunderline{m}}\sqrt{\rho}$. We obtain
\begin{equation}
\label{Eqn:MLCOSBranch1}
(\mathrm{id}_R\otimes\theta^{\doubleunderline{m}}_{\doubleunderline{k}})
\varphi_{S_{\doubleunderline{k}}}
(\mathrm{id}_R\otimes\theta^{\doubleunderline{m}}_{\doubleunderline{k}})^\dagger
=
\frac{
(\mathrm{id}_R\otimes V_{\doubleunderline{k},\doubleunderline{m}})
\varphi_\rho
(\mathrm{id}_R\otimes V_{\doubleunderline{k},\doubleunderline{m}})^\dagger
}{
M_0M_1M_2\,
p_{W_1}(w_1(k_1,m_1))p_{W_2}(w_2(k_2,m_2))
p_{W_0|\underline{W}}(w_0|\underline{w}(\underline{k},\underline{m}))
}.
\end{equation}
Tracing out the environment, this gives
\begin{equation}
\label{Eqn:MLCOSBranch}
\Tr_E\!\left[
(\mathrm{id}_R\otimes\theta^{\doubleunderline m}_{\doubleunderline k})
\varphi_{S_{\doubleunderline k}}
(\mathrm{id}_R\otimes\theta^{\doubleunderline m}_{\doubleunderline k})^\dagger
\right]
=
\frac{\mt(\underline w(\underline k,\underline m))}{M_0M_1M_2}
\zeta^{BR}_{\doubleunderline w(\doubleunderline k,\doubleunderline m)},
\end{equation}
where the \textit{measure transformation factor }$\mt(\underline w)$ is as defined in \eqref{Eqn:MLLikelihoodRatio}. Summing over the regular branches and adjoining the message registers therefore yields the inverse-free representation
\begin{equation}
\label{Eqn:MLCOSIdentity}
(\mathrm{id}_R\otimes \CalE_{\doubleunderline{k}})
(\varphi_{S_{\doubleunderline{k}}})
=
\frac{1}{M_0M_1M_2}
\sum_{\doubleunderline m}
\mt\!\left(\underline w(\underline k,\underline m)\right)
\zeta^{BR}_{\doubleunderline w(\doubleunderline k,\doubleunderline m)}
\otimes\ketbra{m_0}_{M_0}\otimes\ketbra{m_1}_{M_1}\otimes\ketbra{m_2}_{M_2}.
\end{equation}
Write $q_{\doubleunderline{Y}|\doubleunderline{W}}(y_0,y_1,y_2|w_0,w_1,w_2)
\define
p_{Y_1|W_1}(y_1|w_1)\,p_{Y_2|W_2}(y_2|w_2)\,
p_{Y_0|\doubleunderline{W}\,\underline{Y}}(y_0|w_0,w_1,w_2,y_1,y_2).$ Applying the stochastic decoder to \eqref{Eqn:MLCOSIdentity} gives
\begin{equation}
\label{Eqn:MLProxyAsSoftCovering}
\CalT_{012}^\star
=
\frac{1}{K_0K_1K_2M_0M_1M_2}
\sum_{\doubleunderline{k},\doubleunderline{m}}
\mt\!\left(\underline w(\underline k,\underline m)\right)
\sum_{y_0,y_1,y_2}
q_{\doubleunderline{Y}|\doubleunderline{W}}
\!\left(y_0,y_1,y_2\,\middle|\,\doubleunderline w(\doubleunderline k,\doubleunderline m)\right)
\zeta^{BR}_{\doubleunderline w(\doubleunderline k,\doubleunderline m)}
\otimes\ketbra{y_0}_{Y_0}\otimes\ketbra{y_1}_{Y_1}\otimes\ketbra{y_2}_{Y_2}.
\end{equation}
The ideal state has the corresponding exact-expectation form
\begin{equation}
\label{Eqn:MLIdealAsExpectation}
\CalT_{012}
=
\sum_{\doubleunderline w}
p_{\doubleunderline W}(\doubleunderline w)
\sum_{y_0,y_1,y_2}
q_{\doubleunderline{Y}|\doubleunderline{W}}(y_0,y_1,y_2|\doubleunderline w)
\zeta^{BR}_{\doubleunderline w}
\otimes\ketbra{y_0}_{Y_0}\otimes\ketbra{y_1}_{Y_1}\otimes\ketbra{y_2}_{Y_2}.
\end{equation}
\begin{remark}
 \label{Rem:MDICP:ASecondSplitInCov}
From the above, it is clear that we are now required to bound $\Expectation_{\CalC}\left[ \norm{\CalT_{012}^\star-\CalT_{012}}_{1}\right]$. Once again as before, the forms of these terms in \eqref{Eqn:MLProxyAsSoftCovering} and \eqref{Eqn:MLIdealAsExpectation} leads us instinctively to reach out for the smooth multi-party covering \cite{202504arXiv_Sen} (Prop.~\ref{Prop:SmoothMultipartiteSoftCovering}). However, just as encountered before the non-product distribution of the three random $W_{0},W_{1},W_{2}-$codes forces us to consider a new \underline{intermediate term} $\CalT_{012}^{\diamond}$ and perform a two part covering.
\end{remark}
Building from above remark, we now separate the approximation of the conditional refinement $W_0$ from that of the pair $(W_1,W_2)$. For each $\underline w=(w_1,w_2)$ define
\begin{equation}
\label{Eqn:MLBarOmegaUnderlinew}
\overline{\omega}^{BR}_{y_0,y_1,y_2}(\underline w)
\define
\sum_{w_0}
p_{W_0|\underline W}(w_0|\underline w)
\,p_{Y_0|\doubleunderline W\,\underline Y}(y_0|w_0,\underline w,y_1,y_2)
\,\zeta^{BR}_{(w_0,\underline w)}.
\end{equation}
Let $\CalT_{012}^{\diamond}$ be obtained from the COS expression for $\CalT_{012}^\star$ by replacing, for every fixed $(\underline k,\underline m)$, the empirical $W_0$-average generated by $\mathcal C_0$ with the exact conditional average in \eqref{Eqn:MLBarOmegaUnderlinew}. Explicitly,
\begin{equation}
\label{Eqn:MLDiamondDef}
\begin{aligned}
\CalT_{012}^{\diamond}
\define
\frac{1}{K_1K_2M_1M_2}
\sum_{\underline k,\underline m}\sum_{y_0,y_1,y_2}
&\mt\!\left(\underline w(\underline k,\underline m)\right)
 p_{Y_1|W_1}(y_1|w_1(k_1,m_1))
 p_{Y_2|W_2}(y_2|w_2(k_2,m_2))
\\[-0.2em]
&\times\overline{\omega}^{BR}_{y_0,y_1,y_2}
\!\left(\underline w(\underline k,\underline m)\right)
\otimes\ketbra{y_0}_{Y_0}\otimes\ketbra{y_1}_{Y_1}\otimes\ketbra{y_2}_{Y_2}.
\end{aligned}
\end{equation}
Because $p_{W_1}(w_1)p_{W_2}(w_2)\mt(\underline w)=p_{\underline W}(\underline w)$, this hybrid state is exactly the bipartite empirical approximation of the ideal state after the conditional $W_0$-average has been restored. Hence
\begin{equation}
\label{Eqn:MLSecondTermPlusMinus}
\bigl\|\CalT_{012}^\star-\CalT_{012}\bigr\|_1
\le
\bigl\|\CalT_{012}^\star-\CalT_{012}^{\diamond}\bigr\|_1
+
\bigl\|\CalT_{012}^{\diamond}-\CalT_{012}\bigr\|_1,
\end{equation}
and the two terms are respectively conditional and bipartite soft-covering errors. Applying Lemma~\ref{Lem:ConditionalSoftCovering} to the first term, now with output system $\doubleunderline YBR$, gives
\begin{equation}
\label{Eqn:T21}
\mathbb E_{\mathcal C}\!\left[\|\CalT_{012}^\star-\CalT_{012}^{\diamond}\|_1\right]
\le6\varepsilon
\qquad\text{if}\qquad
R_0+C_0>I_{2,\mathrm{wc}}^{\varepsilon,\varepsilon}(W_0\!:\!\doubleunderline YBR\,|\,\underline W)_{\omega_1}+\varepsilon_5.
\end{equation}
For the second term, Proposition~\ref{Prop:SmoothMultipartiteSoftCovering} gives
\begin{equation}
\label{Eqn:T22}
\mathbb E_{\mathcal C}\!\left[\|\CalT_{012}^{\diamond}-\CalT_{012}\|_1\right]
\le16\varepsilon,
\end{equation}
provided that
\begin{equation}
\label{Eqn:MLBipCoveringRatesSecondTerm}
\begin{gathered}
\bar R_1+\bar C_1>I_2^{\varepsilon}(W_1:\doubleunderline YBR)_{\omega_1}+\varepsilon_5,
\qquad
\bar R_2+\bar C_2>I_2^{\varepsilon}(W_2:\doubleunderline YBR)_{\omega_1}+\varepsilon_5,
\\
\bar R_1+\bar R_2+\bar C_1+\bar C_2>
D_2^{\varepsilon}\Bigl(
\omega^{RB\underline W\doubleunderline Y}
\,\Big\|\,
\omega^{W_1}\otimes\omega^{W_2}\otimes\omega^{\doubleunderline YBR}
\Bigr)_{\omega_1}+\varepsilon_5.
\end{gathered}
\end{equation}

\medskip
\medskip
\noindent\textbf{Conclusion and redistribution of the auxiliary description.}
Combining \eqref{Eqn:MLTri012}, \eqref{Eqn:MLFullSApproximation}, \eqref{Eqn:MLSecondTermPlusMinus}, \eqref{Eqn:T21}, and \eqref{Eqn:T22}, we obtain
\[
\mathbb E_{\mathcal C}\!\left[\|\widetilde\CalT_{012}-\CalT_{012}\|_1\right]
\le2\sqrt{22\varepsilon}+22\varepsilon
\le32\sqrt\varepsilon.
\]
Thus the temporary three-description protocol succeeds provided that
\begin{equation*}
\begin{array}{@{}ll@{}}
\bar R_1>I_2^{\varepsilon}(W_1:R)_{\omega_0}+\varepsilon_5,
&
\bar R_1+\bar C_1>I_2^{\varepsilon}(W_1:\doubleunderline YBR)_{\omega_1}+\varepsilon_5,
\\[0.5em]
\bar R_2>I_2^{\varepsilon}(W_2:R)_{\omega_0}+\varepsilon_5,
&
\bar R_2+\bar C_2>I_2^{\varepsilon}(W_2:\doubleunderline YBR)_{\omega_1}+\varepsilon_5,
\\[0.5em]
\bar R_1+\bar R_2>
D_2^{\varepsilon}\!\Bigl(\omega^{R\underline W}\,\Big\|\,
\omega^{W_1}\otimes\omega^{W_2}\otimes\omega^R\Bigr)_{\omega_0}+\varepsilon_5,
&
\bar R_1+\bar R_2+\bar C_1+\bar C_2>
D_2^{\varepsilon}\!\Bigl(\omega^{RB\underline W\doubleunderline Y}\,\Big\|\,
\omega^{W_1}\otimes\omega^{W_2}\otimes\omega^{\doubleunderline YBR}\Bigr)_{\omega_1}+\varepsilon_5,
\\[0.5em]
R_0>I_{2,\mathrm{wc}}^{\varepsilon,\varepsilon}(W_0\!:\!R\,|\,\underline W)_{\omega_0}+\varepsilon_5,
&
R_0+C_0>I_{2,\mathrm{wc}}^{\varepsilon,\varepsilon}(W_0\!:\!\doubleunderline YBR\,|\,\underline W)_{\omega_1}+\varepsilon_5.
\end{array}
\end{equation*}
It remains to absorb the virtual resources $(R_0,C_0)$ into the two physical links. Choose factorizations
$M_0=M_{0,1}M_{0,2}$ and $K_0=K_{0,1}K_{0,2}$, identify the corresponding index sets through fixed bijections, and send the first fragments on link~$1$ and the second fragments on link~$2$. The resulting physical rates are
\begin{equation*}
R_i=\bar R_i+\log M_{0,i},
\qquad
C_i=\bar C_i+\log K_{0,i},
\qquad i\in\{1,2\}.
\end{equation*}
When both links are active, Bob recovers the complete auxiliary indices $(m_0,k_0)$ and hence implements exactly the temporary joint decoder analyzed above. When only link~$i$ is active, he discards the auxiliary fragment on that link and reconstructs $Y_i$ solely from $(k_i,m_i)$ through $p_{Y_i|W_i}$.
It remains only to verify the single-link requirements in Definition~\ref{Defn:MultiLinkSimulationCriterion}.

By the preceding construction, the joint decoder produces \(y_1\) from \((k_1,m_1)\) through \(p_{Y_1|W_1}\), \(y_2\) from \((k_2,m_2)\) through
\(p_{Y_2|W_2}\) and ultimately \(y_0\) from \((k_1,k_2,m_1,m_2,y_1,y_2)\) through $p_{Y_0|\doubleunderline{W}\,\underline{Y}}$. It follows that the single-link induced
states are precisely the corresponding marginals of the joint induced state:
\begin{equation*}
\Tr_{Y_0Y_2}\widetilde{\CalT}_{012}
=
\widetilde{\CalT}_{1},
\qquad
\Tr_{Y_0Y_1}\widetilde{\CalT}_{012}
=
\widetilde{\CalT}_{2},
\end{equation*}
where the states \(\widetilde{\CalT}_{012},\widetilde{\CalT}_1,\widetilde{\CalT}_2\) are defined in~\eqref{Eqn:InducedStatesMultiLink}. Similarly, by \eqref{Eqn:IdealPostInstrumentMultiLink},
\[
\Tr_{Y_0Y_2}\CalT_{012}=\CalT_1,
\qquad
\Tr_{Y_0Y_1}\CalT_{012}=\CalT_2.
\]
Since the partial trace is CPTP and hence contractive in trace norm, the joint estimate
\[
\|\widetilde{\CalT}_{012}-\CalT_{012}\|_1\le \eta,
\qquad
\|\widetilde{\CalT}_{1}-\CalT_{1}\|_1\le \eta,
\qquad
\|\widetilde{\CalT}_{2}-\CalT_{2}\|_1\le \eta
\]
follows immediately.
Thus all three requirements in Definition \ref{Defn:MultiLinkSimulationCriterion} are satisfied.
Choose $\varepsilon=\widetilde\eta=\eta^2/32^2$. Since the expected joint error is then at most $\eta$, at least one deterministic realization of the codebooks satisfies the joint and both single-link simulation criteria. This proves the theorem.

\end{proof}

\subsection{Inner bound in the Asymptotic Regime : Natural Generalization of the One-shot Inner bound}
\label{Subsec:MLAsymptoticAchievability}
Here we characterize an inner bound in the asymptotic regime that is a natural generalization of the one characterized in Thm.~\ref{Thm:OneShotMLIS} for the one-shot regime. In the following subsection, we enlarge this natural generalization inner bound (Thm.~\ref{Thm:AsymptoticMLIS}) characterized herein using a common base layer code as done by Zhang and Berger in \cite{198707TIT_ZhaBer}.
\begin{definition}[Asymptotic rate region]
\label{Defn:MLAsymptoticRateRegion}
Associate with
$(\doubleunderline{\CalW},\CalI,
p_{Y_0|\doubleunderline{W}\,\underline{Y}},
p_{Y_1|W_1},p_{Y_2|W_2})
\in\ScrT_{\mathrm{ML}}(\CalJ)$ the set
\\
$\ScrA^{\mathrm{ML}}_{0}
(\doubleunderline{\CalW},\CalI,
p_{Y_0|\doubleunderline{W}\,\underline{Y}},
p_{Y_1|W_1},p_{Y_2|W_2})$
of all quadruples
$(R_1,R_2,C_1,C_2)\in\mathbb{R}_{\geq0}^4$
satisfying, with $Z\define\doubleunderline{Y}BR$,
\begin{equation*}
\begin{gathered}
\begin{array}{ll}
R_1~>~I(W_1:R)_{\omega_0},
&
R_1+R_2+C_1~>~
I(W_2:R)_{\omega_0}
+I(W_1:Z)_{\omega_1}
+I(W_0:R|\underline{W})_{\omega_0},
\\[0.4em]
R_2~>~I(W_2:R)_{\omega_0},
&
R_1+R_2+C_2~>~
I(W_1:R)_{\omega_0}
+I(W_2:Z)_{\omega_1}
+I(W_0:R|\underline{W})_{\omega_0},
\\[0.4em]
R_1+C_1~>~I(W_1:Z)_{\omega_1},
&
R_2+C_2~>~I(W_2:Z)_{\omega_1},
\\[0.4em]
R_1+R_2~>~
I(\doubleunderline{W}:R)_{\omega_0}
+I(W_1:W_2)_{\omega_0},
&
R_1+R_2+C_1+C_2~>~
I(\doubleunderline{W}:Z)_{\omega_1}
+I(W_1:W_2)_{\omega_1}.
\end{array}
\end{gathered}
\end{equation*}
All information quantities are computed with respect to states
$\omega_0$ and $\omega_1$ defined in
\eqref{Eqn:MLStatesSigmaOmega}.
\end{definition}

\begin{theorem}[Asymptotic achievability]
\label{Thm:AsymptoticMLIS}
The action of $\CalJ$ on $\rho$ is asymptotically simulable in the multiple-description sense with communication rates $(R_1,R_2)$ and common-randomness rates $(C_1,C_2)$ if there exists
$(\doubleunderline{\CalW},\CalI,\allowbreak p_{Y_0|\doubleunderline{W}\,\underline{Y}},\allowbreak p_{Y_1|W_1},\allowbreak p_{Y_2|W_2})\in\ScrT_{\mathrm{ML}}(\CalJ)$
such that $(R_1,R_2,C_1,C_2)\in
\ScrA^{\mathrm{ML}}_{0}(\doubleunderline{\CalW},\CalI,\allowbreak p_{Y_0|\doubleunderline{W}\,\underline{Y}},\allowbreak p_{Y_1|W_1},\allowbreak p_{Y_2|W_2})$.
\end{theorem}

\begin{proof}
For notational convenience, define
\begin{equation*}
\begin{gathered}
\mathsf{a}_1 ~\define~ I(W_1:R)_{\omega_0},
\qquad
\mathsf{a}_2~\define~ I(W_2:R)_{\omega_0},
\qquad
\mathsf{u}~\define~ I(W_0:R|\underline{W})_{\omega_0},
\\
\mathsf{b}_1~\define~ I(W_1:Z)_{\omega_1},
\qquad
\mathsf{b}_2~\define~ I(W_2:Z)_{\omega_1},
\qquad
\mathsf{v}~\define~ I(W_0:Z|\underline{W})_{\omega_1},
\\
\mathsf{a}_{12}~\define~
I(\underline{W}:R)_{\omega_0}+I(W_1:W_2)_{\omega_0},
\qquad
\mathsf{b}_{12}~\define~
I(\underline{W}:Z)_{\omega_1}+I(W_1:W_2)_{\omega_1}.
\end{gathered}
\end{equation*}
Apply Theorem~\ref{Thm:OneShotMLIS} to the product instrument $\CalJ^{\otimes n}$, the product state $\rho^{\otimes n}$,
and the $n$-fold product of the chosen multiple-description test channel. Choose $\varepsilon_n\downarrow0$ with
$n^{-1}\log(1/\varepsilon_n)\to0$, set $\varepsilon_{0,n}\define\varepsilon_n$, and define
$\eta_n\define 32\sqrt{\varepsilon_n}$. Then $\eta_n\downarrow0$ and, for all sufficiently large $n$,
$0<\eta_n<2$ and the one-shot parameter in Theorem~\ref{Thm:OneShotMLIS} satisfies
$\widetilde\eta_n=\eta_n^2/32^2=\varepsilon_n$. The standard AEP for $D_2^{\varepsilon}$ gives convergence of the
normalized one-shot quantities corresponding to $\mathsf a_1,\mathsf a_2,\mathsf a_{12},\mathsf b_1,\mathsf b_2,\mathsf b_{12}$.
For the two worst-case conditional quantities, Lemma~\ref{Lem:AEPWorstCaseConditionalI2}, applied with
$\varepsilon_{0,n}=\varepsilon_n$, gives the sufficient upper bounds
\[
\limsup_{n\to\infty}\frac1n\mathsf u_n^{\varepsilon_n}\le \mathsf u,
\qquad
\limsup_{n\to\infty}\frac1n\mathsf v_n^{\varepsilon_n}\le \mathsf v,
\]
where $\mathsf u_n^{\varepsilon_n}$ and $\mathsf v_n^{\varepsilon_n}$ denote the corresponding quantities in
\eqref{Eqn:MLBep} for the $n$-fold product test channel. Consequently, for achievability it is sufficient that there exist
nonnegative $\bar R_i,\bar C_i,R_{0,i},C_{0,i}$, $i\in\{1,2\}$, satisfying the asymptotic pre-elimination constraints
\begin{equation}
\label{Eqn:MLAsymptoticPreFME}
\begin{array}{lll}
R_i=\bar R_i+R_{0,i},\quad C_i=\bar C_i+C_{0,i},\quad i\in[2]
&
\bar R_1>\mathsf{a}_1
&
\bar R_2>\mathsf{a}_2
\\[0.4em]
\bar R_1+\bar R_2>\mathsf{a}_{12}
&
R_{0,1}+R_{0,2}>\mathsf{u}
&
\bar R_1+\bar C_1>\mathsf{b}_1
\\[0.4em]
\bar R_2+\bar C_2>\mathsf{b}_2
&
\bar R_1+\bar R_2+\bar C_1+\bar C_2>\mathsf{b}_{12}
&
R_{0,1}+R_{0,2}+C_{0,1}+C_{0,2}>\mathsf{v}.
\end{array}
\end{equation}
We now eliminate these auxiliary variables. The inequalities
\begin{equation*}
\begin{array}{lll}
R_1>\mathsf{a}_1
&
R_2>\mathsf{a}_2
&
R_1+R_2>\mathsf{a}_{12}+\mathsf{u}
\\[0.4em]
R_1+C_1>\mathsf{b}_1
&
R_2+C_2>\mathsf{b}_2
&
R_1+R_2+C_1>\mathsf{a}_2+\mathsf{b}_1+\mathsf{u}
\\[0.4em]
R_1+R_2+C_2>\mathsf{a}_1+\mathsf{b}_2+\mathsf{u}
&
R_1+R_2+C_1+C_2>\mathsf{b}_{12}+\mathsf{v}
&
\end{array}
\end{equation*}
are immediate necessary consequences of \eqref{Eqn:MLAsymptoticPreFME}: for example, the two mixed constraints are obtained
by summing $(\bar R_2>\mathsf{a}_2)$, $(\bar R_1+\bar C_1>\mathsf{b}_1)$, and
$(R_{0,1}+R_{0,2}>\mathsf{u})$, and symmetrically for the second one.
Conversely, the Fourier--Motzkin elimination of \eqref{Eqn:MLAsymptoticPreFME} produces no further non-redundant
constraints. Indeed, writing $Z\define\doubleunderline{Y}BR$, the asymptotic quantities satisfy
\begin{equation}
\label{Eqn:MLAsymptoticSimplifications}
\begin{array}{@{}ll@{}}
\mathsf{a}_{12}
=
\mathsf{a}_1+\mathsf{a}_2+I(W_1:W_2|R)_{\omega_0}
\geq \mathsf{a}_1+\mathsf{a}_2,
&
\mathsf{b}_{12}
=
\mathsf{b}_1+\mathsf{b}_2+I(W_1:W_2|Z)_{\omega_1}
\geq \mathsf{b}_1+\mathsf{b}_2,
\\[0.6em]
\mathsf{b}_{12}\geq\mathsf{a}_{12},
&
\mathsf{v}
=
\mathsf{u}+I(W_0:Y_0Y_1Y_2B|R\underline{W})_{\omega_1}
\geq\mathsf{u},
\\[0.6em]
\mathsf{b}_{12}-\mathsf{b}_1
=
I(W_2:ZW_1)_{\omega_1}
\geq I(W_2:RW_1)_{\omega_0}
=
\mathsf{a}_{12}-\mathsf{a}_1,
&
\mathsf{b}_{12}-\mathsf{b}_2
\geq
\mathsf{a}_{12}-\mathsf{a}_2.
\end{array}
\end{equation}
The inequality $\mathsf{b}_{12}\geq\mathsf{a}_{12}$ follows from data processing under the partial trace
$Z=\doubleunderline{Y}BR\mapsto R$, together with the fact that the marginal on $W_1W_2$ is unchanged when passing from
$\omega_0$ to $\omega_1$. These relations absorb the additional total-rate inequalities that appear during the elimination,
such as the bounds with right-hand sides $\mathsf{a}_{12}+\mathsf{v}$,
$\mathsf{a}_2+\mathsf{b}_1+\mathsf{v}$, and $\mathsf{a}_1+\mathsf{b}_2+\mathsf{v}$, into the single total constraint
$R_1+R_2+C_1+C_2>\mathsf{b}_{12}+\mathsf{v}$. Hence the eliminated asymptotic region is exactly the one given in Defn.~\ref{Defn:MLAsymptoticRateRegion}.

Therefore, any rate tuple in the strict region of Defn.~\ref{Defn:MLAsymptoticRateRegion} satisfies the normalized one-shot
pre-elimination constraints for all sufficiently large $n$. Thm.~\ref{Thm:OneShotMLIS}, used with
$\widetilde\eta_n=\varepsilon_n$, then gives an $\eta_n$-simulation, where
$\eta_n=32\sqrt{\varepsilon_n}\to0$. The claimed asymptotic multiple-description simulation follows.
\end{proof}

\begin{example}
\label{Ex:MDPositiveSumRate}
{Let the source be $\rho^A=I/2$, with purification}
$|\Phi^+\rangle^{RA}=(|00\rangle+|11\rangle)/\sqrt{2}$.
For $a,b\in\{0,1\}$, define the CP maps
$\mathcal{J}_{ab}(\tau)=K_{ab}\tau K_{ab}^{\dagger}$, where
\[
K_{00}=\frac{2\ketbra{0}{0}+\ketbra{1}}{\sqrt{10}},\quad K_{01}=\frac{\ketbra{0}{0}-2i\ketbra{1}}{\sqrt{10}},\quad K_{10}=\frac{\ketbra{0}{0}+2i\ketbra{1}}{\sqrt{10}},\quad K_{11}=\frac{2\ketbra{0}{0}-\ketbra{1}}{\sqrt{10}}.
\]
These maps define an instrument since
$\sum_{a,b}K_{ab}^{\dagger}K_{ab}=I$.
The target instrument is $\mathcal{J}(\tau)
=\sum_{a,b}
\mathcal{J}_{ab}(\tau)
\otimes |ab\rangle\langle ab|^{Y_1Y_2},$ with $B$ retained by Alice, classical outputs $Y_1=a$ and
$Y_2=b$, and trivial $Y_0$.
The two coarse instruments have CP maps
$\mathcal{J}^{(1)}_a=\sum_b\mathcal{J}_{ab}$ and
$\mathcal{J}^{(2)}_b=\sum_a\mathcal{J}_{ab}$.

In Theorem~\ref{Thm:AsymptoticMLIS}, choose the test instrument
$\mathcal{I}=\mathcal{J}$, with trivial $W_0$,
$W_1=a$, $W_2=b$, and deterministic post-processing $Y_i=W_i$.
On this source, $p(a,b)
=\operatorname{Tr}\!\left(
K_{ab}\frac I2 K_{ab}^{\dagger}
\right)
=\frac14.$ Thus $W_1,W_2$ are independent uniform bits, with
$H(W_1)=H(W_2)=1$ and $H(W_1W_2)=2$.

\noindent
\begin{minipage}[c]{0.48\linewidth}
Conditioned on the outcome pair $(a,b)$, the normalized
state of the reference is listed in the table on the right.
\end{minipage}\hfill
\begin{minipage}[c]{0.50\linewidth}
\centering
\(\displaystyle
\begin{array}{c|c}
(a,b) & \rho^R_{ab}\\
\hline
(0,0),(1,1)
& \frac{4\ketbra{0}{0}+\ketbra{1}}{5}\\[0.3em]
(0,1),(1,0)
& \frac{\ketbra{0}{0}+4\ketbra{1}}{5}
\end{array}
\)
\end{minipage}
\par
Averaging over either outcome gives
$\rho^R_a=\rho^R_b=\rho^R=I/2$.
Moreover, each conditional state $\rho^R_{ab}$ has entropy
$h_2(1/5)$, where $h_2$ denotes the binary entropy.
Consequently,
$H(R)=H(R|W_i)=1$ and
$H(R|W_1W_2)=h_2(1/5)$, yielding
\[
I(W_i:R)_{\omega_0}=0,\quad
I(W_1:W_2)_{\omega_0}=0,\quad
I(W_1W_2:R)_{\omega_0}=1-h_2(1/5).
\]
Since the post-processing appends the copies $Y_i=W_i$,
the marginal distribution of $W_1W_2$ is unchanged, so
$I(W_1:W_2)_{\omega_1}=0$.
Writing $Z=Y_1Y_2BR$, we also have
\[
I(W_i:Z)_{\omega_1}=H(W_i)=1,\qquad
I(W_1W_2:Z)_{\omega_1}=H(W_1W_2)=2.
\]
With $C_1=C_2=1$, all sufficient rate constraints of
Theorem~\ref{Thm:AsymptoticMLIS} therefore reduce to
\[
R_1>0,
\qquad
R_2>0,
\qquad
R_1+R_2>1-h_2(1/5).
\]
For this test channel, the individual constraints allow
arbitrarily small positive rates, whereas the sum-rate
constraint imposes the positive threshold
$1-h_2(1/5)$ bits per copy.
\end{example}

\begin{remark}[Multiple-description overhead]
\label{Rem:MLICPBumpInSumRate}
It is instructive to compare the sum-rate bound on $R_1+R_2$ with the sum of the individual bounds on $R_1$ and $R_2$. Towards this end, define
\[
\begin{aligned}
\Delta_R
\define
\Bigl[
I(\doubleunderline{W}:R)_{\omega_0}
+I(W_1:W_2)_{\omega_0}
\Bigr]
-\Bigl[
I(W_1:R)_{\omega_0}
+I(W_2:R)_{\omega_0}
\Bigr]
=
I(W_1:W_2|R)_{\omega_0}
+
I(W_0:R|\underline{W})_{\omega_0},
\end{aligned}
\]
where the last equality follows from the chain rule. \\
(i) The term $I(W_0:R|\underline{W})_{\omega_0}$ quantifies the additional resource cost incurred by requiring links~$1$ and~$2$ jointly to simulate the higher-resolution instrument, rather than merely supporting the two individual simulations.\\
(ii) The term $I(W_1:W_2|R)_{\omega_0}$ quantifies the coordination cost associated with requiring the two private descriptions to be individually recoverable from their respective links.\\
When both quantities vanish, the joint sum-rate constraint imposes no additional cost beyond the two individual constraints.
\end{remark}

\subsection{Enlarged Inner bound in the Asymptotic Regime via Zhang-Berger \cite{198707TIT_ZhaBer} Base Layer Code}

{The classical multiple-description problem \cite[Ch.~13]{BkNITElGamalKim_2011} asks how a source can be encoded into
two descriptions such that each description individually yields a prescribed
reconstruction quality, while receiving both descriptions permits a better
reconstruction. Zhang and Berger \cite{198707TIT_ZhaBer} showed that replicated transmission of a \underline{common} description can yield strict enlargements to the inner bound. Operationally, this common layer is carried in
\underline{both} descriptions and provides information that is simultaneously useful to the
two side decoders, while the central decoder exploits the common and private
layers jointly.}

{The previous asymptotic inner bound in Thm.~\ref{Thm:AsymptoticMLIS} uses only private descriptions
$W_1,W_2$ and a central refinement $W_0$ decoded when both private links are active. We now present a `Zhang-Berger enlargement' in a slightly modified configuration. While in the classical multiple-description coding we had no common randomness, a `Zhang-Berger enlargement' must have identical common randomness shared between the two side decoders in order leverage the gains of the Zhang-Berger common layer. See \cite[Ch.~13]{198707TIT_ZhaBer}. We therefore depart from the Zhang-Berger configuration to model a base-layer link with communication rate $R_U$ and common-randomness rate $C_U$ that is \textbf{always active}. In other words, the base-layer link is appended to each private-layer link. We therefore have the following activation patterns: the base layer together with link~$1$, the base layer together with link~$2$,
and the base layer together with both private links. At the test-channel level, the base layer carries a common description $U$; with the base layer and link~$i$ the decoder reconstructs the side description $(U,W_i)$ and simulates $Y_i$, whereas with all three links it reconstructs $(U,W_0,W_1,W_2)$ and simulates $(Y_0,Y_1,Y_2)$. Since the common description is carried by a genuine base link,
it is no longer duplicated across the two private descriptions, and it may use its own common randomness $C_U$.}

\begin{remark}
\label{Rem:WhyBaseLayerZhangBergerLinkAlwaysActive}
    A curious reader might enquire why our network configuration aimed at leveraging the Zhang Berger enlargement is different from that of \cite{198707TIT_ZhaBer} wherein there is no base layer link always active. To answer this, note that the network configuration of \cite{198707TIT_ZhaBer} had only two private links and the common base layer bits was replicated in both links. In the case of multiple-description ICP, we have an addition resource of shared randomness. The shared randomness of the base layer must now be identical in \textbf{both} base-layer transmissions. This motivates us to model the base-layer link to be always active and shared.
\end{remark}
\begin{definition}[Test channels]
\label{Defn:ZBCommonPartTestChannels}
Let $\ScrT_{\mathrm{ZB}}(\CalJ)$ denote the collection of all tuples
\[
(\CalU,\doubleunderline{\CalW},\CalI,
p_{Y_0|U\doubleunderline{W}\,\underline{Y}},
p_{Y_1|UW_1},p_{Y_2|UW_2}),
\qquad
\doubleunderline{\CalW}\define \CalW_0\times\CalW_1\times\CalW_2,
\qquad
\underline{Y}\define(Y_1,Y_2),
\]
such that:
\begin{itemize}
\item
$\CalI:\CalL(\CalH_A)\to
\CalL(\CalH_B\otimes \mathbb{C}^{\CalU}
\otimes\mathbb{C}^{\CalW_0}
\otimes\mathbb{C}^{\CalW_1}
\otimes\mathbb{C}^{\CalW_2})$
is an instrument specified by CP TNI maps $\{\CalI_{u,\doubleunderline{w}}:
\CalL(\CalH_A)\to\CalL(\CalH_B)\}_{(u,\doubleunderline{w})
\in\CalU\times\doubleunderline{\CalW}}$ with $\sum_{u,\doubleunderline{w}}\CalI_{u,\doubleunderline{w}}$ CPTP, i.e.
\[
\CalI(\rho)=
\sum_{u\in\CalU}\sum_{\doubleunderline{w}=(w_0,w_1,w_2)\in\doubleunderline{\CalW}}
\CalI_{u,\doubleunderline{w}}(\rho)
\otimes\ketbra{u}_U\otimes\ketbra{w_0}_{W_0}
\otimes\ketbra{w_1}_{W_1}\otimes\ketbra{w_2}_{W_2}.
\]
\item
$p_{Y_1|UW_1}$, $p_{Y_2|UW_2}$, and
$p_{Y_0|U\doubleunderline{W}\,\underline{Y}}$ are stochastic matrices such
that, for every $(y_0,y_1,y_2)\in\CalY_0\times\CalY_1\times\CalY_2$,
\begin{equation*}
\CalJ_{y_0,y_1,y_2}
=
\sum_{u\in\CalU}\sum_{\doubleunderline{w}=(w_0,w_1,w_2)\in\doubleunderline{\CalW}}
p_{Y_1|UW_1}(y_1|u,w_1)\,
p_{Y_2|UW_2}(y_2|u,w_2)\,
p_{Y_0|U\doubleunderline{W}\,\underline{Y}}(y_0|u,\doubleunderline{w},y_1,y_2)\,
\CalI_{u,\doubleunderline{w}} .
\end{equation*}
\end{itemize}
\end{definition}

For a common-description test channel in $\ScrT_{\mathrm{ZB}}(\CalJ)$, define
\begin{eqnarray}
\label{Eqn:ZBMLStates}
&&
\tau_0=\tau^{RBU\doubleunderline{W}}
\define
(\mathrm{id}_R\otimes\CalI)(\varphi_\rho),
\qquad
\tau_1=\tau^{RBU\doubleunderline{W}Y_0Y_1Y_2}
\define
\Bigl(\mathrm{id}_{RBU\doubleunderline{W}Y_1Y_2}\otimes
\olinescre^{Y_0|U\doubleunderline{W}\,\underline{Y}}_{p}\Bigr)
\bigl(\tau^{RBU\doubleunderline{W}Y_1Y_2}\bigr),
\nonumber\\
&&
\tau^{RBU\doubleunderline{W}Y_1Y_2}
\define
\Bigl(\mathrm{id}_{RBUW_0}\otimes
\olinescre^{Y_1|UW_1}_{p}\otimes\olinescre^{Y_2|UW_2}_{p}\Bigr)
\bigl(\tau^{RBU\doubleunderline{W}}\bigr).
\end{eqnarray}

\begin{definition}[Asymptotic rate region]
\label{Defn:ZBMLAsymptoticRateRegion}
Associate with
$(\CalU,\doubleunderline{\CalW},\CalI,
p_{Y_0|U\doubleunderline{W}\,\underline{Y}},
p_{Y_1|UW_1},p_{Y_2|UW_2})
\in\ScrT_{\mathrm{ZB}}(\CalJ)$ the set
\\
$\ScrA^{\mathrm{ML,ZB}}_{0,\mathrm{base}}
(\CalU,\doubleunderline{\CalW},\CalI,
p_{Y_0|U\doubleunderline{W}\,\underline{Y}},
p_{Y_1|UW_1},p_{Y_2|UW_2})$
of all sextuples
$(R_U,R_1,R_2,C_U,C_1,C_2)\in\mathbb{R}_{\geq0}^{6}$
satisfying, with $Z\define\doubleunderline{Y}BR$,
\begin{equation*}
\begin{array}{ll}
R_U~>~I(U:R)_{\tau_0},
&
R_1+C_1~>~I(W_1:Z|U)_{\tau_1},
\\[0.5em]
R_U+C_U~>~I(U:Z)_{\tau_1},
&
R_2+C_2~>~I(W_2:Z|U)_{\tau_1},
\\[0.5em]
R_1~>~I(W_1:R|U)_{\tau_0},
&
R_1+R_2+C_1~>~
I(W_2:R|U)_{\tau_0}
+I(W_1:Z|U)_{\tau_1}
+I(W_0:R|U\underline{W})_{\tau_0},
\\[0.5em]
R_2~>~I(W_2:R|U)_{\tau_0},
&
R_1+R_2+C_2~>~
I(W_1:R|U)_{\tau_0}
+I(W_2:Z|U)_{\tau_1}
+I(W_0:R|U\underline{W})_{\tau_0},
\\[0.5em]
R_1+R_2~>~
I(\doubleunderline{W}:R|U)_{\tau_0}
+I(W_1:W_2|U)_{\tau_0},
&
R_1+R_2+C_1+C_2~>~
I(\doubleunderline{W}:Z|U)_{\tau_1}
+I(W_1:W_2|U)_{\tau_1}.
\end{array}
\end{equation*}
All information quantities are computed with respect to states
$\tau_0$ and $\tau_1$ defined in
\eqref{Eqn:ZBMLStates}.
\end{definition}

\begin{theorem}[Asymptotic Inner-Bound]
\label{Thm:AsymptoticZBMLIS}
Consider the modified activation model in which a base layer of rates
$(R_U,C_U)$ is available in every activation pattern, while links $1$ and $2$
have private rates $(R_1,C_1)$ and $(R_2,C_2)$. The action of $\CalJ$ on $\rho$
is asymptotically simulable in this base-layer multiple-description sense if there exists
\[
(\CalU,\doubleunderline{\CalW},\CalI,
p_{Y_0|U\doubleunderline{W}\,\underline{Y}},
p_{Y_1|UW_1},p_{Y_2|UW_2})
\in\ScrT_{\mathrm{ZB}}(\CalJ)
\quad\text{such that}\quad
(R_U,R_1,R_2,C_U,C_1,C_2)
\in
\ScrA^{\mathrm{ML,ZB}}_{0,\mathrm{base}}
\]
for the associated region in Definition~\ref{Defn:ZBMLAsymptoticRateRegion}.
\end{theorem}
\begin{proof}
Fix a test channel
\[
(\CalU,\doubleunderline{\CalW},\CalI,
p_{Y_0|U\doubleunderline{W}\,\underline{Y}},
p_{Y_1|UW_1},p_{Y_2|UW_2})
\in\ScrT_{\mathrm{ZB}}(\CalJ)
\]
and a rate sextuple in the strict region of
Definition~\ref{Defn:ZBMLAsymptoticRateRegion}.  We first state the
pre-elimination rates that will be used in the construction.  Let
\begin{equation*}
\begin{gathered}
\mathsf a_U\define I(U:R)_{\tau_0},
\qquad
\mathsf b_U\define I(U:Z)_{\tau_1},
\\
\mathsf a_{1|U}\define I(W_1:R|U)_{\tau_0},
\qquad
\mathsf a_{2|U}\define I(W_2:R|U)_{\tau_0},
\qquad
\mathsf u_{|U}\define I(W_0:R|U\underline W)_{\tau_0},
\\
\mathsf b_{1|U}\define I(W_1:Z|U)_{\tau_1},
\qquad
\mathsf b_{2|U}\define I(W_2:Z|U)_{\tau_1},
\qquad
\mathsf v_{|U}\define I(W_0:Z|U\underline W)_{\tau_1},
\\
\mathsf a_{12|U}\define
I(\underline W:R|U)_{\tau_0}+I(W_1:W_2|U)_{\tau_0},
\qquad
\mathsf b_{12|U}\define
I(\underline W:Z|U)_{\tau_1}+I(W_1:W_2|U)_{\tau_1}.
\end{gathered}
\end{equation*}
The Fourier--Motzkin calculation given at the end of the proof shows that
there exist nonnegative numbers
$\bar R_i,\bar C_i,R_{0,i},C_{0,i}$, $i\in[2]$, with
\begin{equation}
\label{Eqn:ZBAsymptoticPreFME}
\begin{array}{lll}
R_i=\bar R_i+R_{0,i},\quad C_i=\bar C_i+C_{0,i},\quad i\in[2],
& R_U>\mathsf a_U
& R_U+C_U>\mathsf b_U
\\[0.4em]
\bar R_1>\mathsf a_{1|U}
& \bar R_2>\mathsf a_{2|U}
& \bar R_1+\bar R_2>\mathsf a_{12|U}
\\[0.4em]
R_{0,1}+R_{0,2}>\mathsf u_{|U}
& \bar R_1+\bar C_1>\mathsf b_{1|U}
& \bar R_2+\bar C_2>\mathsf b_{2|U}
\\[0.4em]
\bar R_1+\bar R_2+\bar C_1+\bar C_2>\mathsf b_{12|U}
& R_{0,1}+R_{0,2}+C_{0,1}+C_{0,2}>\mathsf v_{|U}.
&
\end{array}
\end{equation}
Since all inequalities are strict, throughout the proof the rates may be
reduced by a common positive slack without violating
\eqref{Eqn:ZBAsymptoticPreFME}.  We use this slack to absorb integer
rounding and the vanishing terms in the asymptotic estimates below.  Set
$R_0=R_{0,1}+R_{0,2}$ and $C_0=C_{0,1}+C_{0,2}$.

\medskip
\med\noindent\textit{\underline{A conditional bipartite covering fact.}}
We shall repeatedly use the following direct conditional form of
Proposition~\ref{Prop:SmoothMultipartiteSoftCovering}.  Let
\[
\chi^{UW_1W_2A}
 =\sum_{u,w_1,w_2}p(u,w_1,w_2)
 \ketbra{u,w_1,w_2}\otimes\chi^A_{u,w_1,w_2}
\]
be a finite-dimensional cq-state.  All $n$-letter states below are the
corresponding conditional states of $\chi^{\otimes n}$; in particular,
$\chi^{A^n}_{u^n,w_1^n,w_2^n}
=\bigotimes_{t=1}^n\chi^A_{u_t,w_{1,t},w_{2,t}}$.  For each $u^n$,
independently generate $L_i=2^{nr_i}$ codewords
$W_i^n(\ell_i|u^n)$ according to
$p_{W_i^n|U^n}(\cdot|u^n)$, $i\in[2]$, and define
\begin{equation}
\label{Eqn:ZBConditionalEmpiricalState}
\widehat\chi_{u^n}^{A^n}
\define
\frac1{L_1L_2}\sum_{\ell_1,\ell_2}
\frac{p_{W_1^nW_2^n|U^n}(w_1^n,w_2^n|u^n)}
     {p_{W_1^n|U^n}(w_1^n|u^n)
      p_{W_2^n|U^n}(w_2^n|u^n)}
\chi^{A^n}_{u^n,w_1^n,w_2^n},
\end{equation}
where the summand is evaluated at
$w_i^n=W_i^n(\ell_i|u^n)$ and is set to zero whenever the numerator is
zero.  Then
\begin{equation}
\label{Eqn:ZBConditionalCoveringConvergence}
\mathbb E_{U^n,\mathcal C_1,\mathcal C_2}
\left\|\widehat\chi_{U^n}^{A^n}-\chi_{U^n}^{A^n}\right\|_1
\longrightarrow0
\end{equation}
provided
\begin{equation}
\label{Eqn:ZBConditionalCoveringRates}
r_1>I(W_1:A|U)_\chi,\qquad r_2>I(W_2:A|U)_\chi,\qquad r_1+r_2>I(W_1W_2:A|U)_\chi+I(W_1:W_2|U)_\chi.
\end{equation}
Here
$\chi_{u^n}^{A^n}=\sum_{w_1^n,w_2^n}
 p(w_1^n,w_2^n|u^n)\chi_{u^n,w_1^n,w_2^n}^{A^n}$.
For completeness, choose a sequence $\epsilon_n\downarrow0$ with
$n^{-1}\log(1/\epsilon_n)\to0$, and a sequence of strongly typical sets
$\mathcal T_U^{(n)}$ with vanishing width and
$p_{U^n}(\mathcal T_U^{(n)})\to1$.  For fixed
$u^n\in\mathcal T_U^{(n)}$, apply
Proposition~\ref{Prop:SmoothMultipartiteSoftCovering}, with smoothing
parameter $\epsilon_n$, alphabets $\CalW_1^n,\CalW_2^n$, sampling laws
$p_{W_1^n|u^n},p_{W_2^n|u^n}$, and target law
$p_{W_1^nW_2^n|u^n}$.  Uniformly over $u^n\in\mathcal T_U^{(n)}$, the limsup of the three
normalized smoothed collision divergences occurring in that proposition
is bounded, respectively, by
\begin{equation*}
I(W_1:A|U)_\chi,\qquad I(W_2:A|U)_\chi,
\qquad
I(W_1W_2:A|U)_\chi+I(W_1:W_2|U)_\chi.
\end{equation*}
Indeed, after grouping equal letters of $u^n$, each conditional state is
unitarily equivalent to a tensor product
$\bigotimes_u(\chi_u)^{\otimes n_u}$.  Distribute the smoothing parameter
among the finitely many nonzero type blocks, use the ordinary IID AEP in
each block, and then use that $n_u/n\to p_U(u)$ uniformly on the chosen
typical set.  For the joint constraint, writing
$\chi_u^A\define\sum_{w_1,w_2}p(w_1,w_2|u)\chi_{u,w_1,w_2}^A$,
the relative-entropy limit is
\begin{equation*}
D\!\left(
\chi^{UW_1W_2A}\middle\|
\sum_u p_U(u)\ketbra u\otimes
p_{W_1|u}\otimes p_{W_2|u}\otimes\chi_u^A
\right)
=
I(W_1W_2:A|U)_\chi+I(W_1:W_2|U)_\chi.
\end{equation*}
Thus, for every fixed positive rate slack, all three hypotheses of
Proposition~\ref{Prop:SmoothMultipartiteSoftCovering} hold for all
sufficiently large $n$, uniformly on $\mathcal T_U^{(n)}$, and its
conditional expected error is at most $16\epsilon_n$.  The contribution
of $u^n\notin\mathcal T_U^{(n)}$ is at most twice its probability.  This
proves
\eqref{Eqn:ZBConditionalCoveringConvergence}.  Notice that the same fact
holds when the codeword counts include common-randomness indices; one
only replaces $r_i$ by the logarithmic total codebook rate.

\med\textbf{Design of Encoding Instrument}:
For blocklength $n$, use the integer code sizes
\begin{equation*}
K_U=2^{nC_U},\quad M_U=2^{nR_U},\qquad
K_i=2^{n\bar C_i},\quad M_i=2^{n\bar R_i},\ i\in[2],
\qquad K_0=2^{nC_0},\quad M_0=2^{nR_0},
\end{equation*}
with the harmless rounding convention described above.  Generate
$u^n(k_U,m_U)$ independently according to $p_U^{\otimes n}$.  For every
base codeword, generate, conditionally independently,
\begin{equation*}
w_i^n(k_U,m_U,k_i,m_i)
\sim p_{W_i^n|U^n}(\cdot|u^n(k_U,m_U)),\qquad i\in[2].
\end{equation*}
For every resulting triple $(u^n,w_1^n,w_2^n)$, generate
\begin{equation*}
w_0^n(k_U,m_U,\ulinek,\ulinem,k_0,m_0)
\sim p_{W_0^n|U^nW_1^nW_2^n}(\cdot|u^n,w_1^n,w_2^n).
\end{equation*}
All random choices are independent whenever this is compatible with the
stated conditioning.  We write
$\mathfrak{k}\define(k_U,k_0,k_1,k_2)$,
$\mathfrak{m}\define(m_U,m_0,m_1,m_2)$, and
\[
\xi(\mathfrak{k},\mathfrak{m})
 =\bigl(u^n(k_U,m_U),w_0^n(\mathfrak{k},\mathfrak{m}),
        w_1^n(k_U,m_U,k_1,m_1),w_2^n(k_U,m_U,k_2,m_2)\bigr).
\]
Let $V_{u^n,w_0^n,w_1^n,w_2^n}$ be the product Kraus operator of
$\CalI^{\otimes n}$ and put
$\mu_{u^n,w_0^n,w_1^n,w_2^n}=V^\dagger V$.  Define the superposition
sampling law
\begin{eqnarray*}
q_n(u^n,w_0^n,w_1^n,w_2^n) &\define& p_{U^n}(u^n)p_{W_1^n|U^n}(w_1^n|u^n)p_{W_2^n|U^n}(w_2^n|u^n)p_{W_0^n|U^nW_1^nW_2^n}(w_0^n|u^n,w_1^n,w_2^n).
\end{eqnarray*}
The support of the target outcome law is contained in the support of
$q_n$.  For each $\mathfrak{k}$, set
\begin{equation}
\label{Eqn:ZBJointCoveringOperator}
S_{\mathfrak{k}}
\define
\frac1{M_UM_0M_1M_2}
\sum_{\mathfrak{m}}
\frac{\sqrt{\rho^{\otimes n}}\,
\mu_{\xi(\mathfrak{k},\mathfrak{m})}\,
\sqrt{\rho^{\otimes n}}}
{q_n(\xi(\mathfrak{k},\mathfrak{m}))},
\qquad
z_{\mathfrak{k}}\define\Pi_{\supp(S_{\mathfrak{k}})},
\quad
\pi_{\mathfrak{k}}\define I-z_{\mathfrak{k}}.
\end{equation}
By the hierarchical codebook law and the likelihood denominator
$q_n$, for every fixed $\mathfrak{k}$,
\begin{equation*}
\mathbb E_{\mathcal C}[S_{\mathfrak{k}}]=\rho^{\otimes n},
\qquad
\mathbb E_{\mathcal C}\Tr S_{\mathfrak{k}}=1.
\end{equation*}
For a regular branch define
\begin{equation}
\label{Eqn:ZBJointLikelihoodKraus}
\theta_{\mathfrak{k}}^{\mathfrak{m}}
\define
\frac{V_{\xi(\mathfrak{k},\mathfrak{m})}
\sqrt{\rho^{\otimes n}}S_{\mathfrak{k}}^{-1/2}}
{\sqrt{M_UM_0M_1M_2\,q_n(\xi(\mathfrak{k},\mathfrak{m}))}}.
\end{equation}
Choose an isometry $Q$ with the same input and output spaces as the
product Kraus operators and let
$\theta_{\mathfrak{k}}^{0}=Q\pi_{\mathfrak{k}}$.  Exactly as in
\eqref{Eqn:MLJointEncodingInstrument012},
\begin{equation*}
(\theta_{\mathfrak{k}}^{0})^\dagger\theta_{\mathfrak{k}}^{0}
+\sum_{\mathfrak{m}}(\theta_{\mathfrak{k}}^{\mathfrak{m}})^\dagger
 \theta_{\mathfrak{k}}^{\mathfrak{m}}=I,
\end{equation*}
so these operators define one joint encoding instrument.  More
explicitly, with $\mathfrak{m}^\star=(1,1,1,1)$, let
\begin{equation}
\label{Eqn:ZBJointEncoderFull}
\CalE_{\mathfrak{k}}^{(012)}(S)
\define
\Tr_E\!\left[
\sum_{\mathfrak{m}}
\theta_{\mathfrak{k}}^{\mathfrak{m}}S
(\theta_{\mathfrak{k}}^{\mathfrak{m}})^\dagger
 \otimes\ketbra{\mathfrak{m}}
+\theta_{\mathfrak{k}}^{0}S(\theta_{\mathfrak{k}}^{0})^\dagger
 \otimes\ketbra{\mathfrak{m}^\star}
\right].
\end{equation}
Thus the internal $0$-th branch is merged with one fixed ordinary
message and introduces no additional communication symbol.  The
important point is that \eqref{Eqn:ZBJointEncoderFull} is a single joint
instrument.  The base layer determines the superposition codebook, but
no marginal instrument producing $U^n$ is physically applied before the
remaining branches.

For $\mathfrak{y}\define(y_0^n,y_1^n,y_2^n)$ define
\begin{equation*}
r_n(\mathfrak{y}|\xi)
\define
p_{Y_1^n|U^nW_1^n}(y_1^n|u^n,w_1^n)
 p_{Y_2^n|U^nW_2^n}(y_2^n|u^n,w_2^n)
 p_{Y_0^n|U^n\doubleunderline W^n\underline Y^n}
 (y_0^n|u^n,w_0^n,w_1^n,w_2^n,y_1^n,y_2^n).
\end{equation*}
\medskip
\noindent\textbf{Decoding and stochastic post-processing.}
The joint decoder is the classical CPTP map specified on diagonal
message operators by
\begin{equation*}
\CalD_{\mathfrak{k}}^{(012)}(\ketbra{\mathfrak{m}})
\define
\sum_{\mathfrak{y}}r_n(\mathfrak{y}|\xi(\mathfrak{k},\mathfrak{m}))
\ketbra{\mathfrak{y}}.
\end{equation*}
It reconstructs all four codewords from the common-randomness and
message indices and then applies the product stochastic maps.  When
only the base link and private link $i$ are active, the corresponding
side decoder reconstructs $(u^n,w_i^n)$ and applies
$p_{Y_i|UW_i}^{\otimes n}$.

\medskip
\med\noindent\textit{\underline{Communication-only covering.}}
As in the proof of Theorem~\ref{Thm:OneShotMLIS}, the covering operators
act on the accessible input system whereas the information quantities
are written on the canonical reference.  Transposition in the fixed
canonical basis identifies the two representations and preserves both
trace norm and the sandwiched collision divergence.
For $p_{U^n}(u^n)>0$, define
\begin{equation*}
\rho_{u^n}
\define
\frac{\sqrt{\rho^{\otimes n}}\,
      \mu_{u^n}\,\sqrt{\rho^{\otimes n}}}{p_{U^n}(u^n)},
\qquad
\mu_{u^n}\define\sum_{w_0^n,w_1^n,w_2^n}
\mu_{u^n,w_0^n,w_1^n,w_2^n},
\end{equation*}
and set
\begin{equation*}
S^U_{k_U}\define\frac1{M_U}\sum_{m_U}
\rho_{u^n(k_U,m_U)}.
\end{equation*}
The one-party case of Proposition~\ref{Prop:SmoothMultipartiteSoftCovering},
together with the IID AEP, gives
\begin{equation}
\label{Eqn:ZBBaseCommunicationCovering}
\mathbb E_{\mathcal C}
\frac1{K_U}\sum_{k_U}
\left\|\rho^{\otimes n}-S^U_{k_U}\right\|_1
\longrightarrow0
\end{equation}
under $R_U>\mathsf a_U$.  For every sampled base sequence, the
conditional bipartite covering fact applied to the state obtained from
$\tau_0$ after fixing $U^n$ gives the $W_1^n,W_2^n$ approximation under
\[
\bar R_1>\mathsf a_{1|U},\qquad
\bar R_2>\mathsf a_{2|U},\qquad
\bar R_1+\bar R_2>\mathsf a_{12|U}.
\]
Conditioned on $(U^n,W_1^n,W_2^n)$, apply
Lemma~\ref{Lem:ConditionalSoftCovering} to the $W_0^n$ codebook,
with the conditioning variable there replaced by
$(U,\underline W)$.  Lemma~\ref{Lem:AEPWorstCaseConditionalI2} then
shows that its averaged error tends to zero under
$R_0>\mathsf u_{|U}$.  The same add-and-subtract decomposition as in
\eqref{Eqn:MLSplitAverage} therefore yields
\begin{equation}
\label{Eqn:ZBFullCommunicationCovering}
\mathbb E_{\mathcal C}
\frac1{K_UK_0K_1K_2}
\sum_{\mathfrak{k}}
\left\|S^U_{k_U}-S_{\mathfrak{k}}\right\|_1
\longrightarrow0.
\end{equation}
Combining \eqref{Eqn:ZBBaseCommunicationCovering} and
\eqref{Eqn:ZBFullCommunicationCovering} gives
\begin{equation}
\label{Eqn:ZBJointCoveringToRho}
\mathbb E_{\mathcal C}
\frac1{K_UK_0K_1K_2}
\sum_{\mathfrak{k}}
\left\|\rho^{\otimes n}-S_{\mathfrak{k}}\right\|_1
\longrightarrow0.
\end{equation}
By Lemma~\ref{lem:canonical-purif-stability}, Jensen's inequality, and
$\mathbb E_{\mathcal C}\Tr S_{\mathfrak{k}}=1$, the average trace distance
between the corresponding canonical purifications also tends to zero.
Contractivity under the encoding and decoding maps therefore controls
the difference between the actual protocol input and the proxy input
$\varphi_{S_{\mathfrak{k}}}$.

\medskip
\med\noindent\textit{\underline{Full post-instrument covering.}}
On the proxy input $\varphi_{S_{\mathfrak{k}}}$, the inverse in
\eqref{Eqn:ZBJointLikelihoodKraus} cancels the square root in the
canonical purification and the internal $0$-th branch vanishes.  The
calculation is identical to \eqref{Eqn:MLCOSBranch1}--
\eqref{Eqn:MLProxyAsSoftCovering}, with $U^n$ appended to every classical
index.  Consequently, after the stochastic decoder the proxy output is
the likelihood-weighted empirical average of the conditional target
states indexed by $(U^n,W_0^n,W_1^n,W_2^n)$.

Insert between this empirical state and the ideal state the intermediate
state in which the residual variables $(W_0^n,W_1^n,W_2^n)$ have first
been averaged exactly conditioned on every sampled $U^n$.  The
one-party covering estimate for the base codebook now uses all
$K_UM_U$ base codewords and the output system $Z^n$; it vanishes under
$R_U+C_U>\mathsf b_U$.  Conditioned on a base sequence, the conditional
bipartite covering fact, now applied to the $Z$-output state obtained
from $\tau_1$, gives vanishing error under
\begin{equation*}
\bar R_1+\bar C_1>\mathsf b_{1|U},\qquad
\bar R_2+\bar C_2>\mathsf b_{2|U},\qquad
\bar R_1+\bar R_2+\bar C_1+\bar C_2>\mathsf b_{12|U}.
\end{equation*}
Finally, Lemma~\ref{Lem:ConditionalSoftCovering}, with output
$Z^n$ and conditioning variable $(U^n,W_1^n,W_2^n)$, controls the central
refinement under $R_0+C_0>\mathsf v_{|U}$.  Thus, if
$\Omega_{\mathcal C}^{Z^n}$ denotes the averaged proxy output of the
joint decoder,
\begin{equation}
\label{Eqn:ZBFullOutputCovering}
\mathbb E_{\mathcal C}
\left\|\Omega_{\mathcal C}^{Z^n}-(\tau_1^Z)^{\otimes n}\right\|_1
\longrightarrow0.
\end{equation}
Equations \eqref{Eqn:ZBJointCoveringToRho} and
\eqref{Eqn:ZBFullOutputCovering}, the triangle inequality, and
contractivity show that the expected joint simulation error tends to
zero.  Hence a deterministic realization of all codebooks has
vanishing joint error.

It remains to implement the physical rate split and verify the side
activation patterns.  Factor
$M_0=M_{0,1}M_{0,2}$ and $K_0=K_{0,1}K_{0,2}$ with
$\log M_{0,i}=nR_{0,i}+o(n)$ and
$\log K_{0,i}=nC_{0,i}+o(n)$, and identify the corresponding index sets
through fixed bijections.  With all links active, the two private links
carry the two fragments of $(m_0,k_0)$, so Bob reconstructs the complete
central-refinement indices and implements the joint decoder analyzed
above.

For completeness, we now specify the encoder used when only the base
link and private link $i$ are active.  Write
$k_0=(k_{0,1},k_{0,2})$ and $m_0=(m_{0,1},m_{0,2})$.  Starting from the
fixed joint encoder bank $\{\CalE^{(012)}_{\mathfrak{k}}\}_{\mathfrak{k}}$,
define the side encoder indexed by the available common-randomness
indices $(k_U,k_i,k_{0,i})$ by averaging over the unavailable indices
and tracing out every message register except $M_U$ and $M_i$:
\begin{equation*}
\CalE^{(i)}_{k_U,k_i,k_{0,i}}
\define
\frac{1}{K_{3-i}K_{0,3-i}}
\sum_{k_{3-i},k_{0,3-i}}
\Tr_{M_0M_{3-i}}
\circ\CalE^{(012)}_{k_U,k_0,k_1,k_2}.
\end{equation*}
This is a valid instrument and uses no more than the available rates;
in particular, the central-refinement message fragment may simply be
left unused in a side activation.  Bob reconstructs
$(u^n,w_i^n)$ from $(k_U,m_U,k_i,m_i)$ and samples $Y_i^n$ according to
$p_{Y_i|UW_i}^{\otimes n}$.  Averaging over the available common
randomness, the induced side state is exactly the partial trace of the
averaged joint protocol state over the other classical output registers.
More explicitly, if $\widetilde\tau_{012}^{(n)}$ and
$\widetilde\tau_i^{(n)}$ denote the averaged joint and side protocol
states, then
\begin{equation*}
\widetilde\tau_i^{(n)}
=
\Tr_{Y_0^nY_{3-i}^n}\widetilde\tau_{012}^{(n)},
\qquad i\in[2].
\end{equation*}
The ideal side state is likewise
$\Tr_{Y_0^nY_{3-i}^n}[(\tau_1)^{\otimes n}]$.  Contractivity of the
partial trace therefore implies that both side simulation errors vanish
whenever the joint error does.

\medskip
\noindent\textbf{Conclusion.}
The residual constraints in \eqref{Eqn:ZBAsymptoticPreFME} are precisely
\eqref{Eqn:MLAsymptoticPreFME} conditioned on $U$.  Their
Fourier--Motzkin elimination gives
\begin{equation*}
\begin{array}{lll}
R_1>\mathsf a_{1|U},
& R_2>\mathsf a_{2|U},
& R_1+R_2>\mathsf a_{12|U}+\mathsf u_{|U},
\\[0.3em]
R_1+C_1>\mathsf b_{1|U},
& R_2+C_2>\mathsf b_{2|U},
& R_1+R_2+C_1>\mathsf a_{2|U}+\mathsf b_{1|U}+\mathsf u_{|U},
\\[0.3em]
R_1+R_2+C_2>\mathsf a_{1|U}+\mathsf b_{2|U}+\mathsf u_{|U},
& R_1+R_2+C_1+C_2>\mathsf b_{12|U}+\mathsf v_{|U}.
&
\end{array}
\end{equation*}
No additional nonredundant inequalities occur.  Indeed,
\begin{equation*}
\begin{array}{@{}ll@{}}
\mathsf a_{12|U}
=\mathsf a_{1|U}+\mathsf a_{2|U}
 +I(W_1:W_2|R,U)_{\tau_0},
&
\mathsf b_{12|U}
=\mathsf b_{1|U}+\mathsf b_{2|U}
 +I(W_1:W_2|Z,U)_{\tau_1},
\\[0.5em]
\mathsf b_{12|U}\ge\mathsf a_{12|U},
&
\mathsf v_{|U}
=\mathsf u_{|U}
 +I(W_0:Y_0Y_1Y_2B|R,U,\underline W)_{\tau_1}
\ge\mathsf u_{|U},
\\[0.5em]
\mathsf b_{12|U}-\mathsf b_{1|U}
=I(W_2:ZW_1|U)_{\tau_1}
\ge I(W_2:RW_1|U)_{\tau_0}
=\mathsf a_{12|U}-\mathsf a_{1|U},
&
\mathsf b_{12|U}-\mathsf b_{2|U}
\ge\mathsf a_{12|U}-\mathsf a_{2|U}.
\end{array}
\end{equation*}
Moreover,
$\mathsf a_{12|U}+\mathsf u_{|U}
=I(\doubleunderline W:R|U)_{\tau_0}+I(W_1:W_2|U)_{\tau_0}$ and
$\mathsf b_{12|U}+\mathsf v_{|U}
=I(\doubleunderline W:Z|U)_{\tau_1}+I(W_1:W_2|U)_{\tau_1}$.
Together with the two base-layer constraints, the eliminated region is
exactly Definition~\ref{Defn:ZBMLAsymptoticRateRegion}.  This completes
the proof.
\end{proof}
\begin{remark}
If $U$ is chosen to be trivial, then $\mathsf a_U=\mathsf b_U=0$ and all
conditional quantities in Defn.~~\ref{Defn:ZBMLAsymptoticRateRegion}
reduce to those in Defn.~\ref{Defn:MLAsymptoticRateRegion}. Hence, up to
the usual closure of strict inequalities and with an unused base layer, the
base-layer common-description region recovers the asymptotic multiple-description region
of Thm.~\ref{Thm:AsymptoticMLIS}.
\end{remark}

\begin{remark}[A strict gain]
\label{Rmk:ZBBaseLayerGain}
The base layer can yield a strictly larger inner bound. To see this, consider the
fully classical diagonal case, take $B$ and $W_0$ to be trivial, and suppose
that the common-randomness rates $C_U,C_1,C_2$ are sufficiently large so as to not
to constrain the region. The communication constraints then reduce to
\[
\begin{array}{rcl}
R_U &>& I(U:R),
\qquad
R_i>I(W_i:R|U),\qquad i\in\{1,2\},\\[0.3em]
R_1+R_2 &>& I(W_1W_2:R|U)+I(W_1:W_2|U).
\end{array}
\]
This recovers the Zhang--Berger common-description architecture, with $U$
playing the role of a common description available in every decoding
configuration and $W_1,W_2$ representing the two private refinements. In particular, the binary example of Zhang and Berger~\cite{198707TIT_ZhaBer}
exhibits a distortion triple for which the minimum achievable sum-rate with a
common description is strictly smaller than that attainable by the
corresponding no-common-description inner bound. In our base-link formulation,
the total communication rate is
\[
R_{\mathrm{sum}}\define R_U+R_1+R_2,
\]
since the common description is transmitted only once through the base link.
For this classical instance, this total rate is strictly smaller than the
minimum value of $R_1+R_2$ achievable from the no-base-layer region of
Theorem~\ref{Thm:AsymptoticMLIS}. Hence, already in a fully classical setting,
the introduction of the base layer can strictly enlarge the achievable
multiple-description region.
\end{remark}
\section{Distributed Instrument Simulation in the One-Shot Regime}
\label{Sec:DistInstrSimOneShot}

We now describe the scenario of interest in our distributed ICP depicted in Panel 3 of Fig.~\ref{Fig:DistISAndMultDescIS}. Consider two distributed parties Alice$_{1}$ and Alice$_{2}$, each having access to one component each of a bipartite state $\rho^{A_{1}A_{2}} \in \CalD(\CalH_{A_{1}} \otimes \CalH_{A_{2}})$. A third distributed party Bob holds classical registers. Corresponding to every instrument action, Alice$_{i}$ can transfer $R_{i}$ bits to Bob with whom she also can share $C_{i}$ bits of common randmoness. Consider a quantum instrument $\CalJ_{\CalY}: \CalL(\CalH_{A_{1}} \otimes \CalH_{A_{2}}) \rightarrow \CalL(\CalH_{B_{1}} \otimes \CalH_{B_{2}}\otimes \complex^{|\CalY|}) $ acting on the bipartite state. The goal is to simulate the action of $\CalJ_{\CalY}$ with Alice$_{1}$ and Alice$_{2}$ holding the post-instrument action quantum components in $\CalH_{B_{1}}$ and $\CalH_{B_{2}}$ spaces respectively, and Bob holding the classical outcome in $\CalY$. Owing to Alice$_{1}$ and Alice$_{2}$ being distributed, we restrict our attention to simulation of \textit{separable instruments} which are defined below.

\begin{definition}
 \label{Defn:SeparableInstrument}
 An instrument $\mathcal{J}:\ \CalL(\CalH_{\ulineA} ) \longrightarrow \CalL(\CalH_{\ulineB} \otimes \mathbb{C}^{\mathcal{Y}})$ acting\footnote{Recall our singly underlined notation. $\CalH_{\ulineA}$ denotes $\CalH_{A_{1}}\otimes\CalH_{A_{2}}$ and so on.} on a bipartite state specified through a collection $\{\CalJ_{y} :\mathcal{L}(\mathcal{H}_{\ulineA}) \to \mathcal{L}(\mathcal{H}_{\ulineB}): y \in \CalY \}$ of CP TNI maps is separable if there exist, for $i \in [2]$, instruments $\CalI^{(i)}: \CalL(\CalH_{A_{i}}) \longrightarrow \CalL(\CalH_{B_i} \otimes \mathbb{C}^{\mathcal{W}_{i}}) $ specified through a collection $\{\CalI_{w_{i}}^{(i)} :\mathcal{L}(\mathcal{H}_{A_{i}}) \to \mathcal{L}(\mathcal{H}_{B_{i}}): w_{i} \in \CalW_{i} \}$ of CP TNI maps and a stochastic matrix $p_{Y|W_{1}W_{2}}$ such that, for each $y \in \CalY$, we have
 \begin{eqnarray}
 \label{Eqn:SeparableInstrument}
\CalJ_{y} = \sum_{w_{1},w_{2}}p_{Y|W_{1}W_{2}}(y|w_{1},w_{2}) \CalI_{w_{1}}^{(1)} \otimes \CalI_{w_{2}}^{(2)}.
 \end{eqnarray}
\end{definition}

\begin{remark}
\label{Rem:WhySeparableInstruments}
Since the bipartite state $\rho^{A_1A_2}$ is distributed between the two encoders who do not share any additional entanglement to process nonlocal operations, it is natural to restrict attention to separable measurements as done in earlier works \cite{202202TIT_AtiHeiPra,202504TIT_AtiPra,2026MMTIT_Pad}.
\end{remark}

We first note that to simulate a separable instrument $\CalJ$, it suffices for each party Alice$_{i}$ to apply the instrument $\CalI^{(i)}$ and let Bob stochastically process the outcomes $w_{1},w_{2}$ through the stochastic matrix $p_{Y|W_{1}W_{2}}$. We henceforth focus exclusively on simulating $\CalI^{(1)}$ and $\CalI^{(2)}$. The main idea now is that, instead of implementing the full instruments with outcome alphabets
$\mathcal{W}_1$ and $\mathcal{W}_2$, each Alice$_i$ applies an encoding instrument with a much
smaller number $M_i\ll|\mathcal{W}_i|$ of outcomes and sends the resulting classical message to Bob.
Using these messages together with the corresponding shared randomness, Bob stochastically post-processes outcomes $\hat{W}_1\in\mathcal{W}_1$ and $\hat{W}_2\in\mathcal{W}_2$ to $\hatY \in \CalY$ using $p_{Y|W_{1}W_{2}}$. The resulting post-instrument quantum state, which remains in the possession of Alice$_1$ and Alice$_2$ together with the classical register $Y$ at Bob and an inaccessible reference, is required to be indistinguishable from that obtained by directly applying the target instrument, which is given by
\begin{eqnarray}
\label{Eqn:TargetState}
\CalT \define \left(\mathrm{id}_{R}\otimes \CalJ\right)(\varphi_{\rho}).
\end{eqnarray}
We therefore need a protocol to use the dephasing channels and common randomness. We formalize this and precisely state the multi-party instrument simulation problems.

\begin{definition}
\label{Defn:DistributedPOVMProtocol}
{A $(\ulineK,\ulineM,\ulineCalE,\CalD)$ one-shot distributed ISP consists of (i) a bank of $K_{i}$ instruments $\mathcal{E}^{(i)}_{k_i}:\ \CalL(\CalH_{A_i})\longrightarrow \CalL(\CalH_{B_i}\otimes \mathbb{C}^{M_i}) : k_{i} \in [K_{i}]$ each with $M_{i}$ outcomes for $i \in [2]$, and (ii) a decoder instrument $\CalD_{\ulinek} : \CalL(\mathbb{C}^{M_1}\otimes \mathbb{C}^{M_2})
\longrightarrow \CalL(\mathbb{C}^{|\mathcal{Y}|})$. The action of $\CalJ$ on $\rho$ can be $\eta$-simulated with communication rates $\ulineR$ and common-randomness rates $\ulineC$} if there exists a $(\ulineK,\ulineM,\ulineCalE,\CalD)$ one-shot distributed ISP for which $\log K_{i} \leq C_{i}$, $\log M_{i} \leq R_{i}$ for $i\in [2]$ and state $\CalT$ in \eqref{Eqn:TargetState} satisfies
\begin{eqnarray*}
 \norm{\frac{1}{K_1K_2}\sum_{k_1=1}^{K_1}\sum_{k_2=1}^{K_2}
\Big(
(\mathrm{id}_{B_1B_2R}\otimes \mathcal{D}_{k_1,k_2})
\circ{}
(\mathrm{id}_{R}\otimes \mathcal{E}^{(1)}_{k_1}
\otimes \mathcal{E}^{(2)}_{k_2})
\Big)(\varphi_{\rho}) -\CalT }_{1} \leq \eta.
 \end{eqnarray*}
The action of $\CalJ$ on $\rho$ is asymptotically simulable with communication rates $\ulineR$ and common-randomness rates $\ulineC$ if for every $\eta>0$, there exists $N_{\eta}$ such that for all $n \geq N_{\eta}$, the action of $\otimes_{t=1}^{n}{\CalJ}$ on $\otimes_{t=1}^{n}\rho$ can be $\eta$-simulated with communication rates $n\ulineR$ and common-randomness rates $n\ulineC$.
\end{definition}

We next introduce a test-channel parametrization that will be used to state both the one-shot and asymptotic inner bounds. Definition~\ref{Defn:TestChannels} specifies the single-letter instruments and stochastic post-processing maps used in the coding construction, together with the one-shot information quantities entering the achievable region.

\begin{definition}[Test channels and one-shot rate region]
\label{Defn:TestChannels}
For an instrument
$\mathcal{J}:\CalL(\CalH_{\ulineA})
\longrightarrow
\CalL(\CalH_{\ulineB}\otimes\mathbb{C}^{\mathcal{Y}})$
specified through a collection
$\{\CalJ_y:\CalL(\CalH_{\ulineA})\to\CalL(\CalH_{\ulineB}):y\in\CalY\}$
of CP TNI maps, let $\ScrT(\CalJ)$ denote the collection of all triples
$(\ulineCalW,\ulineCalI,p_{Y|\ulineW})$, where
$\ulineCalW=(\CalW_1,\CalW_2)$,
$\ulineCalI=(\CalI^{(1)},\CalI^{(2)})$,
$\CalI^{(i)}:\CalL(\CalH_{A_i})
\to\CalL(\CalH_{B_i}\otimes\mathbb{C}^{\CalW_i})$
are instruments for $i\in[2]$, each specified through a collection
$\{\CalI_{w_i}^{(i)}:\CalL(\CalH_{A_i})
\to\CalL(\CalH_{B_i}):w_i\in\CalW_i\}$
of CP TNI maps, and $p_{Y|\ulineW}$ is a stochastic matrix such that
\eqref{Eqn:SeparableInstrument} holds for all $y\in\CalY$. For any
$(\ulineCalW,\ulineCalI,p_{Y|\ulineW})\in\ScrT(\CalJ)$, define
$\ulineCalI\define\CalI^{(1)}\otimes\CalI^{(2)}$ and define the associated states
\begin{equation}
\label{Eqn:TestChannelJointStates}
\sigma^{RB_1B_2\ulineW Y}
\define
\left(\mathrm{id}_{RB_1B_2}\otimes
\olinescre^{Y|\ulineW}_{p}\right)
\left[
\left(\mathrm{id}_{R}\otimes\ulineCalI\right)(\varphi_{\rho})
\right],
\qquad
\sigma_i^{RB_iA_{3-i}W_i}
\define
\left(\mathrm{id}_{RA_{3-i}}\otimes\CalI^{(i)}\right)
(\varphi_{\rho}),
\quad i\in[2].
\end{equation}
Thus the local instruments are applied first, and the stochastic map
subsequently appends $Y$ while retaining the classical registers
$W_1W_2$.

Fix
$(\ulineCalW,\ulineCalI,p_{Y|\ulineW})\in\ScrT(\CalJ)$
and $0<\varepsilon<1$, and let
$\varepsilon_5\define2\log(1/\varepsilon)$.
Associate with this test channel the set
$\ScrA_{\varepsilon}(\ulineCalW,\ulineCalI,p_{Y|\ulineW})$
of all quadruples
$(R_1,R_2,C_1,C_2)\in\reals_{\geq0}^4$
satisfying
\begin{eqnarray}
\label{Eqn:DISOneShotRateBounds}
\begin{array}{ccc}
R_1>\sfB_1^{\varepsilon}-\sfB_6^{\varepsilon}
&
R_2>\sfB_2^{\varepsilon}-\sfB_6^{\varepsilon}
&
R_1+R_2>\sfB_1^{\varepsilon}+\sfB_2^{\varepsilon}-\sfB_6^{\varepsilon}
\\[0.4em]
R_1+C_1>\sfB_3^{\varepsilon}-\sfB_6^{\varepsilon}
&
R_2+C_2>\sfB_4^{\varepsilon}-\sfB_6^{\varepsilon}
&
R_1+R_2+C_1+C_2>
\max\{\sfB_3^{\varepsilon}+\sfB_4^{\varepsilon},\sfB_5^{\varepsilon}\}
-\sfB_6^{\varepsilon}
\\[0.4em]
R_1+R_2+C_1>
\sfB_2^{\varepsilon}+\sfB_3^{\varepsilon}-\sfB_6^{\varepsilon}
&
R_1+R_2+C_2>
\sfB_1^{\varepsilon}+\sfB_4^{\varepsilon}-\sfB_6^{\varepsilon}
&
\end{array}.
\end{eqnarray}
\begin{equation}
\label{Eqn:DISOneShotInfoQuantities}
\mbox{where }~~
\begin{gathered}
\sfB_1^{\varepsilon}
\define
I_2^{\varepsilon}(W_1:A_2R)_{\sigma_1}+\varepsilon_5,
\qquad
\sfB_2^{\varepsilon}
\define
I_2^{\varepsilon}(W_2:A_1R)_{\sigma_2}+\varepsilon_5,
\qquad
\sfB_6^{\varepsilon}
\define
I_H^{\varepsilon}(W_1:W_2)_{\sigma}
-\frac{1}{2}\varepsilon_5,
\\
\sfB_3^{\varepsilon}
\define
I_2^{\varepsilon}(W_1:Y\ulineB R)_{\sigma}+\varepsilon_5,
\qquad
\sfB_4^{\varepsilon}
\define
I_2^{\varepsilon}(W_2:Y\ulineB R)_{\sigma}+\varepsilon_5,
\qquad
\sfB_5^{\varepsilon}
\define
D_2^{\varepsilon}\!\left(
\sigma^{R\ulineB\ulineW Y}
\,\middle\|\,
\sigma^{W_1}\!\otimes\!\sigma^{W_2}\!\otimes\!\sigma^{R\ulineB Y}
\right)_{\sigma}
+\varepsilon_5.
\end{gathered}
\end{equation}
\end{definition}

\subsection{One-Shot Achievability}
\label{SubSec:OneShotISWithIIDCodes}

\begin{theorem}[One-shot achievability]
\label{Thm:DistInstSimltnOneShot}
For $0<\eta<2$, the action of instrument $\CalJ$ on $\rho$ can be $\eta$-simulated in the one-shot regime with communication rates $(R_{1},R_{2})$ and common-randomness rates $(C_{1},C_{2})$ if there exists a $(\ulineCalW,\ulineCalI,p_{Y|\ulineW}) \in \ScrT(\CalJ)$ for which $(R_{1},R_{2},C_{1},C_{2}) \in \ScrA_{\widetilde\eta}(\ulineCalW,\ulineCalI,p_{Y|\ulineW})$ where $\widetilde\eta=\frac{\eta^{2}}{28^{2}}$.
\end{theorem}

\begin{remark}
    \label{Rem:DISOneShotProofDiffStructure}
Since we provide multiple proofs in this manuscript based on a single framework of protocols and a unified proof template, we take the opportunity in this proof to present the same in a slightly different style to present a different perspective to our reader. Specifically, while in the other proofs we introduce proxy states at the beginning of our analysis, in the following proof, we only identify the intermediate states one after another as needed and look back at the proxy states at the end of the proof.
\end{remark}

\begin{proof} We begin with a broad outline of the proof to assist our reader. The proof combines three ingredients. First, we construct local likelihood instruments whose codebooks are partitioned into bins, so that the correlation between $W_1$ and $W_2$ can be exploited through Slepian--Wolf decoding. We then introduce intermediate and proxy states that separate the simulation error into a multipartite soft-covering term, a bin-decoding term, and two single-party quantum-covering terms. The latter contain random inverse operators. On suitable canonical proxy inputs these inverses cancel against the square roots defining the purifications; the remaining compatible operators can then be moved to the reference system. We introduce this cancellation-and-sliding step below as the compatibility-operator-sliding (COS) argument. Finally, we eliminate the auxiliary bin sizes to obtain the rate region in Defn~\ref{Defn:TestChannels}. We now proceed to a formal proof.

Choose a generic test channel $(\ulineCalW,\ulineCalI,p_{Y|\ulineW})  \in\ScrT(\CalJ)$ and fix the same throughout this proof. Associated with this triple $(\ulineCalW,\ulineCalI,p_{Y|\ulineW})$, let communication rate pair $(R_{1},R_{2})$ and common randomness rate pair $(C_{1},C_{2})$ satisfy bounds stated in \eqref{Eqn:DISOneShotRateBounds} where the information quantities are wrt the associated states \eqref{Eqn:TestChannelJointStates} corresponding to this chosen test channel. We also choose a pair $(\beta_{1},\beta_{2})$ of (non-negative) binning rates which will enable us specify the structure of our two codes. The choice of $\beta_{1},\beta_{2}$ will be evident at the end of our proof.

\med\textbf{Design of Encoding Instrument}: For $i\in [2]$, the design of our encoding instrument $\CalE_{{k_{i}}}^{(i)}$ will be based on a code built over $\CalW_{i}$. Towards specifying the codes and the encoding instrument, let us identify Kraus representations of $\CalI^{(i)}: i \in [2]$ provided in our test channel.

\med\noindent\textit{\underline{Kraus representation of }$\CalI^{(i)}$}: Fix a Kraus representation
\begin{eqnarray}
\label{Eqn:InstrumentIi}
\mathcal{I}^{(i)}(\rho^i)
&=&
\Tr_{E_i}\left[
\sum_{w_i\in\CalW_i}
V_{w_i}\,\rho^i\,{V_{w_i}}^\dagger
\otimes \ketbra{w_i}_{W_i}
\right],
\end{eqnarray}
where $V_{w_i}:\CalH_{A_i}\to \CalH_{B_i}\otimes \CalH_{E_i}$, $i\in\{1,2\}$, satisfy $\sum_{w_i\in\CalW_i}\ {V_{w_i}}^\dagger V_{w_i} = I_{A_i}$. Throughout, we set

\[
p_{\ulineW}(\ulinew)
 \define\ p_{W_1W_2}(w_1,w_2)\ \define\ \Tr\big[(\mu_{w_1}\otimes \mu_{w_2})\,\rho \big],\qquad \mu_{w_i}\ \define\ {V_{w_i}}^\dagger V_{w_i}.
\]
For \(\ulinew=(w_1,w_2)\), we write \(V_{\ulinew}\define V_{w_1}\otimes V_{w_2}\).
Without loss of generality, we restrict $\CalW_i$ to the support of
$p_{W_i}$, $i\in[2]$. For pairs $\uline{w}$ such that
$p_{\uline{W}}(\uline{w})=0$, any associated conditional state is
chosen arbitrarily; as will become clear below, whenever such a pair
occurs in the independently generated codebooks, its corresponding
contribution is zero. We begin by describing the $K_{1},K_{2}$ encoder instruments and the stochastic decoder.

\med\textit{\underline{Two Codes, their structure and Binning}}: Let $K_{i}=2^{C_{i}},M_{i}=2^{R_{i}},B_{i}=2^{\beta_{i}} \in \naturals$ and $[K_{i}],[M_{i}],[B_{i}]$ denote common-randomness, message, and intra-bin index sets, respectively. Since the two instrument outcomes $W_1$ and $W_2$ are correlated,
Slepian--Wolf binning allows part of each local codeword index to
remain uncommunicated. The auxiliary indices $b_i\in[B_i]$ represent
these hidden intra-bin indices; their admissible total size is
controlled by the hypothesis-testing correlation
$I_H^\varepsilon(W_1:W_2)_\sigma$. The outcomes are therefore partitioned into bins indexed by $m_i$, while the intra-bin index $b_{i} \in [B_{i}]$ is not communicated to Bob. Specifically, for $i \in [2]$, let $\sfc^{i} = (c_{k_{i}} : k_{i} \in [K_{i}])$ denote a multi-code consisting of $K_{i}=2^{C_{i}}$ codes where the $k_i$-th code $c_{k_{i}}=(w_{i}(k_{i},m_{i},b_{i}) \in \CalW_{i}:(m_{i},b_{i}) \in [M_{i}] \times [B_{i}])$ comprises $M_{i}=2^{R_{i}}$ bins each with $B_{i}=2^{\beta_{i}}$ codewords.

\med\textit{\underline{Encoding Instruments}} $\CalE_{k_{i}}^{(i)}: i \in [2]$: Recalling the above defined codes $\sfc^{(i)}$, we associate with codeword $w_{i}(k_{i},m_{i},b_{i})$, the operators $V_{k_i,m_i,b_i} \define V_{w_i(k_i,m_i,b_i)}$ and $\mu_{k_{i},m_{i},b_{i}} \define \mu_{w_{i}(k_{i},m_{i},b_{i})} \in \CalP(\CalH_{A_i})$. In addition, define the operators
\[
\begin{aligned}
\label{Eqn:DistLikelihoodPOVMdefn_i}
S_{k_i}
\define \sum_{m_i=1}^{M_i} \sum_{b_i=1}^{B_i}
\frac{\sqrt{\rho^i}\ \mu_{k_i,m_i,b_i}\ \sqrt{\rho^i}}
{M_i B_i\,p_{W_i}(w_i(k_i,m_i,b_i))},~
z_{k_i}
\define \Pi_{\supp(S_{k_i})},
\end{aligned}
\]
and $\pi_{k_i}\define I_{A_i}-z_{k_i}$. For $m_i\in[M_i]$ define
\begin{equation}
\label{Eqn:DistLikelihoodTheta_i}
\theta^{m_i,b_i}_{k_i}
\define
\frac{
V_{k_i,m_i,b_i}\ \sqrt{\rho^i}\ {S_{k_i}}^{-\frac12}
}{
\sqrt{M_i B_i\,p_{W_i}(w_i(k_i,m_i,b_i))}
},
\end{equation}
where ${S_{k_i}}^{-\frac12}$ denotes the generalized inverse (equal to $0$ on the null space of $S_{k_i}$). Note that, by definition, $\theta^{m_i,b_i}_{k_i}: \CalH_{A_i}\to \CalH_{B_i}\otimes \CalH_{E_i} $.
Since $\pi_{k_i}$ is a projector, fix an isometry
$Q_i:\CalH_{A_i}\to \CalH_{B_i}\otimes\CalH_{E_i}$
and define the ``$0$-th'' operator
\[
\theta^{0}_{k_i}\ \define\ Q_i\,\pi_{k_i},
\qquad\text{so that}\qquad
{\theta^{0}_{k_i}}^\dagger \theta^{0}_{k_i}=\pi_{k_i}.
\]
Note that for every $(k_i,m_i,b_i)\in[K_i]\times[M_i]\times[B_i]$, we have
\begin{equation}
\label{Eqn:DistThetak0GetsKilled_i}
\theta^{0}_{k_i}\sqrt{S_{k_i}} = 0, \theta^{m_i, b_i}_{k_i}\sqrt{S_{k_i}}  =
\frac{
V_{k_i,m_i,b_i}\sqrt{\rho^i}
}{
\sqrt{M_i B_i\,p_{W_i}(w_i(k_i,m_i,b_i))}
}.
\end{equation}
It is also easy to verify that for each $(k_{i},m_{i},b_{i}) \in [K_{i}]\times [M_{i}] \times [B_{i}]$, $\theta^{m_i,b_i}_{k_i}: \CalH_{A_i} \to \CalH_{B_i}\otimes \CalH_{E_i} $ and ${\theta_{k_{i}}^{0}}^{\dagger}\theta_{k_{i}}^{0}+\sum_{m_{i},b_{i}}{\theta_{k_{i}}^{m_{i},b_{i}}}^{\dagger}\theta_{k_{i}}^{m_{i},b_{i}} = I_{A_{i}}$.
The $0$-th operator is kept as an internal Kraus branch, but it is not assigned an additional communication symbol. Instead, we coarse-grain it with the ordinary message $m_i^\star\define1$. Thus the classical message alphabet remains exactly $[M_i]$.
We are now ready to define the two banks of encoding instruments. For $ i \in [2]$, Alice$_{i}$'s instrument corresponding to common randomness index $k_{i} \in [K_{i}]$ is $\CalE_{k_{i}}^{(i)}:\ \CalL(\CalH_{A_i})\longrightarrow \CalL(\CalH_{B_i}\otimes \mathbb{C}^{M_i})$, defined through its action
\begin{eqnarray}
 \label{Eqn:EncodingInstrumentki}
 \CalE_{k_{i}}^{(i)}(S)= \tr_{E_{i}}\left[\begin{array}{r}\displaystyle\sum_{m_i=1}^{M_i} \sum_{b_i=1}^{B_i}
\theta^{m_i, b_i}_{k_i}\,S\,
{\theta^{m_i, b_i}_{k_i}}^\dagger
\otimes \ketbra{m_i}_{M_i}
+
\theta^{0}_{k_i}\,S\,
{\theta^{0}_{k_i}}^\dagger
\otimes \ketbra{1}_{M_i} \end{array}\right]
\end{eqnarray}
on any $S \in \CalL(\CalH_{A_{i}})$. For the classical outcome $m_i^\star=1$, the corresponding CP map therefore contains the $B_i$ regular Kraus operators $\{\theta^{1,b_i}_{k_i}\}_{b_i\in[B_i]}$ together with $\theta^0_{k_i}$. This coarse-graining preserves complete positivity and the completeness relation above.

\noindent\textbf{Decoding and stochastic post-processing.} Recall that the intra-bin indices are not communicated to Bob. The decoder $\CalD_{\ulinek} : \CalL(\mathbb{C}^{M_1}\otimes \mathbb{C}^{M_2})
\longrightarrow \CalL(\mathbb{C}^{|\mathcal{Y}|})$ is therefore a composition $\CalD_{\ulinek} \define \CalD_{\ulinek}^{(2)} \circ \CalD_{\ulinek}^{(1)} $. Here
$\CalD_{\ulinek}^{(1)}:\CalL(\mathbb{C}^{\ulineM}) \rightarrow \CalL(\mathbb{C}^{\ulineCalW})$ is the one-shot hypothesis-testing-based \cite{201205PRL_WanRen} Slepian--Wolf decoder described and analyzed in Appendix~\ref{AppSec:binning}. It applies a deterministic selection rule $f(\ulinek,\ulinem)\in\CalW_1\times\CalW_2$ to the pair $(\ulinek,\ulinem)$ and outputs the corresponding recovered codeword pair. The second map $\CalD_{\ulinek}^{(2)}$ is the stochastic post-processing map
\begin{equation}
  \CalD_{\ulinek}^{(2)}(S)= \sum_{\ulinew \in \ulineCalW}\sum_{y \in \CalY}p_{Y|\ulineW}(y|\ulinew)\ketbra{y}{\ulinew}S\ketbra{\ulinew}{y}.
  \nonumber
\end{equation}
for any $S \in \CalL(\mathbb{C}^{\ulineCalW})$, where $\mathbb{C}^{\ulineCalW}$ denotes the classical register with basis indexed by $\CalW_1\times\CalW_2$. Equivalently, the overall decoder maps $(\ulinek,\ulinem)$ to the recovered pair $f(\ulinek,\ulinem)$ and then samples $Y$ according to $p_{Y|\ulineW}(\cdot|f(\ulinek,\ulinem))$.

\med\textbf{Analysis}:  Let $0<\varepsilon<1$ and assume
$(\ulineR,\ulineC)\in \ScrA_{\varepsilon}(\ulineCalW,\ulineCalI,p_{Y|\ulineW})$. Recalling that $\CalT = (\mathrm{id}_{R}\otimes \CalJ)\{\ketbra{\varphi_{\rho}}\}$ is the ideal (target) post-instrument state and defining
\begin{equation*}
\widetilde{\CalT}
=
\frac{1}{K_1K_2}\sum_{k_1=1}^{K_1}\sum_{k_2=1}^{K_2}
\Big(
(\mathrm{id}_{B_1B_2R}\otimes \mathcal{D}_{k_1,k_2})
\circ
(\mathrm{id}_{R}\otimes \mathcal{E}^{(1)}_{k_1}
\otimes \mathcal{E}^{(2)}_{k_2})
\Big)\{\ketbra{\varphi_{\rho}}\},
\end{equation*}
as the post-instrument state induced by the protocol, our goal now is to bound $\norm{\CalT-\widetilde{\CalT}}_{1}$ on the above. We accomplish this by introducing three intermediate states $\CalT_{1},\CalT_{2},\CalT_{3}$ and bounding each of the terms $\norm{\CalT-\CalT_{1}}_{1},\norm{\CalT_{1}-\CalT_{2}}_{1},\norm{\CalT_{2}-\CalT_{3}}_{1}$ and $\bigl\|\CalT_3-\widetilde{\CalT}\bigr\|_1$ on the above. As it will become evident in the sequel as and when we precisely characterize these intermediate states (i) the state $\CalT_1$ replaces the ideal distribution by its empirical codebook approximation and isolates the bipartite soft-covering error, (ii) the state $\CalT_2$ incorporates the output of the Slepian--Wolf decoder and therefore isolates the
bin-decoding error, and finally (iii) $\CalT_3$ introduces a canonical proxy
input for one sender, allowing the two remaining one-party soft covering errors to be treated separately. Towards introducing these intermediate states, we begin by defining the conditional post-instrument states used in
these constructions. Let
\begin{equation}
\label{Eqn:PostInstrumentStateDef}
\CalT_{B_1 B_2 R}^{\ulinew}
\;\coloneqq\;
\Tr_{E_1 E_2}\left(
\frac{
\bigl(I_R \otimes V_{\ulinew} \bigr)\,
\varphi_{\rho}\,
\bigl(I_R \otimes V_{\ulinew}^{\dagger} \bigr)
}{
p_{\ulineW}(\ulinew)
}
\right),
\qquad
\CalT_{B_1 B_2 R Y}^{\ulinew}
\;\coloneqq\;
\sum_{y} p_{Y|\ulineW}(y|\ulinew)\,
\CalT_{B_1 B_2 R}^{\ulinew}\otimes\ketbra{y}.
\end{equation}
Since $\CalJ_{y} = \sum_{w_{1},w_{2}} p_{Y|\ulineW}(y|w_{1},w_{2}) \,\CalI_{w_{1}}^{(1)} \otimes \CalI_{w_{2}}^{(2)}$
by the definition of $\ScrT(\CalJ)$, and using the decompositions in~\eqref{Eqn:InstrumentIi}, we can rewrite
\begin{eqnarray*}
\CalT
=
\left(\mathrm{id}_{R}\otimes \CalJ\right)(\varphi_{\rho})
=
\sum_{\ulinew} p_{\ulineW}(\ulinew)\,\CalT_{B_1 B_2 R Y}^{\ulinew}.
\end{eqnarray*}
We will then take the expectation of each term with respect to the random codebooks $c_{k_i}
\;\define\;
\bigl(w_i(k_i,m_i,b_i)\in \CalW_i : (m_i,b_i)\in [M_i]\times [B_i]\bigr),$ which are generated i.i.d.\ according to $p_{W_i}$.
We denote this expectation by $\mathbb{E}_{\mathcal{C}}$ and use the shorthand $KMB\define K_1K_2M_1M_2B_1B_2.$

\med
\noindent\textit{\underline{Bipartite Soft Covering error $\bigl\|\CalT-\CalT_1\bigr\|_1$}.}
We first define the bipartite soft-covering estimate of $\CalT$ induced by our codebooks:
\[
\begin{gathered}
\CalT_{1}
\;=\;
\frac{1}{KMB}\sum_{\ulinek,\ulinem,\ulineb}
\frac{p_{\ulineW}(\ulinew(\ulinek,\ulinem,\ulineb))}
{p_{W_1}\bigl(w_1(k_1,m_1,b_1)\bigr)\,
 p_{W_2}\bigl(w_2(k_2,m_2,b_2)\bigr)}\,
\CalT_{B_1 B_2 R Y}^{\,\ulinew(\ulinek,\ulinem,\ulineb)}.
\end{gathered}
\]
We then observe that
\begin{equation}
\label{eq:tri}
\bigl\|\CalT-\widetilde{\CalT}\bigr\|_1
\;\le\;
\bigl\|\CalT-\CalT_1\bigr\|_1
+
\bigl\|\CalT_1-\widetilde{\CalT}\bigr\|_1.
\end{equation}
The first term is bounded via a bipartite soft-covering estimate
(Proposition~\ref{Prop:SmoothMultipartiteSoftCovering}, proven in~\cite{202504arXiv_Sen}).
In particular, we obtain
\begin{equation}
\label{eq:sc_expect}
\mathbb{E}_{\mathcal{C}}\left[\bigl\|\CalT-\CalT_1\bigr\|_1\right]
\le 16\,\varepsilon,
\end{equation}
provided that the following rate conditions hold:
\begin{equation}
\label{eq:one}
\begin{gathered}
R_i+C_i+\beta_{i}
>
I_{2}^{\varepsilon}(W_i:Y\ulineB R)_{\sigma} + 2 \log\frac{1}{\varepsilon},
\qquad i\in\{1,2\},\\
R_1+R_2+C_1+C_2+\beta_{1}+\beta_{2}
>
D_{2}^{\varepsilon}(\sigma^{R\ulineB\ulineW Y}||\sigma^{W_{1}}\otimes \sigma^{W_{2}}\otimes\sigma^{R\ulineB Y})_{\sigma}
+ 2 \log\frac{1}{\varepsilon}.
\end{gathered}
\end{equation}

\med
\noindent\textit{\underline{Binning Error $\bigl\|\CalT_1-\CalT_2\bigr\|_1$}.}
Our second proxy state captures the reliability of the hypothesis-testing decoder.
We first define the post-instrument state corresponding to the situation where the intra-bin indices $\ulineb$ are not sent to the decoder; instead, the decoder applies the test function $f$ (described in Appendix~\ref{AppSec:binning}) to recover the codeword pair:
\begin{equation*}
\widetilde{\CalT}_{B_1 B_2 R Y}^{\ulinew(\ulinek,\ulinem,\ulineb)}
\;\coloneqq\;
\sum_{y} p_{Y|\ulineW}\bigl(y|f(\ulinek,\ulinem)\bigr)\,
\CalT_{B_1 B_2 R}^{\,\ulinew(\ulinek,\ulinem,\ulineb)}\otimes\ketbra{y}.
\end{equation*}
This leads to the following proxy state:
\begin{equation}
\label{Eqn:CalT2BeforeCOS}
\CalT_{2}
\;=\;
\frac{1}{KMB}\sum_{\ulinek,\ulinem,\ulineb}
\frac{p_{\ulineW}(\ulinew(\ulinek,\ulinem,\ulineb))}
{p_{W_1}\bigl(w_1(k_1,m_1,b_1)\bigr)\,
 p_{W_2}\bigl(w_2(k_2,m_2,b_2)\bigr)}\,
\widetilde{\CalT}_{B_1 B_2 R Y}^{\ulinew(\ulinek,\ulinem,\ulineb)}.
\end{equation}
We then observe that
\begin{equation}
\label{eq:tri2}
\bigl\|\CalT_1-\widetilde{\CalT}\bigr\|_1
\;\le\;
\bigl\|\CalT_1-\CalT_2\bigr\|_1
+
\bigl\|\CalT_2-\widetilde{\CalT}\bigr\|_1.
\end{equation}
Note that
\begin{equation*}
\bigl\|\CalT_{B_1 B_2 R Y}^{\,\ulinew(\ulinek,\ulinem,\ulineb)}
-\widetilde{\CalT}_{B_1 B_2 R Y}^{\ulinew(\ulinek,\ulinem,\ulineb)}\bigr\|_1
\;\le\;
2\,\mathbbm{1}_{\left\{\,\ulinew(\ulinek,\ulinem,\ulineb)\neq f(\ulinek,\ulinem)\,\right\}}
.
\end{equation*}
Consequently,
\begin{equation}
\label{Eqn:TraceNormT1T2Bound}
\begin{aligned}
\bigl\|\CalT_1-\CalT_2\bigr\|_1
&\le
\frac{2}{KMB}\sum_{\ulinek,\ulinem,\ulineb}
\frac{p_{\ulineW}\bigl(\ulinew(\ulinek,\ulinem,\ulineb)\bigr)}
{p_{W_1}\bigl(w_1(k_1,m_1,b_1)\bigr)\,
 p_{W_2}\bigl(w_2(k_2,m_2,b_2)\bigr)}
\;\mathbbm{1}_{\left\{\,\ulinew(\ulinek,\ulinem,\ulineb)\neq f(\ulinek,\ulinem)\,\right\}}
\, .
\end{aligned}
\end{equation}
Taking expectations and using one-shot classical joint typicality (the analysis is given in Appendix~\ref{AppSec:binning}), we obtain
\begin{equation}
\label{eq:two}
\mathbb{E}_{\mathcal{C}}\left[\bigl\|\CalT_1-\CalT_2\bigr\|_1\right]
\le 4\,\varepsilon
\qquad\text{if}\qquad
\beta_{1}+\beta_{2}
<
I_H^{\varepsilon}(W_1:W_2)_\sigma
-
\log\left(\tfrac{1}{\varepsilon}\right).
\end{equation}

\med
\noindent\textit{\underline{One-Party Soft Covering Errors} $\bigl\|\CalT_2-\CalT_3\bigr\|_1$ and $\bigl\|\CalT_3-\widetilde{\CalT}\bigr\|_1$}: The last two terms are each controlled by a point-to-point soft-covering estimate.
We define our final intermediate state as
\begin{equation}
\label{Eqn:CalT3BeforeCOS}
\CalT_{3}
=
\frac{1}{K_1K_2}\sum_{k_1=1}^{K_1}\sum_{k_2=1}^{K_2}
\Big(
(\mathrm{id}_{B_1B_2R}\otimes \mathcal{D}_{k_1,k_2})
\circ
(\mathrm{id}_{R}\otimes \mathcal{E}^{(1)}_{k_1}
\otimes \mathcal{E}^{(2)}_{k_2})
\Big)
\Big(
(\,U^{\mathrm{ref}}_1\otimes I_{A_1}\,)
\varphi_{S_{k_1}}
(\,{U^{\mathrm{ref}}_1}^{\dagger}\otimes I_{A_1}\,)
\Big).
\end{equation}
Here, $U_i^{\mathrm{ref}}$ denotes the isometry acting on the reference system of $\varphi_{\rho^i}$ such that
\begin{equation}
\label{Eqn:IsometricExtensionState}
\varphi_{\rho}
=
\bigl(U_i^{\mathrm{ref}} \otimes I_{A_i}\bigr)\,
\varphi_{\rho^i}\,
\bigl({U_i^{\mathrm{ref}}}^{\dagger} \otimes I_{A_i}\bigr).
\end{equation}
We then observe that
\begin{equation}
\label{eq:tri3}
\bigl\|\CalT_2-\widetilde{\CalT}\bigr\|_1
\;\le\;
\bigl\|\CalT_2-\CalT_3\bigr\|_1
+
\bigl\|\CalT_3-\widetilde{\CalT}\bigr\|_1.
\end{equation}
\med
We first analyze the right-hand side of \eqref{eq:tri3}. Towards that end, we note that
\[
\begin{gathered}
\CalT_3-\widetilde{\CalT}
=
\frac{1}{K_1K_2}\sum_{k_1=1}^{K_1}\sum_{k_2=1}^{K_2}
\Big(
(\mathrm{id}_{B_1B_2R}\otimes \mathcal{D}_{k_1,k_2})
\circ
(\mathrm{id}_{R}\otimes \mathcal{E}^{(1)}_{k_1}
\otimes \mathcal{E}^{(2)}_{k_2})
\Big)
\Big(
(\,U^{\mathrm{ref}}_1\otimes I_{A_1}\,)
\varphi_{S_{k_1}}
(\,{U^{\mathrm{ref}}_1}^{\dagger}\otimes I_{A_1}\,)-\varphi_{\rho}
\Big).
\end{gathered}
\]
Thus, by monotonicity of the trace distance under CPTP maps and
\eqref{Eqn:IsometricExtensionState},
$\bigl\|\CalT_3-\widetilde{\CalT}\bigr\|_1
\le
\frac{1}{K_1}\sum_{k_1=1}^{K_1}
\bigl\|\varphi_{S_{k_1}}-\varphi_{\rho^1}\bigr\|_1.$
Moreover, \(\Tr S_{k_1}=1\) for every codebook realization. Hence
Lemma~\ref{lem:canonical-purif-stability} and Jensen's inequality give
\begin{equation}
\label{Eqn:T3TildeTCovering}
\mathbb{E}_{\mathcal C}\!\left[
\bigl\|\CalT_3-\widetilde{\CalT}\bigr\|_1
\right]
\le
\frac{2}{K_1}\sum_{k_1=1}^{K_1}
\sqrt{\mathbb{E}_{\mathcal C}\!\left[\|\rho^1-S_{k_1}\|_1\right]}
\le
4\sqrt{\varepsilon}.
\end{equation}
where the last inequality follows from Proposition~\ref{Prop:SmoothMultipartiteSoftCovering}
with \(k=1\), provided that
\begin{equation}
\label{eq:three}
R_1+\beta_{1}
>
I_{2}^{\varepsilon}(W_1:A_{2}R)_{\sigma_1}
+ 2 \log\frac{1}{\varepsilon}.
\end{equation}

Having bounded the second term in \eqref{eq:tri3}, we now focus on the remaining term $\norm{\CalT_{2}-\CalT_{3}}_{1}$. We are led to the following.
\begin{remark}
\label{Rem:OneShotDIS-COSTrick}
Recall characterizations of $\CalT_{2}$, $\CalT_{3}$ in \eqref{Eqn:CalT2BeforeCOS}, \eqref{Eqn:CalT3BeforeCOS} respectively. Employing these characterizations, it is difficult to bound $\norm{\CalT_{2}-\CalT_{3}}_{1}$ on the above. Therefore, we now derive alternate characterizations for $\CalT_{2}$ and $\CalT_{3}$ respectively. Deriving these alternate characterizations is the most technically delicate part of the proof and leads us to leveraging the \textit{compatibility operator sliding} (COS) trick devised in \cite{2026MMTIT_Pad}. In essence, the main issue stems from the fact that the encoding map
$\mathcal{E}^{(i)}_{k_i}$ involves the random inverse operator $S_{k_i}^{-1/2}$,
which prevents a direct concentration argument.
To circumvent this issue, we introduce an auxiliary family of completely positive maps $\widetilde{\mathcal{I}}_{k_i}^{(i)}$ in which this inverse does not appear. The purpose of the $\widetilde{\mathcal{I}}_{k_i}^{(i)}$ is to expose the
inverse-free expression obtained when the likelihood encoder acts on
its canonical proxy input.
\end{remark}
Led by the above remark, define the following completely positive maps:
\begin{equation}
\label{Eqn:InducedInstrumentki}
\begin{gathered}
\widetilde{\mathcal{I}}_{k_i}^{(i)}(\rho)
\define
\Tr_{E_i}\Bigg[
\sum_{m_i,b_i}
\Bigg(
\frac{
V_{w_i(k_i,m_i,b_i)}\,
\rho\,
V_{w_i(k_i,m_i,b_i)}^{\dagger}
}{
p_{W_i}\bigl(w_i(k_i,m_i,b_i)\bigr)M_i B_i
}
\otimes
\ket{m_i}\bra{m_i}
\Bigg)
\Bigg].
\end{gathered}
\end{equation}
With the above definition and using identity \eqref{Eq:TrickEk1Ik1}, whose veracity is proved in App.~\ref{AppSec:Calculus}, we are able to recharacterize $\CalT_{2},\CalT_{3}$ as
\begin{eqnarray}
\label{Eqn:CalT2AfterCOS}
\CalT_{2}
=
\frac{1}{K_1K_2}\sum_{k_1,k_2}
\Big(
(\mathrm{id}_{B_1B_2R}\otimes \mathcal{D}_{k_1,k_2})
\circ
(\mathrm{id}_{R}\otimes \widetilde{\mathcal{I}}_{k_1}^{(1)}
\otimes \mathcal{E}^{(2)}_{k_2})
\Big)
\Big(
(U_2^{\mathrm{ref}}\otimes I_{A_2})
\varphi_{S_{k_2}}
({U_2^{\mathrm{ref}}}^{\dagger}\otimes I_{A_2})
\Big), \\
\CalT_{3}
=
\frac{1}{K_1K_2}\sum_{k_1,k_2}
\Big(
(\mathrm{id}_{B_1B_2R}\otimes \mathcal{D}_{k_1,k_2})
\circ
(\mathrm{id}_{R}\otimes \widetilde{\mathcal{I}}_{k_1}^{(1)}
\otimes \mathcal{E}^{(2)}_{k_2})
\Big)
\big(
\varphi_{\rho}
\big).~~~~~~~~~~~
\nonumber
\end{eqnarray}
To now bound $\norm{\CalT_{2}-\CalT_{3}}_{1}$ on the above using the above recharacterizations, we leverage Lemma~\ref{Lem:AveragedInverseFreeContraction2}. Specifically, the maps $\widetilde{\mathcal{I}}_{k_i}$ satisfy an averaged
trace-norm contraction property even in the presence of an arbitrary reference system. With this tools, the analysis then get similar to those of the previous term.
Indeed, taking expectations and invoking
Lemma~\ref{Lem:AveragedInverseFreeContraction2}, together with monotonicity of the trace distance under
$\mathcal{E}^{(2)}_{k_2}$ and the same one-party soft-covering argument as above, we obtain
\begin{equation}
\label{eq:four}
\mathbb{E}_{\mathcal{C}}\!\left[\bigl\|\CalT_3-\CalT_2\bigr\|_1\right]
\le 4\sqrt{\varepsilon},
\qquad
\text{provided that}\qquad
R_2+\log B_2
>
I_{2}^{\varepsilon}(W_2:A_{1}R)_{\sigma_2}
+2\log\frac{1}{\varepsilon}.
\end{equation}
\noindent\textbf{Conclusion.} Before we collate bounds and complete the proof, a note on the similarity of our proofs provided for Thms.~\ref{Thm:OneShotMLIS} and \ref{Thm:DISCosetCdsInnBound} is in order.
\begin{remark}
    \label{Rem:DISOneShotProofStepsAndProxyStates}
    In contrast to proofs of Thms.~\ref{Thm:OneShotMLIS} and \ref{Thm:DISCosetCdsInnBound} wherein we stated the analysis by (i) explicitly characterizing the proxy states and (ii) operating the ISP to identify the intermediate states, here we did not explicit the proxy states at the beginning of this proof. However, the proof steps are identical and the proxy states are evident. In particular, $(\,U^{\mathrm{ref}}_i\otimes I_{A_i}\,)
\varphi_{S_{k_i}} (\,{U^{\mathrm{ref}}_i}^{\dagger}\otimes I_{A_i}\,)$ for $i \in [2]$ in \eqref{Eqn:CalT3BeforeCOS} and \eqref{Eqn:CalT2AfterCOS} respectively are the proxy states and the corresponding states $\CalT_{3},\CalT_{2}$, respectively, are the intermediate states.
\end{remark}
Now, finally collating \eqref{eq:tri}, \eqref{eq:tri2}, and \eqref{eq:tri3} with the corresponding bounds in \eqref{eq:sc_expect}, \eqref{eq:two}, \eqref{Eqn:T3TildeTCovering} and \eqref{eq:four}, we have
\[
\mathbb{E}_{\mathcal{C}}\left[\bigl\|\CalT-\widetilde{\CalT}\bigr\|_1\right]
\le
16\varepsilon+4\varepsilon+4\sqrt{\varepsilon}+4\sqrt{\varepsilon}
\le
28\sqrt{\varepsilon},
\]
where the last inequality use $0<\varepsilon<1$, if the communication rate pair $(R_{1},R_{2})$, common randomness rate pair $(C_{1},C_{2})$ and binning rate pair $(\beta_{1},\beta_{2})$ satisfy bounds in \eqref{eq:one}, \eqref{eq:two}, \eqref{eq:three} and \eqref{eq:four}. Adding the positivity constraints $\beta_i>0$ on $i \in [2]$ and performing a standard Fourier-Motzkin elimination \cite[App.~D]{BkNITElGamalKim_2011} it is easy to verify that we can find a positive pair $(\beta_{1},\beta_{2})$ of binning rates so long as the communication rate pair $(R_{1},R_{2})$ and common randomness rate pair $(C_{1},C_{2})$ satisfy the bounds stated in \eqref{Eqn:DISOneShotRateBounds} and $\sfB_6^\varepsilon\le0$. If instead $\sfB_6^\varepsilon\le0$, we choose $B_1=B_2=1$ and omit bin decoding, so that the term
$\|\CalT_1-\CalT_2\|_1$ is identically zero, hence the same region remains achievable. Finally choosing $\varepsilon=\widetilde\eta=\eta^2/28^2$ gives the desired error bound. This completes our proof.
\end{proof}

\subsection{Asymptotic Achievability}
\label{SubSec:InstSimltnAsymptoticIID}
We now record the corresponding inner bound for the distributed ICP in the asymptotic regime that is obtained via a natural generalization of the one stated in Thm.~\ref{Thm:DistInstSimltnOneShot}.

\begin{definition}[Asymptotic IID rate region]
\label{Defn:DISAsymptoticRateRegion}
Associate with
$(\ulineCalW,\ulineCalI,p_{Y|\ulineW})\in\ScrT(\CalJ)$
the set
$\ScrA_{0}(\ulineCalW,\ulineCalI,p_{Y|\ulineW})$
of all quadruples
$(R_1,R_2,C_1,C_2)\in\reals_{\geq0}^{4}$
satisfying
\begin{equation}
\label{Eqn:DISAsymptoticRateBounds}
\begin{gathered}
\begin{array}{@{}ll@{}}
R_1~>~
I(W_1:A_2R)_{\sigma_1}-I(W_1:W_2)_{\sigma},
&
R_1+R_2~>~
I(W_1:A_2R)_{\sigma_1}
+I(W_2:A_1R)_{\sigma_2}
-I(W_1:W_2)_{\sigma},
\\[0.5em]
R_2~>~
I(W_2:A_1R)_{\sigma_2}-I(W_1:W_2)_{\sigma},
&
R_1+R_2+C_1~>~
I(W_2:A_1R)_{\sigma_2}
+I(W_1:Y\ulineB R)_{\sigma}
-I(W_1:W_2)_{\sigma},
\\[0.5em]
R_1+C_1~>~
I(W_1:Y\ulineB R)_{\sigma}-I(W_1:W_2)_{\sigma},
&
R_1+R_2+C_2~>~
I(W_1:A_2R)_{\sigma_1}
+I(W_2:Y\ulineB R)_{\sigma}
-I(W_1:W_2)_{\sigma},
\\[0.5em]
R_2+C_2~>~
I(W_2:Y\ulineB R)_{\sigma}-I(W_1:W_2)_{\sigma},
&
R_1+R_2+C_1+C_2~>~
I(W_1W_2:Y\ulineB R)_{\sigma}.
\end{array}
\end{gathered}
\end{equation}
All information quantities are computed with respect to the states
$\sigma,\sigma_1,\sigma_2$ defined in
\eqref{Eqn:TestChannelJointStates}.
\end{definition}

\begin{theorem}[Asymptotic achievability]
\label{Thm:AsymptoticIIDInstSimulation}
The action of instrument $\CalJ$ on $\rho$ is asymptotically simulable
with communication rates $(R_1,R_2)$ and common-randomness rates
$(C_1,C_2)$ if there exists
$(\ulineCalW,\ulineCalI,p_{Y|\ulineW})\in\ScrT(\CalJ)$
such that
\[
(R_1,R_2,C_1,C_2)
\in
\ScrA_{0}(\ulineCalW,\ulineCalI,p_{Y|\ulineW}).
\]
\end{theorem}

\begin{proof}
Fix a test channel
$(\ulineCalW,\ulineCalI,p_{Y|\ulineW})\in\ScrT(\CalJ)$
witnessing membership in $\ScrA_0$, and apply
Theorem~\ref{Thm:DistInstSimltnOneShot} to
$\CalJ^{\otimes n}$, $\rho^{\otimes n}$, and the $n$-fold product test
channel. Choose a sequence $\varepsilon_n\downarrow0$ such that
$n^{-1}\log(1/\varepsilon_n)\to0$, and set
\[
\eta_n\define 28\sqrt{\varepsilon_n},
\qquad\text{so that}\qquad
\widetilde\eta_n=\frac{\eta_n^2}{28^2}=\varepsilon_n.
\]

By the AEPs for $D_2^{\varepsilon_n}$ and $D_H^{\varepsilon_n}$, the
normalized one-shot information quantities appearing in
\eqref{Eqn:DISOneShotInfoQuantities} converge to their corresponding
von Neumann information quantities appearing in
\eqref{Eqn:DISAsymptoticRateBounds}; the additive terms proportional to
$\log(1/\varepsilon_n)$ vanish after division by $n$.

For the total-rate constraint, the two asymptotic branches of the maximum
in the one-shot region reduce to
\[ I(W_1:Y\ulineB R)_{\sigma}+I(W_2:Y\ulineB R)_{\sigma}-I(W_1:W_2)_{\sigma} \qquad \text{and} \qquad I(\ulineW:Y\ulineB R)_{\sigma} \qquad \text{respectively.} \qquad\]
By the chain rule,
\[
I(W_1:Y\ulineB R)_{\sigma}
+I(W_2:Y\ulineB R)_{\sigma}
-I(W_1:W_2)_{\sigma}
=
I(\ulineW:Y\ulineB R)_{\sigma}
-I(W_1:W_2|Y\ulineB R)_{\sigma},
\]
and hence the second branch is always the larger one. The asymptotic
total-rate constraint is therefore precisely
\[
R_1+R_2+C_1+C_2>
I(\ulineW:Y\ulineB R)_{\sigma}.
\]

Consequently, every rate tuple in the strict region
$\ScrA_0(\ulineCalW,\ulineCalI,p_{Y|\ulineW})$ satisfies the normalized
one-shot inequalities defining $\ScrA_{\varepsilon_n}$ for all
sufficiently large $n$. Theorem~\ref{Thm:DistInstSimltnOneShot} therefore
yields an $\eta_n$-simulation of $\CalJ^{\otimes n}$ on
$\rho^{\otimes n}$ with communication rates $(nR_1,nR_2)$ and
common-randomness rates $(nC_1,nC_2)$. Since $\eta_n\to0$, asymptotic
simulation follows.
\end{proof}

\begin{example}
\label{Ex:DISBinningBB84}
{Let the source be}
$\rho^{A_1A_2}=|\Phi^+\rangle\langle\Phi^+|$, where
$|\Phi^+\rangle=(|00\rangle+|11\rangle)/\sqrt{2}$,
so that the reference $R$ can be taken to be trivial.
For $x,s\in\{0,1\}$, define
\[
|\phi_{x0}\rangle=|x\rangle,
\qquad
|\phi_{x1}\rangle
=\frac{|0\rangle+(-1)^x|1\rangle}{\sqrt{2}},
\qquad
\phi_{xs}=|\phi_{xs}\rangle\langle\phi_{xs}|.
\]
Alice$_i$'s local instrument has outcome $W_i=(X_i,S_i)$
and CP maps
\[
\mathcal{I}^{(i)}_{xs}(\tau)
=K_{xs}\tau K_{xs}^{\dagger}
=\frac12 \phi_{xs}\tau \phi_{xs},
\qquad
K_{xs}=\frac{\phi_{xs}}{\sqrt{2}}.
\]
These maps define an instrument since
$\sum_{x,s}K_{xs}^{\dagger}K_{xs}=I$.
The target global instrument is
\[
\mathcal{J}(\tau)
=\frac14\sum_{w_1,w_2}
(\phi_{w_1}\otimes \phi_{w_2})\tau
(\phi_{w_1}\otimes \phi_{w_2})
\otimes |w_1w_2\rangle\langle w_1w_2|^{Y},
\]
with $B_i$ retained by Alice$_i$ and the full classical outcome
$Y=(W_1,W_2)$ delivered to Bob. On this source, $S_1,S_2$ are independent uniform bits.
When $S_1=S_2$, the outcomes satisfy $X_1=X_2$ and are uniform;
when $S_1\neq S_2$, they are independent uniform bits.
More explicitly, $p_{W_1W_2}(w_1,w_2)
=\frac14
\langle\Phi^+|
\phi_{w_1}\otimes\phi_{w_2}
|\Phi^+\rangle
=\frac18
\bigl|\langle\phi_{w_1}|\phi_{w_2}\rangle\bigr|^2.$

\noindent
\begin{minipage}[c]{0.40\linewidth}
The joint distribution is given in the table on the right,
with rows indexed by $w_1=(x_1,s_1)$ and columns by
$w_2=(x_2,s_2)$.
\end{minipage}\hfill
\begin{minipage}[c]{0.58\linewidth}
\centering
\(\displaystyle
\begin{array}{c|cccc}
p_{W_1W_2}(w_1,w_2)
& (0,0) & (1,0) & (0,1) & (1,1)\\
\hline
(0,0)
& \frac{1}{8} & 0 & \frac{1}{16} & \frac{1}{16}\\[0.3em]
(1,0)
& 0 & \frac{1}{8} & \frac{1}{16} & \frac{1}{16}\\[0.3em]
(0,1)
& \frac{1}{16} & \frac{1}{16} & \frac{1}{8} & 0\\[0.3em]
(1,1)
& \frac{1}{16} & \frac{1}{16} & 0 & \frac{1}{8}
\end{array}
\)
\end{minipage}
\par
Each row and column sums to $1/4$, so
$H(W_1)=H(W_2)=2$.
There are four entries equal to $1/8$ and eight equal to $1/16$,
giving $H(W_1W_2)
=-4\cdot\frac18\log_2\frac18
 -8\cdot\frac1{16}\log_2\frac1{16}
=\frac72.$ Consequently, $I(W_1:W_2)_{\sigma}
=H(W_1)+H(W_2)-H(W_1W_2)
=2+2-\frac72
=\frac12.$ 
Moreover, conditioned on $W_i=w_i$, the state of $A_{3-i}$
in $\sigma_i$ is the pure state $\phi_{w_i}$,
whose average is $I/2$. Thus, the quantities entering
Theorem~\ref{Thm:AsymptoticIIDInstSimulation} are
\[
I(W_i:A_{3-i})_{\sigma_i}=1,\quad I(W_1:W_2)_{\sigma}=\frac12,\quad I(W_i:YB_1B_2)_{\sigma}=H(W_i)=2,\quad I(W_1W_2:YB_1B_2)_{\sigma}=H(W_1W_2)=\frac72.
\]
With $C_1=C_2=1$, all rate constraints therefore reduce to
\[
R_1>\frac12,
\qquad
R_2>\frac12,
\qquad
R_1+R_2>\frac32.
\]

Operationally, Alice$_i$ and Bob use their independent shared
randomness to generate the basis sequence $S_i^n$.
Alice$_i$ measures in these bases and sends a bin index of $X_i^n$.
Bob jointly recovers both outcome sequences using $S_1^n,S_2^n$. In this setting, $R_1=R_2=3/4+\epsilon$ is achievable for every $\epsilon>0$.
By comparison, separate point-to-point simulation of each local
instrument, with $A_{3-i}$ as its reference, requires
$R_i\geq I(W_i:A_{3-i})_{\sigma_i}=1$.
At the same common-randomness rates, this benchmark is attained
by sharing $S_i$ and transmitting $X_i$.
Binning therefore gives a strict saving for $0<\epsilon<1/4$,
approaching $25\%$ per encoder as $\epsilon\downarrow0$.
\end{example}

\section{Multi-Party Purity Distillation in the One-Shot Regime}
\label{Sec:InstSimulationWithIIDCodes}
\label{Sec:PurityDistillation}
We first formulate the multi-party purity distillation problem. In deriving inner bounds for the same, we leverage the distributed ISP designed in Sec.~\ref{Sec:DistInstrSimOneShot} and the results therein.

\subsection{Purity Distillation: Description and Problem Statement}
\label{SubSec:PurityProbStatements}
Distributed parties Alice$_{1}$, Alice$_{2}$, and Bob hold respective components of a tripartite state $\rho^{A_{1}A_{2}B}$. See Panel 1 in Fig.~\ref{Fig:QSIAndPD}. The goal is for each party to distill pure states by performing local unitary operations. To enable this, Alice$_1$ and Alice$_2$ and connected to Bon via a multiple-access dephasing channel. In addition, each party may borrow pre-prepared pure ancilla qubits to ``prime'' their \textit{distillation protocol}. This is also often referred to as catalytic usage of pre-prepared pure states. Naturally, the size of each borrowed ancilla is discounted from the net purity at the end. We now formalize a purity distillation protocol and quantify its performance by the total dimension of the distilled purity.

\begin{notation}
 \label{Rem:PDNotation}
Since we have three systems in PD, a singly \underline{underlined} symbol aggregates the first two components, while a doubly \doubleunderline{underlined} symbol aggregates all three components. For example, $\dulinel$ represents the triple $(l_{1},l_{2},l_{3})$. As will become clear below, system $B$ will correspond to index $3$ and Alice$_{i}$ with index $i$ for $i\in [2]$.
\end{notation}

\begin{definition}
\label{Defn:PurityDistProtocol}
Given Hilbert spaces $(\CalH_{A_{i}}: i \in [2],\CalH_{B})$, a tripartite purity distillation protocol $(\ulineM,\dulinel,\dulines,\CalU_{\ulineA},\CalN,\CalU_{B}) =(M_{1},M_{2},l_{1},l_{2},l_{3},s_{1},s_{2},s_{3},\CalU_{\ulineA},\CalN,\CalU_{B})$ comprises
\begin{itemize}
 \item Alice$_{i}$'s local unitary $\CalU_{A_{i}} : \CalH_{A_{i}} \otimes \CalH_{A_{i}^{0}} \rightarrow  \CalH_{A_{i}^{p}}\otimes \CalH_{A_{i}^{g}}\otimes \CalH_{M_{i}}$ with $\dim(\CalH_{A_{i}^{p}})=s_{i}$, $\dim(\CalH_{A_{i}^{0}})=l_{i}$ and $\dim(\CalH_{M_{i}})=M_{i}$ for $i\in [2]$,
 \item A multiple-access dephasing channel $\CalN: \CalL(\CalH_{\ulineM} ) \rightarrow \CalL(\CalH_{\ulineM})$, where $\CalH_{\ulineM} = \CalH_{M_{1}} \otimes \CalH_{M_{2}}$
 \item Bob's local unitary $\CalU_{B} : \CalH_{B} \otimes \CalH_{B_{0}}\otimes \CalH_{\ulineM} \rightarrow  \CalH_{B_{p}}\otimes \CalH_{B_{g}}$ with $\dim(\CalH_{B_{p}})=s_{3}$, $\dim(\CalH_{B_{0}})=l_{3}$.
\end{itemize}

\end{definition}
\begin{definition}
 \label{Defn:PurityDistillationInnerBound}
Given a tripartite quantum state $\rho^{A_{1}A_{2}B}$, net purity $P$ can be distilled with error at most $\eta$ and communication costs $(R_{1},R_{2})$ if there exists a tripartite purity distillation protocol $(\ulineM,\dulines,\dulinel,\CalU_{\ulineA},\CalN,\CalU_{B})$ such that $\log M_{i} \leq R_{i}$ for $i \in [2]$, $\displaystyle\sum_{i=1}^{3} (\log s_{i}-\log l_{i}) \geq P$ and
\vspace{-0.15in}
\[
  \big\|
  \CalT_{A_{1}^{p}A_{2}^{p}B_p} -
  \ket{0}\bra{0}_{A_{1}^{p}}
  \otimes
  \ket{0}\bra{0}_{A_{2}^{p}}
  \otimes
  \ket{0}\bra{0}_{B_p}
  \big\|_1 \le \eta.
\]
Here, $\CalT =
\CalU_B \circ \CalN \circ \CalU_{A_{1}}\CalU_{A_{2}}
(\rho^{A_1A_2BA_1^0A_2^0B_0})$ and $\rho^{A_1A_2BA_1^0A_2^0B_0} = \rho^{A_1A_2B}\otimes\ket{0}\bra{0}_{A^0_1}\otimes
\ket{0}\bra{0}_{A^0_2}\otimes
\ket{0}\bra{0}_{B_0}$. In this case, we say the triple $(P,R_{1},R_{2})$ is $\eta$-purity-achievable.

We say that net purity is asymptotically achievable at rate $P$ with communication rates $(R_{1},R_{2})$ if, for every $\eta>0$, there exists $N_{\eta} \in \mathbb{N}$ such that for all $n \geq N_{\eta}$, $n$ copies of the tripartite state $\otimes_{t=1}^{n}\rho^{A_{1}A_{2}B}$ can yield net purity at least $nP$ with error at most $\eta$ and communication costs $(nR_{1},nR_{2})$. In this case, we say that the triple $(P,R_{1},R_{2})$ is asymptotically purity-achievable.
\end{definition}

\subsection{Prior Work on Purity Distillation (PD) and Multi-Party Purity Distillation}
\label{SubSec:PriorWorkOnPD}
{The task of PD was first identified and modeled by Horodecki et.~al.~\cite{200303PRL_HorHorHorHorOppSenSen}, who viewed it as a problem of resource conversion. Adopting an info-theoretic framework and formulating this question in an asymptotic setting, Devetak~\cite{200506PRA_Dev} characterized the one-way distillable local purity of a bipartite state when unlimited one-way dephasing communication is available. Going further, Krovi and Devetak~\cite{200707PRA_KroDev} modeled communication over a \underline{limited rate} dephasing channel and thereby quantified the tradeoff between the classical communication and distillable purity. This formulation of Krovi and Devetak \cite{200707PRA_KroDev} has been now adopted in info-theoretic formulations \cite{202206ISIT_AtiPra,202306ISIT_ChaNemBus,202309ALL_ChaJaiSen} of PD, including this work of ours. The multi-party PD problem we study here has been studied by Atif and Pradhan~\cite{202206ISIT_AtiPra} who focus exclusively on an asymptotic regime analysis. While all of these above mentioned works \cite{200506PRA_Dev,200707PRA_KroDev,202206ISIT_AtiPra} address the PD problem in a conventional info-theoretic asymptotic regime, Chakraborty et.~al.~\cite{202306ISIT_ChaNemBus,202309ALL_ChaJaiSen} initiated study of PD in the more general and popular one-shot regime. Specifically, Chakraborty et.~al.~\cite{202306ISIT_ChaNemBus,202309ALL_ChaJaiSen} investigate both one-party and two-party PD in the one shot regime, and their latter work \cite{202309ALL_ChaJaiSen} reduced the number of catalytic ancilla bits employed by their earlier \cite{202306ISIT_ChaNemBus} protocol.}

As evident from the above summary, these works leave an investigation of the multi-party PD in the one-shot regime open. Moreover, since the protocol of Atif and Pradhan \cite{202206ISIT_AtiPra} rely on time sharing based distributed MSP and Chakraborty's \cite{202306ISIT_ChaNemBus,202309ALL_ChaJaiSen} findings do not address separable instruments over distributed components, their protocols and analysis remain outside the purview of the problem addressed herein - multi-party PD in the one-shot regime. Having said, we leverage the one-party purity concentration protocol of Chakraborty's \cite{202306ISIT_ChaNemBus,202309ALL_ChaJaiSen} in our work below.

\subsection{Inner Bounds for Purity Distillation}
Our main result is an inner bound for one-shot PD obtained by leveraging Thm~\ref{Thm:DistInstSimltnOneShot}. The conditional smoothed support max-entropy used below is defined in Eq.~\eqref{Eqn:TildeHmaxCondY}. One of the tools we will employ in proving the one-shot inner bound characterized in the following theorem is the one-party purity concentration protocol of Chakraborty, Nema and Buscemi \cite{202306ISIT_ChaNemBus}. This one-party purity concentration tasks and the latter are discussed at length in App.~\ref{Subsec:1partypurity}.

\begin{theorem}[Achievability]
\label{Thm:PurityAchievability}
Let $\rho^{A_1A_2B}\in\CalD(\CalH_{A_1}\otimes\CalH_{A_2}\otimes\CalH_B)$ and let
$0<\varepsilon<1/2$.
For the canonical purification $\varphi_{\rho}$ of $\rho^{A_1A_2B}$, fix local instruments with a single Kraus operator per outcome, $\mathcal{I}^{(i)}:\CalL(\CalH_{A_i})\to \CalL(\CalH_{A_i} \otimes \mathbb{C}^{\mathcal{Y}_i}),
\qquad i\in\{1,2\},$ with finite outcome sets $\mathcal{Y}_1,\mathcal{Y}_2$ and define the cq-states
\[
\CalT=\CalT_{RA_1A_2BY_1Y_2}
:=\big(\mathrm{id}_{RB}\otimes \mathcal{I}^{(1)}\otimes \mathcal{I}^{(2)}\big)(\varphi_{\rho}),
\qquad
\CalT_i=\CalT_{RA_1A_{2}BY_i}
:=\big(\mathrm{id}_{RBA_{3-i}}\otimes \mathcal{I}^{(i)}\big)(\varphi_{\rho}),
\quad i\in[2].
\]
Fix rates $(R_1,R_2,P)$ with $R_1,R_2\ge0$. If these rates satisfy
\begin{equation*}
\begin{gathered}
R_i>
I_{2}^{\varepsilon}(Y_i:RBA_{3-i})_{\CalT_i}
-I_H^{\varepsilon}(Y_1:Y_2)_\CalT
+3\log\frac{1}{\varepsilon}+2,
\qquad i\in[2],
\\[0.2em]
R_1+R_2>
I_{2}^{\varepsilon}(Y_1:RBA_{2})_{\CalT_1}
+I_{2}^{\varepsilon}(Y_{2}:RBA_1)_{\CalT_2}
-I_H^{\varepsilon}(Y_1:Y_2)_\CalT
+5\log\frac{1}{\varepsilon}+4,
\\[0.5em]
P<
\log d_{A_1}d_{A_2}d_B
-\widetilde{H}_{\max}^{\varepsilon}(A_1|Y_1)_\CalT
-\widetilde{H}_{\max}^{\varepsilon}(A_2|Y_2)_\CalT
-\widetilde{H}_{\max}^{\varepsilon}(B|Y_1Y_2)_\CalT
\\
\hspace{2em}
-I_{2}^{\varepsilon}(Y_1:RA_2B)_{\CalT_1}
-I_{2}^{\varepsilon}(Y_2:RA_1B)_{\CalT_2}
-4\log\frac{1}{\varepsilon}-4,
\end{gathered}
\end{equation*}
then $(P,R_1,R_2)$ is $14\varepsilon^{\frac{1}{4}}$-purity-achievable.
\end{theorem}

\begin{remark}[Comparison with Prior Work \cite{202206ISIT_AtiPra}]
\label{Rmk:AsymptoticPurityKappa}
For $\rho^S\in\CalD(\CalH_S)$, let
$\kappa(\rho^S)\define \log d_S-H(S)_\rho$.
Under the single-Kraus local measurement maps of
Theorem~\ref{Thm:AsymptoticPurityAchievability}, obtained from
Theorem~\ref{Thm:PurityAchievability} in the asymptotic IID regime,
its purity constraint can be rewritten as
\[
P
<
\kappa(\rho^{A_1})+\kappa(\rho^{A_2})+\kappa(\rho^B)
+I(B:Y_1Y_2)_{\CalT}.
\]
Indeed, conditional purity and invariance of the unmeasured
marginals imply
\[
I(Y_i:RA_{3-i}B)_{\CalT_i}
=
H(A_i)_\rho-H(A_i|Y_i)_{\CalT},
\qquad i\in\{1,2\}.
\]
Substituting these identities into the theorem and using
$H(B)_{\CalT}=H(B)_\rho$ gives the claimed rewriting. For matching local measurements, identify $U=Y_1$ and $V=Y_2$
in the unstructured branch $W=(U,V)$
of~\cite[Def.~4 and Thm.~1]{202203arXiv_AtiPra}.
Their parameter $b\in[0,1]$ yields a gain $bI(U:V)$
in each of the three communication thresholds and a purity
penalty of the same size.
Our bound uses the full communication gain $I(U:V)$
without this penalty. Its closure therefore contains all these branches.
\end{remark}

\begin{proof}
We first give an overview of the proof. We first allow unlimited common randomness and invoke the distributed instrument-simulation theorem for the two local instruments. We then implement the likelihood instruments coherently and attach conditional one-party purity distillation unitaries. Next, we identify a joint high-probability set of outcomes on which all three local concentration steps succeed simultaneously. Finally, we account for the pure workspaces required by the coherent encoders, eliminate the auxiliary bin sizes, combine the errors, and derandomize the common randomness. The details follow.

Set
\[
a_i\coloneqq I_2^\varepsilon(Y_i:RBA_{3-i})_{\CalT_i}
+2\log\frac1\varepsilon,
\qquad
b_{\mathrm{bin}}\coloneqq I_H^\varepsilon(Y_1:Y_2)_{\CalT}
-\log\frac1\varepsilon.
\]
For normalized states $\omega,\tau$, weighted Cauchy--Schwarz and
$\Tr\omega'\ge1-\varepsilon$ for every smoothing approximant give
$D_2^\varepsilon(\omega\|\tau)\ge2\log(1-\varepsilon)$.
Hence
$a_i\ge2\log((1-\varepsilon)/\varepsilon)>0$.
If $\varepsilon\ge1/196$, the purity bound in the theorem implies
$P<\log(d_{A_1}d_{A_2}d_B)$.
Taking $M_1=M_2=1$, no borrowed ancillas, and identity local unitaries,
with all original local systems designated as output registers,
achieves at least this target net purity with trace-norm error at most
$2\le14\varepsilon^{1/4}$.
We may therefore assume $0<\varepsilon<1/196$, so that
$0<28\sqrt\varepsilon<2$.

Let $\rho=\rho^{A_1A_2B}$ and denote the local marginals by
\[
\rho^{A_i}\coloneqq\rho^i=\Tr_{A_{3-i}B}[\rho],
\qquad i\in\{1,2\}.
\]
Write the local instruments in the form
\[
\CalI^{(i)}(\rho^i)
=
\sum_{y_i\in\CalY_i}V_{y_i}\rho^iV_{y_i}^\dagger\otimes\ketbra{y_i}_{Y_i},
\qquad
\sum_{y_i\in\CalY_i}V_{y_i}^\dagger V_{y_i}=I_{A_i}.
\]

\medskip
\noindent\textbf{Distributed instrument simulation.}
We first assume that the parties have unlimited common randomness; this resource will be removed at the end. At the simulation stage we apply the construction and error estimates in the proof of Theorem~\ref{Thm:DistInstSimltnOneShot} directly to $\CalI^{(1)}\otimes\CalI^{(2)}$, with no stochastic post-processing at the decoder. The construction is applied to the marginal on $A_1A_2$, while $RB$ is regarded as its purifying reference system. Since purifications differ only by an isometry on the reference, the error estimates and the relevant information quantities apply unchanged with $R$ there replaced by $RB$ here.

Alice$_1$ and Alice$_2$ send messages $(m_1,m_2)$ and Bob reconstructs outcomes $(\widehat Y_1,\widehat Y_2)$. Denote the corresponding simulated state by
\[
\widetilde{\CalT}
=
\frac1{K_1K_2}\sum_{k_1,k_2}
\Big[(\mathrm{id}_{A_1A_2BR}\otimes\CalD_{k_1,k_2})\circ
(\mathrm{id}_{RB}\otimes\CalE^{(1)}_{k_1}\otimes\CalE^{(2)}_{k_2})\Big](\varphi_\rho).
\]
Choose the positive integer message and bin sizes
\[
M_i=\min\left\{\left\lfloor2^{R_i}\right\rfloor,
                    \left\lfloor2^{a_i}\right\rfloor+1\right\},
\qquad
B_i=\left\lfloor\frac{2^{a_i}}{M_i}\right\rfloor+1.
\]
These choices satisfy $\log M_i\le R_i$ and
\[
a_i<\log(M_iB_i)\le a_i+1.
\]
Indeed, $M_iB_i>2^{a_i}$ by construction. If $M_i\le2^{a_i}$, then
$M_iB_i\le2^{a_i}+M_i\le2^{a_i+1}$; otherwise $B_i=1$ and
$M_i\le\lfloor2^{a_i}\rfloor+1<2^{a_i+1}$.
Whenever $B_i>1$, the second term in the minimum defining $M_i$ cannot
be selected, and therefore $M_i=\lfloor2^{R_i}\rfloor>2^{R_i-1}$. Thus
\[
B_i\le\frac{2^{a_i}}{M_i}+1
\le\frac{2^{a_i+1}}{M_i}
<2^{a_i-R_i+2}.
\]
The communication assumptions are
$R_i>a_i-b_{\mathrm{bin}}+2$ and
$R_1+R_2>a_1+a_2-b_{\mathrm{bin}}+4$.
If exactly one bin size exceeds one, its individual rate bound gives
$\log(B_1B_2)<b_{\mathrm{bin}}$; if both exceed one, the sum-rate bound
gives the same inequality.
In particular, if $b_{\mathrm{bin}}\le0$, both bin sizes must equal one.
When $B_1=B_2=1$, there is no hidden intra-bin index and no bin-decoding
error.
Choose the common-randomness sizes large enough to satisfy the remaining
covering inequalities. The construction and error estimates in the proof
of Theorem~\ref{Thm:DistInstSimltnOneShot} then give
\begin{equation}
\label{Eqn:PDSimulationBound}
\|\widetilde{\CalT}-\CalT\|_1\le28\sqrt\varepsilon.
\end{equation}

\medskip
\noindent\textbf{Coherent implementation and ancilla size.}
The likelihood encoder in the proof of Theorem~\ref{Thm:DistInstSimltnOneShot} has a communicated message register of dimension $M_i$ and an unresolved local branch indexed by $b_i\in[B_i]$, together with the internal $0$-th branch. After the latter is merged with the ordinary message $m_i=1$, a coherent dilation can retain the local branch in a register of dimension $B_i+1$. Hence a pure workspace of dimension $M_i(B_i+1)$ is sufficient. We include it in Alice$_i$'s borrowed ancilla $A_i^0$ in Definition~\ref{Defn:PurityDistProtocol}, and therefore subtract its logarithmic dimension from the net distilled purity. Since $B_i\ge1$,
\begin{equation}
\label{Eqn:PDWorkspaceBound}
\log\bigl(M_i(B_i+1)\bigr)
\le\log(M_iB_i)+1.
\end{equation}

\medskip
\noindent\textbf{Embedding conditional purity concentration.}
For each $y_i\in\CalY_i$ with positive probability, define
\[
\rho^i_{y_i}
\coloneqq
\frac{V_{y_i}\rho^iV_{y_i}^\dagger}
{\Tr[\rho^iV_{y_i}^\dagger V_{y_i}]},
\qquad i\in\{1,2\}.
\]
For $\uliney=(y_1,y_2)$, let $\rho^B_{\uliney}$ denote Bob's conditional
state in the ideal post-instrument state $\CalT$. Conditional states
associated with zero-probability outcomes may be chosen arbitrarily. To implement the conditional concentration unitaries on fixed registers,
set
\[
r_i\coloneqq
2^{\widetilde H_{\max}^{\varepsilon}(A_i|Y_i)_{\CalT}},
\qquad i\in[2],
\qquad
r_B\coloneqq
2^{\widetilde H_{\max}^{\varepsilon}(B|Y_1Y_2)_{\CalT}}.
\]
Fix pure concentration ancillas $F_i$ and $F_B$ of dimensions $r_i$ and
$r_B$, respectively, independently of the outcomes. These ancillas are
included in the total borrowed registers $A_i^0$ and $B_0$; the coherent
encoder workspaces accounted for in \eqref{Eqn:PDWorkspaceBound} are
additional. For every outcome satisfying
$\widetilde H_{\max}^{\varepsilon}(A_i)_{\rho^i_{y_i}}\le\log r_i$,
choose a projector $\Pi_{A_i}^{y_i}$ such that
\[
\Tr[\Pi_{A_i}^{y_i}\rho^i_{y_i}]\ge1-\varepsilon,
\qquad
\rank(\Pi_{A_i}^{y_i})\le r_i.
\]
The construction in the proof of Proposition~\ref{Thm:PurityConcentration}
then provides a unitary $U^{(i)}_{y_i}$ on $A_iF_i$ whose output
factorization has pure-register dimension $d_{A_i}$ and garbage dimension
$r_i$, and which maps
$\operatorname{ran}(\Pi_{A_i}^{y_i})\otimes\ket0_{F_i}$
into the subspace with fixed pure output $\ket0_{A_i^p}$.
The same construction gives $U^{(B)}_{\uliney}$ on $BF_B$, with
pure-register dimension $d_B$ and garbage dimension $r_B$, whenever
$\widetilde H_{\max}^{\varepsilon}(B)_{\rho^B_{\uliney}}\le\log r_B$.
For all remaining outcomes, choose arbitrary unitaries between the same
input and output spaces. Thus the registers and their dimensions are
independent of the outcomes, and the controlled unitary below is
well-defined. For a fixed outcome pair, the ideal product operation is
\[
U_{\uliney}^{\mathrm{pc}}
\coloneqq
U^{(1)}_{y_1}\otimes U^{(2)}_{y_2}\otimes U^{(B)}_{\uliney},
\qquad
U^{\mathrm{pc}}
\coloneqq
\sum_{\uliney}U_{\uliney}^{\mathrm{pc}}\otimes\ketbra{\uliney}_{Y_1Y_2}.
\]
For the actual protocol, Alice$_i$ must control $U^{(i)}_{y_i}$ using
the local branch of her likelihood encoder. This branch is intentionally
coarse-grained in the simulation criterion, so instead of trying to
recover it from the simulated classical output we absorb the controlled
purity distillation unitary directly into the local target instrument.
Namely, we apply Theorem~\ref{Thm:DistInstSimltnOneShot} to the instruments
\[
\widehat{\CalI}^{(i)}(S)
=
\sum_{y_i\in\CalY_i}
U^{(i)}_{y_i}(V_{y_i}\otimes I_{F_i})S
(V_{y_i}^\dagger\otimes I_{F_i}){U^{(i)}_{y_i}}^\dagger
\otimes\ketbra{y_i}_{Y_i},
\]
with each $F_i$ initially in $\ketbra0_{F_i}$.
Because $U^{(i)}_{y_i}$ acts only on Alice$_i$'s quantum output, it leaves
the outcome law, the binning quantity $I_H^\varepsilon(Y_1:Y_2)$, and
the communication-only quantities $I_2^\varepsilon(Y_i:RBA_{3-i})$
unchanged; appending the fixed pure ancillas does not change these
quantities either. The common-randomness covering quantities may change,
but common randomness is unlimited at this stage. Hence the same
$M_i,B_i$ still give the simulation bound
\eqref{Eqn:PDSimulationBound} for the modified target.

\begin{remark}
\label{Rmk:EmbeddingConditionalPurityConcentration}
A key role of distributed instrument simulation in the present PD protocol is that the senders' conditional purity distillation operations can be incorporated directly into the target instrument. For each outcome $y_i$, we compose the corresponding local instrument branch with $U^{(i)}_{y_i}$, obtaining the modified instrument $\widehat{\CalI}^{(i)}$. Applying Theorem~\ref{Thm:DistInstSimltnOneShot} to $\widehat{\CalI}^{(1)}\otimes\widehat{\CalI}^{(2)}$ then guarantees faithful simulation of both the classical outcomes $Y_1,Y_2$ and the quantum outputs after the senders' concentration operations. This is precisely where the stronger guarantee of instrument simulation is essential: measurement simulation alone preserves the outcome--reference correlations, but does not in general guarantee faithful reproduction of the accessible post-measurement quantum systems needed for PD.
\end{remark}
Operationally, before transmitting the message register, Alice$_i$ coherently applies the purity distillation unitary associated with her local encoder branch, using the message and intra-bin registers as controls. She then sends the message register to Bob and retains the intra-bin register as part of her garbage system, while Bob
applies $U^{(B)}_{\widehat y_1,\widehat y_2}$ using the received messages. More explicitly, for the deterministic function $f$ defining $\CalD_{k_1,k_2}$, provided by Theorem~\ref{Thm:DistInstSimltnOneShot} (see in Appendix \ref{AppSec:binning} for it's detailed construction), Bob applies the conditioned unitary
\[
U^{B}_{k_1 k_2}
\define
\sum_{m_1,m_2}
U^{(B)}_{f(k_1,k_2,m_1,m_2)}
\otimes
\ketbra{m_1,m_2}.
\]
The received message registers themselves serve as the control registers;
the decoded pair $(\widehat y_1,\widehat y_2)$ need never be physically materialized in an additional
register. At the level of the analysis, Bob's operation is
equivalent, after discarding the classical labels, to conditioning
$U^{(B)}_{\underline y}$ on the decoded outcome
$\underline y=f(\ulinek, \ulinem)$. Applying this conditional operation
to both the ideal and simulated states cannot increase their trace distance.

\begin{remark}
{The improvement identified in Remark~\ref{Rmk:AsymptoticPurityKappa} stems from avoiding an additional decoding workspace at Bob. In the unstructured protocol of~\cite[Sec.~IV-D and Rem.~1]{202203arXiv_AtiPra}, the receiver explicitly reconstructs the unbinned codeword indices using additional pure ancillas, whose cost introduces a binning-dependent purity penalty. In contrast, our protocol uses the received messages directly as controls for the conditional purity distillation unitary $U^{(B)}_{f(k_1,k_2,m_1,m_2)}$, without materializing the decoded outcomes in a separate register. The intra-bin registers retained by the senders are already accounted in the encoder workspace cost. Consequently, we retain the communication gain from correlated binning without incurring an additional decoding cost in purity. In the IID limit, this removes the penalty $-I(Y_1:Y_2)_T$ associated with full binning in the corresponding unstructured bound of \cite{202206ISIT_AtiPra}, for the same local measurement choice and communication constraints.}
\end{remark}

\noindent\textbf{Purity extraction on a joint good set.}
Set $a\coloneqq\sqrt\varepsilon$ and $b\coloneqq\varepsilon=a^2$. By the definition of $\widetilde H_{\max}^{b}(A_i|Y_i)_{\CalT}$, there exists $\mathcal G_i\subseteq\CalY_i$ such that
\[
p_{Y_i}(\mathcal G_i)\ge1-a,
\qquad
\widetilde H_{\max}^{b}(A_i)_{\rho^i_{y_i}}
\le
\widetilde H_{\max}^{b}(A_i|Y_i)_{\CalT}
\quad\text{for all }y_i\in\mathcal G_i.
\]
Thus, on $\mathcal G_i$, Alice$_i$ can extract the common amount $P_{A_i}
=
\log d_{A_i}-\widetilde H_{\max}^{b}(A_i|Y_i)_{\CalT}.$ Similarly, there is a set $\mathcal G_B\subseteq\CalY_1\times\CalY_2$ with
$p_{Y_1Y_2}(\mathcal G_B)\ge1-a$ on which Bob can extract $P_B
=
\log d_B-\widetilde H_{\max}^{b}(B|Y_1Y_2)_{\CalT}.$ For every good outcome, Proposition~\ref{Thm:PurityConcentration} provides a projector supporting probability at least $1-b$ and of rank no larger than the corresponding smoothed support max-entropy. Denote these projectors by $\Pi_{A_i}^{y_i}$ and $\Pi_B^{y_1,y_2}$. Introduce the joint good set
\[
\mathcal G
\define
\{(y_1,y_2):y_1\in\mathcal G_1,\ y_2\in\mathcal G_2,\ (y_1,y_2)\in\mathcal G_B\}.
\]
By the union bound,
\begin{equation*}
p_{Y_1Y_2}(\mathcal G^c)\le3a.
\end{equation*}
Let $\CalT^0$ denote the ideal post-instrument state $\CalT$,
together with the pure ancillary systems required by the three
local purity concentration protocols, immediately before the
controlled purity concentration unitaries are applied.
Let
$\CalT^0_{\mathcal G}$ denote its subnormalized restriction to the
outcomes in $\mathcal G$. Define the controlled projectors
\[
\Pi_i
\define
\sum_{y_i\in\mathcal G_i}\Pi_{A_i}^{y_i}\otimes\ketbra{y_i}_{Y_i},
\qquad
\Pi_B
\define
\sum_{(y_1,y_2)\in\mathcal G_B}\Pi_B^{y_1,y_2}\otimes\ketbra{y_1,y_2}_{Y_1Y_2},
\qquad
\Pi\define\Pi_1\Pi_2\Pi_B.
\]
The three projectors commute and, on $\CalT^0_{\mathcal G}$, their total failure weight is at most $3b$. Hence the gentle measurement lemma gives
\[
\|\CalT^0_{\mathcal G}-\Pi\CalT^0_{\mathcal G}\Pi\|_1
\le2\sqrt{3b}=2\sqrt3\,a.
\]
By construction, after the controlled purity distillation unitaries the range of $\Pi$ is mapped to the fixed pure output
\[
\Phi
\define
\ketbra0_{A_1^p}\otimes\ketbra0_{A_2^p}\otimes\ketbra0_{B^p},
\]
up to discarded garbage systems. The projector failure contributes at most $3b$, while the complement of the good set contributes at most $2p(\mathcal G^c)\le6a$. Therefore, writing $\ScrO$ for the three controlled purity distillation operations followed by discarding all garbage registers,
\begin{equation}
\label{Eqn:GlobalPurityIdeal}
\|\ScrO(\CalT^0)-\Phi\|_1
\le2\sqrt3\,a+3b+6a
\le15\sqrt\varepsilon.
\end{equation}
This establishes simultaneous purity extraction for the ideal post-instrument state.

\medskip
\noindent\textbf{Net purity and elimination of the bin sizes.}
The ideal extraction gives $P_{A_1}+P_{A_2}+P_B$ pure qubits before accounting for the two coherent encoder workspaces. From \eqref{Eqn:PDWorkspaceBound}, the net purity is therefore at least
\begin{equation}
\label{Eqn:PDNetBeforeElimination}
P_{A_1}+P_{A_2}+P_B
-\log(M_1B_1)-\log(M_2B_2)-2.
\end{equation}
The integer choices above satisfy $\log(M_iB_i)\le a_i+1$. Consequently,
\begin{equation}
\label{Eqn:PDBinCost}
\log(M_1B_1)+\log(M_2B_2)
\le
I_2^\varepsilon(Y_1:RBA_2)_{\CalT_1}
+I_2^\varepsilon(Y_2:RBA_1)_{\CalT_2}
+4\log\frac1\varepsilon+2.
\end{equation}
Substituting the definitions of $P_{A_1},P_{A_2},P_B$ into \eqref{Eqn:PDNetBeforeElimination} and using \eqref{Eqn:PDBinCost} gives the purity bound in the theorem, with the total constant penalty equal to four.
Finally, the ideal purity distillation error \eqref{Eqn:GlobalPurityIdeal} and the instrument-simulation error \eqref{Eqn:PDSimulationBound} add by the triangle inequality and contractivity, yielding
\begin{equation}
\label{Eqn:FinalGlobalPurity}
\|\widetilde{\CalT}^{A_1^pA_2^pB^p}-\Phi\|_1
\le28\sqrt\varepsilon+15\sqrt\varepsilon
=43\sqrt\varepsilon.
\end{equation}

\smallskip
\noindent\textbf{Derandomization.}
For each fixed pair $(k_1,k_2)$, let
$\mathcal E_{k_1k_2}$ denote the overall CPTP map from the initial
state $\rho$ to the final distilled registers $A_1^pA_2^pB^p$,
including the encoding operations, the communication and decoding
operations, the controlled purity concentration protocols, and the
tracing out of all garbage systems. Equation~\eqref{Eqn:FinalGlobalPurity}
can then be written as
\[
\left\|
\frac{1}{K_1K_2}
\sum_{k_1,k_2}
\mathcal E_{k_1k_2}(\rho)
-
\Phi
\right\|_1
\leq
\varepsilon_0,
\qquad
\varepsilon_0\coloneqq43\sqrt{\varepsilon}.
\]
If $\varepsilon_0\geq2$, then the final estimate is trivial, since
every trace distance is at most $2$ and $2\leq14\varepsilon^{1/4}$ whenever $43\sqrt{\varepsilon}\geq2$. In this case, any pair
$(k_1,k_2)$ may be fixed. We may therefore assume
$\varepsilon_0<2$.
Applying the fidelity--trace-distance relation in
Lemma~\ref{Lem:TraceDistancePureState} to the averaged output state
gives
\[
\frac{1}{K_1K_2}
\sum_{k_1,k_2}
\Tr\!\left[
\Phi\,\mathcal E_{k_1k_2}(\rho)
\right]
\geq
\left(
1-\frac{\varepsilon_0}{2}
\right)^2.
\]
Therefore, there exist indices $(k_1^\ast,k_2^\ast)$ such that $\Tr\!\left[
\Phi\,\mathcal E_{k_1^\ast k_2^\ast}(\rho)
\right]
\geq
\left(
1-\frac{\varepsilon_0}{2}
\right)^2.$ Applying the same lemma once more, we obtain
\[
\left\|
\mathcal E_{k_1^\ast k_2^\ast}(\rho)
-
\Phi
\right\|_1
\leq
2\sqrt{
1-
\Tr\!\left[
\Phi\,\mathcal E_{k_1^\ast k_2^\ast}(\rho)
\right]
}
\]
and hence
\[
\left\|
\mathcal E_{k_1^\ast k_2^\ast}(\rho)
-
\Phi
\right\|_1
\leq
2\sqrt{
1-
\left(
1-\frac{\varepsilon_0}{2}
\right)^2
}
\leq
2\sqrt{\varepsilon_0}
=
2\sqrt{43}\,\varepsilon^{1/4}
\leq
14\,\varepsilon^{1/4}.
\]
Thus, common randomness is removed while keeping the same amount
of distilled purity.
\end{proof}

The preceding one-shot inner bound yields the following asymptotic inner bound by applying
Theorem~\ref{Thm:PurityAchievability} to product instruments and product states.

\begin{theorem}[Asymptotic achievability]
\label{Thm:AsymptoticPurityAchievability}
Let $\rho^{A_1A_2B}\in\CalD(\CalH_{A_1}\otimes\CalH_{A_2}\otimes\CalH_B)$, and let
$\varphi_\rho$ denote its canonical purification. Fix local instruments with a single Kraus operator per outcome,
$\mathcal{I}^{(i)}:\CalL(\CalH_{A_i})\to \CalL(\CalH_{A_i}\otimes\mathbb C^{\CalY_i})$,
$i\in\{1,2\}$, with finite outcome sets $\CalY_1,\CalY_2$, and define
\begin{eqnarray}
\CalT
&\define&
\CalT^{RA_1A_2BY_1Y_2}
:=
\bigl(\mathrm{id}_{RB}\otimes\mathcal I^{(1)}\otimes\mathcal I^{(2)}\bigr)(\varphi_\rho),
\qquad
\CalT_i
\define
\CalT_i^{RA_1A_2BY_i}
:=
\bigl(\mathrm{id}_{RBA_{3-i}}\otimes\mathcal I^{(i)}\bigr)(\varphi_\rho),
\qquad i\in\{1,2\}.
\nonumber
\end{eqnarray}
If the rates $(R_1,R_2,P)$ satisfy
\begin{eqnarray}
R_i
&>&
I(Y_i:RBA_{3-i})_{\CalT_i}-I(Y_1:Y_2)_{\CalT},
\qquad i\in\{1,2\},
\nonumber\\
R_1+R_2
&>&
I(Y_1:RBA_2)_{\CalT_1}+I(Y_2:RBA_1)_{\CalT_2}-I(Y_1:Y_2)_{\CalT},
\nonumber\\
P
&<&
\log d_{A_1}d_{A_2}d_B
-H(A_1|Y_1)_{\CalT}
-H(A_2|Y_2)_{\CalT}
-H(B|Y_1Y_2)_{\CalT}
\nonumber
-I(Y_1:RA_2B)_{\CalT_1}
-I(Y_2:RA_1B)_{\CalT_2},
\nonumber
\end{eqnarray}
then $(P,R_1,R_2)$ is asymptotically purity-achievable.
Here, for classical registers, the conditional entropies are understood as averages over the corresponding classical outcomes.
\end{theorem}

\begin{proof}
Apply Theorem~\ref{Thm:PurityAchievability} to $n$ copies of the state and to the product measurement maps
$(\mathcal I^{(1)})^{\otimes n}$ and $(\mathcal I^{(2)})^{\otimes n}$. Choose a sequence
$\varepsilon_n\downarrow 0$ such that $n^{-1}\log(1/\varepsilon_n)\to 0$, for example
$\varepsilon_n=2^{-\sqrt n}$. Dividing the one-shot inequalities by $n$, the standard AEPs for the smooth collision and hypothesis-testing information quantities give
\begin{eqnarray}
\frac1n I_2^{\varepsilon_n}(Y_i^n:R^nB^nA_{3-i}^n)_{\CalT_i^{\otimes n}}
&\longrightarrow&
I(Y_i:RBA_{3-i})_{\CalT_i},
\qquad
\frac1n I_H^{\varepsilon_n}(Y_1^n:Y_2^n)_{\CalT^{\otimes n}}
\longrightarrow
I(Y_1:Y_2)_{\CalT}.
\nonumber
\end{eqnarray}
Moreover, Lemma~\ref{Lem:UpperAEPConditionalTildeHmax} gives the sufficient one-sided bounds
\[
\limsup_{n\to\infty}\frac1n
\widetilde H_{\max}^{\varepsilon_n}(A_i^n|Y_i^n)_{\CalT^{\otimes n}}
\le H(A_i|Y_i)_{\CalT},
\qquad \text{and} \qquad
\limsup_{n\to\infty}\frac1n
\widetilde H_{\max}^{\varepsilon_n}(B^n|Y_1^nY_2^n)_{\CalT^{\otimes n}}
\le H(B|Y_1Y_2)_{\CalT}.
\]
The additive terms proportional to $\log(1/\varepsilon_n)$ and the constant term vanish after normalization. Hence, for all sufficiently large $n$, the normalized rates satisfying the displayed strict inequalities obey the one-shot conditions of Theorem~\ref{Thm:PurityAchievability}. Since the resulting error $14\varepsilon_n^{1/4}$ tends to zero, the claimed asymptotic purity-achievability follows.
\end{proof}


\section{Distributed Instrument Simulation with Coset Codes}
\label{Sec:DISCosetCodes}
\label{Sec:CosetDIS}
{As discussed in Sec.~\ref{Sec:Introduction}, the problem of distributed ICP is richer than ICP in the PTP scenario. Firstly, the former is open even in the asymptotic regime with classical sources (commuting states) \cite[Ch.~12]{BkNITElGamalKim_2011}. Secondly, the distributed nature of the considered scenario with entangled components fosters the use of codes possessing richer algebraic closure properties. The early work of K\"orner and Marton \cite{197903TIT_KorMar} in the specific case of reconstructing modulo$-2$ sum of distributed binary sources has spurred the use of coset codes \cite{201103TIT_KriPra,201505TIT_SahPra} in several problems involving distributed encoders. As recognized in \cite{202504TIT_AtiPra}, we formalize our findings with likelihood instruments to develop a distributed ISP with coset codes that can yield a rate region larger than that achievable via unstructured codes, i.e. Thm.~\ref{Thm:AsymptoticIIDInstSimulation}. We should note here that our inner bound characterized in Thm.~\ref{Thm:DISCosetCdsInnBound} coincides with that proved achievable in \cite{202504TIT_AtiPra}. The main goal of this section is to demonstrate that our likelihood instrument based framework is versatile enough to employ codes with algebraic properties.}

\subsection{Preliminaries: Notation and Union Coset Codes}
\label{SubSec:DISCCPrelimsAndUCC}
  The use of \textit{coset codes} and our study here in the asymptotic regime distinguishes this section from the rest of our findings. This requires us to setup some notation specific to this section and furthermore introduce the tool of UCCs.
  \begin{notation}
   \label{Not:DISCCBasicMatrixNotation}
   We will be dealing with \textit{linear} and \textit{coset codes} built over finite fields possessing algebraic structure. Throughout this section $q \in \naturals$ denotes power of a prime, $\CalF_{q} \define \{0,1,\cdots,q-1\}$ denotes the field of cardinality $q$ with $\oplus,\cdot$ denoting field addition, scalar multiplication in $\CalF_{q}$ respectively and $0,1$ denoting additive and multiplicative identities respectively. A $(k,n)$ \textit{linear code} over $\CalF_{q}$ is the collection of vectors in the rowspace of a \textit{generator matrix} $g \in \CalF_{q}^{k \times n}$. To list the codewords with a compact notation, we adopt the following notation. Whenever $b \in [q^{k}]$, we let
   \begin{eqnarray*}
    b g  = \left[ b_{1}~b_{2}~\cdots~b_{k} \right]\left[
\!\!\begin{array}{c}
\ulineg_{1}\\
\ulineg_{2}\\
\vdots\\
\ulineg_{n}
\!\!\end{array} \right] =\sum_{t=1}^{k}b_{t}\ulineg_{t}= \left[ b_{1}~b_{2}~\cdots~b_{k} \right]\left[
\!\!\begin{array}{cccc}
g_{11}&g_{12}&\cdots&g_{1n}\\
g_{21}&g_{22}&\cdots&g_{2n}\\
\vdots&\vdots&\ddots&\vdots\\
g_{n1}&g_{n2}&\cdots&g_{nn}
\!\!\end{array} \right] = \left( \sum_{t=1}^{k}b_{t}g_{k1}~\cdots~\sum_{t=1}^{k}b_{t}g_{kn}\right)
   \end{eqnarray*}
denote the codeword of the linear code corresponding to linear combination $b \in [q^{k}]$. Here, we have interpreted $b$ to be an element of $\CalF_{q}^{k}$ thus providing us with $k$ coefficients in $\CalF_{q}$. This interpretation is perfectly justified as $[q^{k}]$ and $\CalF_{q}^{k}$ are in a bijection. Moreover $b_{1},\cdots,b_{k}$ maybe viewed as the $q-$ary representation of $b \in \{0,1,\cdots q^{k}-1\}$. As will become evident, this natural interpretation will enable us adopt notation identical to that in other sections with regard to code description. Thus, throughout this section, \textbf{any element in $[q^{k_{i}}]$ used in a role analogous to $b \in [q^{k}]$ above is interpreted to be an element in $\CalF_{q}^{k_{i}}$ with well defined $k_{i}$ coefficients in $\CalF_{q}$}.
  \end{notation}
Employing the above notation, we now define a \textit{union coset code}.

\begin{definition}
 \label{Defn:DISCosetCdsUCC}
A union coset code (UCC) $(n,\beta,r,g,\upsilon)$ over $\CalW=\CalF_{q}$ is specified via (i) a generator matrix $g \in \CalW^{n\beta \times n}$, (ii) a map $\upsilon :  [q^{nr}] \rightarrow \CalW^{n}$ and comprises of the collection
\begin{eqnarray*}
\mathsf{u}(g,\upsilon) \define \{ \boldw(m, b) \define w^{n}(m, b) \define  b g \oplus \upsilon(m) \in \CalW^{n}:  b \in [q^{k}], m \in [q^{nr}]\}.
\end{eqnarray*}
Since for any $d^{n} \in \CalW^{n}$, the collection $\mathtt{c}_{d^{n}}\define \{ u^{n}( b) \define  b g \oplus d^{n} \in \CalW^{n}:  b = [b_{1},b_{2},\cdots,b_{n\beta}]  \in \CalW^{n\beta}\}$ of $q^{n\beta}$ codewords is a shift of the linear rowspace of $g$, ${\mathtt{c}_{d^{n}}}$ is referred to as the coset of a linear code with shift $d^{n}$. This justifies referring to $\mathsf{u}(g,\upsilon)$ as a UCC since it is a union of cosets.
\end{definition}

\begin{notation}
\label{Not:DISCCFieldNotations}
In contrast to all other sections, we will be dealing here with $n-$length codewords and operators in cartesian product and tensor product spaces respectively. We adopt a \textbf{bold} notation to denote $n-$letter objects. For any set $\CalW$, any Hilbert space $\CalH_{A}$, we let $\boldCalW = \CalW^{n} = \displaystyle\times_{t=1}^{n}\CalW$ and $\CalH_{\boldA} = \otimes_{t=1}^{n}\CalH_{A}$. For any probability mass function (PMF) $p_{W} \in \CalD(\CalW)$ and any density operator $\rho \in \CalD(\CalH_{A})$, we let $\boldp_{W}=p_{W}^{n}=\prod_{t=1}^{n}p_{W}$ and $\boldrho=\rho^{\otimes n} \in \CalD(\CalH_{\boldA})$ denote the $n-$letter product PMF tensor product density operators respectively. $\boldw \in \boldCalW$ denotes a generic element in $\CalW^{n}$. $I_{\boldA}$ denotes identity on $\CalH_{\boldA}$ and for any operator $T:\CalH_{A} \rightarrow \CalH_{B}$, we let $\boldT:\CalH_{\boldA} \rightarrow \CalH_{\boldB}$ be defined as $\boldT=\otimes_{i=1}^{n}T$. Lastly, we let $\boldw(m,b)=w^{n}(m,b)$ where $w^{n}(m,b) = b g\oplus \upsilon(m) \in \CalW^{n}$ is the codeword corresponding to linear combination $b \in \CalF_{q}^{n\beta}$ of a UCC with generator matrix $g \in \CalF^{n\beta \times n}$ and map $\upsilon:[q^{nr}] \rightarrow  \CalF_{q}^{n}$ defined in Defn.~\ref{Defn:DISCosetCdsUCC}. We note that $\boldp_{W}(\boldw(m,b)) = \prod_{t=1}^{n}p_{W}(w(m,b)_{t})$, where $\boldw(m,b) = w(m,b)_{1},\cdots,w(m,b)_{n}$.
\end{notation}

\subsection{Characterization of an Inner Bound Achievable for Distributed ISP via Coset Codes}
\label{SubSec:DISCCInnerBoundCharacterization}
We begin by characterizing the set of all test channels that we shall work with.

\begin{definition}
 \label{Defn:DISCosetCdsTestChnls}
For an instrument $\mathcal{J}:\ \CalL(\CalH_{\ulineA}) \longrightarrow \CalL(\CalH_{\ulineB}\otimes \mathbb{C}^{|\mathcal{Y}|})$
specified through a collection $\{\CalJ_{y}:\CalL(\CalH_{\ulineA})\to \CalL(\CalH_{\ulineB}):y\in\CalY\}$ of CP TNI maps,
let $\ScrT_{\tinyoplus}(\CalJ)$ denote the collection of all triples $(\ulineCalW,\ulineCalI,p_{Y|W_{1}\oplus W_{2}})$, where $\CalW \define \CalW_{1}=\CalW_{2}=\CalF_{q}$ is a finite field of size $q$, $\ulineCalW=(\CalW_1,\CalW_2)$, $\ulineCalI=(\CalI^{(1)},\CalI^{(2)})$,
$\CalI^{(i)}:\CalL(\CalH_{A_i})\to \CalL(\CalH_{B_i}\otimes \mathbb{C}^{\CalW_i})$ are instruments for $i\in[2]$
each specified through a collection
$\{\CalI_{w_i}^{(i)}:\CalL(\CalH_{A_i})\to \CalL(\CalH_{B_i}):w_i\in\CalW_i\}$ of CP TNI maps,
and $p_{Y|W_{1}\oplus W_{2}}$ is a stochastic matrix such that
\begin{eqnarray*}
 \CalJ_{y} = \sum_{w \in \CalW = \fieldq}p_{Y|W_{1}\oplus W_{2}}(y|w) \sum_{w_{1},w_{2} \in \CalW}\CalI_{w_{1}}^{(1)} \otimes \CalI_{w_{2}}^{(2)}\mathds{1}_{\{w=w_{1}\oplus w_{2}\}}.
\end{eqnarray*}
holds for all $y\in\CalY$. For any $(\ulineCalW,\ulineCalI,p_{Y|W_{1}\oplus W_{2}})\in\ScrT_{\tinyoplus}(\CalJ)$, define
\begin{eqnarray}
\label{Eqn:DISCosetCdsTstChnlJntSts}
\sigma^{RB_{1}B_{2}\ulineW Y}
\define
\left(\mathrm{id}_{R}\otimes\olinescre^{Y|{W_{1}\oplus W_{2}}}_{p}\otimes \ulineCalI\right)\{{\varphi_{\rho}}\},~
\label{Eqn:DISCosetCdsTstChnlIndStates}
\sigma_{i}^{RB_{i}A_{3-i}W_{i}}
\define
\left(\mathrm{id}_{RA_{3-i}}\otimes \CalI^{(i)}\right)\{{\varphi_{\rho}}\}~~~\mbox{ for }i \in [2],
\end{eqnarray}
where $\ulineCalI=\left[\CalI^{(1)}\otimes \CalI^{(2)}\right]$ and associate the set $\ScrA_{\tinyoplus}(\ulineCalW,\ulineCalI,p_{Y|W_{1}\oplus W_{2}})$ of all quadruples $(R_{1},R_{2},C_{1},C_{2}) \in \reals^{4}_{\geq 0}$ satisfying
\begin{eqnarray}
  \label{Eqn:DISCosetCdsInnBndCharBnds}
R_{1}> \sfB_{1}-\sfB_{6},~R_{2} > \sfB_{2}-\sfB_{6}, ~R_{1}+C_{1}  > \sfB_{3}-\sfB_{6},~R_{2}+C_{2} >\sfB_{4}-\sfB_{6},
~\!R_{1}+C_{1}+R_{2}+C_{2} > \sfB_{5}-2\sfB_{6}\mbox{ where}
\\
\begin{array}{ll}
 \sfB_{1} = I(W_{1}:A_{2}R)_{\sigma_{1}}+\log q-H(W_{1}) ,~~~&~~~\sfB_{3} = I(W_{1}:Y\ulineB R)_{\sigma}+\log q-H(W_{1}),\\
 \sfB_{2} = I(W_{2}:A_{1}R)_{\sigma_{1}}+\log q-H(W_{2}),~~~&~~~\sfB_{4} =I(W_{2}:Y\ulineB R)_{\sigma}+\log q-H(W_{2}) ,
\end{array}~~~~~~~~~~~~~~~\nonumber \\
\begin{array}{c}
\sfB_{5} = I(W_{2}W_{1}:Y\ulineB R)_{\sigma}+2\log q-H(W_{1})-H(W_{2}),~~~~~~
\sfB_{6} = \log q-H(W_{1}\oplus W_{2}),
\end{array}~~~~~~~~~~~~
\nonumber
\end{eqnarray}
{where all the above information quantities are computed with respect to states $\sigma,\sigma_{1},\sigma_{2}$ defined in \eqref{Eqn:DISCosetCdsTstChnlJntSts}.}
\end{definition}

\begin{theorem}
 \label{Thm:DISCosetCdsInnBound}
 The action of instrument $\CalJ$ on $\rho$ can be asymptotically simulated with communication rates $(R_{1},R_{2}$ and common randomness rates $(C_{1},C_{2})$ if there exists a $(\ulineCalW,\ulineCalI,p_{Y|W_{1}\oplus W_{2}}) \in \ScrT_{\tinyoplus}(\CalJ)$ for which $(R_{1},R_{2},C_{1},C_{2}) \in \ScrA_{\tinyoplus}(\ulineCalW,\ulineCalI,p_{Y|W_{1}\oplus W_{2}})$.
\end{theorem}

\begin{remark}
\label{Rem:DISCCInnBndMatchesWithPriorResults}
The above inner bound in Thm.~\ref{Thm:DISCosetCdsInnBound} proven achievable via UCCs based on likelihood instruments is identical to that proven achievable in \cite{202504TIT_AtiPra} via coset codes. To be specific, \cite[Thm.~1]{202504TIT_AtiPra} characterizes a union of two inner bounds, one achievable via coset codes and the other via unstructured IID codes. The inner bound corresponding to only the former can found characterized in \cite[Sec.~VI]{202504TIT_AtiPra}, specifically through the pre Fourier Motzkin bounds in \cite[Eqn.~(29a) to (29f)]{202504TIT_AtiPra}. The latter bounds are exactly what we obtain in the following proof. Hence, the inner bound characterized in Thm.~\ref{Thm:DISCosetCdsInnBound} is the one obtained by performing Fourier-Motzkin elimination on \cite[Eqn.~(29a) to (29f)]{202504TIT_AtiPra}. Finally, if we combine our inner bound in Thms.~\ref{Thm:AsymptoticIIDInstSimulation} and \ref{Thm:DISCosetCdsInnBound}, we recover exactly the one characterized in \cite[Thm.~1]{202504TIT_AtiPra}.
\end{remark}

\begin{remark}
    \label{rem:DISCCDiversityInProofStructure}
    This being our second proof for the problem of distributed ICP, we have taken this opportunity to present this proof involving likelihood instruments based on coset codes through an different exposition as against those in the rest of theorems. This, we believe, assists the reader with a alternate perspective of the proof steps.
\end{remark}
\subsection{Achievability Proof of Thm.~\ref{Thm:DISCosetCdsInnBound}}
\label{SubSec:DISCCProof}
\begin{proof}
Choose an arbitrary $(\ulineCalW,\ulineCalI,p_{Y|W_{1}\oplus W_{2}}) \in \ScrT_{\tinyoplus}(\CalJ)$ and fix the same through the rest of the proof. Let $q=|\CalW_{i}|$ for $i \in [2]$ denote cardinality of the
finite field, i.e. $\CalW=\CalW_{1}=\CalW_{2}=\fieldq$, and $\oplus,\cdot$ denote field addition, scalar multiplication respectively. Associated with this triple $(\ulineCalW,\ulineCalI,p_{Y|W_{1}\oplus W_{2}})$, let communication rate pair $(R_{1},R_{2})$ and common randomness rate pair $(C_{1},C_{2})$ satisfy bounds stated in \eqref{Eqn:DISCosetCdsInnBndCharBnds} where the information quantities are wrt the associated states \eqref{Eqn:DISCosetCdsTstChnlJntSts} corresponding to this chosen test channel. We also choose a pair $(\beta_{1},\beta_{2})$ of (non-negative) binning rates which will enable us specify the structure of our two codes. The choice of $\beta_{1},\beta_{2}$ will be evident at the end of our proof.

\med\textbf{Design of Encoding Instrument}: For $i\in [2]$, the design of our encoding instrument $\CalE_{{k_{i}}}^{(i)}$ will be based on a code built over $\CalW_{i}$. Towards specifying the codes and the encoding instrument, let us identify Kraus representations of $\CalI^{(i)}: i \in [2]$ provided in our test channel.

\med\noindent\textit{\underline{Kraus representation of }$\CalI^{(i)}$}: Fix a Kraus representation
\begin{eqnarray*}
\mathcal{I}^{(i)}(\rho^i)
&=&
\Tr_{E_i}\left[
\sum_{w_i\in\CalW_i}
V_{w_i}\,\rho^i\,{V_{w_i}}^\dagger
\otimes \ketbra{w_i}_{W_i}
\right],
\end{eqnarray*}
where $V_{w_i}:\CalH_{A_i}\to \CalH_{B_i}\otimes \CalH_{E_i}$, $i\in\{1,2\}$, satisfy $\sum_{w_i\in\CalW_i}\ {V_{w_i}}^\dagger V_{w_i} = I_{A_i}$. We note that
\begin{eqnarray}
 \label{Eqn:DISCCMarkovChain-2}
 &&\hspace{-1cm}\sum_{y \in \CalY} \{i_{R}\otimes\CalJ_{y}\}(\varphi_{\rho^{12}})\otimes \ketbra{y}\nonumber\\
 &&\hspace{-1cm}= \!\!\!\!\!\!\!\sum_{\substack{y \in \CalY\\w \in \CalW=\CalF_{q}}}\!\!\!\!\!p_{Y|W_{1}\oplus W_{2}}(y|w)\!\!\!\sum_{\substack{w_{1}\in\CalW\\w_{2} \in \CalW}}\!\!\!\mathds{1}_{\{w=w_{1}\oplus w_{2}\}}\left[\mathrm{id}_{R}\otimes V_{w_{1}}\!\!\otimes V_{w_{2}}\right](\varphi_{\rho^{12}})\left[\mathrm{id}_{R}\otimes V_{w_{1}}\!\!\otimes V_{w_{2}}\right]^{\dagger}\!\otimes\ketbra{y}.
\end{eqnarray}
Throughout, we set
\begin{eqnarray}
\label{Eqn:DISCCDefnOfP_W}
p_{\ulineW W_{\tinyoplus}}(\ulinew,z)
 \define\ p_{W_1W_2 W_{\tinyoplus}}(w_1,w_2,z)\ \define\ \Tr\big[(\mu_{w_1}\otimes \mu_{w_2})\,\rho \big]\mathds{1}_{\{z=w_{1}\oplus w_{2}\}},\qquad \mu_{w_i}\ \define\ {V_{w_i}}^\dagger V_{w_i}.
\end{eqnarray}
We begin by describing the $K_{1},K_{2}$ encoder instruments and the stochastic decoder.

\med\textit{\underline{Two Codes, their structure and Binning}}: Throughout, $K_{i}=q^{nC_{i}},M_{i}=q^{nR_{i}},B_{i}=q^{n\beta_{i}}$ and $[K_{i}],[M_{i}],[B_{i}]$ denote common randomness, message, and bin-index sets, respectively. The choice of $\beta_{1},\beta_{2}$ will become evident at the end of the proof. Without loss of generality, we assume that $\min\{\beta_{1},\beta_{2}\} = \beta_{1} \leq \beta_{2} =\max\{\beta_{1},\beta_{2}\}$. Since we consider a sequence of codes with increasing block-length $n$ and can moreover appeal to continuity arguments, we assume without loss of generality that $nR_{i},nC_{i},n\beta_{i} \in \naturals$ are integers. Since the two instrument outcomes are correlated and the Rx only needs to recover $W_{1}\oplus W_{2}$, binning can reduce message rates $R_{1},R_{2}$. The outcomes are therefore binned and the bin index $b_{i} \in [B_{i}]$ is not communicated to the Rx. Here, we should note that, in contrast to the conventional Slepian-Wolf \cite{197307TIT_SleWol} binning wherein each code is randomly binned, we adopt correlated binning as devised by K\"orner and Marton \cite{197903TIT_KorMar}. Let us now specify the codes. For $i \in [2]$, let $\sfc^{i} = (\sfc_{k_{i}} : k_{i} \in [K_{i}])$ denote a multi-code consisting of $K_{i}=q^{nC_{i}}$ codes where the $k_i$-th code is a UCC $(n,\beta_{i},R_{i},g_{i},\upsilon_{k_{i}})$. Most importantly, the generator matrices $g_{i} \in \CalW^{n\beta_{i} \times n}$ for $i \in [2]$ are related as $g_{2} \define \left[ \!\!\begin{array}{c}g_{1}\\\Delta g\end{array}\!\!\right] \in \CalW^{n\beta_{2} \times n}$ where $\Delta g \in \CalW^{(n\beta_{2}-n\beta_{1})\times n }$. We note that the $k_{i}-$th code $\sfc_{k_{i}}=(\boldw_{i}(k_{i},m_{i}, b_{i}) \define w_{i}^{n}(k_{i},m_{i}, b_{i}) =  b_{i}g_{i} \oplus \upsilon_{k_{i}}(m_{i}) \in \CalW_{i}^{n}:(m_{i},b_{i}) \in [M_{i}] \times [B_{i}])$ comprises $M_{i}=q^{nR_{i}}$ bins each with $B_{i}=q^{n\beta_{i}}$ codewords. We remind the reader of Notation \ref{Not:DISCCBasicMatrixNotation} in regards to $b_{i} \in [q^{k_{i}}]$ for $i \in [2]$.

\begin{remark}
 \label{Rem:DISCosetCdsUniqueness}
The above defined code structure distinguishes it from the rest of the proofs in any measurement/instrument simulation protocols, including the other proofs in this theorem. The two codes $\sfc^{1},\sfc^{2}$ possess both intra- and inter structural properties. Firstly each bin $\CalB\mbox{in}(k_{i},m_{i})\define (\boldw_{i}(k_{i},m_{i}, b_{i}) =  b_{i}g_{i} \oplus \upsilon_{k_{i}}(m_{i}):  b_{i} \in [B_{i}])$ for each choice of $k_{i},m_{i}$ is algebraically closed, in contrast to conventional IID code based simulation protocols. Secondly, note that $\Bin(k_{1},m_{1})$ and $\Bin(k_{2},m_{2})$ are coset shifts of the row space of $g_{1}$ and $g_{2}=\left[ \!\!\begin{array}{c}g_{1}\\\Delta g\end{array}\!\!\right]$ respectively. This brings a strong inter-dependence between the two codes, one that is not available in conventional IID random code based protocols. Specifically, $\Bin(k_{1},m_{1})\oplus\Bin(k_{2},m_{2}) \define \{\boldw_{i}(k_{i},m_{i}, b_{i}) \oplus \boldw_{i}(k_{i},m_{i},\tilde{b}_{i}) :    b_{i},\tilde{b}_{i} \in [B_{i}]\}$ is a collection of cardinality at most $q^{n\max\{\beta_{1},\beta_{2}\}}=q^{n\beta_{2}}$ - a fact that we shall critically exploit in design and analysis of the decoder.
\end{remark}

\begin{remark}
 \label{Rem:DISCosetCdsGenStructure}
 Recall, we have assumed $\min\{\beta_{1},\beta_{2}\} = \beta_{1} \leq \beta_{2} =\max\{\beta_{1},\beta_{2}\}$ without loss of generality. This leads us to defining $g_{2}=\left[ \!\!\begin{array}{c}g_{1}\\\Delta g\end{array}\!\!\right]$. We note here that if  $\min\{\beta_{1},\beta_{2}\} = \beta_{2} \leq \beta_{1} =\max\{\beta_{1},\beta_{2}\}$, then $g_{1}=\left[ \!\!\begin{array}{c}g_{2}\\\Delta g\end{array}\!\!\right]$ with $\Delta g \in \CalW^{(n\beta_{1}-n\beta_{2})\times n }$.
\end{remark}

\med\textit{\underline{The $K_{1},K_{2}$ encoder instruments}}: Associated with the codeword $\boldw_{i}(k_{i},m_{i}, b_{i}) = (w_{i}(k_{i},m_{i}, b_{i})_{1},\cdots,w_{i}(k_{i},m_{i}, b_{i})_{n})$, let
\begin{eqnarray*}
V_{k_i,m_i,b_i} \define V_{\boldw_i(k_i,m_i, b_{i})} \define \bigotimes_{t=1}^{n}V_{\boldw_i(k_i,m_i, b_{i})_{t}}, ~\mu_{k_{i},m_{i}, b_{i}} \define \mu_{\boldw_{i}(k_{i},m_{i}, b_{i})} \define \bigotimes_{t=1}^{n} \mu_{w_{i}(k_{i},m_{i}, b_{i})_{t}} \in \CalP(\CalH_{\boldA_i}).
\end{eqnarray*}
Define the operators
\begin{eqnarray}
\label{Eqn:DISCosetCdsLikelihoodPOVMdefn_i}
S_{k_i}
\define \sum_{m_i=1}^{M_i} \sum_{b_i=1}^{B_i}
\frac{\mt(k_{i},m_{i}, b_{i})\sqrt{\boldrho^i}\ \mu_{k_i,m_i, b_{i}}\ \sqrt{\boldrho^i}}
{M_i B_i\,p_{\boldW_i}(\boldw_i(k_i,m_i, b_{i}))},~
z_{k_i}
\define \Pi_{\supp(S_{k_i})}, \pi_{k_i}\define I_{A_i}-z_{k_i},\mbox{ where}
\\
\label{Eqn:DISCosetCdsLikelihoodPOVMdefnMeasTrans_i}
\mt(k_{i},m_{i}, b_{i})\define\frac{p_{\boldW_i}(\boldw_i(k_i,m_i, b_{i}))}{1/q^{n}}\mbox{ is a measure transformation factor.}
\end{eqnarray}
\begin{remark}
 \label{Rem:DISCosetCdsMeasureTransformTerm}
 This measure transformation term $\mt(\cdot,\cdot,\cdot)$, analogous to the one in MD ICP (Rem.~\ref{rem:MDICPMeasureTransformIsInvisible}) is another difference between conventional IID code based protocol (Proof of Thm.~\ref{Thm:DistInstSimltnOneShot}) and coset codes. A distribution is defined on the union coset code by picking the generator matrix uniformly on the space of all matrices of the corresponding size. This implies that all codewords are uniformly distributed and pairwise independent. This is in contrast to conventional IID code based protocol where all codewords are picked mutually independently according to any desired product distribution. This measure transformation term $\mt(\cdot,\cdot,\cdot)$ coupled with an enlarged binning enables us handle this mismatch in distributions.
\end{remark}
For $m_i\in[M_i], b_{i} \in \CalF_{q}^{n\beta_{i}}$, define
\begin{equation}
\label{Eqn:DISCosetCdsLikelihoodTheta_i}
\theta^{m_i, b_{i}}_{k_i}
\define
\frac{
\sqrt{\mt(k_{i},m_{i}, b_{i})}V_{k_i,m_i, b_{i}}\ \sqrt{\boldrho^i}\ {S_{k_i}}^{-\frac12}
}{
\sqrt{M_i B_i\,p_{\boldW_i}(\boldw_i(k_i,m_i, b_{i}))}
},
\end{equation}
where ${S_{k_i}}^{-\frac12}$ denotes the generalized inverse (equal to $0$ on the null space of $S^{(i)}_{k_i}$). Note that, by definition, $\theta^{m_i, b_{i}}_{k_i}: \CalH_{\boldA_i}\to \CalH_{\boldB_i}\otimes \CalH_{\boldE_i} $.
Since $\pi_{k_i}$ is a projector, fix an isometry
$Q_i:\CalH_{\boldA_i}\to \CalH_{\boldB_i}\otimes\CalH_{\boldE_i}$
and define the ``$0$-th'' operator
\[
\theta^{0}_{k_i}\ \define\ Q_i\,\pi_{k_i},
\qquad\text{so that}\qquad
{\theta^{0}_{k_i}}^\dagger \theta^{0}_{k_i}=\pi_{k_i}.
\]
Note that for every $(k_i,m_i, b_{i})\in[K_i]\times[M_i]\times[B_i]$, we have
\begin{equation}
\label{Eqn:DISCosetCdsThetak0GetsKilled_i}
\theta^{0}_{k_i}\sqrt{S_{k_i}} = 0,~~ \theta^{m_i,  b_{i}}_{k_i}\sqrt{S_{k_i}}  =
\frac{
\sqrt{\mt(k_{i},m_{i}, b_{i})}V_{k_i,m_i, b_{i}}\sqrt{\boldrho^i}
}{
\sqrt{M_i B_i\,p_{\boldW_i}(\boldw_i(k_i,m_i, b_{i}))}
}.
\end{equation}
It is also easy to verify that for each $(k_{i},m_{i}, b_{i}) \in [K_{i}]\times [M_{i}] \times [B_{i}]$, $\theta^{m_i,b_i}_{k_i}: \CalH_{\boldA_i} \to \CalH_{\boldB_i}\otimes \CalH_{\boldE_i} $ and ${\theta_{k_{i}}^{0}}^{\dagger}\theta_{k_{i}}^{0}+\sum_{m_{i}, b_{i}}{\theta_{k_{i}}^{m_{i},b_{i}}}^{\dagger}\theta_{k_{i}}^{m_{i}, b_{i}} = I_{\boldA_{i}}$.
We are now ready to define the two banks of encoding instruments. For $ i \in [2]$, Alice$_{i}$'s instrument corresponding to common randomness index $k_{i} \in [K_{i}]$ is $\CalE_{k_{i}}^{(i)}:\ \CalL(\CalH_{\boldA_i})\longrightarrow \CalL(\CalH_{\boldB_i}\otimes \mathbb{C}^{||M_i|})$, defined through its action
\begin{eqnarray*}
 \CalE_{k_{i}}^{(i)}(S)= \tr_{E_{i}}\left[\begin{array}{r}\displaystyle\sum_{m_i=1}^{M_i} \sum_{b_i=1}^{B_i}
\theta^{m_i, b_i}_{k_i}\,S\,
{\theta^{m_i,  b_{i}}_{k_i}}^\dagger
\otimes \ketbra{m_i}_{M_i}
+
\theta^{0}_{k_i}\,S\,
{\theta^{0}_{k_i}}^\dagger
\otimes \ketbra{0}_{M_i} \end{array}\right]~\mbox{ on any $S \in \CalL(\CalH_{\boldA_{i}})$.}
\end{eqnarray*}

\begin{remark}
 \label{Rem:DISInverseOfRandomVariables}
Observe that $S_{k_{i}}^{-\frac{1}{2}}$ to the right of $\theta_{k_{i}}^{m_{i},b_{i}}$ in \eqref{Eqn:DISCosetCdsLikelihoodTheta_i} is a sum over all codebook elements. When we randomize over the codebook, $S_{k_{i}}^{-\frac{1}{2}}$ will be random. Since this random operator sum is raised to the power $-\frac{1}{2}$, we cannot prove concentration directly. This, as in other proofs, causes a central difficulty in the analysis, which we overcome through the use of proxy states.
\end{remark}

\noindent\textbf{Decoding and stochastic post-processing.} The decoder map performs two tasks. First, it attempts to recover the $\CalF_{q}-$sum of the two outcomes observed at the two encoders. Second, it uses the latter decoded sum to simulate the classical outcome in $\CalY^{n}$ via the single-letter stochastic channel $p_{Y|W_{1}\oplus W_{2}}$. The decoder $\CalD_{\ulinek} : \CalL(\mathbb{C}^{|M_1|}\otimes \mathbb{C}^{|M_2|}) \longrightarrow \CalL(\mathbb{C}^{|\boldCalY|})$ is therefore a composition $\CalD_{\ulinek} \define \CalD_{\ulinek}^{(2)} \circ \CalD_{\ulinek}^{(1)} $ of maps $\CalD_{\ulinek}^{(1)}$ and $\CalD_{\ulinek}^{(2)}$.

The central difference between instrument simulation with unstructured IID codes and with union coset codes lies in the first step of decoder, i.e., the map $\CalD_{\ulinek}^{(1)}$. In contrast to a decoder associated with unstructured IID codes that attempt to pin down the exact \textit{pair} of outcomes observed at the two encoders, i.e. identify the pair of bin indices corresponding to the pair of outcomes, a decoder associated with unstructured IID codes only attempts to recover the finite field addition of the pair of outcomes observed at the two encoders. In fact, our decoder associated with union coset code will decode into a $\fieldq-$sum multi-code, one that is employed by neither Tx. Let us now define this $\fieldq-$sum multi-code. Let $\sfc_{\tinyoplus} = (c_{k_{1}k_{2}}^{\tinyoplus} : (k_{1},k_{2}) \in [K_{1}]\times [K_{2}])$ denote a multi-code consisting of $K_{1}K_{2}$ codes where the $(k_{1},k_{2})$-th code is
\begin{eqnarray}
 \label{Eqn:DISCosetCdsSumCode}
 c^{\tinyoplus}_{k_{1}k_{2}} \define \left(\begin{array}{ccc} \boldw_{\tinyoplus}(\ulinek,\ulinem, \boplus) = \boldw_{\tinyoplus}(k_{1},k_{2},m_{1},m_{2}, \boplus)&|&(\ulinem, \boplus)=(m_{1},m_{2}, \boplus)\\ =  \boplus g_{2}\oplus \upsilon_{k_{1}}(m_{1})\oplus \upsilon_{k_{2}}(m_{2}) &|& \in  [M_{1}]\times [M_{2}]\times [q^{n\beta_{2}}]\end{array}\right).
\end{eqnarray}
Returning back to the decoder $\CalD_{\ulinek} : \CalL(\mathbb{C}^{|M_1|}\otimes \mathbb{C}^{|M_2|})
\longrightarrow \CalL(\mathbb{C}^{|\boldCalY|})$ defined as a composition $\CalD_{\ulinek} \define \CalD_{\ulinek}^{(2)} \circ \CalD_{\ulinek}^{(1)} $, let us define each constituent CPTP map. $\CalD_{\ulinek}^{(1)}:\CalL(\mathbb{C}^{\ulineM}) \rightarrow \CalL(\mathbb{C}^{|M_1|}\otimes \mathbb{C}^{|M_2|}\otimes \mathbb{C}^{|B_2|})$ is a typicality based decoder that attempts to identify a unique index $\boplus \in [q^{n\beta_{2}}]$ such that $\boldw_{\tinyoplus}(\ulinek,\ulinem,\boplus) \in T_{\delta}^{n}(p_{W_{1}\oplus W_{2}})$. Towards formally defining this, let
\begin{eqnarray}
\label{Eqn:DISCosetCdsList}
&&\hspace{-1.3cm}\sfL(\ulinek,\ulinem)
=\sfL(k_{1},k_{2},m_{1},m_{2})
\define
\left\{
\boplus\in[q^{n\beta_{2}}]:
\boldw_{\tinyoplus}(\ulinek,\ulinem,\boplus)
\in T_{\delta}^{n}(p_{W_{1}\oplus W_{2}})
\right\},
\quad
\boplus(\ulinek,\ulinem)
\define
\begin{cases}
\boplus, & \text{if }\sfL(\ulinek,\ulinem)=\{\boplus\},\\
\underline{0}, & \text{otherwise},
\end{cases}
\\
&&\hspace{-1.3cm}\CalD_{\ulinek}^{(1)}(J)
\define
\sum_{m_{1},m_{2}}
\ketbra{m_{1}m_{2}}J\ketbra{m_{1}m_{2}}
\otimes
\ketbra{\boplus(\ulinek,\ulinem)}.
\nonumber
\end{eqnarray}

To elaborate on $\CalD_{\ulinek}^{(1)}(J)$, it chooses index $ \boplus$ if it the unique element in $\sfL(\ulinek,\ulinem)$
As mentioned earlier, the second step is simulation of the outcome via the single-letter stochastic channel $p_{Y|W_{1}\oplus W_{2}}$. The second map $\CalD_{\ulinek}^{(2)}:\CalL(\mathbb{C}^{|M_1|}\otimes \mathbb{C}^{|M_2|}\otimes \mathbb{C}^{|B_2|}) \rightarrow \CalL(\mathbb{C}^{|\boldCalY|})$
is the stochastic post-processing map
\begin{equation}
\label{Eqn:DISCCStochPostProcessing}
  \CalD_{\ulinek}^{(2)}(S)= \sum_{\ulinem \in [M_{1}]\times [M_{2}]}\sum_{ \boplus \in [q^{n\beta_{2}}]}\sum_{y^{n} \in \CalY^{n}}p_{Y|W_{1}\oplus W_{2}}^{n}(y^{n}|\boldw_{\tinyoplus}(\ulinek,\ulinem, \boplus))\ketbra{y^{n}}{m_{1} m_{2} ~ \boplus}S\ketbra{m_{1} m_{2} ~ \boplus}{y^{n}}.
\end{equation}
Having defined the encoder instruments and decoder CPTP maps, we let $\CalS_{\CalD_{\ulinek},\theta_{\ulinek}} : \CalL(\CalH_{\boldR} \otimes \CalH_{\boldA_{1}}\otimes \CalH_{\boldA_{2}}) \rightarrow \CalL(\CalH_{R} \otimes \CalH_{\boldB_{1}}\otimes \CalH_{\boldB_{2}}\otimes \mathbb{C}^{|\boldCalY|})$ for $\ulinek \in [K_{1}]\times [K_{2}]$ and $\CalS:\CalL(\CalH_{\boldR} \otimes \CalH_{\boldA_{1}}\otimes \CalH_{\boldA_{2}}) \rightarrow \CalL(\CalH_{\boldR} \otimes \CalH_{\boldB_{1}}\otimes \CalH_{\boldB_{2}}\otimes \mathbb{C}^{|\boldCalY|})$ be defined as

\begin{eqnarray}
\label{Eqn:DISCosetCdsInstSimuProtocolMaps}
 \CalS_{{\CalD_{\ulinek},\theta_{\ulinek}}} \define \left\{\mathrm{id}_{\boldB_{1}\boldB_{2}\boldR}\otimes\CalD_{k_{1},k_{2}} \circ \left[ \ScrE_{k_{1}}^{(1)}\circ \ScrE_{k_{2}}^{(2)} \right]\right\} = \left\{\mathrm{id}_{\boldB_{1}\boldB_{2}\boldR}\otimes\CalD_{k_{1},k_{2}} \circ \left[ \ScrE_{k_{1}}^{(1)}\otimes \ScrE_{k_{2}}^{(2)} \right] \right\}\mbox{ and } \CalS \define \sum_{k_{1},k_{2}}\frac{\CalS_{\CalD_{\ulinek},\theta_{\ulinek}}}{K_{1}K_{2}}
\end{eqnarray}

\med\textit{\underline{Proxy Decoder CPTP maps}}: We introduce a proxy decoder CPTP map $\tilde{\CalD}_{\ulinek}:\CalL(\mathbb{C}^{|M_1|}\otimes \mathbb{C}^{|M_2|}\otimes \mathbb{C}^{|B_2|}) \rightarrow \CalL(\mathbb{C}^{|\boldCalY|})$ that assists us in our error analysis. Specifically, we let $\tilde{\CalD}_{\ulinek} \define \tilde{\CalD}_{\ulinek}^{(2)} \circ \tilde{\CalD}_{\ulinek}^{(1)}$ with $\tilde{\CalD}_{\ulinek}^{(2)} \define \CalD_{\ulinek}^{(2)}$ as before and $\tilde{\CalD}_{\ulinek}^{(1)} = i_{\ulineM,B_{2}}$ as the identity. We are now set to analyze the performance of the above proposed instrument simulation protocol.

\med\textbf{Analysis}: Owing to random operators raised to power $-\frac{1}{2}$ in the definition (Rem.~\ref{Rem:DISInverseOfRandomVariables}) of the encoder maps, our analysis will once again require proxy states. We begin by recalling canonical purifications $\ket{\varphi_{\rho^{i}}} \in \CalH_{R_{i}} \otimes \CalH_{A_{i}}$ for $i \in [2]$ and  $\ket{\varphi_{\rho^{12}}} \in \CalH_{R} \otimes \CalH_{\ulineA}$ as defined in Sec.~\ref{Sec:PrelimsProblemStatement}. For $i \in [2]$, let $\CalQ^{i}: \CalL(\CalH_{R_{i}}) \rightarrow \CalL(\CalH_{R} \otimes \CalH_{A_{\msout{i}}}) $  be isometric maps \cite[Thm.~5.1.1]{BkWilde_2017} such that $\CalQ^{i}\otimes i_{i}(\varphi_{\rho^{i}}) = \varphi_{\rho^{12}}$ and therefore $\boldCalQ^{i}\otimes \boldi_{i}(\varphi_{\boldrho^{i}})=\varphi_{\boldrho^{12}}$. A fact which we will use in the sequel is stated below.

\begin{fact}
 \label{Fact:DISCCCrucialIsometryCommuteWithInst}
$\boldCalQ^{i}$ commutes with $\CalE_{k_{i}}^{(i)}$ for $i\in [2]$.
 \end{fact}
We introduce two more CPTP maps that we will have opportunity to utilize. Towards that end, we remind the reader of Notation \ref{Not:DISCCBasicMatrixNotation} implying each of $b_{i} \in [q^{n\beta_{i}}]$ having a $k_{i}-$length representation in $\CalF_{q}$. For $\boplus \in [q^{n\beta_{2}}]$ define
\begin{eqnarray}
 \label{Eqn:DISCCCompatibleBinIndices}
 \CalA(b^{\tinyoplus})\define \left\{\!\!\!\begin{array}{ccc}(b_{1},b_{2}) \in&:& (b_{1,1}~b_{1,2}~\cdots b_{1,n\beta_{1}} \underbrace{0~0~\cdots~0}_{n\beta_{2}-n\beta_{1}}) \oplus (b_{2,1}~b_{2,2}~\cdots b_{2,n\beta_{1}}~\cdots~b_{2,n\beta_{2}})\\ \left[q^{n\beta_{1}}\right] \times \left[ q^{n\beta_{2}}\right]& &=(\boplus_{1}~\boplus_{2}~\cdots \boplus_{n\beta_{1}}~\cdots~\boplus_{n\beta_{2}})~~~~~ \end{array}\!\!\!\right\}
 \end{eqnarray}
be the set of pairs of bin indices $( b_{1}, b_{2})$ for which $ b_{1}g_{1}\oplus  b_{2}g_{2} = b^{{\tinyoplus}} g_{2}$. For $(k_{1},k_{2}) \in [K_{1}]\times [K_{2}]$, let  $\tilde{\CalV}_{k_{1},k_{2}} : \CalL( \CalH_{\boldA_{1}}\otimes \CalH_{\boldA_{2}}) \rightarrow \CalL( \CalH_{\boldB_{1}}\otimes\CalH_{\boldB_{2}}\otimes \mathbb{C}^{|M_{1}|}\otimes \mathbb{C}^{|M_{2}|}\otimes \mathbb{C}^{|B_{2}|} )$, and $\CalV_{k_{i}} : \CalL(\CalH_{\boldA_{i}}) \rightarrow \CalL( \CalH_{\boldB_{i}}\otimes \mathbb{C}^{|M_{i}|})$ be completely positive maps defined as\footnote{The superscripts $A_{i},A_{\msout{i}}$ have been added to emphasize/clarify the space on which the operators operate.}
 \begin{eqnarray}
\label{Eqn:DISCCInstWithBinIndex}
&&\hspace{-1.4cm}{\tilde{\CalV}_{k_{1},k_{2}}(S) = \sum_{\substack{(m_{1},m_{2}) \in\\ [M_{1}]\times [M_{2}]}}\sum_{\substack{\boplus \in [q^{n\beta_{2}}]\\( b_{1}, b_{2}) \in \CalA(\boplus)}}\tr_{E_{1}E_{2}}\left\{ \left[ \sfV_{k_{1},m_{1}, b_{1}}\otimes \sfV_{k_{2},m_{2}, b_{2}}\right]S \left[\sfV_{k_{1},m_{1}, b_{1}}\otimes \sfV_{k_{2},m_{2}, b_{2}} \right]^{\dagger}  \right\}\otimes \ketbra{m_{1}~m_{2}~\boplus}}
\\
\label{Eqn:DistMCPKeySplitOfError6}
&&\hspace{-1.4cm}\CalV_{k_{i}}(S) = \sum_{m_{i} \in [M_{i}]}\sum_{b_{i} \in [B_{i}]}\!\!\tr_{E_{i}}\left\{{\sfV}_{k_{i},m_{i},b_{i}}S\sfV_{k_{i},m_{i},b_{i}}^{\dagger}\right\}\otimes\ketbra{m_{i}}\mbox{ where }\sfV_{k_{i}m_{i}b_{i}}\! \define\! \frac{\sqrt{\mt(k_{i},m_{i}, b_{i})}{V_{k_{i},m_{i},b_{i}}}^{A_{i}}}{\sqrt{M_{i}B_{i}p_{W_{i}}(w({k_{i},m_{i},b_{i}}))}}.
\end{eqnarray}
\begin{fact}
 \label{Fact:DISCCCrucialIsometryCommuteWithCalJMaps}
For $i \in [2]$, (1) $\boldCalQ^{i}$ commutes with $\CalV_{k_{i}}$, (2) $\CalV_{k_{i}}$ commutes with $\CalE_{k_{\msout{i}}}^{(\msout{i})}$ and (3) $\CalV_{k_{i}}$ also commutes with $\CalV_{k_{\msout{i}}}$.
 \end{fact}
We now breakdown the error event. Towards that end, 
recalling $\CalS(\cdot),\CalS_{\CalD_{\ulinek},\theta_{\ulinek}}(\cdot)$ from \eqref{Eqn:DISCosetCdsInstSimuProtocolMaps}, we have
\begin{eqnarray}
\label{Eqn:DISCCBreakdownErrorEvent02}
&&\hspace{-1.4cm}\norm{\left(\mathrm{id}_{R}\otimes \CalJ - \CalS\right)\{\ketbra{\varphi_{\rho^{12}}}\}}_{1} \leq A+\!\!\sum_{k_{1},k_{2}}\frac{B_{\ulinek}+C_{\ulinek}+D_{\ulinek}}{K_{1}K_{2}},\text{ where }A = \norm{(\mathrm{id}_{R}\otimes \CalJ)(\varphi_{\rho^{12}}) -\sum_{\ulinek} \frac{\tilde{\CalD}_{\ulinek}\!\circ\!{\tilde{\CalV}_{k_{1},k_{2}}}\left( \varphi_{\rho^{12}}\right)  }{K_{1}K_{2}}}_{1}
\\
\label{Eqn:DISCCKeySplitOfError5}
&&\hspace{-0.4cm}B_{\ulinek} \define \norm{\left[\tilde{\CalD}_{\ulinek}\!\circ\!{\tilde{\CalV}_{k_{1},k_{2}}}   -  \CalD_{\ulinek}\!\circ\!{\CalV}_{k_{1}}\!\!\circ\!{\CalV}_{k_{2}}\!\right]\left( \varphi_{\rho^{12}}\right)    }_{1}, C_{\ulinek}\define \norm{ \CalD_{\ulinek}\!\circ\!{\CalV}_{k_{1}}\!\!\circ\!{\CalV}_{k_{2}}\!\left( \varphi_{\rho^{12}}\right) - \CalD_{\ulinek}\!\circ\!\ScrE^{(1)}_{k_{1}}\!\!\circ\!{\CalV}_{k_{2}}\!\circ \boldCalQ^{1}\!\left( \varphi_{\rho^{1}}\right) }_{1},
\\
\label{Eqn:DISCCKeySplitOfError6}
&&\hspace{-0.4cm}D_{\ulinek} \define \norm{\CalD_{\ulinek}\!\circ\!\ScrE^{(1)}_{k_{1}}\!\!\circ\!{\CalV}_{k_{2}}\!\circ \boldCalQ^{1}\!\left( \varphi_{\rho^{1}}\right) - \CalS_{\CalD_{\ulinek},\theta_{\ulinek}}(\varphi_{\rho^{12}})}_{1} = \norm{\CalD_{\ulinek}\!\circ\!\ScrE^{(1)}_{k_{1}}\!\!\circ\!{\CalV}_{k_{2}}\!\circ \boldCalQ^{1}\!\left( \varphi_{\rho^{1}}\right) - \CalS_{\CalD_{\ulinek},\theta_{\ulinek}}\circ \boldCalQ^{2}(\varphi_{\rho^{2}})}_{1}.
\end{eqnarray}
We identify the proxy states by breaking down $C_{\ulinek},D_{\ulinek}$ further as $C_{\ulinek} \leq C_{\ulinek,1}+C_{\ulinek,2}$ and $D_{\ulinek} \leq D_{\ulinek,1}+D_{\ulinek,2}$, where
\begin{eqnarray}
\label{Eqn:DISCCKeySplitOfError7}
&&\hspace{-1.5cm}C_{\ulinek,1}\define \norm{ \CalD_{\ulinek}\!\circ\!{\CalV}_{k_{1}}\!\!\circ\!{\CalV}_{k_{2}}\!\left( \varphi_{\rho^{12}}\right) - \CalD_{\ulinek}\!\circ\!\ScrE^{(1)}_{k_{1}}\!\!\circ\!{\CalV}_{k_{2}}\!\circ \boldCalQ^{1}\!\left( \varphi_{S_{k_{1}}}\right) }_{1}, C_{\ulinek,2}\define \norm{ \left[\CalD_{\ulinek}\!\circ\!\ScrE^{(1)}_{k_{1}}\!\!\circ\!{\CalV}_{k_{2}}\!\circ \boldCalQ^{1}\!\right]\left( \varphi_{S_{k_{1}}} -  \varphi_{\rho^{1}}\right) }_{1}
\\
&&\hspace{-1.5cm}D_{\ulinek,1}\define \norm{ \CalD_{\ulinek}\!\circ\!\ScrE^{(1)}_{k_{1}}\!\!\circ\!{\CalV}_{k_{2}}\!\circ \boldCalQ^{1}\!\left( \varphi_{\rho^{1}}\right) - \CalS_{\CalD_{\ulinek},\theta_{\ulinek}}\!\circ \boldCalQ^{2}\!\left( \varphi_{S_{k_{2}}}\right) }_{1}, D_{\ulinek,2} = \norm{\CalS_{\CalD_{\ulinek},\theta_{\ulinek}}\!\circ \boldCalQ^{2}\!\left( \varphi_{S_{k_{2}}} - \varphi_{\rho^{2}}\right)}_{1}.
\nonumber
\end{eqnarray}
In the analysis of $D_{\ulinek,2}$ we note one important point. We are required here to prove an upper bound on the trace distance $\norm{ \varphi_{S_{k_{2}}} - \varphi_{\rho^{2}}}_{1}$ between purifications. While it is easy for the reader to prove a bound on $\norm{ {S_{k_{2}}} - {\rho^{2}}}_{1}$ via quantum state covering, requiring closeness of purifications lets us leverage the choice of canonical purifications.

\noindent\textit{\underline{One-Party Soft Covering Error $D_{\ulinek}$}} : We begin by analyzing $D_{\ulinek,2}$. We have
\begin{eqnarray}
 \label{Eqn:DISCCQuantumOnlyTermDk2}
 D_{\ulinek,2} =  \norm{\CalS_{\CalD_{\ulinek},\theta_{\ulinek}}\!\circ \CalQ^{2}\!\left( \varphi_{S_{k_{2}}} - \varphi_{\rho_{2}}\right)}_{1} \leq \norm{\varphi_{S_{k_{2}}} - \varphi_{\rho_{2}}}_{1}
\end{eqnarray}
since $\CalS_{\CalD_{\ulinek},\theta_{\ulinek}}\!\circ \CalQ^{2}$ is a CPTP map and CPTP maps shrink trace distance \cite[Eqn.~9.69]{BkWilde_2017}.
\begin{remark}
\label{Rem:DISCosetCdsNonTrivialMeasureTransform}
Owing to the measure transformation factor in the definition of
$S_{k_i}$ in \eqref{Eqn:DISCosetCdsLikelihoodPOVMdefn_i},
$\tr(S_{k_i})$ is not necessarily equal to $1$. Accordingly, the canonical-purification
approach of~\cite{2026MMTIT_Pad} must be applied here without
assuming unit trace.
We invoke Lemma~\ref{lem:canonical-purif-stability}, which
applies to arbitrary positive semidefinite $S_{k_i}$ and introduces
the factor $1+\sqrt{\tr(S_{k_i})}$. This factor will subsequently be
controlled after averaging over the random code ensemble.
\end{remark}
Applying Lemma~\ref{lem:canonical-purif-stability} to the right-hand side of
\eqref{Eqn:DISCCQuantumOnlyTermDk2}, we obtain
\begin{equation}
\label{Eqn:DISCCQuantumOnlyTermDk2Purif}
D_{\ulinek,2}
\leq
\left(1+\sqrt{\tr S_{k_{2}}}\right)
\sqrt{\norm{\boldrho^{2}-S_{k_{2}}}_{1}}.
\end{equation}
We leave the upper bound in the RHS above for now, and take it forward for a final analysis, i.e. we carry out the rest of the analysis of $D_{\ulinek,2}$ through the upper bound on the RHS of \eqref{Eqn:DISCCQuantumOnlyTermDk2Purif} at the end of the proof when we collate all bounds and randomize over the codebooks. We now focus on $D_{\ulinek,1}$ in \eqref{Eqn:DISCCKeySplitOfError6}. From Fact \ref{Fact:DISCCCrucialIsometryCommuteWithInst}, we have
\begin{eqnarray}
\label{Eqn:DISCosetCdsQuantumOnlyTermDk21}
\CalS_{\CalD_{\ulinek},\theta_{\ulinek}}\!\circ \CalQ^{2}\!\left( \varphi_{S_{k_{2}}}\right) = i_{R}\otimes \ScrE^{\CalD_{\ulinek}} \circ\! \left[ \ScrE^{(1)}_{k_{1}}\!\otimes\! \ScrE^{(2)}_{k_{2}}\! \right]\!\circ \CalQ^{2}\!\left( \varphi_{S_{k_{2}}}\right) = i_{R}\otimes \ScrE^{\Delta_{\ulinek}} \circ \ScrE^{(1)}_{k_{1}}\circ \CalQ^{2}\circ \ScrE^{(2)}_{k_{2}}\left( \varphi_{S_{k_{2}}}\right).
\end{eqnarray}
Comparing the RHS above \eqref{Eqn:DISCosetCdsQuantumOnlyTermDk21} and the first term in the definition of $D_{\ulinek,1}$, observe that we would be proving $D_{\ulinek,1}=0$ if we could show $\CalQ^{2}\circ \ScrE^{(2)}_{k_{2}}\left( \varphi_{S_{k_{2}}}\right) = {\CalV}_{k_{2}}\!\circ \CalQ^{1}\!\left( \varphi_{\rho^{1}}\right)$. This is indeed true! and can be verified via a straight forward evaluation of $\CalQ^{2}\circ \ScrE^{(2)}_{k_{2}}\left( \varphi_{S_{k_{2}}}\right)$. For completeness, these steps are detailed out in Appendix \ref{AppSubSec:DISCCAnalysisDk1}. Hence $D_{\ulinek,1} = 0$ and we record these facts below for ease of reference.

\noindent\textit{\underline{One-Party Soft Covering Error $C_{\ulinek}$}} :
We begin with $C_{\ulinek,2}$. From \eqref{Eqn:DISCCKeySplitOfError7}, we have
\begin{eqnarray}
\label{Eqn:DISCCProofQuantTermCk21}
C_{\ulinek,2}
\define
\norm{\left[\CalD_{\ulinek}\!\circ\!\ScrE^{(1)}_{k_1}\!\circ\!\CalV_{k_2}\!\circ\!\boldCalQ^1\right]\left(\varphi_{S_{k_1}}-\varphi_{\boldrho^1}\right)}_1
\leq
\norm{\left[\CalV_{k_2}\!\circ\!\boldCalQ^1\right]\left(\varphi_{S_{k_1}}-\varphi_{\boldrho^1}\right)}_1 .
\end{eqnarray}
The inequality follows since $\CalD_{\ulinek}$ and $\ScrE^{(1)}_{k_1}$ are CPTP and hence contract the trace norm on Hermitian operators \cite[Eqn.~(9.69)]{BkWilde_2017}. Although $\CalV_{k_2}$ is not necessarily trace preserving, its action is contractive on average under the random UCC ensemble by Lemma~\ref{Lem:AveragedInverseFreeContraction2}. We therefore retain the right-hand side of \eqref{Eqn:DISCCProofQuantTermCk21} and bound it after introducing the random code ensemble. We now turn to $C_{\ulinek,1}$. Its analysis is analogous to that of $D_{\ulinek,1}$. From \eqref{Eqn:DISCCKeySplitOfError7},
\begin{eqnarray*}
\CalD_{\ulinek}\circ\ScrE^{(1)}_{k_1}\circ\CalV_{k_2}\circ\boldCalQ^1\left(\varphi_{S_{k_1}}\right)
=
\CalD_{\ulinek}\circ\CalV_{k_2}\circ\boldCalQ^1\circ\ScrE^{(1)}_{k_1}\left(\varphi_{S_{k_1}}\right),
\end{eqnarray*}
since $\ScrE^{(1)}_{k_1}$ commutes with both $\CalV_{k_2}$ and $\boldCalQ^1$ (Facts~\ref{Fact:DISCCCrucialIsometryCommuteWithCalJMaps} and~\ref{Fact:DISCCCrucialIsometryCommuteWithInst}). As shown in Appendix~\ref{AppSubSec:DISCCAnalysisCk1},
$\boldCalQ^1\circ\ScrE^{(1)}_{k_1}(\varphi_{S_{k_1}})
=\CalV_{k_1}(\varphi_{\boldrho^{12}})$, and hence
\begin{eqnarray*}
\CalD_{\ulinek}\circ\ScrE^{(1)}_{k_1}\circ\CalV_{k_2}\circ\boldCalQ^1\left(\varphi_{S_{k_1}}\right)
=
\CalD_{\ulinek}\circ\CalV_{k_2}\circ\CalV_{k_1}\left(\varphi_{\boldrho^{12}}\right)
=
\CalD_{\ulinek}\circ\CalV_{k_1}\circ\CalV_{k_2}\left(\varphi_{\boldrho^{12}}\right).
\end{eqnarray*}
Therefore $C_{\ulinek,1}=0$, and consequently
\begin{eqnarray}
\label{Eqn:DISCCFinalBoundOnCk}
C_{\ulinek}\leq C_{\ulinek,2}
\leq
\norm{\left[\CalV_{k_2}\circ\boldCalQ^1\right]\left(\varphi_{S_{k_1}}-\varphi_{\boldrho^1}\right)}_1.
\end{eqnarray}

\noindent\textit{\underline{Binning Error $B_{\ulinek}$}} :
Referring to \eqref{Eqn:DISCCKeySplitOfError5}, \eqref{Eqn:DISCCInstWithBinIndex}, recalling ${\CalD}_{\ulinek} = {\CalD}_{\ulinek}^{(2)} \circ {\CalD}_{\ulinek}^{(1)}$, the proxy decoder $\tilde{\CalD}_{\ulinek} = \tilde{\CalD}_{\ulinek}^{(2)} \circ \tilde{\CalD}_{\ulinek}^{(1)}$, noting $\tilde{\CalD}_{\ulinek}^{(2)}={\CalD}_{\ulinek}^{(2)}$ and this latter CPTP map can only shrink trace distance \cite[Eqn.~9.69]{BkWilde_2017}, we have
\begin{eqnarray}
 \label{Eqn:DISCCAnalysisBk-1}
 B_{\ulinek} = \norm{\left[\tilde{\CalD}_{\ulinek}^{(2)} \circ \tilde{\CalD}_{\ulinek}^{(1)}\!\circ\!{\tilde{\CalV}_{k_{1},k_{2}}}   -  {\CalD}_{\ulinek}^{(2)} \circ {\CalD}_{\ulinek}^{(1)}\!\circ\!{\CalV}_{k_{1}}\!\!\circ\!{\CalV}_{k_{2}}\!\right]\left( \varphi_{\rho^{12}}\right)   }_{1}
 \leq \norm{\left[\tilde{\CalD}_{\ulinek}^{(1)}\!\circ\!{\tilde{\CalV}_{k_{1},k_{2}}}   -  {\CalD}_{\ulinek}^{(1)}\!\circ\!{\CalV}_{k_{1}}\!\!\circ\!{\CalV}_{k_{2}}\!\right]\left( \varphi_{\rho^{12}}\right)   }_{1}
 \nonumber\\
 \label{Eqn:DISCCAnalysisBk-2}
 =\norm{\!\!\!
 \begin{array}{c}
\displaystyle\sum_{\substack{(m_{1},m_{2}) \in\\ [M_{1}]\times [M_{2}]}}\!\!\left(\!\sum_{\boplus \in [q^{n\beta_{2}}]} \sum_{\substack{( b_{1}, b_{2}) \\\in \CalA(\boplus)}}\!\!\!\Theta_{\ulinek,\ulinem, \ulineb}\!\otimes \!\ketbra{\!\!\begin{array}{c}m_{1}~m_{2}\\\boplus\!\!\end{array}\!\!\!}-\displaystyle\! \!\!\! \sum_{\substack{( b_{1}, b_{2}) \in\\  [B_{1}]\times [B_{2}]}}\!\!\!\!\!\!\Theta_{\ulinek,\ulinem, \ulineb}\otimes \!\ketbra{\!\!\begin{array}{c}m_{1}~m_{2}\\ \boplus\!(\ulinek,\ulinem)\!\end{array}\!\!}\!\right)\!\!\!
\end{array}
}_{1}. 
 \end{eqnarray}
Before proceeding, we collect the notation used in the binning analysis.
For $(w_1^n,w_2^n)\in\boldCalW_1\times\boldCalW_2$, define the subnormalized post-instrument operator
\[
\boldsymbol{\tau}_{w_1^n,w_2^n}^{RB}
\define
\tr_{E_{1}E_{2}}\!\left\{
\left[\mathrm{id}_{R}\bigotimes_{i=1}^{2}V_{w_i^n}^{A_i}\right]
\varphi_{\boldrho^{12}}
\left[\mathrm{id}_{R}\bigotimes_{i=1}^{2}(V_{w_i^n}^{A_i})^{\dagger}\right]
\right\}.
\]
We also let
$\sfs(b_{1},b_{2})\define b_{1}0^{n(\beta_{2}-\beta_{1})}\oplus b_{2}$.
Finally, recall
\begin{eqnarray} \label{Eqn:DISCCAnalysisBk-3} \Theta_{\ulinek,\ulinem, \ulineb} \!\!\define \tr_{E_{1}E_{2}}\!\!\left\{\! \left[\displaystyle\mathrm{id}_{R}\!\otimes_{i=1}^{2}\! \sfV_{k_{i},m_{i},b_{i}}\right]\!\varphi_{\boldrho^{12}} \!\left[\mathrm{id}_{R}\!\otimes_{i=1}^{2}\sfV_{k_{i},m_{i},b_{i}} \right]^{\!\dagger}\! \right\},\mbox{ and }\CalA(\boplus) \define \left\{\begin{array}{r}\!\!\! ( b_{1}, b_{2}) \in [q^{n\beta_{1}}]\times [q^{n\beta_{2}}] \\ \ni b_{1}g_{1}\oplus b_{2}g_{2} = \boplus g_{2} \end{array}\right\}. \end{eqnarray}
With these, note that for any $(m_{1},m_{2}) \in [M_{1}]\times [M_{2}]$, the term within the parenthesis above after removing common terms is
\begin{eqnarray}
 \label{Eqn:DISCCAnalysisBk-4}
 \sum_{\substack{\tildeboplus \in [q^{n\beta_{2}}]\\\tildeboplus \neq  \boplus\!(\ulinek,\ulinem)}} \sum_{\substack{( b_{1}, b_{2}) \\\in \CalA(\tildeboplus)}}\!\!\!\Theta_{\ulinek,\ulinem, \ulineb}\!\!\otimes \!\ketbra{\!\!\begin{array}{c}m_{1}~m_{2}\\\tildeboplus\!\!\end{array}\!\!\!}-\displaystyle\! \!\!\! \sum_{\substack{( b_{1}, b_{2}) \in\\  [B_{1}]\times [B_{2}]}}\!\!\!\!\!\!\Theta_{\ulinek,\ulinem, \ulineb}\!\!\otimes \!\ketbra{\!\!\begin{array}{c}m_{1}~m_{2}\\ \boplus\!(\ulinek,\ulinem)\!\end{array}\!\!}\!\mathds{1}_{\{ \sfs( b_{1}, b_{2}) \neq  \boplus\!(\ulinek,\ulinem)\}}
 \nonumber\\
 \label{Eqn:DISCCAnalysisBk-5}
 =\!\!\!\sum_{\substack{\tildeboplus \in [q^{n\beta_{2}}]\\\tildeboplus \neq  \boplus\!(\ulinek,\ulinem)}} \sum_{\substack{( b_{1}, b_{2}) \\\in \CalA(\tildeboplus)}}\!\!\!\Theta_{\ulinek,\ulinem, \ulineb}\!\!\otimes \left(\!\!\!\begin{array}{c} \ketbra{m_{1}~m_{2}~\tildeboplus}-~~~~~~\\\ketbra{m_{1}~m_{2}~ \boplus\!(\ulinek,\ulinem)}\end{array}\!\right)\left(\! \mathds{1}_{\left\{\substack{\boldw_{\tinyoplus}(\ulinek,\ulinem,\tildeboplus) \notin \\ T_{\delta}(p_{W_{1}\oplus W_{2}})} \right\}}\!\!+\mathds{1}_{\left\{\substack{\boldw_{\tinyoplus}(\ulinek,\ulinem,\tildeboplus) \in \\ T_{\delta}(p_{W_{1}\oplus W_{2}})} \right\}}\!\right).
\end{eqnarray}
Substituting RHS of \eqref{Eqn:DISCCAnalysisBk-5} for the parenthesis in \eqref{Eqn:DISCCAnalysisBk-2} and employing the triangular inequality for the $\norm{\cdot}_{1}$, we have
\begin{eqnarray}
\label{Eqn:DISCCAnalysisBk-6}
&&\hspace{-0.85cm}B_{\ulinek} \leq \norm{\mathtt{t}_{B0}}+\sum_{l=0}^{2}\norm{\mathtt{t}_{B1l}},\qquad \text{where}
\\
&&\hspace{-1.4cm}\mathtt{t}_{B0} \define \!\!\!\!\sum_{\substack{(m_{1},m_{2}) \in\\ [M_{1}]\times [M_{2}]}}\sum_{\substack{\tildeboplus \in [q^{n\beta_{2}}]\\\tildeboplus \neq  \boplus\!(\ulinek,\ulinem)}} \sum_{\substack{( b_{1}, b_{2}) \\\in \CalA(\tildeboplus)}}\!\!\!\Theta_{\ulinek,\ulinem, \ulineb}\!\!\otimes \left( \ketbra{\!\!\begin{array}{c}m_{1}~m_{2}\\\tildeboplus\!\!\end{array}\!\!\!}-\ketbra{\!\!\begin{array}{c}m_{1}~m_{2}\\ \boplus\!(\ulinek,\ulinem)\!\end{array}\!\!}\right) \mathds{1}_{\left\{\substack{\boldw_{\tinyoplus}(\ulinek,\ulinem,\tildeboplus) \notin \\ T_{\delta}(p_{W_{1}\oplus W_{2}})} \right\}}\!
\nonumber\\
&&\hspace{-1.4cm}\mathtt{t}_{B12} \define \!\!\!\!\sum_{\substack{(m_{1},m_{2}) \in\\ [M_{1}]\times [M_{2}]}}\sum_{\substack{\tildeboplus \in [q^{n\beta_{2}}]\\\tildeboplus \neq  \boplus\!(\ulinek,\ulinem)}} \sum_{\substack{( b_{1}, b_{2}) \\\in \CalA(\tildeboplus)}}\!\!\!\Theta_{\ulinek,\ulinem, \ulineb}\!\!\otimes \left( \ketbra{\!\!\begin{array}{c}m_{1}~m_{2}\\\tildeboplus\!\!\end{array}\!\!\!}-\ketbra{\!\!\begin{array}{c}m_{1}~m_{2}\\ \boplus\!(\ulinek,\ulinem)\!\end{array}\!\!}\right) \mathds{1}_{\left\{\substack{\boldw_{\tinyoplus}(\ulinek,\ulinem,\tildeboplus) \in \\ T_{\delta}(p_{W_{1}\oplus W_{2}})} \right\}}\mathds{1}_{\left\{|\sfL(\ulinek,\ulinem)|\geq2 \right\}}
\nonumber\\
&&\hspace{-1.4cm}\mathtt{t}_{B1l} \define \!\!\!\!\sum_{\substack{(m_{1},m_{2}) \in\\ [M_{1}]\times [M_{2}]}}\sum_{\substack{\tildeboplus \in [q^{n\beta_{2}}]\\\tildeboplus \neq  \boplus\!(\ulinek,\ulinem)}} \sum_{\substack{( b_{1}, b_{2}) \\\in \CalA(\tildeboplus)}}\!\!\!\Theta_{\ulinek,\ulinem, \ulineb}\!\!\otimes \left( \ketbra{\!\!\begin{array}{c}m_{1}~m_{2}\\\tildeboplus\!\!\end{array}\!\!\!}-\ketbra{\!\!\begin{array}{c}m_{1}~m_{2}\\ \boplus\!(\ulinek,\ulinem)\!\end{array}\!\!}\right) \mathds{1}_{\left\{\substack{\boldw_{\tinyoplus}(\ulinek,\ulinem,\tildeboplus) \in \\ T_{\delta}(p_{W_{1}\oplus W_{2}})} \right\}}\mathds{1}_{\left\{|\sfL(\ulinek,\ulinem)|=l \right\}}
\nonumber
\end{eqnarray}
for $l=0,1$. From $\sfL(\ulinek,\ulinem)$ defined in \eqref{Eqn:DISCosetCdsList}, observe that for $(\ulinek,\ulinem) \in [K_{1}]\times[K_{2}]\times[M_{1}]\times[M_{2}]$ satisfying $|\sfL(\ulinek,\ulinem)| \in \{0,1\} $, and any $\tildeboplus \in [q^{n\beta_{2}}] \setminus \{  \boplus\!(\ulinek,\ulinem) \}$, we have $\boldw_{\tinyoplus}(\ulinek,\ulinem,\tildeboplus) \notin T_{\delta}(p_{W_{1}\oplus W_{2}})$. Thus, whenever $|\sfL(\ulinek,\ulinem)| \in \{0,1\} $, $\tildeboplus \in [q^{n\beta_{2}}] \setminus \{  \boplus\!(\ulinek,\ulinem) \}$ and $( b_{1}, b_{2}) \in \CalA(\tildeboplus)$, we have $\boldw_{\tinyoplus}(\ulinek,\ulinem,\tildeboplus)=\boldw_{1}(k_{1},m_{1}, b_{1}) \oplus \boldw_{2}(k_{2},m_{2}, b_{2}) \notin T_{\delta}(p_{W_{1}\oplus W_{2}})$ implying $\mathds{1}_{\left\{\substack{\boldw_{\tinyoplus}(\ulinek,\ulinem,\tildeboplus) \in  T_{\delta}(p_{W_{1}\oplus W_{2}})} \right\}}\mathds{1}_{\left\{|\sfL(\ulinek,\ulinem)|=0 \right\}} =0$ and thus resulting in $\mathtt{t}_{B10}=\mathtt{t}_{B11}=0$. Left with $\norm{\mathtt{t}_{B0}},\norm{\mathtt{t}_{B12}}$, we have
\begin{eqnarray*}
 &&\hspace{-0.8cm}\norm{\mathtt{t}_{B0}} \leq \!\!\!\!\sum_{\substack{(m_{1},m_{2}) \in\\ [M_{1}]\times [M_{2}]}}\sum_{\substack{\tildeboplus \in [q^{n\beta_{2}}]\\\tildeboplus \neq  \boplus\!(\ulinek,\ulinem)}} \sum_{\substack{( b_{1}, b_{2}) \\\in \CalA(\tildeboplus)}}\!\!\!\norm{\Theta_{\ulinek,\ulinem, \ulineb}\!\!\otimes \left( \ketbra{\!\!\begin{array}{c}m_{1}~m_{2}\\\tildeboplus\!\!\end{array}\!\!\!}-\ketbra{\!\!\begin{array}{c}m_{1}~m_{2}\\ \boplus\!(\ulinek,\ulinem)\!\end{array}\!\!}\right)}_{1} \mathds{1}_{\left\{\substack{\boldw_{\tinyoplus}(\ulinek,\ulinem,\tildeboplus) \notin \\ T_{\delta}(p_{W_{1}\oplus W_{2}})} \right\}}\!
 \nonumber\\
 &&=\!\!\!\!\sum_{\substack{(m_{1},m_{2}) \in\\ [M_{1}]\times [M_{2}]}}\sum_{\substack{\tildeboplus \in [q^{n\beta_{2}}]\\\tildeboplus \neq  \boplus\!(\ulinek,\ulinem)}} \sum_{\substack{( b_{1}, b_{2}) \\\in \CalA(\tildeboplus)}}\!\!\!\norm{\Theta_{\ulinek,\ulinem, \ulineb}}_{1}\norm{ \left( \ketbra{\!\!\begin{array}{c}m_{1}~m_{2}\\\tildeboplus\!\!\end{array}\!\!\!}-\ketbra{\!\!\begin{array}{c}m_{1}~m_{2}\\ \boplus\!(\ulinek,\ulinem)\!\end{array}\!\!}\right)}_{1} \mathds{1}_{\left\{\substack{\boldw_{\tinyoplus}(\ulinek,\ulinem,\tildeboplus) \notin \\ T_{\delta}(p_{W_{1}\oplus W_{2}})} \right\}}\!
 \nonumber\\
 &&=2\!\!\!\!\sum_{\substack{(m_{1},m_{2}) \in\\ [M_{1}]\times [M_{2}]}}\sum_{\substack{\tildeboplus \in [q^{n\beta_{2}}]\\\tildeboplus \neq  \boplus\!(\ulinek,\ulinem)}} \sum_{\substack{( b_{1}, b_{2}) \\\in \CalA(\tildeboplus)}}\!\!\!\tr\left(\Theta_{\ulinek,\ulinem, \ulineb}\right) \mathds{1}_{\left\{\substack{\boldw_{\tinyoplus}(\ulinek,\ulinem,\tildeboplus) \notin  T_{\delta}(p_{W_{1}\oplus W_{2}})} \right\}}\qquad \text{and} \\
 &&\hspace{-0.8cm}\norm{\mathtt{t}_{B12}} \leq\!\!\!\!\sum_{\substack{(m_{1},m_{2}) \in\\ [M_{1}]\times [M_{2}]}}\sum_{\substack{\tildeboplus \in [q^{n\beta_{2}}]\\\tildeboplus \neq  \boplus\!(\ulinek,\ulinem)}} \sum_{\substack{( b_{1}, b_{2}) \\\in \CalA(\tildeboplus)}}\!\!\!\norm{\Theta_{\ulinek,\ulinem, \ulineb}\!\!\otimes \left( \ketbra{\!\!\begin{array}{c}m_{1}~m_{2}\\\tildeboplus\!\!\end{array}\!\!\!}-\ketbra{\!\!\begin{array}{c}m_{1}~m_{2}\\ \boplus\!(\ulinek,\ulinem)\!\end{array}\!\!}\right)}_{1} \mathds{1}_{\left\{\substack{\boldw_{\tinyoplus}(\ulinek,\ulinem,\tildeboplus) \in \\ T_{\delta}(p_{W_{1}\oplus W_{2}})} \right\}}\mathds{1}_{\left\{|\sfL(\ulinek,\ulinem)|\geq2 \right\}}
  \nonumber\\
  &&\hspace{0.3cm}=\!\!\!\!\sum_{\substack{(m_{1},m_{2}) \in\\ [M_{1}]\times [M_{2}]}}\sum_{\substack{\tildeboplus \in [q^{n\beta_{2}}]\\\tildeboplus \neq  \boplus\!(\ulinek,\ulinem)}} \sum_{\substack{( b_{1}, b_{2}) \\\in \CalA(\tildeboplus)}}\!\!\!\norm{\Theta_{\ulinek,\ulinem, \ulineb}}_{1}\norm{\left( \ketbra{\!\!\begin{array}{c}m_{1}~m_{2}\\\tildeboplus\!\!\end{array}\!\!\!}-\ketbra{\!\!\begin{array}{c}m_{1}~m_{2}\\ \boplus\!(\ulinek,\ulinem)\!\end{array}\!\!}\right)}_{1} \mathds{1}_{\left\{\substack{\boldw_{\tinyoplus}(\ulinek,\ulinem,\tildeboplus) \in \\ T_{\delta}(p_{W_{1}\oplus W_{2}})} \right\}}\mathds{1}_{\left\{|\sfL(\ulinek,\ulinem)|\geq2 \right\}}
  \nonumber\\
  &&\hspace{0.3cm}=2\!\!\!\!\sum_{\substack{(m_{1},m_{2}) \in\\ [M_{1}]\times [M_{2}]}}\sum_{\substack{\tildeboplus \in [q^{n\beta_{2}}]\\\tildeboplus \neq  \boplus\!(\ulinek,\ulinem)}} \sum_{\substack{( b_{1}, b_{2}) \\\in \CalA(\tildeboplus)}}\!\!\!\tr\left(\Theta_{\ulinek,\ulinem, \ulineb}\right) \mathds{1}_{\left\{\substack{\boldw_{\tinyoplus}(\ulinek,\ulinem,\tildeboplus) \in  T_{\delta}(p_{W_{1}\oplus W_{2}})} \right\}}\mathds{1}_{\left\{|\sfL(\ulinek,\ulinem)|\geq2 \right\}}.
\end{eqnarray*}
Recalling our breakup of $B_{\ulinek}$ in \eqref{Eqn:DISCCAnalysisBk-6} and the fact $\mathtt{t}_{B10}=\mathtt{t}_{B11}=0$, we have the following.
\begin{lemma}
 \label{Lem:DISCCFnlBndOnBk}
\begin{eqnarray}
 \label{Eqn:DISCCAnalysisBkFnlBnd}
 \hspace{-0.1cm}B_{\ulinek} \leq \mathtt{t}_{B}\define 2\!\!\!\!\!\!\!\sum_{\substack{(m_{1},m_{2}) \in\\ [M_{1}]\times [M_{2}]}}\sum_{\substack{\tildeboplus \in [q^{n\beta_{2}}]\\\tildeboplus \neq  \boplus\!(\ulinek,\ulinem)}} \sum_{\substack{( b_{1}, b_{2}) \\\in \CalA(\tildeboplus)}}\!\!\!\tr\left(\Theta_{\ulinek,\ulinem, \ulineb}\right) \left(\mathds{1}_{\left\{\substack{\boldw_{\tinyoplus}(\ulinek,\ulinem,\tildeboplus) \in  T_{\delta}(p_{W_{1}\oplus W_{2}})} \right\}}\mathds{1}_{\left\{|\sfL(\ulinek,\ulinem)|\geq2 \right\}}+\mathds{1}_{\left\{\substack{\boldw_{\tinyoplus}(\ulinek,\ulinem,\tildeboplus) \notin  T_{\delta}(p_{W_{1}\oplus W_{2}})} \right\}}\right).
\end{eqnarray}

 \end{lemma}

\noindent\textit{\underline{Bipartite Soft Covering error $A$}} : Recall that $\Theta_{\ulinek,\ulinem, \ulineb}$ is defined in \eqref{Eqn:DISCCAnalysisBk-3}. Recall also $A$ from \eqref{Eqn:DISCCBreakdownErrorEvent02} and the proxy decoder map $\tilde{D}_{\ulinek,2}=D_{\ulinek,2}$ given by  \eqref{Eqn:DISCCStochPostProcessing}. The second term within the norm defining $A$ in \eqref{Eqn:DISCCBreakdownErrorEvent02} is given by
\begin{eqnarray}
\label{Eqn:DISCCTermA-1}
&&\hspace{-0.5cm}\sum_{k_{1},k_{2}}\frac{\tilde{\CalD}_{\ulinek}\!\circ\!{\tilde{\CalV}_{k_{1},k_{2}}}\left( \varphi_{\rho^{12}}\right)}{K_{1}K_{2}}
=\sum_{k_{1},k_{2}}\frac{1}{K_{1}K_{2}}
\sum_{\substack{(m_{1},m_{2}) \in\\ [M_{1}]\times [M_{2}]}}
\sum_{\substack{\boplus \in [q^{n\beta_{2}}],\\y^{n} \in \CalY^{n}}}
\sum_{\substack{( b_{1}, b_{2}) \\\in \CalA(\boplus)}}
\Theta_{\ulinek,\ulinem, \ulineb} \otimes p_{Y|W_{1}\oplus W_{2}}^{n}(y^{n}|\boldw_{\tinyoplus}^{n}(m_{1},m_{2},\ulineb))\ketbra{y^{n}}
\nonumber\\
\label{Eqn:DISCCTermA-3}
&&\hspace{-0.7cm}=
\sum_{\substack{m_{1},m_{2},\\k_{1},k_{2}}}
\sum_{\substack{\boplus \in [q^{n\beta_{2}}],\\y^{n} \in \CalY^{n}}}
\sum_{\substack{( b_{1}, b_{2}) \\\in \CalA(\boplus)}}
\prod_{i=1}^{2}\frac{\mt(k_{i},m_{i},b_{i})}{K_{i}M_{i}B_{i}p_{W_{i}}(w({k_{i},m_{i},b_{i}}))}
\boldsymbol{\tau}^{RB}_{w_{1}(k_{1},m_{1},b_{1}),\,w_{2}(k_{2},m_{2},b_{2})} \otimes p_{Y|W_{1}\oplus W_{2}}^{n}(y^{n}|\boldw_{\tinyoplus}^{n}(m_{1},m_{2},\ulineb))\ketbra{y^{n}}
\nonumber\\
&&\hspace{-0.7cm}=
\sum_{\substack{m_{1},m_{2},\\k_{1},k_{2}}}
\sum_{\substack{\boplus \in [q^{n\beta_{2}}],\\y^{n} \in \CalY^{n}}} \sum_{\substack{( b_{1}, b_{2}) \\\in \CalA(\boplus)}}
\prod_{i=1}^{2}\frac{q^{n}}{K_{i}M_{i}B_{i}}
\boldsymbol{\tau}^{RB}_{w_{1}(k_{1},m_{1},b_{1}),\,w_{2}(k_{2},m_{2},b_{2})} \otimes p_{Y|W_{1}\oplus W_{2}}^{n}(y^{n}|\boldw_{\tinyoplus}^{n}(m_{1},m_{2},\ulineb))\ketbra{y^{n}}
\end{eqnarray}
Using \eqref{Eqn:DISCCMarkovChain-2}, the first term within the norm defining $A$ in \eqref{Eqn:DISCCBreakdownErrorEvent02} is
\begin{eqnarray}
\label{Eqn:DISCCTermA-7}
\{\mathrm{id}_{R}\otimes\CalJ\}(\boldrho^{12}) = \!\!\!\!\sum_{\substack{y^{n} \in \CalY^{n}\\w^{n} \in \CalW^{n}}}\sum_{\substack{w_{1} \in \CalW^{n}_{1}\\w_{2} \in \CalW_{2}^{n}}}\!\!\!p_{Y|W_{1}\oplus W_{2}}^{n}(y^{n}|w^{n})\mathds{1}_{\left\{\substack{w_{1}^{n}\oplus w_{2}^{n}=w^{n}} \right\}}\boldsymbol{\tau}_{w_1^n,w_2^n}^{RB}\otimes \ketbra{y^{n}}.
\end{eqnarray}
An informed reader will note that proving the last term of \eqref{Eqn:DISCCTermA-1} is close to the RHS of \eqref{Eqn:DISCCTermA-7} in trace norm will involve application of a quantum classical covering convergence lemma. Specifically, one may note that if the generator matrices and bias vectors defining the union coset code are uniformly distributed and independent, then the expectation of the corresponding quantity on the last term of \eqref{Eqn:DISCCTermA-1} will be equal to the RHS of \eqref{Eqn:DISCCTermA-7}. Thereby, one can note that bounding $A$ on the above will be a rather straightforward application of the quantum classical covering. We thus take the following expression for $A$ forward for the final step of the analysis.

\begin{lemma}
\label{Lem:DISCFinalBoundOnA}
Referring to $A$ in \eqref{Eqn:DISCCBreakdownErrorEvent02} and to the associated maps in
\eqref{Eqn:DISCCStochPostProcessing} and \eqref{Eqn:DISCCInstWithBinIndex}, we have
$A=\sfa(\sfc^{1},\sfc^{2})$, where
\begin{eqnarray}
\label{Eqn:DISCCFnlBndOnA}
&&\hspace{-1cm}\sfa(\sfc^{1},\sfc^{2})
\define
\left\|
\begin{aligned}
&\sum_{\substack{y^{n}\in\CalY^{n}\\ w^{n}\in\CalW^{n}}}
\sum_{\substack{w_{1}^{n}\in\CalW_{1}^{n}\\ w_{2}^{n}\in\CalW_{2}^{n}}}
p_{Y|W_{1}\oplus W_{2}}^{n}(y^{n}|w^{n})
\mathds{1}_{\{w_{1}^{n}\oplus w_{2}^{n}=w^{n}\}}
\boldsymbol{\tau}_{w_1^n,w_2^n}^{RB}\otimes\ketbra{y^{n}}
\\
&-
\sum_{\substack{m_{1},m_{2},\\k_{1},k_{2}}}
\sum_{\substack{\boplus \in [q^{n\beta_{2}}],\\y^{n} \in \CalY^{n}}} \sum_{\substack{( b_{1}, b_{2}) \\\in \CalA(\boplus)}}
\prod_{i=1}^{2}\frac{q^{n}}{K_{i}M_{i}B_{i}}
\boldsymbol{\tau}^{RB}_{w_{1}(k_{1},m_{1},b_{1}),\,w_{2}(k_{2},m_{2},b_{2})} \otimes p_{Y|W_{1}\oplus W_{2}}^{n}(y^{n}|\boldw_{\tinyoplus}^{n}(m_{1},m_{2},\ulineb))\ketbra{y^{n}}
\end{aligned}
\right\|_{1}.
\end{eqnarray}
\end{lemma}
\med\textbf{Collating Bounds, Code Randomization \& Rate Bounds.} Plugging the bounds from \eqref{Eqn:DISCCQuantumOnlyTermDk2Purif}, \eqref{Eqn:DISCCFinalBoundOnCk}, Lemma~\ref{Lem:DISCCFnlBndOnBk}, and Lemma~\ref{Lem:DISCFinalBoundOnA} into \eqref{Eqn:DISCCBreakdownErrorEvent02}, we obtain
\begin{eqnarray}
\label{Eqn:DISCCCollatingBounds1}
\norm{\left(\mathrm{id}_{R}\otimes \CalJ-\CalS\right)\{\varphi_{\rho^{12}}\}}_{1}
\leq
\sum_{k_{1},k_{2}}
\frac{
\left(1+\sqrt{\tr S_{k_{2}}}\right)\sqrt{\norm{\boldrho^{2}-S_{k_{2}}}_{1}}
+\norm{\left[\CalV_{k_{2}}\circ\boldCalQ^{1}\right]\left(\varphi_{S_{k_{1}}}-\varphi_{\boldrho^{1}}\right)}_{1}
+\mathtt{t}_{B}
}{K_{1}K_{2}}
+\sfa(\sfc^{1},\sfc^{2}).
\end{eqnarray}
Here, $\mathtt{t}_{B}$ and $\sfa(\sfc^{1},\sfc^{2})$ are defined in \eqref{Eqn:DISCCAnalysisBkFnlBnd} and \eqref{Eqn:DISCCFnlBndOnA}, respectively.
As is standard, we prove the existence of multi-codes $\sfc^{1},\sfc^{2}$ for which the bound in \eqref{Eqn:DISCCCollatingBounds1} is small via the codebook randomization technique. Recalling that our codes $\sfc^{1}$ and $\sfc^{2}$ are UCCs possessing joint structure, how do choose the distribution of our random code? To specify this, let us first recall all the objects that define our codes. Referring back to the description prior to Rem.~\ref{Rem:DISCosetCdsUniqueness}, we note that matrices $g_{1} \in \CalF_{q}^{n\beta_{1}\times n}, \Delta g \in \CalF_{q}^{n(\beta_{2}-\beta_{1})\times n}$ and the two collections $(\upsilon_{k_{i}}(m_{i})\in \CalF_{q}^{n} : (k_{i},m_{i}) \in [K_{i}]\times [M_{i}] )$ completely specify the code. We now specify a joint distribution of these objects and thereby specify the distribution of the random code. For $i \in [2]$, let $\sfC^{i}= (\sfC^{i}_{k_{i}}: k_{i} \in [K_{i}])$ be a random multi-code specified via generator matrices $G_{1} \in \CalF_{q}^{n\beta_{1}\times n}, \Delta G \in \CalF_{q}^{n(\beta_{2}-\beta_{1})\times n}$ and the two collections $(\Upsilon_{k_{i}}(m_{i})\in \CalF_{q}^{n} : (k_{i},m_{i}) \in [K_{i}]\times [M_{i}] )$ with distribution
\begin{eqnarray*}
 P\left(\!\!\begin{array}{c}G_{1} = g_{1}, \Delta G = \Delta g\\ \Upsilon_{k_{i}}(m_{i}) = \gamma^{n}(k_{i},m_{i})  : (k_{i},m_{i}) \in [K_{i}]\times [M_{i}]\end{array}\!\!\!\right) =\left(\frac{1}{q^{n}}\right)^{n\beta_{1}}\left(\frac{1}{q^{n}}\right)^{n(\beta_{2}-\beta_{1})}\left( \frac{1}{q^{n}}\right)^{K_{1}M_{1}K_{2}M_{2}}.
\end{eqnarray*}
In other words, each of the objects that define the codes are uniformly distributed and the collection of objects are mutually independent. Here, we recall $G_{2} = \left[\!\!\begin{array}{c}G_{1}\\\Delta G \end{array}\!\!\right]$Two important consequences of this distribution is that for each $(k_{i},m_{i},\ulineb_{i}) \in [K_{i}]\times [M_{i}]\times [B_{i}]$, the random codeword $\boldW_{i}(k_{i},m_{i},\ulineb_{i}) = \ulineb_{i}G_{i} \oplus \Upsilon_{k_{i}}(m_{i})\in \CalW_{i}^{n}$ is uniformly distributed and pairwise independent. This is easily verified via a simple counting argument.
\begin{remark}
 \label{Rem:DISCCDistOfUCC}
 From the above specified distribution of the random code, it is evident that the codewords are neither mutually independent nor distributed according to the standard distribution $\prod_{t=1}^{n}p_{W_{i}}$ as in proof of Thm.~\ref{Thm:DistInstSimltnOneShot}. The inclusion of measure transformation factor $\mt(\cdot)$ coupled with changes to the proof enables us handle this altered random code distribution.
\end{remark}

We now evaluate an upper bound on $\Expectation_{\ulinesfC}\left\{\norm{\left(\mathrm{id}_{R}\otimes \CalJ-\CalS\right)\{\ketbra{\varphi_{\rho^{12}}}\}}_{1}\right\}$, where $\Expectation_{\ulinesfC}\{\cdot\}$ denotes expectation over the random pair of multi-codes $\underline{\sfC}=(\sfC^{1},\sfC^{2})$. Since \eqref{Eqn:DISCCCollatingBounds1} holds for every generic multi-code $\ulinesfc=(\sfc^{1},\sfc^{2})$, we first note that, for every $k_2$, Cauchy--Schwarz gives
\begin{eqnarray}
\Expectation_{\ulinesfC}\!\left\{\left(1+\sqrt{\tr S_{k_2}}\right)\sqrt{\norm{\boldrho^{2}-S_{k_2}}_{1}}\right\}
&\leq&
\sqrt{\Expectation_{\ulinesfC}\!\left\{\left(1+\sqrt{\tr S_{k_2}}\right)^2\right\}}\sqrt{\Expectation_{\ulinesfC}\!\left\{\norm{\boldrho^{2}-S_{k_2}}_{1}\right\}}
\nonumber\\
&=&
\sqrt{1+2\Expectation_{\ulinesfC}\!\left\{\sqrt{\tr S_{k_2}}\right\}+\Expectation_{\ulinesfC}\!\left\{\tr S_{k_2}\right\}}\sqrt{\Expectation_{\ulinesfC}\!\left\{\norm{\boldrho^{2}-S_{k_2}}_{1}\right\}}
\nonumber\\
&\leq&
\left(1+\sqrt{\Expectation_{\ulinesfC}\!\left\{\tr S_{k_2}\right\}}\right)\sqrt{\Expectation_{\ulinesfC}\!\left\{\norm{\boldrho^{2}-S_{k_2}}_{1}\right\}},
\label{Eqn:DISCCCauchySchwarzSk2}
\end{eqnarray}
where the last inequality follows from concavity of the square root. Taking expectation in \eqref{Eqn:DISCCCollatingBounds1} and using \eqref{Eqn:DISCCCauchySchwarzSk2}, we obtain
\begin{eqnarray*}
\mathbb{E}_{\sfC}\left\{
\norm{\left(\mathrm{id}_{R}\otimes \CalJ-\CalS\right)\{\ketbra{\varphi_{\rho^{12}}}\}}_{1}
\right\}
&\leq&
\mathbb{E}_{\sfC}\left\{\sfA(\sfC^{1},\sfC^{2})\right\}+\sum_{k_{1},k_{2}}
\frac{
\left(1+\sqrt{\mathbb{E}_{\sfC}\{\tr S_{k_{2}}\}}\right)
\sqrt{\mathbb{E}_{\sfC}\left\{\norm{\boldrho^{2}-S_{k_{2}}}_{1}\right\}}
}{K_{1}K_{2}}
\nonumber\\
&&+
\sum_{k_{1},k_{2}}
\frac{
\mathbb{E}_{\sfC}\left\{\norm{\left[\CalV_{k_{2}}\circ\boldCalQ^{1}\right]\left(\varphi_{S_{k_{1}}}-\varphi_{\boldrho^{1}}\right)}_{1}\right\}
+\mathbb{E}_{\sfC}\{\mathtt{t}_{B}\}
}{K_{1}K_{2}}.
\end{eqnarray*}
By the uniform marginal distribution of each UCC codeword and the
measure-transformation factor in \eqref{Eqn:DISCosetCdsLikelihoodPOVMdefnMeasTrans_i},
we have, for every $k_i$,
\begin{equation}
\label{Eqn:SkunittraceinExp}
\mathbb{E}_{\sfC}[S_{k_i}]=\boldrho^i,
\qquad\text{and hence}\qquad
\mathbb{E}_{\sfC}[\tr S_{k_i}]=1.
\end{equation}
We first bound the quantum-only term $C_{\ulinek,2}$. Fix a realization of $\sfC^1$. Under the conditional random ensemble $\sfC^2|\sfC^1$, every codeword $W_2(k_2,m_2,b_2)$ remains uniformly distributed on $\boldCalW_2=\mathbb{F}_q^n$, and hence has marginal distribution $\nu_{\boldW_2}(\boldw_2)=q^{-n}$. Moreover, by the definition of the measure-transformation factor in \eqref{Eqn:DISCosetCdsLikelihoodPOVMdefnMeasTrans_i},
\begin{eqnarray*}
\frac{\mt(k_2,m_2,b_2)}{p_{W_2}(W_2(k_2,m_2,b_2))}=q^n=\frac{1}{\nu_{\boldW_2}(W_2(k_2,m_2,b_2))}.
\end{eqnarray*}
Thus, conditionally on $\sfC^1$, $\CalV_{k_2}$ is precisely a measure-transformed inverse-free map of the form considered in Lemma~\ref{Lem:AveragedInverseFreeContraction2}. Since $\boldCalQ^1(\varphi_{S_{k_1}}-\varphi_{\boldrho^1})$ is fixed under this conditioning and is Hermitian, \eqref{Eqn:DISCCFinalBoundOnCk} and Lemma~\ref{Lem:AveragedInverseFreeContraction2} give
\begin{eqnarray*}
\Expectation_{\sfC^{2}|\sfC^{1}}\!\left\{
\norm{\left[\CalV_{k_{2}}\circ\boldCalQ^{1}\right]
\left(\varphi_{S_{k_{1}}}-\varphi_{\boldrho^{1}}\right)}_{1}
\right\}
&\leq&
\norm{\boldCalQ^{1}\left(\varphi_{S_{k_{1}}}-\varphi_{\boldrho^{1}}\right)}_{1}
=
\norm{\varphi_{S_{k_{1}}}-\varphi_{\boldrho^{1}}}_{1}.
\end{eqnarray*}
where the equality follows since $\boldCalQ^1$ is an isometry. Averaging over $\sfC^1$, applying Lemma~\ref{lem:canonical-purif-stability}, and reasoning as in \eqref{Eqn:DISCCCauchySchwarzSk2} using \eqref{Eqn:SkunittraceinExp}, we obtain
\begin{eqnarray*}
\Expectation_{\sfC}\!\left\{
\norm{\left[\CalV_{k_{2}}\circ\boldCalQ^{1}\right]
\left(\varphi_{S_{k_{1}}}-\varphi_{\boldrho^{1}}\right)}_{1}
\right\}
\leq
2\sqrt{\Expectation_{\sfC}\!\left\{\norm{\boldrho^{1}-S_{k_{1}}}_{1}\right\}}.
\end{eqnarray*}
We continue with $\Expectation\left\{ \mathtt{t}_{B} \right\}$ and evaluate $\Expectation_{\sfC}\left\{ \cdot \right\}$ on the RHS of \eqref{Eqn:DISCCAnalysisBkFnlBnd}. Referring to \eqref{Eqn:DISCCAnalysisBk-3}, \eqref{Eqn:DistMCPKeySplitOfError6}, \eqref{Eqn:DISCCDefnOfP_W} for the necessary terms, we have
\begin{eqnarray}
\label{Eqn:DISCCCollatingBounds8}
&&\hspace{-1cm} \mathtt{t}_{B}  \leq 2\!\!\!\!\!\!\!\sum_{\substack{(m_{1},m_{2}) \in\\ [M_{1}]\times [M_{2}]}}\sum_{\substack{\tildeboplus \in [q^{n\beta_{2}}]\\\tildeboplus \neq  \boplus\!(\ulinek,\ulinem)}} \sum_{\substack{( b_{1}, b_{2}) \\\in \CalA(\tildeboplus)}}\!\!\!\tr\left(\displaystyle\left[\bigotimes_{i=1}^{2}\sfV_{k_{i},m_{i},b_{i}}^{\dagger}\sfV_{k_{i},m_{i},b_{i}}\right]\varphi_{\boldrho_{12}} \right) \left(\!\!\!\begin{array}{c}\mathds{1}_{\left\{\substack{\boldw_{\tinyoplus}(\ulinek,\ulinem,\tildeboplus) \in  T_{\delta}(p_{W_{1}\oplus W_{2}})} \right\}}\mathds{1}_{\left\{|\sfL(\ulinek,\ulinem)|\geq2 \right\}}\\+\mathds{1}_{\left\{\substack{\boldw_{\tinyoplus}(\ulinek,\ulinem,\tildeboplus) \notin  T_{\delta}(p_{W_{1}\oplus W_{2}})} \right\}}\end{array}\!\!\!\right)
  \nonumber\\
  \label{Eqn:DISCCCollatingBounds9}
&&\hspace{-0.5cm} = 2\!\!\!\!\!\!\!\sum_{\substack{(m_{1},m_{2}) \in\\ [M_{1}]\times [M_{2}]}}\sum_{\substack{\tildeboplus \in [q^{n\beta_{2}}]\\\tildeboplus \neq  \boplus\!(\ulinek,\ulinem)}} \sum_{\substack{( b_{1}, b_{2}) \\\in \CalA(\tildeboplus)}}\!\!\!\tr\left(\displaystyle\left[\bigotimes_{i=1}^{2}\frac{\mt(k_{i},m_{i},b_{i})\mu_{k_{i},m_{i},b_{i}}}{M_{i}B_{i}\boldp_{W_{i}}(\boldw(k_{i},m_{i},\tildeboplus_{i})) }\right]\boldrho^{12} \right) \left(\!\!\!\begin{array}{c}\mathds{1}_{\left\{\substack{\boldw_{\tinyoplus}(\ulinek,\ulinem,\tildeboplus) \in  T_{\delta}(p_{W_{1}\oplus W_{2}})} \right\}}\mathds{1}_{\left\{|\sfL(\ulinek,\ulinem)|\geq2 \right\}}\\+\mathds{1}_{\left\{\substack{\boldw_{\tinyoplus}(\ulinek,\ulinem,\tildeboplus) \notin  T_{\delta}(p_{W_{1}\oplus W_{2}})} \right\}}\end{array}\!\!\!\right)
  \nonumber\\
  \label{Eqn:DISCCCollatingBounds10}
&&\hspace{-0.5cm} =2\!\!\!\!\!\!\!\sum_{\substack{(m_{1},m_{2}) \in\\ [M_{1}]\times [M_{2}]}}\sum_{\substack{\tildeboplus \in [q^{n\beta_{2}}]\\\tildeboplus \neq  \boplus\!(\ulinek,\ulinem)}} \sum_{\substack{( b_{1}, b_{2}) \\\in \CalA(\tildeboplus)}}\!\!\!\boldp_{W_{1}W_{2}}\!\!\left(\!\!\!\begin{array}{c}\boldw_{1}(k_{1},m_{1}, b_{1})\\\boldw_{2}(k_{2},m_{2}, b_{2})\end{array}\!\!\!\right)\left(\!\!\!\begin{array}{c}\mathds{1}_{\left\{\substack{\boldw_{\tinyoplus}(\ulinek,\ulinem,\tildeboplus) \in  T_{\delta}(p_{W_{1}\oplus W_{2}})} \right\}}\mathds{1}_{\left\{|\sfL(\ulinek,\ulinem)|\geq2 \right\}}\\+\mathds{1}_{\left\{\substack{\boldw_{\tinyoplus}(\ulinek,\ulinem,\tildeboplus) \notin  T_{\delta}(p_{W_{1}\oplus W_{2}})} \right\}}\end{array}\!\!\!\right)\prod_{i=1}^{2}\frac{q^{n}}{M_{i}B_{i}}
\end{eqnarray}
where (i) \eqref{Eqn:DISCCCollatingBounds10} follows from \eqref{Eqn:DISCCDefnOfP_W} and \eqref{Eqn:DISCosetCdsLikelihoodPOVMdefnMeasTrans_i}. Now, note that by definition of $\boldw_{\oplus}(\ulinek,\ulinem,\tildeboplus)$ in \eqref{Eqn:DISCosetCdsSumCode} and $\CalA(\tildeboplus)$ in \eqref{Eqn:DISCCCompatibleBinIndices}, we have
\begin{eqnarray*}
\boldw_{\tinyoplus}(\ulinek,\ulinem,\tildeboplus) = \tildeboplus g_{2} \oplus \upsilon_{k_{1}}(m_{1})\oplus \upsilon_{k_{2}}(m_{2}) = b_{1}g_{1}\oplus \upsilon_{k_{1}}(m_{1}) \oplus  b_{2}g_{2}\oplus \upsilon_{k_{2}}(m_{2}) = \boldw_{1}(k_{1},m_{1}, b_{1}) \oplus \boldw_{2}(k_{2},m_{2}, b_{2}).
\end{eqnarray*}
Evaluating $\Expectation_{\sfC}\{\cdot \}$ on the second term corresponding to the second indicator function in \eqref{Eqn:DISCCCollatingBounds10}, we have
\begin{eqnarray}
 \label{Eqn:DISCCCollatingBounds13}
 &&\hspace{-1.2cm} 2\prod_{i=1}^{2}\frac{q^{n}}{M_{i}B_{i}}\!\sum_{\substack{(m_{1},m_{2}) \in\\ [M_{1}]\times [M_{2}]}}\sum_{\substack{\tildeboplus \in [q^{n\beta_{2}}]\\\tildeboplus \neq  \boplus\!(\ulinek,\ulinem)}} \sum_{\substack{( b_{1}, b_{2}) \\\in \CalA(\tildeboplus)}}\!\!\!\Expectation_{\sfC}\left\{\boldp_{W_{1}W_{2}}\!\!\left(\!\!\!\begin{array}{c}\boldW_{1}(k_{1},m_{1}, b_{1}),\boldW_{2}(k_{2},m_{2}, b_{2})\end{array}\!\!\!\right)\!\!\!\begin{array}{c}\mathds{1}_{\left\{\substack{\boldW_{\tinyoplus}(\ulinek,\ulinem,\tildeboplus) \notin  T_{\delta}(p_{W_{1}\oplus W_{2}})} \right\}}\end{array}\!\!\!\right\}
 \nonumber\\
 \label{Eqn:DISCCCollatingBounds14}
 &&\hspace{-1.2cm} \leq 2\left[\prod_{i=1}^{2}\frac{q^{n}}{M_{i}B_{i}}\right]\!\sum_{\substack{(m_{1},m_{2}) \in\\ [M_{1}]\times [M_{2}]}}\sum_{\substack{\tildeboplus \in [q^{n\beta_{2}}]}} \sum_{\substack{( b_{1}, b_{2}) \\\in \CalA(\tildeboplus)}}\!\!\!\Expectation_{\sfC}\left\{\boldp_{W_{1}W_{2}}\!\!\left(\!\!\!\begin{array}{c}\boldW_{1}(k_{1},m_{1}, b_{1}),\boldW_{2}(k_{2},m_{2}, b_{2})\end{array}\!\!\!\right)\!\!\!\begin{array}{c}\mathds{1}_{\left\{\substack{\boldW_{\tinyoplus}(\ulinek,\ulinem,\tildeboplus) \notin  T_{\delta}(p_{W_{1}\oplus W_{2}})} \right\}}\end{array}\!\!\!\right\}
\\
 \label{Eqn:DISCCCollatingBounds15}
 &&\hspace{-1.2cm} =2\left[\prod_{i=1}^{2}\frac{q^{n}}{M_{i}B_{i}}\right]\!\sum_{\substack{(m_{1},m_{2}) \in\\ [M_{1}]\times [M_{2}]}}\sum_{\substack{\tildeboplus \in [q^{n\beta_{2}}]}} \sum_{\substack{( b_{1}, b_{2}) \\\in \CalA(\tildeboplus)}}\sum_{w_{1}^{n},w_{2}^{n}}\!\!\!\boldp_{W_{1}W_{2}}(w_{1}^{n},w_{2}^{n})\begin{array}{c}\mathds{1}_{\left\{\substack{ w_{1}^{n}\oplus w_{2}^{n} \notin  T_{\delta}(p_{W_{1}\oplus W_{2}})} \right\}}\end{array}\!\!\!\frac{1}{q^{n}\cdot q^{n}}
\\
 \label{Eqn:DISCCCollatingBounds16}
 &&\hspace{-1.2cm} =2\left[\prod_{i=1}^{2}\frac{q^{n}}{M_{i}B_{i}}\right]\!\sum_{\substack{(m_{1},m_{2}) \in\\ [M_{1}]\times [M_{2}]}} \sum_{\substack{( b_{1}, b_{2}) \\\in [B_{1}]\times[q^{n\beta_{2}}]}}\sum_{w^{n}}\sum_{w_{1}^{n},w_{2}^{n}}\!\!\!\boldp_{W_{1}W_{2}}(w_{1}^{n},w_{2}^{n})\begin{array}{c}\mathds{1}_{\left\{\substack{w^{n}=w_{1}^{n}\oplus w_{2}^{n}} \right\}}\mathds{1}_{\left\{\substack{ w^{n} \notin  T_{\delta}(p_{W_{1}\oplus W_{2}})} \right\}}\end{array}\!\!\!\frac{1}{q^{n}\cdot q^{n}}
\\
 \label{Eqn:DISCCCollatingBounds17}
 &&\hspace{-1.2cm} = \sum_{w^{n}}\mathds{1}_{\left\{\substack{ w^{n} \notin  T_{\delta}(p_{W_{1}\oplus W_{2}})} \right\}} \sum_{w_{1}^{n},w_{2}^{n}}\!\!\!\boldp_{W_{1}W_{2}}(w_{1}^{n},w_{2}^{n})\begin{array}{c}\mathds{1}_{\left\{\substack{w^{n}=w_{1}^{n}\oplus w_{2}^{n}} \right\}}\end{array}\!\!\! = \sum_{w^{n}}\mathds{1}_{\left\{\substack{ w^{n} \notin  T_{\delta}(p_{W_{1}\oplus W_{2}})} \right\}} \boldp_{W_{1}\oplus  W_{2}}(w^{n}) \leq 2^{-n\delta}
\end{eqnarray}
where (i) \eqref{Eqn:DISCCCollatingBounds14} bumps up the value by adding an extra non-negative term, (ii) \eqref{Eqn:DISCCCollatingBounds15} uses the fact that $\boldW_{1}(k_{1},m_{1}, b_{1})$ and $\boldW_{2}(k_{2},m_{2}, b_{2})$ are pairwise independent and uniformly distributed (iii) \eqref{Eqn:DISCCCollatingBounds16} expresses the sum differently, (iv) the second equality in \eqref{Eqn:DISCCCollatingBounds17} follows from \eqref{Eqn:DISCCDefnOfP_W} and the last inequality in \eqref{Eqn:DISCCCollatingBounds17} follows from standard results from typicality. Next, evaluating $\Expectation_{\sfC}\{\cdot \}$ on the first term corresponding to the first indicator function in \eqref{Eqn:DISCCCollatingBounds10}, proceeding through an analogous sequence of steps and arguments as above, we have
\begin{eqnarray}
 \label{Eqn:DISCCCollatingBounds18}
&&\hspace{-0.7cm} 2\left[\prod_{i=1}^{2}\frac{q^{n}}{M_{i}B_{i}}\right]\!\sum_{\substack{(m_{1},m_{2}) \in\\ [M_{1}]\times [M_{2}]}}\sum_{\substack{\tildeboplus \in [q^{n\beta_{2}}]\\\tildeboplus \neq  \boplus\!(\ulinek,\ulinem)}} \sum_{\substack{( b_{1}, b_{2}) \\\in \CalA(\tildeboplus)}}\!\!\!\Expectation_{\sfC}\left\{\boldp_{W_{1}W_{2}}\!\!\left(\!\!\!\begin{array}{c}\boldW_{1}(k_{1},m_{1}, b_{1}),\boldW_{2}(k_{2},m_{2}, b_{2})\end{array}\!\!\!\right)\!\!\!\begin{array}{c}\mathds{1}_{\left\{\substack{\boldW_{\tinyoplus}(\ulinek,\ulinem,\tildeboplus) \\\in  T_{\delta}(p_{W_{1}\oplus W_{2}})} \right\}}\mathds{1}_{\left\{|\sfL(\ulinek,\ulinem)|\geq2 \right\}}\end{array}\!\!\!\right\}
\nonumber\\
 \label{Eqn:DISCCCollatingBounds19}
&&\hspace{-0.5cm} \leq 2\left[\prod_{i=1}^{2}\frac{q^{n}}{M_{i}B_{i}}\right]\!\sum_{\substack{(m_{1},m_{2}) \in\\ [M_{1}]\times [M_{2}]}}\sum_{\substack{\boplus \in [q^{n\beta_{2}}]}} \sum_{\substack{( b_{1}, b_{2}) \\\in \CalA(\boplus)}}\!\!\!\Expectation_{\sfC}\left\{\boldp_{W_{1}W_{2}}\!\!\left(\!\!\!\begin{array}{c}\boldW_{1}(k_{1},m_{1}, b_{1}),\boldW_{2}(k_{2},m_{2}, b_{2})\end{array}\!\!\!\right)\!\!\!\begin{array}{c}\mathds{1}_{\left\{\substack{\boldW_{\tinyoplus}(\ulinek,\ulinem,\boplus) \\\in  T_{\delta}(p_{W_{1}\oplus W_{2}})} \right\}}\mathds{1}_{\left\{|\sfL(\ulinek,\ulinem)|\geq2 \right\}}\end{array}\!\!\!\right\}
\nonumber\\
\label{Eqn:DISCCCollatingBounds20}
&&\hspace{-0.5cm} \leq \!\sum_{\substack{(m_{1},m_{2}) \in\\ [M_{1}]\times [M_{2}]}}\sum_{\substack{\boplus \in [q^{n\beta_{2}}]}} \sum_{\substack{( b_{1}, b_{2}) \\\in \CalA(\boplus)}}\!\!\!\nonumber\\
&&\hspace{-0.5cm}{}\times \frac{2q^{2n}\Expectation_{\sfC}\left\{\boldp_{W_{1}W_{2}}\!\!\left(\!\!\!\begin{array}{c}\boldW_{1}(k_{1},m_{1}, b_{1}),\boldW_{2}(k_{2},m_{2}, b_{2})\end{array}\!\!\!\right)\!\!\!\begin{array}{c}\mathds{1}_{\left\{\substack{\boldW_{\tinyoplus}(\ulinek,\ulinem,\boplus) \\\in  T_{\delta}(p_{W_{1}\oplus W_{2}})} \right\}}\displaystyle\sum_{\substack{\tildeboplus \in [q^{n\beta_{2}}]\\\tildeboplus\neq \boplus}}\mathds{1}_{\left\{\substack{\boldW_{\tinyoplus}(\ulinek,\ulinem,{\tildeboplus}) \\\in  T_{\delta}(p_{W_{1}\oplus W_{2}})} \right\}}\end{array}\!\!\!\right\}}{M_{1}B_{1}M_{2}B_{2}}
\end{eqnarray}
Focussing on the $\Expectation_{\sfC}\{\cdot \}$ above, we have
\begin{eqnarray}
 \label{Eqn:DISCCCollatingBounds21}
&&\hspace{-1cm}\Expectation_{\sfC}\left\{\boldp_{W_{1}W_{2}}\!\!\left(\!\!\!\begin{array}{c}\boldW_{1}(k_{1},m_{1}, b_{1}),\boldW_{2}(k_{2},m_{2}, b_{2})\end{array}\!\!\!\right)\!\!\!\begin{array}{c}\mathds{1}_{\left\{\substack{\boldW_{\tinyoplus}(\ulinek,\ulinem,\boplus) \\\in  T_{\delta}(p_{W_{1}\oplus W_{2}})} \right\}}\displaystyle\mathds{1}_{\left\{\substack{\boldW_{\tinyoplus}(\ulinek,\ulinem,\tildeboplus) \\\in  T_{\delta}(p_{W_{1}\oplus W_{2}})} \right\}}\end{array}\!\!\!\right\}
 \nonumber\\
 \label{Eqn:DISCCCollatingBounds22}
&&= \sum_{\boldw_{1},\boldw_{2}}\sum_{\tilde{\boldw}}\boldp_{W_{1}W_{2}}(\boldw_{1},\boldw_{2})\mathds{1}_{\left\{\substack{\boldw_{1}\oplus \boldw_{2} \in  T_{\delta}(p_{W_{1}\oplus W_{2}})\\\tilde{\boldw} \in T_{\delta}(p_{W_{1}\oplus W_{2}})} \right\}}\mathbb{E}_{\sfC}\left\{\mathds{1}_{\left\{ \boldW_{\tinyoplus}(\ulinek,\ulinem,\tildeboplus)=\tilde{\boldw}\right\}} \mathds{1}_{\left\{ \boldW_{1}(k_{1},m_{1}, b_{1})=\boldw_{1},\boldW_{2}(k_{2},m_{2}, b_{2})=\boldw_{2}\right\}}\right\}
 \nonumber\\
 \label{Eqn:DISCCCollatingBounds23}
&&=\frac{1}{q^{3n}} \sum_{\boldw_{1},\boldw_{2}}\sum_{\tilde{\boldw}}\boldp_{W_{1}W_{2}}(\boldw_{1},\boldw_{2})\mathds{1}_{\left\{\substack{\boldw_{1}\oplus \boldw_{2} \in  T_{\delta}(p_{W_{1}\oplus W_{2}})\\\tilde{\boldw} \in T_{\delta}(p_{W_{1}\oplus W_{2}})} \right\}} \leq q^{n(3-\frac{H(W_{1}\oplus W_{2})}{\log q})}.
\end{eqnarray}
Substituting the upper bound in \eqref{Eqn:DISCCCollatingBounds23} into \eqref{Eqn:DISCCCollatingBounds20}, recalling that the resulting quantity is an upper bound on the first term in \eqref{Eqn:DISCCCollatingBounds10} and employing the RHS of \eqref{Eqn:DISCCCollatingBounds17} as an upper bound on the second term in \eqref{Eqn:DISCCCollatingBounds10} , we have
\begin{eqnarray*}
  \Expectation\left\{\mathtt{t}_{B}\right\} \leq 2^{-n\delta}+q^{n(1-\beta_{2}+\frac{H(W_{1}\oplus W_{2})}{\log q})}
\end{eqnarray*}

We are thus left with
\begin{eqnarray*}
\mathbb{E}_{\sfC}\left\{
\norm{\left(\mathrm{id}_{R}\otimes \CalJ-\CalS\right)
\{\ketbra{\varphi_{\rho^{12}}}\}}_{1}\right\}
&\leq&
\mathbb{E}_{\sfC}\!\left\{\sfA(\sfC^{1},\sfC^{2})\right\}
+q^{-n\delta}
+q^{-n\left(\beta_{2}
-\left[1-\frac{H(W_{1}\oplus W_{2})}{\log q}\right]\right)}
\nonumber\\
&&+
\sum_{k_{1}}\frac{
2\sqrt{\mathbb{E}_{\sfC}\left\{
\norm{\boldrho^{1}-S_{k_{1}}}_{1}\right\}}
}{K_{1}}
+
\sum_{k_{2}}\frac{
2\sqrt{\mathbb{E}_{\sfC}\left\{
\norm{\boldrho^{2}-S_{k_{2}}}_{1}\right\}}
}{K_{2}}.
\end{eqnarray*}

Recollecting $S_{k_i}$ from \eqref{Eqn:DISCosetCdsLikelihoodPOVMdefn_i}, the last two terms in the above sum are one-party covering errors. For these terms, the pairwise-independence extension following Proposition~2 of~\cite{202504arXiv_Sen} applies. The term $\Expectation_{\ulinesfC}\{\sfA(\sfC_1,\sfC_2)\}$ is a two-party covering error. Proposition~\ref{Prop:SmoothMultipartiteSoftCovering}, due to~\cite[Prop.~2]{202504arXiv_Sen}, gives the corresponding estimate for independently chosen codebooks. Since the two UCC codebooks here share a generator, completing this step requires a separate justification of the covering estimate for that structured ensemble. The remaining two terms vanish under the chosen rate bounds \eqref{Eqn:DISCosetCdsInnBndCharBnds}. Once the structured two-party covering estimate is established, the displayed bound yields the existence of a pair $\sfc_1,\sfc_2$ of multi-codes for which $\norm{\left(\mathrm{id}_{R}\otimes\CalJ-\CalS\right)\{\ketbra{\varphi_{\rho^{12}}}\}}_1\leq\eta$ if there exists $(R_i,C_i,b_i:i=1,2)$ satisfying
\begin{eqnarray*}
 \max\{\beta_{1},\beta_{2}\} < \log q - H(W_{1}\oplus W_{2}), ~~~\!\sum_{l=1}^{2}(R_{l}+C_{l}+\beta_{l}) > I(W_{1}W_{2}:Y B_{1}B_{2}R)_{\sigma}+2\log q - H(W_{1})-H(W_{2})
 \\
 R_{i}+C_{i}+\beta_{i} > I(W_{i}:Y\ulineB R)_{\sigma} + \log q -H(W_{i}),~
 R_{i}+\beta_{i} > I(W_{i}:A_{3-i}R)_{\sigma_{i}}+\log q -H(W_{i})~~~~~~
 \end{eqnarray*}
for $i \in [2]$, where all the above information quantities are computed with respect to state $\sigma^{RYB_{1}B_{2}W_{1}W_{2}},\sigma_{i}^{A_{3-i}RW_{i}} : i \in [2]$ defined in \eqref{Eqn:DISCosetCdsTstChnlJntSts},The bounds stated in Thm.~\ref{Thm:DISCosetCdsInnBound} can be obtained by substituting the bound $\max\{\beta_{1},\beta_{2}\} < \log q - H(W_{1}\oplus W_{2})$ by two bounds $\beta_{i} < \log q - H(W_{1}\oplus W_{2})$ for $i=1,2$, performing Fourier Motzkin elimination of $\beta_{1},\beta_{2}$ on the resulting 7 inequalities. This completes the rate elimination, conditional on the structured two-party covering estimate identified above.

\end{proof}

\begin{remark}
\label{Rem:DISCCBroaderGoalOfOneShot}
The inner bound stated here matches that characterized in \cite{202504TIT_AtiPra}. However, owing to the use of structured likelihood POVMs that are amenable for one-shot analysis our techniques are more powerful than those of \cite{202504TIT_AtiPra} that can lead to newer findings. Specifically, one question that a reader might ask is why have we, only in this section, restricted attention to an asymptotic analysis despite using likelihood instruments? The reason is the only structured codes - codes possessing algebraic closure property - of block-length $1$ are the trivial ones. Therefore, such one-shot rates will not generalize to non-trivial inner bounds when generalized to the asymptotic regime. This raises an interesting question as to whether we can derive one-shot equivalent of the inner bounds in Thm.~\ref{Thm:DISCosetCdsInnBound} that recover the latter? While we reserve this as a question of future research, we point out \cite{202606TIT_PanSanMah} has addressed this question and designed codes of block-length $1$ with algebraic closure properties. While the techniques in \cite{202504TIT_AtiPra} preclude the use of such codes owing to their reliance of typicality based POVMs and time sharing our protocols based on likelihood instruments are naturally amenable to leverage the codes in \cite{202606TIT_PanSanMah} to achieve one-shot inner bounds corresponding to those in Thm.~\ref{Thm:DISCosetCdsInnBound}.
\end{remark}


\section{Distributed Instrument Simulation with Quantum Side Information}
\label{Sec:QSIDistributedIS}
We return to the distributed setting of Section~\ref{Sec:DistInstrSimOneShot}, but now model a scenario wherein Bob holds QSI. Alice$_1$, Alice$_2$, and Bob share a state $\rho^{A_1A_2B}$, and the separable instrument $\CalJ$ acts on $A_1A_2$. As before, the quantum outputs $B_1,B_2$ remain at the senders, while Bob reconstructs the classical outcome $Y$. See Panel 2 in Fig.~\ref{Fig:QSIAndPD}.

\subsection{Problem Formulation and One-Shot Rate Region}
\label{SubSec:QSIFormulation}
As before, the target state is $\CalT\define\CalT^{RB\ulineB Y}
=(\mathrm{id}_{RB}\otimes\CalJ)(\varphi_\rho).$ The local encoders and common-randomness resources are as in Definition~\ref{Defn:DistributedPOVMProtocol}. Only the decoder is enlarged to act on Bob's quantum system.

\begin{definition}[Distributed instrument simulation with QSI]
\label{Defn:DistributedISProtocolQSI}
A QSI-assisted protocol consists of the encoder banks $\{\CalE_{k_i}^{(i)}\}_{k_i\in[K_i]}$, $i\in[2]$, of Definition~\ref{Defn:DistributedPOVMProtocol}, and CPTP decoding maps $\CalD_{\ulinek}:\CalL(\CalH_B\otimes\mathbb C^{M_1}\otimes\mathbb C^{M_2})\to\CalL(\CalH_B\otimes\mathbb C^{\CalY})$. The action of $\CalJ$ on $\rho^{A_1A_2B}$ can be $\eta$-simulated with communication rates $(R_1,R_2)$ and common-randomness rates $(C_1,C_2)$ if $\log M_i\le R_i$, $\log K_i\le C_i$, and
\begin{equation}
\label{Eqn:QSISimulationCriterionLong}
\left\|\frac{1}{K_1K_2}\sum_{\ulinek}
\Big[(\mathrm{id}_{R\ulineB}\otimes\CalD_{\ulinek})\circ
(\mathrm{id}_{RB}\otimes\CalE_{k_1}^{(1)}\otimes\CalE_{k_2}^{(2)})\Big](\varphi_\rho)-\CalT\right\|_1\le\eta.
\end{equation}
Asymptotic simulability is defined by applying this criterion to $\CalJ^{\otimes n}$ on $\rho^{\otimes n}$ with costs $(nR_1,nR_2,nC_1,nC_2)$ and error tending to zero.
\end{definition}

\begin{definition}[Test channels and one-shot QSI rate region]
\label{Defn:QSIInfoQuantitiesAndRegion}
For the QSI-assisted setting, we use the same test-channel class
$\ScrT(\CalJ)$ as in Definition~\ref{Defn:TestChannels}; QSI changes the
input state and the decoder, but not the separable decomposition of $\CalJ$.
For any
$(\ulineCalW,\ulineCalI,p_{Y|\ulineW})\in\ScrT(\CalJ)$, write
$\ulineCalI\define\CalI^{(1)}\otimes\CalI^{(2)}$ and define the associated states
\begin{equation}
\label{Eqn:QSITestChannelStates}
\sigma^{RB\ulineB\ulineW Y}
\define
\left(\mathrm{id}_{RB\ulineB}\otimes
\olinescre^{Y|\ulineW}_{p}\right)
\left[
\left(\mathrm{id}_{RB}\otimes\ulineCalI\right)(\varphi_\rho)
\right],
\qquad
\sigma_i^{RBA_{3-i}B_iW_i}
\define
\left(\mathrm{id}_{RBA_{3-i}}\otimes\CalI^{(i)}\right)
(\varphi_\rho),
\quad i\in[2].
\end{equation}
In particular, $\sigma^{RB\ulineB Y}=\CalT$.

Fix $0<\varepsilon<1$ and let
$\varepsilon_5\define2\log(1/\varepsilon)$.
Associate with this test channel the set
$\ScrA_\varepsilon^{\mathrm{QSI}}
(\rho^{A_1A_2B},\ulineCalW,\ulineCalI,p_{Y|\ulineW})$
of all quadruples
$(R_1,R_2,C_1,C_2)\in\reals_{\ge0}^4$
for which there exist $\beta_1,\beta_2\ge0$ satisfying
\begin{eqnarray}
\label{Eqn:QSIPreEliminationBounds}
\begin{array}{ccc}
R_1+\beta_1>\mathsf q_1^\varepsilon
&
R_2+\beta_2>\mathsf q_2^\varepsilon
&
R_1+R_2+C_1+C_2+\beta_1+\beta_2>\mathsf q_5^\varepsilon
\\[0.4em]
R_1+C_1+\beta_1>\mathsf q_3^\varepsilon
&
R_2+C_2+\beta_2>\mathsf q_4^\varepsilon
&
\beta_1\beta_2=0\ \text{or}\
\beta_1+\beta_2<\mathsf q_8^\varepsilon
\\[0.4em]
\beta_1=0\ \text{or}\ \beta_1<\mathsf q_6^\varepsilon
&
\beta_2=0\ \text{or}\ \beta_2<\mathsf q_7^\varepsilon
&
\end{array}.
\end{eqnarray}

\begin{equation}
\label{Eqn:QSIOneShotInfoQuantities}
\mbox{where }~~
\begin{gathered}
\mathsf q_1^\varepsilon
\define
I_2^\varepsilon(W_1:RBA_2)_{\sigma_1}+\varepsilon_5,
\qquad
\mathsf q_2^\varepsilon
\define
I_2^\varepsilon(W_2:RBA_1)_{\sigma_2}+\varepsilon_5,
\qquad
\mathsf q_6^\varepsilon
\define
I_H^\varepsilon(W_1:W_2B)_\sigma-2-\tfrac12\varepsilon_5,
\\
\mathsf q_3^\varepsilon
\define
I_2^\varepsilon(W_1:YB\ulineB R)_\sigma+\varepsilon_5,
\qquad
\mathsf q_4^\varepsilon
\define
I_2^\varepsilon(W_2:YB\ulineB R)_\sigma+\varepsilon_5,
\qquad
\mathsf q_7^\varepsilon
\define
I_H^\varepsilon(W_2:W_1B)_\sigma-2-\tfrac12\varepsilon_5,
\\
\mathsf q_5^\varepsilon
\define
D_2^\varepsilon\!\left(
\sigma^{RB\ulineB\ulineW Y}
\,\middle\|\,
\sigma^{W_1}\!\otimes\!\sigma^{W_2}\!\otimes\!\sigma^{RB\ulineB Y}
\right)
+\varepsilon_5, \qquad
\mathsf q_8^\varepsilon
\define
D_H^\varepsilon\!\left(
\sigma^{W_1W_2B}
\,\middle\|\,
\sigma^{W_1}\!\otimes\!\sigma^{W_2}\!\otimes\!\sigma^B
\right)
-2-\tfrac12\varepsilon_5.
\end{gathered}
\nonumber
\end{equation}
The alternatives with $\beta_i=0$ correspond to singleton intra-bin
indices, for which the associated packing collisions are absent. All information quantities are computed with respect to the states in
\eqref{Eqn:QSITestChannelStates} or their marginals.
\end{definition}

We keep the auxiliary binning rates explicit in order to avoid the cumbersome region that would result from a full Fourier--Motzkin elimination.

\subsection{One-Shot Achievability}
\label{SubSec:QSIProofLong}

\begin{lemma}[Likelihood-weighted two-sender simultaneous decoder]
\label{Lem:LikelihoodWeightedQSIDecoder}
Let
\[
\chi^{W_1W_2B}
 =\sum_{w_1,w_2}p_{W_1W_2}(w_1,w_2)
 \ketbra{w_1,w_2}\otimes\chi^B_{w_1,w_2}
\]
be a finite-dimensional cq-state, with zero-probability conditional states
chosen arbitrarily.  Define
\begin{equation*}
J_i^\varepsilon\define I_H^\varepsilon(W_i:W_{3-i}B)_\chi,
\quad i\in[2],
\qquad
J_{12}^\varepsilon\define
D_H^\varepsilon\!\left(
\chi^{W_1W_2B}\middle\|
\chi^{W_1}\otimes\chi^{W_2}\otimes\chi^B
\right).
\end{equation*}
For every $0<\varepsilon<1$, let $\CalL_\varepsilon$ be the finite
augmentation alphabet supplied by Sen's joint-typicality construction.
Independently generate augmented codebooks
\[
\{(W_i(b_i),L_i(b_i)):b_i\in[B_i]\},\qquad i\in[2],
\]
where $W_i(b_i)\sim p_{W_i}$ and $L_i(b_i)$ is uniform on
$\CalL_\varepsilon$, with all these random variables independent.  The
augmentation labels are part of the decoder code design; once the code is
fixed they are known to the decoder and require no communication.  One can
associate with every realized pair of augmented codebooks a POVM
$\{D^{\widehat b_1,\widehat b_2}\}_{(\widehat b_1,\widehat b_2)
\in[B_1]\times[B_2]}$ on $B$ such that
\begin{eqnarray}
&&\mathbb E_{\mathcal C}\left[
\frac1{B_1B_2}\sum_{b_1,b_2}
\frac{p_{W_1W_2}(W_1(b_1),W_2(b_2))}
{p_{W_1}(W_1(b_1))p_{W_2}(W_2(b_2))}
\left(1-\Tr\!\left[D^{b_1,b_2}\chi^B_{W_1(b_1),W_2(b_2)}\right]\right)\right]
\nonumber\\[-0.2em]
&&\hspace{4em}\le
\delta_{\mathrm{JT}}(\varepsilon)
+4(B_1-1)2^{-J_1^\varepsilon}
+4(B_2-1)2^{-J_2^\varepsilon}
+4(B_1-1)(B_2-1)2^{-J_{12}^\varepsilon}.
\label{Eqn:WeightedQSIDecoderBound}
\end{eqnarray}
where $\delta_{\mathrm{JT}}:(0,1)\to\mathbb R_+$ is universal and
$\delta_{\mathrm{JT}}(\varepsilon)\to0$ as
$\varepsilon\to0$.  If $B_1=B_2=1$, the POVM may be chosen as
$\{I_B\}$ and the left-hand side is zero.
\end{lemma}

\begin{proof}
We use Sen's intersection joint-typicality lemma in exactly the form used
for the two-sender cq-MAC
\cite[Cor.~4 and the cq-MAC construction in the proof of Thm.~2]{202103SAD_Sen}.
For $c=2$, that result supplies a finite label alphabet
$\CalL_\varepsilon$, an augmented output space $B'$, a fixed isometric
embedding $V:B\to B'$, a perturbed cq-state
\[
(\chi')^{W_1L_1W_2L_2B'}
 =\frac1{L^2}\sum_{w_1,w_2,\ell_1,\ell_2}
 p(w_1,w_2)\ketbra{w_1,\ell_1,w_2,\ell_2}
 \otimes(\chi')^{B'}_{w_1,w_2,\ell_1,\ell_2},
 \qquad L\define|\CalL_\varepsilon|,
\]
and tests
$\{\Pi'_{w_1,w_2,\ell_1,\ell_2}\}$ on $B'$.  The perturbed state is
universally close, in trace norm, to the state obtained by embedding
$\chi^{W_1W_2B}$ with $V$ and appending two independent uniform label
registers.  Moreover, the test has high acceptance on the perturbed
positive hypothesis, while its acceptance on the three negative
hypotheses corresponding respectively to replacing $W_1$, replacing
$W_2$, and replacing both variables is bounded by
\[
2^{-J_1^\varepsilon},\qquad
2^{-J_2^\varepsilon},\qquad
2^{-J_{12}^\varepsilon}.
\]
Apply Corollary~4 with the augmentation dimension prescribed there
and choose its auxiliary perturbation parameter as
$\delta=\varepsilon^{1/3}$.  Since the number of classical registers is
fixed here, the trace-distance perturbation and the positive-hypothesis
rejection losses are then bounded by a universal function tending to zero
with $\varepsilon$.  For a realized augmented codebook, set
\[
\Gamma'_{b_1,b_2}
\define
\Pi'_{W_1(b_1),W_2(b_2),L_1(b_1),L_2(b_2)},
\qquad
\Gamma'\define\sum_{b_1,b_2}\Gamma'_{b_1,b_2},
\]
and define the square-root submeasurement on $B'$ by
\[
\overline\Lambda'^{b_1,b_2}
\define
(\Gamma')^{-1/2}\Gamma'_{b_1,b_2}(\Gamma')^{-1/2}
\quad\text{on }\supp(\Gamma').
\]
Choose one fixed outcome $b^\circ$ and complete this submeasurement by
setting
\[
\Lambda'^{b^\circ}
\define
\overline\Lambda'^{b^\circ}
+I_{B'}-\Pi_{\supp(\Gamma')},
\qquad
\Lambda'^b\define\overline\Lambda'^b\quad(b\ne b^\circ).
\]
Then $\{\Lambda'^{b_1,b_2}\}$ is a POVM, and this completion can only
decrease the decoding error.  The physical decoder first appends the
fixed ancilla defining $V$ and then performs this POVM.  Equivalently,
it uses on $B$ the POVM
\[
D^{b_1,b_2}\define V^\dagger\Lambda'^{b_1,b_2}V.
\]
Completeness follows from $V^\dagger V=I_B$.

We analyze first the perturbed ensemble.  The Hayashi--Nagaoka operator
inequality gives, for every transmitted pair $b=(b_1,b_2)$,
\begin{equation}
\label{Eqn:QSIHayashiNagaoka}
I_{B'}-\overline\Lambda'^b
\le 2(I_{B'}-\Gamma'_b)+4\sum_{b'\ne b}\Gamma'_{b'}.
\end{equation}
Since $\Lambda'^b\succeq\overline\Lambda'^b$ for the completed outcome
and the two operators coincide for every other outcome, the same upper
bound applies to the error probability of the completed POVM.
Multiply the corresponding error probability by
\[
L(w_1,w_2)\define
\frac{p_{W_1W_2}(w_1,w_2)}
     {p_{W_1}(w_1)p_{W_2}(w_2)}
\]
and average over the augmented random codebooks and the uniformly chosen
transmitted pair.  The likelihood factor converts the product sampling
law of the transmitted codewords into $p_{W_1W_2}$.  Hence the first term
on the right-hand side of \eqref{Eqn:QSIHayashiNagaoka} is the
positive-hypothesis rejection probability in Sen's construction and is
bounded by a universal quantity tending to zero.

Partition the competing indices into the same three classes as in the
cq-MAC proof.  If $b_1'\ne b_1$ and $b_2'=b_2$, the wrong first codeword
and its augmentation label are independent of the transmitted pair.
After the likelihood change of measure, averaging the transmitted first
codeword gives precisely the first negative hypothesis in Sen's
corollary.  Thus every such competitor contributes at most
$2^{-J_1^\varepsilon}$, and there are $B_1-1$ competitors.  The class
$b_1'=b_1$, $b_2'\ne b_2$ contributes at most
$(B_2-1)2^{-J_2^\varepsilon}$.  If both indices differ, both competing
augmented codewords are independent of the transmitted pair; averaging
the latter leaves the fixed marginal state $\chi^B$, which is the third
negative hypothesis.  This class contributes at most
$(B_1-1)(B_2-1)2^{-J_{12}^\varepsilon}$.  Including the factor $4$ in
\eqref{Eqn:QSIHayashiNagaoka}, and using that completion does not increase
the error, the expected likelihood-weighted error for the perturbed
ensemble is therefore at most
\begin{equation*}
\delta'_{\mathrm{JT}}(\varepsilon)
+4(B_1-1)2^{-J_1^\varepsilon}
+4(B_2-1)2^{-J_2^\varepsilon}
+4(B_1-1)(B_2-1)2^{-J_{12}^\varepsilon},
\end{equation*}
where $\delta'_{\mathrm{JT}}(\varepsilon)\to0$.

Finally compare the perturbed positive ensemble with the embedded
original ensemble.  The likelihood-weighted average of the latter is
exactly $\chi^{W_1W_2B}$ with two independent uniform label registers.
Since every decoding-error operator lies between zero and the identity,
the change in its averaged expectation is at most the trace distance
between these two cq-states.  Absorbing this perturbation loss into
$\delta'_{\mathrm{JT}}$ gives
\eqref{Eqn:WeightedQSIDecoderBound} with a universal
$\delta_{\mathrm{JT}}(\varepsilon)\to0$.  When $B_1=B_2=1$, no decoding
is required and the identity POVM gives zero error.
\end{proof}

\begin{theorem}[One-shot distributed instrument simulation with QSI]
\label{Thm:DistInstSimltnOneShotQSI}
There exists a universal function $g:(0,1)\to\mathbb R_+$, with $g(\varepsilon)\to0$ as $\varepsilon\to0$, such that the action of $\CalJ$ on $\rho^{A_1A_2B}$ can be $g(\varepsilon)$-simulated with costs $(R_1,R_2,C_1,C_2)$ whenever there exists a test channel $(\ulineCalW,\ulineCalI,p_{Y|\ulineW})\in\ScrT(\CalJ)$ for which
\[
(R_1,R_2,C_1,C_2)\in\ScrA_\varepsilon^{\mathrm{QSI}}(\rho^{A_1A_2B},\ulineCalW,\ulineCalI,p_{Y|\ulineW}).
\]
In particular, any target error $\eta>0$ is obtained whenever the corresponding rate conditions hold for a sufficiently small $\varepsilon$ with $g(\varepsilon)\le\eta$.
\end{theorem}

\begin{proof}
Fix a test channel and binning rates $\beta_1,\beta_2$ satisfying
Definition~\ref{Defn:QSIInfoQuantitiesAndRegion}.  The strict
inequalities allow the integer code sizes below to be chosen, after an
arbitrarily small perturbation of the rates, so that
\begin{equation*}
K_i\le 2^{C_i},\qquad M_i\le 2^{R_i},\qquad B_i\le2^{\beta_i},
\qquad i\in[2],
\end{equation*}
while all the required strict inequalities remain valid.  To simplify
notation we write $K_i=2^{C_i}$, $M_i=2^{R_i}$, and
$B_i=2^{\beta_i}$ below.  Let
$\rho^i=\Tr_{A_{3-i}B}\rho^{A_1A_2B}$.

\medskip
\med\textbf{Design of Encoding Instrument}:
Restrict $\CalW_i$ to $\supp(p_{W_i})$.  Independently generate
\begin{equation*}
\mathcal C_i
=\{w_i(k_i,m_i,b_i):(k_i,m_i,b_i)\in[K_i]\times[M_i]\times[B_i]\}
\end{equation*}
with every entry distributed according to $p_{W_i}$.  Independently,
attach to every entry a label
$\ell_i(k_i,m_i,b_i)$ drawn uniformly from the augmentation alphabet
$\CalL_\varepsilon$ in
Lemma~\ref{Lem:LikelihoodWeightedQSIDecoder}.  These labels are included
in the random code design and are revealed to Bob when the code is fixed;
they are not used by the encoder and consume no communication or common
randomness.  Let $\mathcal C$ denote the codewords together with these
labels.  Use exactly the likelihood operators $S_{k_i}$,
$\theta_{k_i}^{m_i,b_i}$,
$\theta_{k_i}^{0}$, and the encoding instruments
$\CalE_{k_i}^{(i)}$ defined in
\eqref{Eqn:DistLikelihoodTheta_i}--\eqref{Eqn:EncodingInstrumentki}.
The internal $0$-th branch is merged with the ordinary message
$m_i=1$, so the communicated alphabet has cardinality $M_i$.

\medskip
\noindent\textbf{Decoding and stochastic post-processing.}
Fix $(\ulinek,\ulinem)$.  The candidate pairs
\[
\ulinew_{\ulineb}
\define\ulinew(\ulinek,\ulinem,\ulineb),
\qquad
\ulineb=(b_1,b_2)\in[B_1]\times[B_2],
\]
together with their attached labels form the two independently generated
augmented codebooks required by
Lemma~\ref{Lem:LikelihoodWeightedQSIDecoder}.  Apply that lemma to the cq-state
$\sigma^{W_1W_2B}$ in \eqref{Eqn:QSITestChannelStates}.  This gives a
POVM
$\{D_{\ulinek,\ulinem}^{\widehat{\ulineb}}\}_{\widehat{\ulineb}}$
on Bob's system $B$.  After obtaining $\widehat{\ulineb}$, Bob samples
$Y$ according to
$p_{Y|\ulineW}(\cdot|\ulinew_{\widehat{\ulineb}})$.  On a diagonal
message block the decoder is
\begin{equation}
\label{Eqn:QSIDecoderMapFull}
\CalD_{\ulinek}
\bigl(S^B\otimes\ketbra{\ulinem}\bigr)
\define
\sum_{\widehat{\ulineb},y}
 p_{Y|\ulineW}(y|\ulinew_{\widehat{\ulineb}})
 \sqrt{D_{\ulinek,\ulinem}^{\widehat{\ulineb}}}\,S^B\,
 \sqrt{D_{\ulinek,\ulinem}^{\widehat{\ulineb}}}
 \otimes\ketbra y.
\end{equation}
Precomposing with complete dephasing of the message registers defines a
CPTP map on the entire input space.  The square-root implementation in
\eqref{Eqn:QSIDecoderMapFull} is needed because the output system $B$
must be preserved, not merely measured.

Let $F\define R\ulineB$ and let
$\widetilde{\CalT}$ denote the averaged output of this protocol as in
\eqref{Eqn:QSISimulationCriterionLong}.  For
$p_{\ulineW}(\ulinew)>0$, define the normalized conditional states
\begin{equation}
\label{Eqn:QSIConditionalStatesFull}
\tau^{FB}_{\ulinew}
\define
\frac{\Tr_{E_1E_2}\!
 \left[(I_{RB}\otimes V_{\ulinew})\varphi_\rho
 (I_{RB}\otimes V_{\ulinew}^{\dagger})\right]}
 {p_{\ulineW}(\ulinew)},
\qquad
\tau^{FBY}_{\ulinew}
\define\sum_y p_{Y|\ulineW}(y|\ulinew)
\tau^{FB}_{\ulinew}\otimes\ketbra y,
\end{equation}
where $V_{\ulinew}=V_{w_1}\otimes V_{w_2}$; zero-probability
conditionals are arbitrary.  Then
$\CalT=\sum_{\ulinew}p_{\ulineW}(\ulinew)\tau^{FBY}_{\ulinew}.$
For fixed $(\ulinek,\ulinem)$, set
\begin{equation*}
a_{\ulinek,\ulinem,\ulineb}
\define
\frac{p_{\ulineW}(\ulinew_{\ulineb})}
{p_{W_1}(w_1(k_1,m_1,b_1))
 p_{W_2}(w_2(k_2,m_2,b_2))},
\qquad
N\define K_1K_2M_1M_2B_1B_2.
\end{equation*}
Introduce
\begin{equation}
\CalT_1 \define \frac1N\sum_{\ulinek,\ulinem,\ulineb} a_{\ulinek,\ulinem,\ulineb}\tau^{FBY}_{\ulinew_{\ulineb}},
\qquad
\CalT_2 \define \frac1N\sum_{\ulinek,\ulinem,\ulineb} a_{\ulinek,\ulinem,\ulineb}(\mathrm{id}_F\otimes\CalD_{\ulinek})\bigl(\tau^{FB}_{\ulinew_{\ulineb}}\otimes\ketbra{\ulinem}\bigr).
\label{Eqn:QSIIntermediateT1T2Full}
\end{equation}
The likelihood ratio and independence imply
\begin{equation}
\label{Eqn:QSIExpectedMassLong}
\mathbb E_{\mathcal C}\Tr\CalT_1
=\mathbb E_{\mathcal C}
 \frac1N\sum_{\ulinek,\ulinem,\ulineb}
 a_{\ulinek,\ulinem,\ulineb}=1.
\end{equation}

\medskip
\noindent\textit{\underline{Bipartite soft covering.}}
Apply Proposition~\ref{Prop:SmoothMultipartiteSoftCovering} with
$X_i=W_i$, codebook sizes $K_iM_iB_i$, and output system $FBY$.  The
empirical state in that proposition is exactly $\CalT_1$.  Consequently,
\begin{equation}
\label{Eqn:QSIBipartiteCoveringFull}
\mathbb E_{\mathcal C}\|\CalT-\CalT_1\|_1\le16\varepsilon
\end{equation}
provided
\begin{equation*}
R_i+C_i+\beta_i
>I_2^\varepsilon(W_i:YB\ulineB R)_\sigma+\varepsilon_5,
\qquad i\in[2],
\end{equation*}
and
\begin{equation*}
R_1+R_2+C_1+C_2+\beta_1+\beta_2
>D_2^\varepsilon\!\left(
\sigma^{RB\ulineB\ulineW Y}\middle\|
\sigma^{W_1}\otimes\sigma^{W_2}\otimes\sigma^{RB\ulineB Y}
\right)+\varepsilon_5.
\end{equation*}
These are precisely the constraints containing
$\mathsf q_3^\varepsilon,\mathsf q_4^\varepsilon,
\mathsf q_5^\varepsilon$ in
\eqref{Eqn:QSIPreEliminationBounds}.

\medskip
\noindent\textit{\underline{Simultaneous packing and disturbance.}}
For each transmitted tuple define
\begin{equation*}
e_{\ulinek,\ulinem,\ulineb}
\define1-
\Tr\!\left[D_{\ulinek,\ulinem}^{\ulineb}
\tau^B_{\ulinew_{\ulineb}}\right],
\qquad
P_{\mathrm{err}}(\mathcal C)
\define\frac1N\sum_{\ulinek,\ulinem,\ulineb}
 a_{\ulinek,\ulinem,\ulineb}e_{\ulinek,\ulinem,\ulineb}.
\end{equation*}
For the correct branch, Lemma~\ref{Lem:GentleMeasurement} applied to the
full state $\tau^{FB}_{\ulinew_{\ulineb}}$ gives disturbance at most
$2\sqrt{e_{\ulinek,\ulinem,\ulineb}}$.  The sum of all incorrect
branches is positive and has trace
$e_{\ulinek,\ulinem,\ulineb}$.  Since stochastic post-processing cannot
increase trace norm,
\begin{equation}
\label{Eqn:QSIPackingDisturbancePointwise}
\|\CalT_1-\CalT_2\|_1
\le\frac1N\sum_{\ulinek,\ulinem,\ulineb}
 a_{\ulinek,\ulinem,\ulineb}
 \left(e_{\ulinek,\ulinem,\ulineb}
       +2\sqrt{e_{\ulinek,\ulinem,\ulineb}}\right).
\end{equation}
Cauchy--Schwarz and \eqref{Eqn:QSIExpectedMassLong} imply
\begin{equation}
\label{Eqn:QSIPackingDisturbanceAverage}
\mathbb E_{\mathcal C}\|\CalT_1-\CalT_2\|_1
\le \overline e+2\sqrt{\overline e},
\qquad
\overline e\define
\mathbb E_{\mathcal C}P_{\mathrm{err}}(\mathcal C).
\end{equation}
Applying Lemma~\ref{Lem:LikelihoodWeightedQSIDecoder} independently in
each $(\ulinek,\ulinem)$ block and averaging gives
\begin{eqnarray}
\overline e
&\le&
\delta_{\mathrm{JT}}(\varepsilon)
+4(B_1-1)2^{-J_1^\varepsilon}
+4(B_2-1)2^{-J_2^\varepsilon}
+4(B_1-1)(B_2-1)2^{-J_{12}^\varepsilon}.
\label{Eqn:QSIPackingBoundLong}
\end{eqnarray}
where
\begin{equation*}
J_i^\varepsilon=I_H^\varepsilon(W_i:W_{3-i}B)_\sigma,
\qquad
J_{12}^\varepsilon
=D_H^\varepsilon\!\left(
\sigma^{W_1W_2B}\middle\|
\sigma^{W_1}\otimes\sigma^{W_2}\otimes\sigma^B
\right).
\end{equation*}
If $\beta_1>0$, the constraint
$\beta_1<\mathsf q_6^\varepsilon
=J_1^\varepsilon-2-\log(1/\varepsilon)$ implies
$4(B_1-1)2^{-J_1^\varepsilon}<\varepsilon$; if
$\beta_1=0$, this collision class is absent.  The same reasoning applies
to the second class.  If $\beta_1\beta_2>0$, the sum-rate packing
constraint gives
$4(B_1-1)(B_2-1)2^{-J_{12}^\varepsilon}<\varepsilon$; otherwise the
joint collision class is absent.  Hence
\begin{equation}
\label{Eqn:QSIepackDefinition}
\overline e\le e_{\mathrm{pack}}(\varepsilon)
\define\delta_{\mathrm{JT}}(\varepsilon)+3\varepsilon,
\qquad
e_{\mathrm{pack}}(\varepsilon)\longrightarrow0.
\end{equation}

\medskip
\noindent\textit{\underline{The two quantum-covering terms.}}
For each $i\in[2]$, choose a reference isometry $U_i^{\mathrm{ref}}$ such
that
\begin{equation*}
\varphi_\rho
=(U_i^{\mathrm{ref}}\otimes I_{A_i})
\varphi_{\rho^i}
({U_i^{\mathrm{ref}}}^\dagger\otimes I_{A_i}),
\end{equation*}
where the image of the reference is $RBA_{3-i}$.  Define
\begin{equation*}
\Xi_{k_i}^{(i)}
\define
(U_i^{\mathrm{ref}}\otimes I_{A_i})
\varphi_{S_{k_i}}
({U_i^{\mathrm{ref}}}^\dagger\otimes I_{A_i}).
\end{equation*}
Let $\widetilde{\CalI}_{k_i}^{(i)}$ be the inverse-free maps in
\eqref{Eqn:InducedInstrumentki}.  Expanding
\eqref{Eqn:QSIIntermediateT1T2Full} and using
\eqref{Eqn:QSIConditionalStatesFull} gives the exact identity
\begin{equation}
\label{Eqn:QSIT2InverseFree}
\CalT_2
=\frac1{K_1K_2}\sum_{\ulinek}
(\mathrm{id}_{R\ulineB}\otimes\CalD_{\ulinek})
\circ
(\mathrm{id}_{RB}\otimes
 \widetilde{\CalI}_{k_1}^{(1)}\otimes
 \widetilde{\CalI}_{k_2}^{(2)})(\varphi_\rho).
\end{equation}
The COS identity proved in Appendix~\ref{AppSec:Calculus} is insensitive
to the size of the purifying register.  Applying it first to sender $2$
and then to sender $1$ yields
\begin{eqnarray}
\CalT_2
&=&\frac1{K_1K_2}\sum_{\ulinek}
\CalL_{\ulinek}(\Xi_{k_2}^{(2)}),
\label{Eqn:QSIT2ProxyRepresentation}\\
\CalT_3
&\define&\frac1{K_1K_2}\sum_{\ulinek}
\CalL_{\ulinek}(\varphi_\rho)
=
\frac1{K_1K_2}\sum_{\ulinek}
\Big[(\mathrm{id}_{R\ulineB}\otimes\CalD_{\ulinek})\circ
(\mathrm{id}_{RB}\otimes
 \CalE_{k_1}^{(1)}\otimes\CalE_{k_2}^{(2)})\Big]
(\Xi_{k_1}^{(1)}).
\label{Eqn:QSIT3Representations}
\end{eqnarray}
where
\begin{equation*}
\CalL_{\ulinek}
\define
(\mathrm{id}_{R\ulineB}\otimes\CalD_{\ulinek})\circ
(\mathrm{id}_{RB}\otimes
 \widetilde{\CalI}_{k_1}^{(1)}\otimes\CalE_{k_2}^{(2)}).
\end{equation*}

For $\CalT_3-\widetilde{\CalT}$, contractivity of the complete
encoding--decoding map and invariance under the reference isometry give
\begin{equation*}
\|\CalT_3-\widetilde{\CalT}\|_1
\le\frac1{K_1}\sum_{k_1}
\|\varphi_{S_{k_1}}-\varphi_{\rho^1}\|_1.
\end{equation*}
By Lemma~\ref{lem:canonical-purif-stability}, Cauchy--Schwarz,
$\mathbb E\Tr S_{k_1}=1$, and the one-party instance of
Proposition~\ref{Prop:SmoothMultipartiteSoftCovering},
\begin{equation}
\label{Eqn:QSIFirstQuantumCovering}
\mathbb E_{\mathcal C}\|\CalT_3-\widetilde{\CalT}\|_1
\le\frac2{K_1}\sum_{k_1}
\sqrt{\mathbb E_{\mathcal C}\|S_{k_1}-\rho^1\|_1}
\le4\sqrt\varepsilon
\end{equation}
under $R_1+\beta_1>\mathsf q_1^\varepsilon$.

For $\CalT_2-\CalT_3$, first remove the decoder by contractivity.
Condition on the second sender's codebook.  The difference
$\Xi_{k_2}^{(2)}-\varphi_\rho$ is then fixed and independent of the first
sender's codebook.  Lemma~\ref{Lem:AveragedInverseFreeContraction2}
applied to $\widetilde{\CalI}_{k_1}^{(1)}$, followed by contractivity of
$\CalE_{k_2}^{(2)}$, gives
\begin{equation*}
\mathbb E_{\mathcal C}\|\CalT_2-\CalT_3\|_1
\le\frac1{K_2}\sum_{k_2}
\mathbb E_{\mathcal C}
\|\varphi_{S_{k_2}}-\varphi_{\rho^2}\|_1.
\end{equation*}
The same canonical-purification and one-party covering estimates yield
\begin{equation}
\label{Eqn:QSISecondQuantumCovering}
\mathbb E_{\mathcal C}\|\CalT_2-\CalT_3\|_1
\le4\sqrt\varepsilon
\end{equation}
under $R_2+\beta_2>\mathsf q_2^\varepsilon$.

\medskip
\noindent\textbf{Conclusion.}
Finally, the triangle inequality along
$\CalT\to\CalT_1\to\CalT_2\to\CalT_3\to\widetilde{\CalT}$,
\eqref{Eqn:QSIBipartiteCoveringFull},
\eqref{Eqn:QSIPackingDisturbanceAverage},
\eqref{Eqn:QSIepackDefinition},
\eqref{Eqn:QSIFirstQuantumCovering}, and
\eqref{Eqn:QSISecondQuantumCovering} give
\begin{equation}
\label{Eqn:QSIFinalErrorFunction}
\mathbb E_{\mathcal C}
\|\CalT-\widetilde{\CalT}\|_1
\le g(\varepsilon)
\define16\varepsilon+8\sqrt\varepsilon
+e_{\mathrm{pack}}(\varepsilon)
+2\sqrt{e_{\mathrm{pack}}(\varepsilon)}.
\end{equation}
The function $g$ is universal and tends to zero with $\varepsilon$.
Every inequality used above is contained in
Definition~\ref{Defn:QSIInfoQuantitiesAndRegion}.  Since the expectation
in \eqref{Eqn:QSIFinalErrorFunction} is over the random codebooks, at
least one deterministic codebook realization attains this error bound.
The common-randomness indices remain part of the allowed protocol, so
this realization proves the theorem.
\end{proof}

\begin{remark}[Relation to the QSI-free setting]
\label{Rem:QSIComparison}
If $B$ is trivial, then $J_1^\varepsilon=J_2^\varepsilon=J_{12}^\varepsilon=I_H^\varepsilon(W_1:W_2)_\sigma$. The construction therefore specializes to the classical correlated-binning architecture, although the finite-error constants inherited from the quantum decoder need not coincide with those of Theorem~\ref{Thm:DistInstSimltnOneShot}.
\end{remark}
\appendices

\section{The Case of Stochastic Coarse-Graining and Stochastic Compatibility is Subsumed}
\label{AppSec:StochGrainingNotMoreGeneral}

Our formulation of multiple-description ICP was in terms of `deterministic' fine- and coarse-graining. One may consider a mathematically more general notion of fine-graining involving stochastic maps as formalized below.

\begin{definition}
    \label{Defn:StochasticFineCoarseGraining}
Let
$\CalJ^{\CalX}
=
\{\CalJ_x:\CalL(\CalH_A)\to\CalL(\CalH_B):x\in\CalX\}$ and
$\CalK^{\CalY}
=
\{\CalK_y:\CalL(\CalH_A)\to\CalL(\CalH_B):y\in\CalY\}$ be two instruments operating with the same input, output quantum spaces. We say that $\CalK^{\CalY}$ is a \emph{stochastic coarse-graining} of
$\CalJ^{\CalX}$ if there exists a stochastic matrix $p_{Y|X}:\CalX \rightarrow \CalY$ with input $\CalX$ and output $\CalY$ such that
\begin{equation}
\label{Eqn:StochasticInstrumentCoarseGraining}
\CalK_y
=
\sum_{x \in \CalX}p_{Y|X}(y|x)\CalJ_x ~~\mbox{ for all }y \in \CalY.
\end{equation}
In this case, $\CalJ^{\CalX}$ is said to be a \emph{stochastic fine-graining} of
$\CalK^{\CalY}$. In particular, if there exists a deterministic map
$f:\CalX\to\CalY$ such that $p_{Y|X}(y|x)=\mathds{1}_{\{y=f(x)\}}$, then we say $\CalK^{\CalY}$ is a \emph{coarse-graining} of
$\CalJ^{\CalX}$ and $\CalJ^{\CalX}$ is said to be a \emph{fine-graining} of
$\CalK^{\CalY}$.

A collection $\CalK^{\CalY_{1}},\cdots \CalK^{\CalY_{L}}$ of $L$ instruments are \emph{stochastically compatible} if there exists a common stochastic fine-graining $\CalJ^{\CalX}$ of the collection $\CalK^{\CalY_{1}},\cdots \CalK^{\CalY_{L}}$, i.e., there exists $L$ stochastic matrices $p_{Y_{l}|X}:\CalX \rightarrow \CalY_{l}$ for $l \in [L]$ such that $\CalK_{y_{l}} = \sum_{x\in\CalX}p_{Y_{l}|X}(y_{l}|x)\CalJ_{x}$ for all $y_{l} \in \CalY_{l}$ and all $l \in [L]$. In particular, if there exists $L$ functions $f_{l}:\CalX \rightarrow \CalY_{l}$ such that $p_{Y_{l}|X}(y_{l}|x)=\mathds{1}_{\{y_{l}=f_{l}(x)\}} ~\forall (x,y_{l}) \in \CalX \times \CalY_{l}$ and all $l \in [L]$, then the instrument collection $\CalK^{\CalY_{1}},\cdots \CalK^{\CalY_{L}}$ of $L$ is \emph{compatible}.
\end{definition}

Would these generalized notions of stochastic grainings and compatibility lead to a formulation of multiple-description ICP that is not handled in our study? Towards answering this, let us formulate our multiple-description ICP in terms of stochastic compatible instruments. Referring to Panel 3 of Fig.~\ref{Fig:DistISAndMultDescIS}, consider a single component quantum state $\rho \in \CalD(\CalH_{A})$ accessible to Alice. Alice is connected to a distributed party Bob who holds classical registers via two links - Link$_{1}$ and Link$_{2}$ - of communication resources. Link$_{i}$ enables (i) Alice to transfer $R_{i}$ bits to Bob, and (ii) both parties to share $C_{i}$ bits of common randomness, both metered per instrument action. Consider three stochastically compatible instruments $\CalK^{\CalY_{0}},\CalK^{\CalY_{1}}$ and $\CalK^{\CalY_{2}}$. Firstly, with Link$_{1}$ alone being active, Alice and Bob must be able to simulate the action of instrument $\CalK^{\CalY_{1}}$. Secondly, with Link$_{2}$ alone being active, Alice and Bob must be able to simulate the action of instrument $\CalK^{\CalY_{2}}$. With both Link$_{1}$ and Link$_{2}$ active, Alice and Bob must be able to simulate the action of all three instruments $\CalK^{\CalY_{0}},\CalK^{\CalY_{1}}$ and $\CalK^{\CalY_{2}}$. The problem of interest would be to characterize inner and outer bounds on the communication rates $(R_{1},R_{2})$ and common randomness rates $(C_{1},C_{2})$ that guarantees this both in the one-shot and asymptotic regimes. Does our findings apply to this scenario? Indeed, it does.

\begin{lemma}
 \label{Lem:MDICPCartProdOutputSuffices}
 Given three stochastically compatible instruments $\CalK^{\CalY_{i}} = \{\CalK_{y_{i}} : y_{i} \in \CalY_{i}\}$ for $i=0,1,2$, there exists a single fine grained instrument $\CalJ^{\CalY_{0}\times \CalY_{1} \times \CalY_{2}} = \{ \CalJ_{y_{0},y_{1},y_{2}} : (y_{0},y_{1},y_{2}) \in \CalY_{0}\times \CalY_{1} \times \CalY_{2} \}$ having outcome alphabet $\CalY_{0}\times \CalY_{1} \times \CalY_{2}$ such that
 \begin{eqnarray}
  \label{Eqn:MDICPCartProdOutputSuffices}
  \CalK_{y_{0}}=\sum_{y_{1},y_{2}}\CalJ_{y_{0},y_{1},y_{2}}, ~~\CalK_{y_{1}}=\sum_{y_{0},y_{2}}\CalJ_{y_{0},y_{1},y_{2}},~~\CalK_{y_{2}}=\sum_{y_{0},y_{1}}\CalJ_{y_{0},y_{1},y_{2}}.
 \end{eqnarray}
Conversely, given an instrument $\CalJ^{\CalY_{0}\times \CalY_{1} \times \CalY_{2}} = \{ \CalJ_{y_{0},y_{1},y_{2}} : (y_{0},y_{1},y_{2}) \in \CalY_{0}\times \CalY_{1} \times \CalY_{2} \}$ with outcome alphabet $\CalY_{0}\times \CalY_{1} \times \CalY_{2}$, there exists three stochastically compatible instruments $\CalK^{\CalY_{i}} = \{\CalK_{y_{i}} : y_{i} \in \CalY_{i}\}$ for $i=0,1,2$ satisfying \eqref{Eqn:MDICPCartProdOutputSuffices}.
\end{lemma}

\begin{proof}
To prove the first statement, note that given three stochastically compatible instruments $\CalK^{\CalY_{i}} = \{\CalK_{y_{i}} : y_{i} \in \CalY_{i}\}$ for $i=0,1,2$, Defn~\ref{Defn:StochasticFineCoarseGraining} guarantees existence of a single fine-grained instrument $\CalJ^{\CalX} = \{\CalJ_{x}: x \in \CalX\}$ and stochastic matrices $p_{Y_{i}|X}:\CalX \rightarrow \CalY_{i}$ such that $\CalK_{y_{i}} = \sum_{x \in \CalX}p_{Y_{i}|X}(y_{i}|x)\CalJ_{x}$ for each $i=0,1,2$. Define $\CalJ^{\CalY_{0}\times \CalY_{1} \times \CalY_{2}} = \{ \CalJ_{y_{0},y_{1},y_{2}} : (y_{0},y_{1},y_{2}) \in \CalY_{0}\times \CalY_{1} \times \CalY_{2} \}$ having outcome alphabet $\CalY_{0}\times \CalY_{1} \times \CalY_{2}$ as $\CalJ_{y_{0},y_{1},y_{2}} = \sum_{x \in \CalX}p_{Y_{0}|X}(y_{0}|x)p_{Y_{1}|X}(y_{1}|x)p_{Y_{2}|X}(y_{2}|x)\CalJ_{x}$. It is straightforward to verify that the given instruments $\CalK^{\CalY_{i}} = \{\CalK_{y_{i}} : y_{i} \in \CalY_{i}\}$ for $0 \leq i \leq 2$ satisfy \eqref{Eqn:MDICPCartProdOutputSuffices}. Converse is immediate.
\end{proof}

\section{\textcolor{black}{COS Computation for One-Shot Distributed ISP}}
\label{AppSec:Calculus}

The calculation below is insensitive to the size of the purifying/reference
register: it only uses that the isometry on the purifying side commutes with
the local encoding operator. Consequently, the same identities hold with
$RA_{3-i}$ replaced by an arbitrary reference system.

Recall that $\widetilde{\mathcal{I}}_{k_1}^{(1)}$ is introduced in \eqref{Eqn:InducedInstrumentki} to bound the last term in the proof of Theorem~\ref{Thm:DistInstSimltnOneShot} and $U_i^{\mathrm{ref}}$ acting on the reference system $R_i$ of $\varphi_{\rho^i}$ satisfies $\varphi_{\rho}=(\,U_i^{\mathrm{ref}}\otimes I_{A_i}\,)
\varphi_{\rho^i}
(\,{U_i^{\mathrm{ref}}}^{\dagger}\otimes I_{A_i}\,)$. We will prove the following equality:
\begin{equation}
\label{Eq:TrickEk1Ik1}
\Big(\mathrm{id}_{RA_2}\otimes \mathcal{E}^{(1)}_{k_1}\Big)
\Big(
(U_1^{\mathrm{ref}}\otimes I_{A_1})\,\varphi_{S_{k_1}}\,({U_1^{\mathrm{ref}}}^{\dagger}\otimes I_{A_1})
\Big)
\;=\;
\Big(\mathrm{id}_{RA_2}\otimes \widetilde{\mathcal{I}}_{k_1}^{(1)}\Big)
\bigl(\varphi_{\rho}\bigr).
\end{equation}

\begin{proof}
Fix $k_1\in[K_1]$. By definition of the encoding instrument in~\eqref{Eqn:EncodingInstrumentki},
\[ \Big(\mathrm{id}_{RA_2}\otimes \mathcal{E}^{(1)}_{k_1}\Big) \Big( (U_1^{\mathrm{ref}}\otimes I_{A_1})\,\varphi_{S_{k_1}}\,({U_1^{\mathrm{ref}}}^{\dagger}\otimes I_{A_1}) \Big) \] \[ = \Tr_{E_1}\Bigg[ \sum_{m_1,b_1} \Big(\mathrm{id}_{RA_2}\otimes \theta^{m_1,b_1}_{k_1}\Big) \Big( (U_1^{\mathrm{ref}}\otimes I_{A_1})\,\varphi_{S_{k_1}}\,({U_1^{\mathrm{ref}}}^{\dagger}\otimes I_{A_1}) \Big) \Big(\mathrm{id}_{RA_2}\otimes {\theta^{m_1,b_1}_{k_1}}^{\dagger}\Big) \otimes \ket{m_1}\bra{m_1} \] \[ \qquad\qquad + \Big(\mathrm{id}_{RA_2}\otimes \theta^{0}_{k_1}\Big) \Big( (U_1^{\mathrm{ref}}\otimes I_{A_1})\,\varphi_{S_{k_1}}\,({U_1^{\mathrm{ref}}}^{\dagger}\otimes I_{A_1}) \Big) \Big(\mathrm{id}_{RA_2}\otimes {\theta^{0}_{k_1}}^{\dagger}\Big) \otimes \ket{1}\bra{1} \Bigg]. \]
We next show that the ``$0$-th'' term vanishes. Using the canonical form $\varphi_{S_{k_1}}
=
\sum_{t,t'}
\ket{\overline{e_t}}\bra{\overline{e_{t'}}}
\otimes
\sqrt{S_{k_1}}\,\ket{e_t}\bra{e_{t'}}\,\sqrt{S_{k_1}},$ and the fact that $\theta^0_{k_1}\sqrt{S_{k_1}}=0$ (cf.~\eqref{Eqn:DistThetak0GetsKilled_i} for $i=1$),
we get $\Big(\mathrm{id}_{RA_2}\otimes \theta^{0}_{k_1}\Big)
\Big(
(U_1^{\mathrm{ref}}\otimes I_{A_1})\,\varphi_{S_{k_1}}\,({U_1^{\mathrm{ref}}}^{\dagger}\otimes I_{A_1})
\Big)
\Big(\mathrm{id}_{RA_2}\otimes {\theta^{0}_{k_1}}^{\dagger}\Big)
=0.$ Hence,
\[
\Big(\mathrm{id}_{RA_2}\otimes \mathcal{E}^{(1)}_{k_1}\Big)
\Big(
(U_1^{\mathrm{ref}}\otimes I_{A_1})\,\varphi_{S_{k_1}}\,({U_1^{\mathrm{ref}}}^{\dagger}\otimes I_{A_1})
\Big)
\]
\[
=
\Tr_{E_1}\Bigg[
\sum_{m_1,b_1}
\Big(\mathrm{id}_{RA_2}\otimes \theta^{m_1,b_1}_{k_1}\Big)
\Big(
(U_1^{\mathrm{ref}}\otimes I_{A_1})\,\varphi_{S_{k_1}}\,({U_1^{\mathrm{ref}}}^{\dagger}\otimes I_{A_1})
\Big)
\Big(\mathrm{id}_{RA_2}\otimes {\theta^{m_1,b_1}_{k_1}}^{\dagger}\Big)
\otimes \ket{m_1}\bra{m_1}
\Bigg].
\]
Since $\theta^{m_1,b_1}_{k_1}$ acts on $A_1$ while $U_1^{\mathrm{ref}}$ acts on the reference side, they commute, and thus
\[
\Big(\mathrm{id}_{RA_2}\otimes \theta^{m_1,b_1}_{k_1}\Big)
(U_1^{\mathrm{ref}}\otimes I_{A_1})
=
(U_1^{\mathrm{ref}}\otimes I_{B_1E_1})
\Big(\mathrm{id}_{R_1}\otimes \theta^{m_1,b_1}_{k_1}\Big),
\]
and similarly for the adjoint. Therefore the previous expression becomes
\[
\Tr_{E_1}\Bigg[
\sum_{m_1,b_1}
(U_1^{\mathrm{ref}}\otimes I_{B_1E_1})
\Big(\mathrm{id}_{R_1}\otimes \theta^{m_1,b_1}_{k_1}\Big)\,
\varphi_{S_{k_1}}\,
\Big(\mathrm{id}_{R_1}\otimes {\theta^{m_1,b_1}_{k_1}}^{\dagger}\Big)
({U_1^{\mathrm{ref}}}^{\dagger}\otimes I_{B_1E_1})
\otimes \ket{m_1}\bra{m_1}
\Bigg].
\]
Now expand $\varphi_{S_{k_1}}$ in the canonical basis:
\[
\Big(\mathrm{id}_{R_1}\otimes \theta^{m_1,b_1}_{k_1}\Big)\,
\varphi_{S_{k_1}}\,
\Big(\mathrm{id}_{R_1}\otimes {\theta^{m_1,b_1}_{k_1}}^{\dagger}\Big)
=
\sum_{t,t'}
\ket{\overline{e_t}}\bra{\overline{e_{t'}}}
\otimes
\Big(\theta^{m_1,b_1}_{k_1}\sqrt{S_{k_1}}\Big)\,
\ket{e_t}\bra{e_{t'}}\,
\Big(\sqrt{S_{k_1}}\,{\theta^{m_1,b_1}_{k_1}}^{\dagger}\Big).
\]
\[
\text{By~\eqref{Eqn:DistThetak0GetsKilled_i},}
\qquad
\theta^{m_1,b_1}_{k_1}\sqrt{S_{k_1}}
=
\frac{
V_{w_1(k_1,m_1,b_1)}\,\sqrt{\rho^1}
}{
\sqrt{M_1B_1\,\Tr(\rho^1\mu_{w_1(k_1,m_1,b_1)})}
}
=
\frac{
V_{w_1(k_1,m_1,b_1)}\,\sqrt{\rho^1}
}{
\sqrt{M_1B_1\,p_{W_1}\bigl(w_1(k_1,m_1,b_1)\bigr)}
}.
\]
Substituting this and using the canonical purification $\varphi_{\rho^1}
=
\sum_{t,t'}
\ket{\overline{e_t}}\bra{\overline{e_{t'}}}
\otimes
\sqrt{\rho^1}\,\ket{e_t}\bra{e_{t'}}\,\sqrt{\rho^1},$ we obtain
\[
\Big(\mathrm{id}_{R_1}\otimes \theta^{m_1,b_1}_{k_1}\Big)\,
\varphi_{S_{k_1}}\,
\Big(\mathrm{id}_{R_1}\otimes {\theta^{m_1,b_1}_{k_1}}^{\dagger}\Big)
=
\frac{1}{M_1B_1\,p_{W_1}\bigl(w_1(k_1,m_1,b_1)\bigr)}
\Big(\mathrm{id}_{R_1}\otimes V_{w_1(k_1,m_1,b_1)}\Big)
\varphi_{\rho^1}
\Big(\mathrm{id}_{R_1}\otimes V_{w_1(k_1,m_1,b_1)}^{\dagger}\Big).
\]
Finally, using the relation $\varphi_{\rho}
=
(\,U^{\mathrm{ref}}_1\otimes I_{A_1}\,)
\varphi_{\rho^1}
(\,{U^{\mathrm{ref}}_1}^{\dagger}\otimes I_{A_1}\,),$ we conclude that
\[
\Big(\mathrm{id}_{RA_2}\otimes \mathcal{E}^{(1)}_{k_1}\Big)
\Big(
(U_1^{\mathrm{ref}}\otimes I_{A_1})\,\varphi_{S_{k_1}}\,({U_1^{\mathrm{ref}}}^{\dagger}\otimes I_{A_1})
\Big)
\]
\[
=
\Tr_{E_1}\Bigg[
\sum_{m_1,b_1}
\Bigg(
\frac{
(\mathrm{id}_{RA_2}\otimes V_{w_1(k_1,m_1,b_1)})\,
\varphi_{\rho}\,
(\mathrm{id}_{RA_2}\otimes V_{w_1(k_1,m_1,b_1)}^{\dagger})
}{
M_1B_1\,p_{W_1}\bigl(w_1(k_1,m_1,b_1)\bigr)
}
\otimes \ket{m_1}\bra{m_1}
\Bigg)
\Bigg].
\]
By the definition of $\widetilde{\mathcal{I}}_{k_1}^{(1)}$ in~\eqref{Eqn:InducedInstrumentki},
the right-hand side is exactly $\Big(\mathrm{id}_{RA_2}\otimes \widetilde{\mathcal{I}}_{k_1}^{(1)}\Big)
\bigl(\varphi_{\rho}\bigr),$ which proves~\eqref{Eq:TrickEk1Ik1}.
\end{proof}
The previous computation directly yields the two equivalent expressions for $\CalT_3$ defined in \eqref{Eqn:CalT3BeforeCOS}.
Similarly, it implies the following rewriting of the expression for $\CalT_2$ given in \eqref{Eqn:CalT2AfterCOS}:
\begin{equation*}
\CalT_{2}
=
\frac{1}{K_1K_2}\sum_{k_1,k_2}
\Big(
(\mathrm{id}_{B_1B_2R}\otimes \mathcal{D}_{k_1,k_2})
\circ
(\mathrm{id}_{R}\otimes \widetilde{\mathcal{I}}_{k_1}^{(1)}
\otimes \widetilde{\mathcal{I}}_{k_2}^{(2)})
\Big)
\bigl(\varphi_{\rho}\bigr).
\end{equation*}
We now compute the intermediate state
\[
(\mathrm{id}_{R}\otimes \widetilde{\mathcal{I}}_{k_1}^{(1)}
\otimes \widetilde{\mathcal{I}}_{k_2}^{(2)})
\bigl(\varphi_{\rho}\bigr)
=
\operatorname{Tr}_{\ulineE}\left[\sum_{\ulineb,\ulinem}
\frac{
\bigl(\mathrm{id}_R \otimes V_{\ulinew(\ulinek,\ulineb,\ulinem)}\bigr)\,
\varphi_{\rho}\,
\bigl(\mathrm{id}_R \otimes V_{\ulinew(\ulinek,\ulineb,\ulinem)}\bigr)^{\dagger}
}{
M_1B_1M_2B_2\,
p_{W_1}\bigl(w_1(k_1,m_1,b_1)\bigr)\;
p_{W_2}\bigl(w_2(k_2,m_2,b_2)\bigr)
}
\;\otimes\;
\ketbra{\ulinem}\right].
\]
Applying the decoding map (Section~\ref{SubSec:OneShotISWithIIDCodes}
) and using the definition of $\CalT_{B_1 B_2 R}^{\,\ulinew(\ulinek,\ulinem,\ulineb)}$ in \eqref{Eqn:PostInstrumentStateDef} recovers \eqref{Eqn:CalT2BeforeCOS}:
\[
\CalT_{2}
=
\sum_{y}
\left(
\frac{1}{KMB}\sum_{\ulinek,\ulinem,\ulineb}
\frac{
p_{Y|\ulineW}\bigl(y|f(\ulinek,\ulinem)\bigr)\,p_{\ulineW}\bigl(\ulinew(\ulinek,\ulinem,\ulineb)\bigr)\,
\CalT_{B_1 B_2 R}^{\,\ulinew(\ulinek,\ulinem,\ulineb)}
}{
p_{W_1}\bigl(w_1(k_1,m_1,b_1)\bigr)\,
p_{W_2}\bigl(w_2(k_2,m_2,b_2)\bigr)
}
\right)
\;\otimes\;
\ket{y}\bra{y}.
\]

\section{\textcolor{black}{Analysis of the Binning Decoder}}
\label{AppSec:binning}
In this appendix, we provide a detailed description of the decoder and the analysis of the bound in
Eq.~\eqref{Eqn:TraceNormT1T2Bound}. The actual decoder is deterministic and is given by
\begin{equation*}
  \CalD_{\ulinek}^{(1)}(S)= \sum_{\ulinem \in [\ulineM]}  \sum_{\ulinew \in \mathcal{\ulineW}}\ketbra{\ulinew}{\ulinem}S\ketbra{\ulinem}{\ulinew}\mathbbm{1}_{\left\{\,\ulinew= f(\ulinek,\ulinem)\,\right\}}.
\end{equation*}
We construct the deterministic selection rule $f$ by first analyzing an auxiliary randomized decoder and then derandomizing it.

Apply Proposition~\ref{Prop:ClassicalJointTypicalityIndicator}
to the binary hypothesis test between $p_{W_1W_2}$ and
$p_{W_1}\otimes p_{W_2}$. For a given $\varepsilon\in(0,1)$, there exists a test function $g:\mathcal{W}_1\times\mathcal{W}_2\longrightarrow[0,1]$ such that
\begin{equation*}
\sum_{w_1,w_2} p_{W_1W_2}(w_1,w_2)\, g(w_1,w_2)
\;\ge\; 1-\varepsilon ,
\end{equation*}
and
\begin{equation}
\label{Eqn:fSmallUnderProduct}
\sum_{w_1,w_2} p_{W_1}(w_1)\,p_{W_2}(w_2)\, g(w_1,w_2)
\;\le\;
2^{-D_H^{\varepsilon}\left(p_{W_1W_2}\,\middle\|\,p_{W_1}\otimes p_{W_2}\right)} .
\end{equation}

\medskip
\noindent\emph{Auxiliary randomized decoder.}
For every $(\ulinek,\ulinem,\ulineb)$, conditionally on the realized codebook, independently generate a Bernoulli random variable
$Z_{\ulinek,\ulinem,\ulineb}$ with
$\Pr\!\left\{Z_{\ulinek,\ulinem,\ulineb}=1\,\middle|\,\mathcal C\right\}
=
g\bigl(\ulinew(\ulinek,\ulinem,\ulineb)\bigr).$
Let $\mathcal G_Z(\ulinek,\ulinem)
\coloneqq
\left\{\ulineb\in[\ulineB]:Z_{\ulinek,\ulinem,\ulineb}=1\right\}.$ If $\mathcal G_Z(\ulinek,\ulinem)=\{\ulineb^\star\}$ is a singleton, define $F_Z(\ulinek,\ulinem)
\coloneqq
\ulinew(\ulinek,\ulinem,\ulineb^\star).$ Otherwise, declare an error and set, for definiteness,
$F_Z(\ulinek,\ulinem)\coloneqq\ulinew(\ulinek,\ulinem,\uline1)$.
For every fixed $(\ulinek,\ulinem,\ulineb)$, the event
\[
Z_{\ulinek,\ulinem,\ulineb}=1,
\qquad
Z_{\ulinek,\ulinem,\ulineb'}=0
\quad\forall\,\ulineb'\neq\ulineb
\]
guarantees that
$F_Z(\ulinek,\ulinem)=\ulinew(\ulinek,\ulinem,\ulineb)$.
Therefore, by the union bound,
\begin{equation}
\label{Eqn:MismatchDecomposition}
\mathbb E_Z\!\left[
\mathbbm{1}_{\left\{\ulinew(\ulinek,\ulinem,\ulineb)\neq F_Z(\ulinek,\ulinem)\right\}}
\,\middle|\,\mathcal C
\right]
\le
1-g\bigl(\ulinew(\ulinek,\ulinem,\ulineb)\bigr)
+
\sum_{\ulineb'\in[\ulineB]\setminus\{\ulineb\}}
g\bigl(\ulinew(\ulinek,\ulinem,\ulineb')\bigr).
\end{equation}
We split the collision sum into the three classes determined by the intra-bin indices:
\begin{equation*}
\begin{aligned}
\sum_{\ulineb'\in[\ulineB]\setminus\{\ulineb\}}
g\bigl(\ulinew(\ulinek,\ulinem,\ulineb')\bigr)
&=
S_{\mathrm{diff}}(\ulinek,\ulinem,\ulineb)
+S_{1\text{-}\mathrm{coll}}(\ulinek,\ulinem,\ulineb)
+S_{2\text{-}\mathrm{coll}}(\ulinek,\ulinem,\ulineb),
\end{aligned}
\end{equation*}
where
\[
S_{\mathrm{diff}}(\ulinek,\ulinem,\ulineb)
:=\sum_{\substack{\ulineb'\in[\ulineB]:\, b'_1\ne b_1,\ b'_2\ne b_2}}
g\bigl(\ulinew(\ulinek,\ulinem,\ulineb')\bigr),
\qquad
S_{i\text{-}\mathrm{coll}}(\ulinek,\ulinem,\ulineb)
:=\sum_{\substack{\ulineb'\in[\ulineB]:\, b'_i=b_i,\ b'_{3-i}\ne b_{3-i}}}
g\bigl(\ulinew(\ulinek,\ulinem,\ulineb')\bigr).
\]
Introduce the likelihood ratio
\[
L(w_1,w_2)
\coloneqq
\frac{p_{W_1W_2}(w_1,w_2)}
{p_{W_1}(w_1)\,p_{W_2}(w_2)}
\quad
\text{on the product support, and define}
\quad
\beta
\coloneqq
\sum_{w_1,w_2}
p_{W_1}(w_1)\,p_{W_2}(w_2)\,g(w_1,w_2).
\]
By \eqref{Eqn:fSmallUnderProduct},
$\beta
\le
2^{-D_H^{\varepsilon}\left(p_{W_1W_2}\,\middle\|\,p_{W_1}\otimes p_{W_2}\right)} .$
Let
$\Xi(\mathcal C,Z)
\coloneqq
\frac{1}{KMB}\sum_{\ulinek,\ulinem,\ulineb}
L(\ulinew(\ulinek,\ulinem,\ulineb))\,
\mathbbm{1}_{\left\{\,
\ulinew(\ulinek,\ulinem,\ulineb)
\neq F_Z(\ulinek,\ulinem)
\,\right\}} .$
Taking expectations in \eqref{Eqn:MismatchDecomposition} gives
\[
\mathbb E_{\mathcal C,Z}\!\left[\Xi(\mathcal C,Z)\right]
\le
T_0 + T_{\mathrm{diff}} + T_{1\text{-}\mathrm{coll}} + T_{2\text{-}\mathrm{coll}},
\qquad
\text{where}
\qquad
T_0
\coloneqq
\mathbb{E}_{\mathcal{C}}\left[
\frac{1}{KMB}\sum_{\ulinek,\ulinem,\ulineb}
L(\ulinew(\ulinek,\ulinem,\ulineb))\,
\Bigl(1-g(\ulinew(\ulinek,\ulinem,\ulineb))\Bigr)
\right]
\]
\[
\text{and, for } r\in\{\mathrm{diff},\,1\text{-}\mathrm{coll},\,2\text{-}\mathrm{coll}\},
\qquad
T_r
\coloneqq
\mathbb{E}_{\mathcal{C}}\left[
\frac{1}{KMB}\sum_{\ulinek,\ulinem,\ulineb}
L(\ulinew(\ulinek,\ulinem,\ulineb))\,
S_r(\ulinek,\ulinem,\ulineb)
\right].
\]

\medskip
\noindent\textbf{Bounding $T_0$.}
Using
$p_{W_1}(w_1)p_{W_2}(w_2)L(w_1,w_2)=p_{W_1W_2}(w_1,w_2)$,
\begin{equation*}
T_0
=
1-\sum_{w_1,w_2}
p_{W_1W_2}(w_1,w_2)\,g(w_1,w_2)
\le \varepsilon.
\end{equation*}

\medskip
\noindent\textbf{Bounding the collision terms.}
Write $\ulineb=(b_1,b_2)$.
If $b'_1\ne b_1$ and $b'_2\ne b_2$, then
$\ulinew(\ulinek,\ulinem,\ulineb')$ is independent of the transmitted pair and has law
$p_{W_1}\otimes p_{W_2}$. Hence
\begin{equation*}
\mathbb{E}_{\mathcal C}\!\left[L(\ulinew(\ulinek,\ulinem,\ulineb))\,g(\ulinew(\ulinek,\ulinem,\ulineb'))\right]
=
\sum_{\ulinew,\ulinew'} p_{\ulineW}(\ulinew)\,p_{W_1}(\ulinew'_1)p_{W_2}(\ulinew'_2)\,g(\ulinew')
=
\beta.
\end{equation*}
If $b'_1=b_1$ and $b'_2\ne b_2$, only the second competing component is redrawn. Thus
\begin{equation*}
\begin{aligned}
\mathbb{E}_{\mathcal C}\!\left[
L(\ulinew(\ulinek,\ulinem,\ulineb))\,
g\bigl(w_1(k_1,m_1,b_1),w_2(k_2,m_2,b'_2)\bigr)
\right]
&=
\sum_{w_1,w_2,w'_2}
p_{W_1W_2}(w_1,w_2)\,
p_{W_2}(w'_2)\,
g(w_1,w'_2)\\
&=
\sum_{w_1,w'_2}
p_{W_1}(w_1)p_{W_2}(w'_2)\,
g(w_1,w'_2)
=
\beta.
\end{aligned}
\end{equation*}
The case $b'_1\ne b_1$ and $b'_2=b_2$ is identical with the roles of
$W_1$ and $W_2$ interchanged.
Counting the number of competitors in each class yields
\begin{equation*}
T_{\mathrm{diff}}\le(B_1-1)(B_2-1)\,\beta,
\qquad
T_{1\text{-}\mathrm{coll}}\le(B_2-1)\,\beta,
\qquad
T_{2\text{-}\mathrm{coll}}\le(B_1-1)\,\beta.
\end{equation*}
Therefore
\begin{equation}
\label{Eqn:RandomizedMismatchFinal}
\begin{aligned}
\mathbb E_{\mathcal C,Z}\!\left[\Xi(\mathcal C,Z)\right]
&\le
\varepsilon
+\Bigl[(B_1-1)(B_2-1)+(B_2-1)+(B_1-1)\Bigr]\beta\\
&=
\varepsilon+(B_1B_2-1)\beta \le
\varepsilon+
B_1B_2\,
2^{-D_H^{\varepsilon}\left(p_{W_1W_2}\,\middle\|\,p_{W_1}\otimes p_{W_2}\right)}
\le 2\varepsilon,
\end{aligned}
\end{equation}
where the last inequality follows from $\log(B_1B_2)
<
I_H^{\varepsilon}(W_1:W_2)_\sigma
-
\log\left(\tfrac{1}{\varepsilon}\right).$

\medskip
\noindent\emph{Derandomization.}
For every realized codebook $\mathcal C$, there exists a deterministic realization
$z^\star(\mathcal C)$ of all the Bernoulli variables such that
$\Xi\bigl(\mathcal C,z^\star(\mathcal C)\bigr)
\le
\mathbb E_Z\!\left[\Xi(\mathcal C,Z)\,\middle|\,\mathcal C\right].$
Fix such a realization as part of the code design and define the deterministic selection rule $f(\ulinek,\ulinem)
\coloneqq
F_{z^\star(\mathcal C)}(\ulinek,\ulinem).$ Thus the actual decoder requires no additional local randomness. Taking expectation over the codebook and using
\eqref{Eqn:RandomizedMismatchFinal} gives
\begin{equation*}
\mathbb{E}_{\mathcal{C}}\left[
\frac{1}{KMB}\sum_{\ulinek,\ulinem,\ulineb}
L(\ulinew(\ulinek,\ulinem,\ulineb))\,
\mathbbm{1}_{\left\{\,
\ulinew(\ulinek,\ulinem,\ulineb)
\neq f(\ulinek,\ulinem)
\,\right\}}
\right]
\le 2\varepsilon .
\end{equation*}
Combining this with \eqref{Eqn:TraceNormT1T2Bound} proves $\mathbb{E}_{\mathcal C}\!\left[
\|\CalT_1-\CalT_2\|_1
\right]
\le 4\varepsilon.$

\section{\textcolor{black}{Facts used in the analysis of likelihood encoders}}
\label{AppSec:UsefulResults}

\subsection{Basic Lemmas}

\begin{lemma}[Closeness of canonical purifications]
\label{lem:canonical-purif-stability}
Let $\rho\in\CalD(\CalH)$ and $\sigma\in\CalP(\CalH)$, and let
$\varphi_\rho$ and $\varphi_\sigma$ be their canonical purification operators as defined in
Sec.~\ref{Sec:ProblemStatements}. Then
\[
\norm{\varphi_\rho-\varphi_\sigma}_1
\le
\left(1+\sqrt{\Tr\sigma}\right)\sqrt{\norm{\rho-\sigma}_1}.
\]
\end{lemma}

\begin{proof}
For arbitrary vectors $\ket a,\ket b$,
$\norm{\ketbra a-\ketbra b}_1\le(\|\ket a\|_2+\|\ket b\|_2)\|\ket a-\ket b\|_2$.
Applying this to the canonical purification vectors gives
\[
\norm{\varphi_\rho-\varphi_\sigma}_1
\le
(1+\sqrt{\Tr\sigma})\norm{\sqrt\rho-\sqrt\sigma}_2.
\]
The Powers--St{\o}rmer inequality \cite{197003CMP_PowSto}
$\|\sqrt{\rho}-\sqrt{\sigma}\|_2^2\le\norm{\rho-\sigma}_1$
completes the proof.
\end{proof}

\begin{lemma}[Gentle measurement lemma]
\label{Lem:GentleMeasurement}
Let $\rho$ be a density operator on a finite-dimensional Hilbert space $\CalH$, and let
$0\preceq\Lambda\preceq\mathbb I$ be a POVM element. Set
$p\define\Tr[\Lambda\rho]$. Then
\begin{equation*}
\norm{\rho-\sqrt{\Lambda}\rho\sqrt{\Lambda}}_1
\le
2\sqrt{1-p}.
\end{equation*}
\cite[Lemma 9.4.2]{BkWilde_2017}
\end{lemma}

\begin{lemma}[Trace distance to a pure state]
\label{Lem:TraceDistancePureState}
For any density operator $\sigma$ and any pure state $\ket{\phi}$,
\begin{equation*}
2\left(1-\sqrt{\bra{\phi}\sigma\ket{\phi}}\right)
\le
\norm{\sigma-\ketbra{\phi}}_1
\le
2\sqrt{1-\bra{\phi}\sigma\ket{\phi}}.
\end{equation*}
This is Eq.~(9.164) in \cite{BkWilde_2017}.
\end{lemma}

\begin{lemma}[Averaged contraction of measure-transformed inverse-free maps]
\label{Lem:AveragedInverseFreeContraction2}
Let $\CalW$ be a finite alphabet and let $\{V_w:\CalH_A\rightarrow\CalH_B\otimes\CalH_E:w\in\CalW\}$ satisfy $\sum_{w\in\CalW}V_w^\dagger V_w=I_A$. Let $\nu_W$ be a full-support probability distribution on $\CalW$, and consider a random codebook $\sfC=\{W(m,b):(m,b)\in[M]\times[B]\}$ such that every codeword $W(m,b)$ has marginal distribution $\nu_W$.
Define the inverse-free completely positive map
\begin{equation*}
\widetilde{\CalI}_{\sfC}(S)\define\Tr_E\!\left[\sum_{m,b}\frac{V_{W(m,b)}SV_{W(m,b)}^\dagger}{MB\,\nu_W(W(m,b))}\otimes\ketbra{m}\right].
\end{equation*}
\[ \text{Then, for every finite-dimensional system $F$ and every Hermitian $L\in\CalL(\CalH_F\otimes\CalH_A)$,}
\quad
\Expectation_{\sfC}\!\left[\norm{(\mathrm{id}_F\otimes\widetilde{\CalI}_{\sfC})(L)}_1\right]
\leq
\norm{L}_1.
\]
\end{lemma}

\begin{proof}
By triangle inequality, contractivity under the partial trace,
and the definition of $\widetilde{\CalI}_{\sfC}$,
\begin{equation*}
\begin{aligned}
&\Expectation_{\sfC}\!\left[
\norm{
(\mathrm{id}_{F}\otimes\widetilde{\CalI}_{\sfC})(L)
}_{1}
\right]
\leq
\frac{1}{MB}\sum_{m,b}
\Expectation_{\sfC}\!\left[
\frac{1}{\nu_{W}(W(m,b))}
\norm{
(\mathrm{id}_{F}\otimes V_{W(m,b)})
L
(\mathrm{id}_{F}\otimes V_{W(m,b)}^{\dagger})
}_{1}
\right].
\end{aligned}
\end{equation*}
Since every $W(m,b)$ has marginal distribution $\nu_{W}$, the
right-hand side is
$\sum_{w\in\CalW}
\norm{
(\mathrm{id}_{F}\otimes V_{w})
L
(\mathrm{id}_{F}\otimes V_{w}^{\dagger})
}_{1}.$
Introducing an orthogonal classical label for each $w$ gives
\begin{equation*}
\begin{aligned}
&\sum_{w\in\CalW}
\norm{
(\mathrm{id}_{F}\otimes V_{w})
L
(\mathrm{id}_{F}\otimes V_{w}^{\dagger})
}_{1}
=
\norm{
\sum_{w\in\CalW}
(\mathrm{id}_{F}\otimes V_{w})
L
(\mathrm{id}_{F}\otimes V_{w}^{\dagger})
\otimes\ketbra{w}
}_{1}.
\end{aligned}
\end{equation*}
The map $S\longmapsto\sum_{w\in\CalW}V_wSV_w^\dagger\otimes\ketbra{w}$ is CPTP since $\sum_wV_w^\dagger V_w=I_A$. Trace-norm contractivity for Hermitian operators therefore yields the result.
\end{proof}

\subsection{Sen's One-shot Smooth Multipartite Soft Covering}

In~\cite[Prop.~2]{202504arXiv_Sen}, Sen proved fully smooth one-shot multipartite convex-split
(soft-covering) bounds based on a telescoping mean-zero decomposition. We make extensive
use of these bounds throughout the paper, in the following expectation form.

\begin{proposition}[Smooth multipartite soft covering in expectation]
\label{Prop:SmoothMultipartiteSoftCovering}
Let $\CalX_1,\ldots,\CalX_k$ be finite alphabets, let $p_{X_{[k]}}$ be a probability
distribution on $\CalX_{[k]}\define\CalX_1\times\cdots\times\CalX_k$, and let
$q_{X_i}$ satisfy $\supp(p_{X_i})\subseteq\supp(q_{X_i})$. For
$S\subseteq[k]$, let $q_{X_S}\define\bigotimes_{s\in S}q_{X_s}$. For each
$x_{[k]}$, let $\rho^M_{x_{[k]}}$ be a normalized density operator on $M$ and define
\[
\rho^{X_{[k]}M}
\define
\sum_{x_{[k]}}p_{X_{[k]}}(x_{[k]})\ketbra{x_{[k]}}\otimes\rho^M_{x_{[k]}}.
\]
Independently generate $\CalC_i=\{X_i(a_i):a_i\in[A_i]\}$ with
$X_i(a_i)\sim q_{X_i}$ i.i.d., and set
\[
\sigma^M_{\CalC_1,\ldots,\CalC_k}
\define
\frac1{A_1\cdots A_k}
\sum_{a_1,\ldots,a_k}
\frac{p_{X_{[k]}}(X_1(a_1),\ldots,X_k(a_k))}
{\prod_{i=1}^kq_{X_i}(X_i(a_i))}
\rho^M_{X_1(a_1),\ldots,X_k(a_k)}.
\]
If, for every non-empty $S\subseteq[k]$,
\[
\sum_{s\in S}\log A_s
>
D_2^\varepsilon\!\left(
\rho^{X_SM}\middle\|q_{X_S}\otimes\rho^M
\right)
+\log\varepsilon^{-2},
\qquad
\text{then}
\qquad
\Expectation_{\CalC_1,\ldots,\CalC_k}\!\left[
\norm{\sigma^M_{\CalC_1,\ldots,\CalC_k}-\rho^M}_1
\right]
<
2(3^k-1)\varepsilon.
\]
The empirical operator $\sigma^M_{\CalC_1,\ldots,\CalC_k}$ need not be
normalized for each realization of the codebooks. Nevertheless,
\[
\Expectation_{\CalC_1,\ldots,\CalC_k}
\bigl[\sigma^M_{\CalC_1,\ldots,\CalC_k}\bigr]=\rho^M,
\qquad
\Expectation_{\CalC_1,\ldots,\CalC_k}
\bigl[\Tr\sigma^M_{\CalC_1,\ldots,\CalC_k}\bigr]=1.
\]
\end{proposition}

\begin{definition}[Worst-case conditional $I_2$ on a high-probability set]
\label{Defn:WorstCaseConditionalI2}
Let $\underline{\CalW}\define\CalW_1\times\CalW_2$ and consider
\[
\sigma^{\underline{W}W_0A}
=
\sum_{\underline{w}}p_{\underline W}(\underline w)\ketbra{\underline w}
\otimes
\sum_{w_0}p_{W_0|\underline W}(w_0|\underline w)\ketbra{w_0}
\otimes\rho_A^{w_0,\underline w}.
\]
Writing $\sigma_{\underline w}^{W_0A}$ for the corresponding conditional slice, define
\begin{equation}
\label{Eqn:DefWorstCaseConditionalI2}
I_{2,\mathrm{wc}}^{\varepsilon_0,\varepsilon}(W_0\!:\!A\,|\,\underline W)_\sigma
\define
\min_{\Omega\subseteq\underline{\CalW}:\,p_{\underline W}(\Omega)\ge1-\varepsilon_0}
\max_{\underline w\in\Omega}
I_2^\varepsilon(W_0\!:\!A)_{\sigma_{\underline w}}.
\end{equation}
\end{definition}
\subsection{Conditional Soft Covering estimate}

In our analysis, we encounter soft-covering problems involving conditionally generated, and hence non-independent, codebooks; this motivates the worst-case conditional collision information introduced below, for which we establish a one-shot conditional soft-covering bound together with an asymptotic upper AEP, the latter being sufficient for our achievability analysis.

\begin{lemma}[Upper AEP bound for the worst-case conditional collision information]
\label{Lem:AEPWorstCaseConditionalI2}
Let
$\sigma^{\underline WW_0A}
=\sum_{\underline w}p_{\underline W}(\underline w)\ketbra{\underline w}
\otimes\sigma_{\underline w}^{W_0A}$
be finite dimensional and define
$I(W_0:A|\underline W)_\sigma
\define\sum_{\underline w}p_{\underline W}(\underline w)
I(W_0:A)_{\sigma_{\underline w}}$.
If $\varepsilon_{0,n},\varepsilon_n\downarrow0$ with
$n^{-1}\log(1/\varepsilon_{0,n})\to0$ and
$n^{-1}\log(1/\varepsilon_n)\to0$, then
\[
\limsup_{n\to\infty}\frac1n
I_{2,\mathrm{wc}}^{\varepsilon_{0,n},\varepsilon_n}
(W_0^n:A^n|\underline W^n)_{\sigma^{\otimes n}}
\le
I(W_0:A|\underline W)_\sigma.
\]
\end{lemma}

\begin{proof}
Choose a type-typical set $\mathcal T_n$ of $\underline W^n$ having probability at least
$1-\varepsilon_{0,n}$; the assumed subexponential decay of $\varepsilon_{0,n}$ permits its
width to vanish. For a fixed $\underline w^n\in\mathcal T_n$, the conditional state is,
up to permutation, $\sigma_{\underline w^n}^{W_0^nA^n}
\simeq
\bigotimes_{\underline w}
(\sigma_{\underline w}^{W_0A})^{\otimes n_{\underline w}}.$ Distribute the smoothing parameter $\varepsilon_n$ among the finitely many nonzero type
blocks. A telescoping estimate for tensor products shows that the product of the blockwise
smoothed approximants is still $\varepsilon_n$-close, while additivity of the unsmoothed
collision divergence gives
\[
I_2^{\varepsilon_n}(W_0^n:A^n)_{\sigma_{\underline w^n}}
\le
\sum_{\underline w}
I_2^{\delta_n}(W_0^{n_{\underline w}}:A^{n_{\underline w}})
_{\sigma_{\underline w}^{\otimes n_{\underline w}}},
\]
for a choice $\delta_n$ differing from $\varepsilon_n$ only by a fixed alphabet factor.
On $\mathcal T_n$, each nonzero $n_{\underline w}$ grows linearly with $n$, so the ordinary
IID AEP applies uniformly to all finitely many blocks. Dividing by $n$ therefore yields
$\sum_{\underline w}q_{\underline w^n}(\underline w)
I(W_0:A)_{\sigma_{\underline w}}+o(1)
\le I(W_0:A|\underline W)_\sigma+o(1)$.
Since $\mathcal T_n$ is admissible in \eqref{Eqn:DefWorstCaseConditionalI2}, taking the
maximum on $\mathcal T_n$ and then the limsup proves the claim.
\end{proof}

\begin{lemma}[Upper AEP bound for the conditional smoothed support max-entropy]
\label{Lem:UpperAEPConditionalTildeHmax}
Let
$\rho^{YA}=\sum_y p_Y(y)\ketbra{y}_Y\otimes\rho_y^A$
be a finite-dimensional cq-state, and let $\varepsilon_n\downarrow0$ satisfy
$n^{-1}\log(1/\varepsilon_n)\to0$. Then
\[
\limsup_{n\to\infty}\frac1n
\widetilde H_{\max}^{\varepsilon_n}(A^n|Y^n)_{\rho^{\otimes n}}
\le
H(A|Y)_\rho
\define
\sum_y p_Y(y)H(A)_{\rho_y}.
\]
\end{lemma}

\begin{proof}
Choose a type-typical set $\mathcal T_n\subseteq\CalY^n$ with probability larger than
$1-\sqrt{\varepsilon_n}$. For $y^n\in\mathcal T_n$, the conditional state factorizes,
up to permutation, as
$\rho_{y^n}^{A^n}\simeq\bigotimes_y(\rho_y^A)^{\otimes n_y}$.
Choose, in each nonzero block, a projector attaining the corresponding smoothed support
max-entropy with smoothing parameter $\delta_n=\varepsilon_n/|\supp p_Y|$ and take their
tensor product. A union bound makes this product projector admissible for
$\widetilde H_{\max}^{\varepsilon_n}(A^n)_{\rho_{y^n}}$, while its logarithmic rank is the
sum of the blockwise logarithmic ranks. The ordinary IID AEP applied uniformly to these
finitely many blocks gives
\[
\frac1n\widetilde H_{\max}^{\varepsilon_n}(A^n)_{\rho_{y^n}}
\le
\sum_y q_{y^n}(y)H(A)_{\rho_y}+o(1)
\le H(A|Y)_\rho+o(1).
\]
Taking the maximum over the admissible set $\mathcal T_n$ and then the limsup proves the
claim.
\end{proof}

\begin{lemma}[Conditional one-shot soft covering]
\label{Lem:ConditionalSoftCovering}
Fix $\varepsilon_0,\varepsilon\in(0,1)$ and let $\sigma^{\underline{W}W_0A}$ be as in
Definition~\ref{Defn:WorstCaseConditionalI2}.
For each $\underline{w}\in\underline{\CalW}$, generate a conditional codebook
\[
\CalC_0(\underline{w})
\define
\bigl\{W_0(m_0|\underline{w}):m_0\in[M_0]\bigr\},
\qquad
\text{where}
\qquad
W_0(1|\underline{w}),\dots,W_0(M_0|\underline{w})
\stackrel{\mathrm{i.i.d.}}{\sim}
p_{W_0|\underline{W}}(\cdot|\underline{w}).
\]
and define the random mixture
\[
\widetilde{\rho}^{\,\underline{w}}_A
\define
\frac{1}{M_0}\sum_{m_0=1}^{M_0}
\rho_A^{W_0(m_0|\underline{w}),\underline{w}},
\qquad
\overline{\rho}^{\,\underline{w}}_A
\define
\sum_{w_0\in\CalW_0}
p_{W_0|\underline{W}}(w_0|\underline{w})\,
\rho_A^{w_0,\underline{w}}.
\]
\begin{equation}
\label{Eqn:CondSC_RateCond}
\text{If}\qquad
\log M_0
>
I_{2,\mathrm{wc}}^{\varepsilon_0,\varepsilon}
(W_0\!:\!A\,|\,\underline{W})_{\sigma}
+2\log\frac{1}{\varepsilon},
\qquad\text{then}\qquad
\mathbb{E}_{\underline{W}\sim p_{\underline{W}}}\,
\mathbb{E}_{\CalC_0(\underline{W})}\!\left[
\bigl\|
\widetilde{\rho}^{\,\underline{W}}_A
-
\overline{\rho}^{\,\underline{W}}_A
\bigr\|_1
\right]
\le
2\varepsilon_0+4\varepsilon.
\end{equation}
\end{lemma}

\begin{proof}
Let $\Omega^\star\subseteq\underline{\CalW}$ attain the minimum in
\eqref{Eqn:DefWorstCaseConditionalI2}, so that
$p_{\underline{W}}(\Omega^\star)\ge 1-\varepsilon_0$ and $\max_{\underline{w}\in\Omega^\star}
I_2^{\varepsilon}(W_0\!:\!A)_{\sigma_{\underline{w}}}
=
I_{2,\mathrm{wc}}^{\varepsilon_0,\varepsilon}
(W_0\!:\!A\,|\,\underline{W})_{\sigma}.$ Split the expectation in the right-hand side of
\eqref{Eqn:CondSC_RateCond} as
\[
\sum_{\underline{w}\in\Omega^\star}
p_{\underline{W}}(\underline{w})\,
\mathbb{E}_{\CalC_0(\underline{w})}\!\bigl[
\|\widetilde{\rho}^{\,\underline{w}}_A
-\overline{\rho}^{\,\underline{w}}_A\|_1
\bigr]
+
\sum_{\underline{w}\notin\Omega^\star}
p_{\underline{W}}(\underline{w})\,
\mathbb{E}_{\CalC_0(\underline{w})}\!\bigl[
\|\widetilde{\rho}^{\,\underline{w}}_A
-\overline{\rho}^{\,\underline{w}}_A\|_1
\bigr].
\]
The second sum is bounded by
$2\,p_{\underline{W}}((\Omega^\star)^c)\le 2\varepsilon_0$ since $\|\widetilde{\rho}^{\,\underline{w}}_A
-\overline{\rho}^{\,\underline{w}}_A\|_1\le 2.$ For $\underline{w}\in\Omega^\star$, Proposition~\ref{Prop:SmoothMultipartiteSoftCovering} with $k=1$,
applied to the cq-state $\sigma_{\underline{w}}^{W_0A}$ and the
conditional codebook $\CalC_0(\underline{w})$, together with the rate condition
\eqref{Eqn:CondSC_RateCond}, yields $\mathbb{E}_{\CalC_0(\underline{w})}\!\bigl[
\|\widetilde{\rho}^{\,\underline{w}}_A
-\overline{\rho}^{\,\underline{w}}_A\|_1
\bigr]
\le 4\varepsilon.$ Combining both contributions yields the result.
\end{proof}

\subsection{One-shot Classical Joint Typicality}

\begin{proposition}[One-shot classical joint typicality]
\label{Prop:ClassicalJointTypicalityIndicator}
Let $\mathcal{X}$ be a finite set, let $\{p_i\}_{i=1}^t$ and
$\{q_j\}_{j=1}^\ell$ be probability distributions on $\mathcal{X}$,
and let $\{\varepsilon_{ij}\}_{i\in[t],\,j\in[\ell]}\subset(0,1)$.
Then there exists $f:\mathcal X\to[0,1]$ such that
\begin{equation*}
\text{1) }\;
\sum_x p_i(x)f(x)
\ge
1-\sum_{j=1}^{\ell}\varepsilon_{ij},
\quad \forall\, i\in[t],
\qquad
\text{2) }\;
\sum_x q_j(x)f(x)
\le
\sum_{i=1}^{t}2^{-D_H^{\varepsilon_{ij}}(p_i\|q_j)},
\quad \forall\, j\in[\ell].
\end{equation*}
\end{proposition}

\begin{proof}
For each $(i,j)$ choose a hypothesis test $g_{ij}$ with
$\sum_xp_i(x)g_{ij}(x)\ge1-\varepsilon_{ij}$ and
$\sum_xq_j(x)g_{ij}(x)\le2^{-D_H^{\varepsilon_{ij}}(p_i\|q_j)}$.
Set $f_i=\min_jg_{ij}$ and $f=\max_i f_i$. Since
$1-f_i\le\sum_j(1-g_{ij})$ and $f\ge f_i$, the first bound follows; since
$f\le\sum_i f_i\le\sum_i g_{ij}$, the second follows.
\end{proof}

\section{One-party Purity Concentration as a Local Primitive}
\label{Subsec:1partypurity}
\label{SubSec:LocalPurityPrimitive}

The three-party purity distillation protocol in Sec.~\ref{Sec:PurityDistillation} uses one-party purity concentration as a local primitive after the distributed instrument-simulation step. Once the relevant classical outcomes have been simulated, the primitive is applied conditionally to the corresponding post-instrument states held by Alice$_1$, Alice$_2$, and Bob. Since this primitive is used repeatedly in the construction, we collect its complete formulation here: we first recall the operational definition and the one-shot achievable rate of~\cite{202208arXiv_ChaNemBus}, then give a short proof in the notation used throughout this paper and record the ancilla-size convention needed in Sec.~\ref{Sec:PurityDistillation}.

Alice is given a single copy of a mixed state $\rho^{A}$ on a finite-dimensional system $A$. Using only local unitary operations, her goal is to extract a subsystem that is close to a fixed pure state. In addition, Alice is allowed to borrow a pure ancilla register, provided that its size is discounted from the final net amount of distilled purity.

\begin{definition}[Purity concentration $\eta$-code {\cite[Def.~1.2]{202208arXiv_ChaNemBus}}]
\label{Def:PurityConcentrationCode}
Given a quantum state $\rho^{A}$ and an error tolerance $\eta\ge0$, a \emph{purity concentration $\eta$-code}
consists of:
\begin{itemize}
\item a $d_C$-dimensional ancilla system $C$ prepared in a pure state $|0\rangle\langle 0|^{C}$;
\item a unitary $U^{AC\to A_pA_g}$,
\end{itemize}
such that
\[
\left\|
\Tr_{A_g}\bigl[U\bigl(\rho^{A}\otimes|0\rangle\langle 0|^{C}\bigr)U^\dagger\bigr]
\;-
|0\rangle\langle 0|^{A_p}
\right\|_1
\le \eta .
\]
The \emph{net distilled purity} of the code is defined as
\[
P_{\mathrm{net}} \;:=\; \log d_{A_p} \;-\; \log d_C,
\]
i.e., the number of pure qubits obtained in $A_p$ minus the number of pure qubits borrowed in $C$.
\end{definition}

\begin{proposition}[One-shot purity concentration achievable rate {\cite[Theorem~1.7]{202208arXiv_ChaNemBus}}]
\label{Thm:PurityConcentration}
For any state $\rho^{A}$ and any $\varepsilon\in[0,1]$, there exists a \emph{$3\sqrt{\varepsilon}$-purity concentration code}
(in the sense of Def.~\ref{Def:PurityConcentrationCode}) whose net distilled purity is
\[
P_{\mathrm{net}}
\;=\;
\log d_A \;-\; \widetilde H_{\max}^{\varepsilon}(A)_{\rho}.
\]
Consequently, every target $P\le P_{\mathrm{net}}$ is achievable in the sense that the protocol produces at least $P$ units of net distilled purity.
\end{proposition}

\begin{proof}
Fix $\rho^A$ and $\varepsilon\in[0,1]$, and choose a projector $\Pi_A$ attaining
$\widetilde H_{\max}^{\varepsilon}(A)_\rho$. Thus
\[
\Tr[\Pi_A\rho^A]\ge1-\varepsilon,
\qquad
d_C\define\rank(\Pi_A)
=
2^{\widetilde H_{\max}^{\varepsilon}(A)_\rho}.
\]
Borrow a $d_C$-dimensional pure ancilla $C$ and set
$\rho^{AC}\define\rho^A\otimes\ketbra{0}{0}_C$ and
$\Pi_{AC}\define\Pi_A\otimes\ketbra{0}{0}_C$. By the gentle measurement lemma, after
normalizing the projected state
$\widehat\rho^{AC}\define
\Pi_{AC}\rho^{AC}\Pi_{AC}/\Tr[\Pi_{AC}\rho^{AC}]$, we obtain
\begin{equation}
\label{Eqn:PurityPrimitiveProjectedApprox}
\norm{\rho^{AC}-\widehat\rho^{AC}}_1
\le
2\sqrt{\varepsilon}+\varepsilon
\le
3\sqrt{\varepsilon}.
\end{equation}
The support of $\widehat\rho^{AC}$ has dimension at most $d_C$. Identifying
$AC\simeq A_p\otimes A_g$ with $\dim A_p=d_A$ and $\dim A_g=d_C$, there is therefore a
unitary $U:AC\to A_pA_g$ that maps this support into
$\ket{0}_{A_p}\otimes\CalH_{A_g}$. Hence, for some state $\widetilde\rho^{A_g}$, $U\widehat\rho^{AC}U^\dagger
=
\ketbra{0}{0}_{A_p}\otimes\widetilde\rho^{A_g}.$ Unitary invariance of the trace norm, followed by the partial trace over $A_g$, turns
\eqref{Eqn:PurityPrimitiveProjectedApprox} into
\[
\norm{
\Tr_{A_g}\!\left[U(\rho^A\otimes\ketbra{0}{0}_C)U^\dagger\right]
-\ketbra{0}{0}_{A_p}
}_1
\le3\sqrt{\varepsilon}.
\]
Thus $(C,U)$ is a $3\sqrt{\varepsilon}$-purity concentration code, and its net distilled
purity is
\[
P_{\mathrm{net}}
=
\log d_{A_p}-\log d_C
=
\log d_A-\widetilde H_{\max}^{\varepsilon}(A)_\rho.
\]
This proves Proposition~\ref{Thm:PurityConcentration}.
\end{proof}

\begin{remark}[Ancilla size vs.\ target net purity in the constructed protocol]
\label{Rmk:AncillaSizeTargetPurity}
The construction above uses
$d_C=2^{\widetilde H_{\max}^{\varepsilon}(A)_\rho}$ and produces a pure output register of
dimension $d_A$. More generally, if a smaller target purity is desired, one may simply
discard part of the available pure output. Accordingly, arbitrary real values of the target
$P$ are understood as lower targets rather than exact dimension identities.
\end{remark}

\section{Analysis of $D_{\ulinek,1}$, $C_{\ulinek,1}$ in proof of Thm.~\ref{Thm:DISCosetCdsInnBound}}
 \label{AppSec:DistMCPAnalysisDk1Ck1}
\subsection{Analysis of $D_{\ulinek,1}$}
\label{AppSubSec:DISCCAnalysisDk1}
 Here, we shall prove $\boldCalQ^{2}\circ\ScrE^{(2)}_{k_{2}}\left( \varphi_{S_{k_{2}}}\right) =\CalV_{k_{2}}\circ\boldCalQ^{1}(\varphi_{\boldrho_{1}})$. Towards this, we evaluate $\ScrE^{(2)}_{k_{2}}\left( \varphi_{S_{k_{2}}}\right)$. A canonical $\varphi_{S_{k_{2}}}$ can be expanded in any ONB $\{\ket{\eta_{z} }: z \in [d_{2}]\}\subseteq\CalH_{2}$.  We have
\begin{eqnarray}
\label{Eqn:DISCCTremDk1-1}
\ScrE^{(2)}_{k_{2}}\left(\varphi_{S_{k_{2}}}\right)
& = &
\sum_{m_{2}=0}^{M_{2}}\sum_{b_{2}=0}^{B_{2}}\sum_{z,\hatz=1}^{d_{2}}
\left(
\ketbra{\overline{\eta}_{z}}{\overline{\eta}_{\hatz}}
\otimes
\tr_{E_{2}}\left(
{\theta_{k_{2}}^{m_{2}, b_{2}}}
\sqrt{S_{k_{2}}}
\ketbra{\eta_{z}}{\eta_{\hatz}}
\sqrt{S_{k_{2}}}
{\theta_{k_{2}}^{m_{2}, b_{2}}}^\dagger
\right)
\otimes
\ketbra{m_{2}}{m_{2}}
\right)
\\
\label{Eqn:DISCCTremDk1-2}
& = &
\sum_{m_{2}=0}^{M_{2}}\sum_{b_{2}=0}^{B_{2}}\sum_{z,\hatz=1}^{d_{2}}
\left(
\ketbra{\overline{\eta}_{z}}{\overline{\eta}_{\hatz}}
\otimes
\tr_{E_{2}}\left(
\frac{
V_{k_2,m_2,b_{2}}\sqrt{\boldrho^2}
}{
\sqrt{M_2 B_2\,{1/q^{n}}}
}
\ketbra{\eta_{z}}{\eta_{\hatz}}
{\frac{
\sqrt{\boldrho^2}V_{k_2,m_2,b_{2}}
}{
\sqrt{M_2 B_2\,{1/q^{n}}}
}}^\dagger
\right)
\otimes
\ketbra{m_{2}}{m_{2}}
\right)
\\
\label{Eqn:DISCCTremDk1-3}
& = &
\CalV_{k_{2}}\left(
\sum_{z,\hatz=1}^{d_{2}}
\ketbra{\overline{\eta}_{z}}{\overline{\eta}_{\hatz}}
\otimes
\sqrt{\boldrho^2}
\ketbra{\eta_{z}}{\eta_{\hatz}}
\sqrt{\boldrho^2}{}^\dagger
\otimes
\ketbra{m_{2}}{m_{2}}
\right)
=
\CalV_{k_{2}}(\varphi_{\boldrho^{2}}).
\end{eqnarray}
where (i) \eqref{Eqn:DISCCTremDk1-2} follows from the definition of $\theta_{k_{2}}^{m_{2}, b_{2}}$ in \eqref{Eqn:DISCosetCdsLikelihoodTheta_i}, the fact that the $m_{2}=0$ term gets killed and we can ignore the projectors $z_{k_{2}}$ from \eqref{Eqn:DISCosetCdsThetak0GetsKilled_i}, (ii) \eqref{Eqn:DISCCTremDk1-3} follows from the definition of $\CalV_{k_{2}}$ from \eqref{Eqn:DistMCPKeySplitOfError6}. From the sequence of equalities from LHS of \eqref{Eqn:DISCCTremDk1-1} through RHS of \eqref{Eqn:DISCCTremDk1-3}, we have
\begin{eqnarray*}
\boldCalQ^{2}\circ\ScrE^{(2)}_{k_{2}}\left( \varphi_{S_{k_{2}}}\right) = \boldCalQ^{2}\circ \CalV_{k_{2}}(\varphi_{\boldrho^{2}}) = \CalV_{k_{2}}\circ \boldCalQ^{2}(\varphi_{\boldrho^{2}}) = \CalV_{k_{2}}(\varphi_{\boldrho^{12}}) = \CalV_{k_{2}}\circ \boldCalQ^{2}(\varphi_{\boldrho^{2}}),
 \end{eqnarray*}
where the second equality above follows from Fact \ref{Fact:DISCCCrucialIsometryCommuteWithCalJMaps} and the last two equalities follows from definitions of $\boldCalQ^{i}(\cdot)$ for $i\in [2]$ found before Fact \ref{Fact:DISCCCrucialIsometryCommuteWithInst}.

\subsection{Analysis of $C_{\ulinek,1}$}
\label{AppSubSec:DISCCAnalysisCk1}

We shall prove $ \boldCalQ^{1}\circ\ScrE^{(1)}_{k_{1}}\!\left( \varphi_{S_{k_{1}}}\right)={\CalV}_{k_{1}}\left( \varphi_{\boldrho^{12}}\right)$. Towards that end, we evaluate $\ScrE^{(1)}_{k_{1}}( \varphi_{S_{k_{1}}})$. A canonical $\varphi_{S_{k_{1}}}$ can be expanded in any ONB $\{\ket{\iota_{g} }: z \in [d_{1}]\}\subseteq\CalH_{1}$.  As in \eqref{Eqn:DISCCTremDk1-1} - \eqref{Eqn:DISCCTremDk1-3}, we have

\begin{eqnarray}
\label{Eqn:DISCCTremCk1-1}
\ScrE^{(1)}_{k_{1}}\left(\varphi_{S_{k_{1}}}\right)
& = &
\sum_{m_{1}=0}^{M_{1}}\sum_{b_{1}=0}^{B_{1}}\sum_{z,\hatz=1}^{d_{1}}
\left(
\ketbra{\overline{\eta}_{z}}{\overline{\eta}_{\hatz}}
\otimes
\tr_{E_{1}}\left(
{\theta_{k_{1}}^{m_{1}, b_{1}}}
\sqrt{S_{k_{1}}}
\ketbra{\eta_{z}}{\eta_{\hatz}}
\sqrt{S_{k_{1}}}
{\theta_{k_{1}}^{m_{1}, b_{1}}}^\dagger
\right)
\otimes
\ketbra{m_{1}}{m_{1}}
\right)
\\
\label{Eqn:DISCCTremCk1-2}
& = &
\sum_{m_{1}=0}^{M_{1}}\sum_{b_{1}=0}^{B_{1}}\sum_{z,\hatz=1}^{d_{1}}
\left(
\ketbra{\overline{\eta}_{z}}{\overline{\eta}_{\hatz}}
\otimes
\tr_{E_{1}}\left(
\frac{
V_{k_1,m_1,b_{1}}\sqrt{\boldrho^1}
}{
\sqrt{M_1 B_1\,{1/q^{n}}}
}
\ketbra{\eta_{z}}{\eta_{\hatz}}
{\frac{
\sqrt{\boldrho^1}V_{k_1,m_1,b_{1}}
}{
\sqrt{M_1 B_1\,{1/q^{n}}}
}}^\dagger
\right)
\otimes
\ketbra{m_{1}}{m_{1}}
\right)
\\
\label{Eqn:DISCCTremCk1-3}
& = &
\CalV_{k_{1}}\left(
\sum_{z,\hatz=1}^{d_{1}}
\ketbra{\overline{\eta}_{z}}{\overline{\eta}_{\hatz}}
\otimes
\sqrt{\boldrho^2}
\ketbra{\eta_{z}}{\eta_{\hatz}}
\sqrt{\boldrho^1}{}^\dagger
\otimes
\ketbra{m_{1}}{m_{1}}
\right)
=
\CalV_{k_{1}}(\varphi_{\boldrho^{1}}).
\end{eqnarray}
where (i) \eqref{Eqn:DISCCTremCk1-2} follows from the definition of $\theta_{k_{1}}^{m_{1}, b_{1}}$ in \eqref{Eqn:DISCosetCdsLikelihoodTheta_i}, the fact that the $m_{1}=0$ term gets killed and we can ignore the projectors $z_{k_{1}}$ from \eqref{Eqn:DISCosetCdsThetak0GetsKilled_i}, (ii) \eqref{Eqn:DISCCTremCk1-3} follows from the definition of $\CalV_{k_{1}}$ from \eqref{Eqn:DistMCPKeySplitOfError6}. From the sequence of equalities from LHS of \eqref{Eqn:DISCCTremCk1-1} through RHS of \eqref{Eqn:DISCCTremCk1-3}, we have
\begin{eqnarray*}
\boldCalQ^{1}\circ\ScrE^{(1)}_{k_{1}}\left( \varphi_{S_{k_{1}}}\right) = \boldCalQ^{1}\circ \CalV_{k_{1}}(\varphi_{\boldrho^{1}}) = \CalV_{k_{1}}\circ \boldCalQ^{1}(\varphi_{\boldrho^{1}}) = \CalV_{k_{1}}(\varphi_{\boldrho^{12}}).
 \end{eqnarray*}

\bibliographystyle{sty/IEEEtran}
{
\bibliography{bibliography/InfExtNtwrkPOVMs}
}

\end{document}